%% file: gesamtIIa4.tex
\documentstyle[aps,epsfig,a4,tighten,twoside,psfig]{revtex}

\def\v1{\vspace{1cm}} 
\def\be{\begin{equation}} 
\def\ee{\end{equation}}
\newcommand{\en}{\end{equation}} 
\def\eeas{\end{eqnarray*}}
\def\beas{\begin{eqnarray*}}
\newcommand{\ena}{\end{eqnarray}}

\newcommand{\hbo}{\hbox to 1 true cm {\hfill } }
\newcommand{\tr}{\hbox{tr}}
\def\tr{\mbox{Tr}\,}

\def\bc{\begin{center}} 
\def\ec{\end{center}} 
\def\vh{\varphi} 
 
\newcommand{\bea}{\begin{eqnarray}} 
\newcommand{\eea}{\end{eqnarray}}

\newcommand{\pr}{\partial}

\renewcommand{\deg}{^{\circ}}

\newcommand{\eps}{\varepsilon}

\newcommand{\pip}[1]{\left( #1 \right)} %put in parentheses
\newcommand{\pib}[1]{\left[ #1 \right]}%put in braces
\newcommand{\ov}[1]{\overline {#1}}

\newcommand{\wsep}[2]{\hspace{#1 in} \mbox{#2} \hspace{#1in}}

\newcommand{\w}[1]{\mbox{#1}}

\newcommand{\eqnb}{\begin{equation}}
\newcommand{\eqne}{\end{equation}}

\newcommand{\question}[1]{}
\newcommand{\lsim}[1]{
\setlength{\unitlength}{12pt}
\begin{picture}(1.4,1.)
\put(.7,-0.3){\makebox(0.0,1.)[t]{$<$}}
\put(.7,-0.3){\makebox(0.0,1.)[b]{$\sim$}}
\end{picture}#1}
\newcommand{\NSt}{{\mbox{\scriptsize\it NS}}}
\newcommand{\St}{{\mbox{\scriptsize\it S}}}
\def\Ref#1{(\ref{#1})}
\def\Cite#1{\protect\cite{#1}}
\def\Caption#1{\begin{quotation}\caption{\sl #1}\end{quotation}}

\def\BA{\begin{eqnarray}}
\def\BE{\begin{equation}}
\def\BF{\begin{figure}[htb]}
\def\BT{\begin{table}[htb]}
\def\EA{\end{eqnarray}}
\def\EE{\end{equation}}
\def\EF{\end{figure}}
\def\ET{\end{table}}
\def\vr{\vec r\,}  \def\rT{r_T} \def\brT{{\vec r_T}}
\def\vp{\vec p\,}  \def\pT{p_T} \def\bpT{{\vec p_T}}
   
\def\la{\langle}
\def\ra{\rangle}
\def\mb{\,\mbox{mb}}
\def\fm{\,\mbox{fm}}
\def\GeV{\,\mbox{GeV}}
\def\eps{\varepsilon}

\def\Jpsi{J\!/\!\psi}

\newcommand{\psibar}{\bar\psi}
\newcommand{\chibar}{\bar\chi}
\newcommand{\bra}{\langle}
\newcommand{\ket}{\rangle}
\newcommand{\pbp}{\langle\bar\psi\psi\rangle}
\newcommand{\cbc}{\langle\bar\chi\chi\rangle}
\newcommand{\U}[1]{\mathrm{U}(#1)}

\newcommand{\One}{1\kern-4.5pt1}
\newcommand{\half}{\frac{1}{2}}
\newcommand{\Tr}{\mathrm{Tr}}
\newcommand{\cc}[1]{#1}
\newlength{\colw}
\newcommand{\m}{\hphantom{$-$}}
\newcommand{\ga}{\alpha}
\newcommand{\gb}{\beta}
\newcommand{\gc}{\gamma}
\newcommand{\gd}{\delta}
\newcommand{\gk}{\kappa}

\newcommand{\gs}{\sigma}

\def\Journal#1#2#3#4{{#1} {\bf #2} (#4) #3}

\def\AP{{\em Ann. Phys.}}

\def\EPJC{{\em Eur. Phys. J.} C}

\def\NPB{{\em Nucl. Phys.} B}

\def\PLB{{\em Phys. Lett.} B}
\def\PRL{\em Phys. Rev. Lett.}

\def\PRD{{\em Phys. Rev.} D}

\def\PREPC{{\em Phys. Rep.} C}

\def\RMP{{\em Rev. Mod. Phys.}}
\def\RNC{{\em Riv. Nuovo Cimento}}

\def\YA{{\em Yad. Fiz.}}
\newcommand{\hf} {{1\over2}}
\newcommand{\nonu}{\nonumber\\}
\def\eq#1{(\ref{#1})}
\def\ra{\rangle}
\def\la{\langle}

\newcommand{\zr}[1]{\mbox{\hspace*{#1em}}}
 \newcommand{\ID}{\mbox{{\sf 1}\zr{-0.14}\rule{0.04em}{1.55ex}\zr{0.1}}}

\newcommand{\pfig}[3]{
 \refstepcounter{dafigcounter}
 \begin{minipage}[t]{#2}
  \begin{center}
   {\epsfxsize=#2 \mbox{\epsffile{#1.eps}}}
  \end{center}
  \label{#1}
  \small \bf Fig.~\thedafigcounter\rm\ #3
 \end{minipage}
}

\newcounter{dafigcounter}

\renewcommand{\thefootnote}{\fnsymbol{footnote}}

\def\trans{\mbox{\tiny$\bot$}} % Transverse component
\def\longi{\mbox{\tiny$\|$}}   % Longitudinal component
\newcommand{\nablaleftright}{\stackrel{\leftrightarrow}{\nabla}}

\newcommand{\wvp}{{\vec P}}
\begin{document}

%\title{}
%\author{}
%\maketitle

\setcounter{tocdepth}{1}

\input{SOURCE/proctitle2.tex}
\input{SOURCE/ganzleer.tex}
\input{SOURCE/proctitle}

\input{SOURCE/ganzleer.tex}
\input{SOURCE/preface.tex}

\input{SOURCE/ganzleer.tex}
\tableofcontents
\newpage
\input{SOURCE/QCD.tex}
\input{SOURCE/ganzleer1.tex}
\input{SOURCE/alkoferNEU.tex}     %Alkofer%
\input{SOURCE/ganzleer1.tex}
\input{SOURCE/g1NEU.tex}          %Langfeld%
\input{SOURCE/topqcd.tex}         %Pervushin%
\input{SOURCE/ganzleer1.tex}
\input{SOURCE/polnoiNEU.tex}      %Polonyi%
\input{SOURCE/rostock00NEU.tex}   %Skullerud%

\input{SOURCE/ganzleer1.tex}
\input{SOURCE/HIC.tex}

\input{SOURCE/ganzleer1.tex}
\input{SOURCE/arne.tex}           %Hoell%
\input{SOURCE/ganzleer1.tex}
\input{SOURCE/trentoNEU.tex}      %Nayak%
\input{SOURCE/002eproc.tex}       %Schmidt%
\input{SOURCE/ganzleer1.tex}
\input{SOURCE/paper.tex}          %Serreau%
\input{SOURCE/ganzleer1.tex}
\input{SOURCE/psi-A5.tex}         %Huefner%
\input{SOURCE/ganzleer1.tex}
\input{SOURCE/textyura.tex}       %Ivanov%
\input{SOURCE/ganzleer1.tex}
\input{SOURCE/proceed.tex}        %Burau%
\input{SOURCE/ganzleer1.tex}
\input{SOURCE/procAPP.tex}        %Beyer%
\input{SOURCE/main.tex}           %Bugaev%

\input{SOURCE/ganzleer1.tex}
\input{SOURCE/DSE.tex}

\input{SOURCE/ganzleer1.tex}
\input{SOURCE/shortRdr.tex}       %Klabucar%
\input{SOURCE/ganzleer1.tex}
\input{SOURCE/ruivo1NEU.tex}      %Ruivo%
\input{SOURCE/tre01rev.tex}       %Scoccola%
\input{SOURCE/sqm_sepNEU.tex}     %Gocke%
\newpage
\input{SOURCE/ganzleer1.tex}
\input{SOURCE/dgh2.tex}           %Blaschke%
\input{SOURCE/ganzleer1.tex}
\input{SOURCE/proc_1412.tex}      %Behnke%
\input{SOURCE/ganzleer1.tex}
\newpage
\input{SOURCE/PARTLIST.tex}

\input{SOURCE/ganzleer1.tex}
\input{SOURCE/participant.tex}

\end{document}

%% file: SOURCE/proctitle2.tex
\thispagestyle{empty}
$\left. \right.$
\vspace*{0.5cm}
\begin{center}
\large
{\LARGE{\bf Exploring Quark Matter}}\\
\vspace{1.5cm}
{\rm Editors:}\\
\vspace{0.5cm}
{\bf Gerhard R.G. Burau}\\[.3cm]
{\bf David B. Blaschke}\\[.3cm]
{\bf Sebastian M. Schmidt}\\
\vspace{1cm}
\begin{minipage}[htbp]{5cm}
\psfig{figure=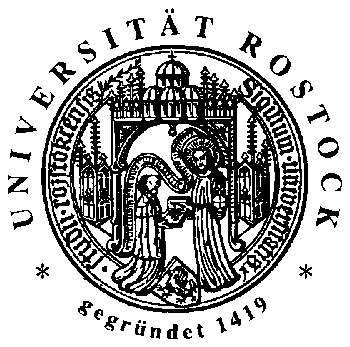,width=5cm}
\end{minipage}

\vspace{1cm}
{\Large{\bf Universit\"at Rostock}}\\
Fachbereich Physik
\end{center}
\newpage

%% file: SOURCE/ganzleer.tex
\pagestyle{empty}
\hspace*{7cm}

$\left. \right.$

\newpage

%% file: SOURCE/proctitle.tex
\thispagestyle{empty}
$\left. \right.$
\vspace*{0.5cm}
\begin{center}
\large
{\LARGE{\bf Exploring Quark Matter}}\\
\vspace{1.5cm}
{\Large{\rm Proceedings of the Workshops}}\\
\vspace{1cm}
{\bf Quark Matter in Astro- and Particle Physics}\\
\vspace{0.3cm}
{\rm Rostock (Germany), November 2000}\\
\vspace{1cm}
{\bf Dynamical Aspects of the QCD Phase Transition }\\
\vspace{0.3cm}
{\rm Trento (Italy), March 2001}\\
\vspace{1cm}
{\Large{\rm Dedicated to \\
J\"org H\"ufner on the occasion of his 65th birthday\\
and \\
Gerd R{\"o}pke on the occasion of his 60th birthday}}\\
\vspace{1.5cm}
{\rm Editors}\\
\vspace{0.5cm}
{\bf Gerhard R.G. Burau} (Universit\"at Rostock)\\[.3cm]
{\bf David B. Blaschke} (Universit\"at Rostock)\\[.3cm]
{\bf Sebastian M. Schmidt} (Universit\"at T\"ubingen)
\end{center}
\newpage

%% file: SOURCE/preface.tex
\pagestyle{empty}
%\pagenumbering{roman}
\begin{center}
\section*{\bf Preface}
%\vspace*{2mm}
\end{center}
\newcounter{kapitel00}[section]
\setcounter{section}{-1}
\stepcounter{section}
\vspace*{-0.2cm}
These proceedings contain research results  presented during the
Workshops {\bf "Quark Matter in Astro- and Particle Physics"} held at the
University of Rostock, November 27 - 29, 2000 and {\bf "Dynamical Aspects
of the QCD Phase Transition"} held at the ECT* in Trento, March 12 - 15, 2001. 
These collaboration meetings are a
continuation of a long term tradition of short workshops.
After the spring and fall meetings on Quantum Statistics held in
Ahrenshoop during the eighties, there was a number of workshops in the
nineties on strongly interacting many-particle systems in- and out-of
equilibrium where the groups of Dubna, Heidelberg and Rostock 
have rotated organization. 
The most recent sequence of collaboration meetings is devoted to the 
applications of thermal field theory and many-particle aspects of
quark matter formation for hot and dense matter systems in nuclear
collisions 
and in astrophysics.
The research on these challenging problems is stimulated by
various collaborations, e.g. with Heidelberg, the JINR Dubna and the ANL Argonne.
We are glad that new collaborators have joined us and that we can
present a compilation of contributions to these fascinating subjects.\\
\noindent
\newline
%\vspace{0.1cm}
\noindent 
Rostock \& T\"ubingen, April 2001\hfill D. Blaschke, G. Burau, S. Schmidt

\vspace*{0.6cm}

\section*{Acknowledgements}
\vspace*{-0.2cm}
We would like to thank all those who helped in organzing the
workshops. We are grateful to the local crew in Rostock: Marina Hertzfeldt,
Christian Gocke, Danilo Behnke and also to Hannelore Gellert from the 
International Office of the University of Rostock. We acknowledge the support 
of the ECT* in Trento in  hosting the collaboration meeting in March 2001 and
thank heartily Ines Campo for the local organization. The workshops
were supported in part by the DFG Graduiertenkolleg ``Stark korrelierte 
Vielteilchensysteme'' at the University of Rostock, 
Deutscher Akademischer Austauschdienst (DAAD), 
Deutsche Forschungsgemeinschaft (DFG 436 114/202/00), Heisenberg - Landau 
program of the BMBF, the Ministry of education, science and culture of
Mecklenburg - Vorpommern and the ECT* in Trento/Italy.
\newpage

%% file: SOURCE/QCD.tex
\pagestyle{plain}
\addcontentsline{toc}{section}{$\left. \right.$\hspace{0.3cm} \bf Quantum Chromodynamics}
$\left. \right.$
\vspace*{5cm}
\begin{center}
{\Large Contributions on}
\end{center}
\begin{center}
{\Large {\bf Quantum Chromodynamics}}
\end{center}
\newpage

%% file: SOURCE/ganzleer1.tex
\pagestyle{empty}
\hspace*{7cm}

$\left. \right.$

\newpage

\pagestyle{plain}

%% file: SOURCE/alkoferNEU.tex
\section*{\bf The Kugo--Ojima Confinement Criterion from Dyson--Schwinger Equations}
\addcontentsline{toc}{section}{\protect\numberline{}{The Kugo--Ojima Confinement Criterion from Dyson--Schwinger Equations \\ \mbox{\it R. Alkofer, L. von Smekal and P. Watson}}}
\begin{center}
\vspace*{2mm}
{Reinhard Alkofer$^a$, Lorenz von Smekal$^b$ and Peter Watson$^a$}\\[0.3cm]
{\small\it $^a$Universit\"at T\"ubingen, Institut f\"ur Theoretische Physik,\\
         Auf der Morgenstelle 14, 72076 T\"ubingen, Germany\\
         E-mail: Reinhard.Alkofer@uni-tuebingen.de\\
         watson@pion01.tphys.physik.uni-tuebingen.de\\ ~\\
         $^b$Universit\"at Erlangen-N\"urnberg,
         Institut f\"ur Theoretische Physik III,\\
         Staudtstr.~7, 91058 Erlangen, Germany\\
         E-mail: smekal@theorie3.physik.uni-erlangen.de}
\end{center}

\newcounter{eqn16}[equation]
\setcounter{equation}{-1}
\stepcounter{equation}

\newcounter{bild16}[figure]
\setcounter{figure}{-1}
\stepcounter{figure}

\newcounter{tabelle16}[table]
\setcounter{table}{-1}
\stepcounter{table}

\newcounter{unterkapitel16}[subsection]
\setcounter{subsection}{-1}
\stepcounter{subsection}

\newcounter{unterunterkapitel16}[subsubsection]
\setcounter{subsubsection}{-1}
\stepcounter{subsubsection}

\newcounter{paragraph16}[paragraph]
\setcounter{paragraph}{-1}
\stepcounter{paragraph}

\begin{abstract}

Prerequisites of confinement in the covariant and local description of QCD are
reviewed. In particular, the Kugo--Ojima confinement criterion, the positivity
violations of transverse gluon and quark states, and the conditions necessary
to avoid  the decomposition property for colored clusters are discussed. In
Landau gauge QCD, the Kugo--Ojima confinement criterion follows from the 
ghost Dyson--Schwinger equation if the corresponding Green's functions 
can be expanded in an asymptotic series. Furthermore, the infrared behaviour
of the  propagators in Landau gauge QCD as extracted from solutions to
truncated Dyson--Schwinger equations and lattice simulations is discussed
in the light of these issues.  

\end{abstract}

\vskip 5mm

\noindent
{\bf Prerequisites of a Covariant Description of Confinement\\}

The confinement phenomenon in QCD cannot be accommodated within the  standard
framework of quantum field theory. Thereby it is known that covariant quantum
theories of gauge fields require indefinite metric spaces. Maintaining the
much stronger principle of locality, great emphasis has been put on the idea of
relating confinement to the violation of positivity in QCD. Just as in QED,
where the Gupta-Bleuler prescription is to enforce the Lorentz condition on
physical states, a semi-definite {\em physical subspace} can be defined as the
kernel of an operator.   The physical states then correspond to equivalence
classes of states  in this subspace differing by zero norm components.  
Besides transverse photons covariance implies the existence of  longitudinal
and scalar photons in QED. The latter two form metric partners in the
indefinite space. The Lorentz condition eliminates half of these leaving
unpaired states of zero norm which do not contribute to observables. Since the
Lorentz condition commutes with the $S$-Matrix, physical states scatter into
physical ones exclusively. 

Due to the gluon self-interactions the corresponding mechanism is more 
complicated in QCD. Here, the Becchi--Rouet--Stora (BRS) symmetry of the
gauge fixed action proves to be helpful.
Within the framework of BRS algebra, in the simplest version for
the BRS-charge $Q_B$ and the ghost number $Q_c$ given by,
\begin{equation} 
           Q_B^2 = 0 \; , \quad \left[ iQ_c , Q_B \right] = Q_B \; ,
\end{equation}
completeness of the nilpotent BRS-charge $Q_B$ in a state space $\mathcal{V}$
of indefinite metric is assumed. This charge generates the BRS 
transformations by ghost number graded commutators $\{ \, , \}$,
{\it i.e.}, by commutators or anticommutators for fields with even
or odd ghost number, respectively.
The semi-definite subspace 
${\mathcal{V}}_{\mbox{\tiny p}}  = \mbox{Ker}\, Q_B  $
is defined on the basis of this algebra by those states which are annihilated
by the BRS charge $Q_B$. 
Since $Q_B^2 =0 $, this subspace contains the space $ \mbox{Im}\, Q_B $
of so-called daughter
states which are images of others, their parent states in $\mathcal{V}$.
A {\em physical} Hilbert space is then obtained as the 
covariant space of equivalence classes, the BRS-cohomology of states in the
kernel modulo those in the image of $Q_B$,
\begin{equation}
     {\mathcal{H}}(Q_B,{\mathcal{V}}) = {\mbox{Ker}\, Q_B}/{\mbox{Im}\, Q_B} 
       \simeq  {\mathcal{V}}_s \; , 
\end{equation}
which is isomorphic to the space ${\mathcal{V}}_s$ of BRS singlets.    
Completeness is thereby important in the proof of positivity for physical
states \cite{Kug79,Nak90} because it assures the absence of metric
partners of BRS-singlets.

With completeness, all states in $\mathcal{V}$ are either BRS
singlets in ${\mathcal{V}}_s$ or belong to quartets which are 
metric-partner pairs of BRS-doublets (of parent with daughter states).
This exhausts all possibilities. The generalization of the
Gupta--Bleuler condition on physical states, $Q_B |\psi\rangle = 0$ in
$\mathcal{V}_{\mbox{\tiny p}}$, eliminates half of these metric partners
leaving unpaired states of zero norm  which do not contribute to any
observable. This essentially is the quartet mechanism: 
\begin{itemize}
\item[] 
Just as in QED, one such quartet, the elementary quartet, is formed by
the massless asymptotic states of longitudinal and timelike gluons together 
with ghosts and antighosts which are thus all unobservable. 
\item[] 
In contrast to QED, however, one expects the quartet mechanism also 
to apply to transverse gluon and quark states, as far as they exist
asymptotically. A violation of positivity for such states then entails
that they have to be unobservable also. 
\end{itemize}

Asymptotic transverse gluon and quark states
may or may not exist in the indefinite metric space $\mathcal{V}$. If either 
of them do exist and the Kugo--Ojima criterion  (see below) is realized, they
belong to unobservable quartets. In that case, the BRS-transformations of their
asymptotic fields entail that they form these quartets together with
ghost-gluon and/or ghost-quark bound states, respectively \cite{Nak90}.
It is furthermore  crucial for confinement, however, to have a mass gap in
transverse gluon correlations, {\it i.e.}, the massless transverse gluon
states of perturbation theory have to dissappear even though they should
belong to quartets due to color antiscreening \cite{Oeh80,Nis94,Alk00}. 

The interpretation  in terms of transition probabilities  holds between
physical states. For a local operator $A$ to be observable it is necessary to
be BRS-closed, $\{ iQ_B , A \} \, = 0 $, which coincides with the requirement
of its local gauge invariance. It then follows that all states generated
from the vacuum  $|\Omega\rangle$ by any such observable fulfill positivity: 
On the other hand, unobservable, {\it i.e.}, confined, states violate 
positivity.

The remaining dynamical aspect of confinement in this
formulation resides in the cluster decomposition property \cite{Haa96}. 
Including the indefinite metric spaces of covariant gauge
theories it can be summarized as
follows: For the vacuum expectation values of
two local operators $A$ and $B$, translated to a large spacelike
separaration $R$ of each other one obtains the following bounds depending on
the existence of a finite gap $M$ in the spectrum of the 
mass operator in $\mathcal{V}$ \cite{Nak90}   
\begin{eqnarray} 
        \Big|  \langle  \Omega | A(x) B(0) |\Omega \rangle  &-&         
 \langle  \Omega | A(x) |\Omega \rangle  \,  \langle  \Omega
             |  B(0) |\Omega \rangle  \Big|  \\
   && \hskip -.2cm \le  \;  \bigg\{  \begin{array}{ll} 
   \mbox{\small Const.} \, \times \, R^{-3/2 + 2N} \, e^{-MR} \!\!, \quad 
                      & \mbox{mass gap } M \; , \\
   \mbox{\small Const.} \, \times \, R^{-2 + 2N} \,, \;\; 
                      & \mbox{no mass gap} \; ,  \end{array}  \nonumber  
\end{eqnarray}
for $R^2 = - x^2 \to \infty $. Herein, positivity entails that $N = 0$, but a
positive integer $N$ is possible for the indefinite inner product structure in
$\mathcal{V}$. Therefore, in order to avoid the decomposition property
for products of unobservable operators $A$ and $B$ which 
together with the Kugo-Ojima criterion (see below) 
is equivalent to avoiding the
decomposition property for colored clusters, there should 
be no mass gap in the indefinite space $\mathcal{V}$. 
Of course, this implies nothing on the physical spectrum of the mass operator
in $\mathcal{H}$ which certainly should have a mass gap. 
In fact, if the cluster decomposition property holds for a product $A(x) B(0)$ 
forming an observable, it can be shown that both 
$A$ and $B$ are observables themselves. 
This then eliminates the possibility of scattering a physical state into
color singlet states consisting of widely separated colored clusters (the
``behind-the-moon'' problem) \cite{Nak90}. 

Confinement depends on the 
realization of the unfixed global gauge symmetries in this formulation.
The identification of the 
BRS-singlets in the physical Hilbert space $\mathcal{H}$ with
color singlets is possible only if the charge of global gauge transformations
is BRS-exact {\em and} unbroken. The sufficent conditions for this are
provided by the Kugo-Ojima criterion: Considering the 
globally conserved current     
\begin{equation} 
    J^a_\mu = \partial_\nu F_{\mu\nu}^a  + \{ Q_B , D_{\mu}^{ab} \bar c^b \} 
    \qquad (\mbox{with} \; \partial_\mu J^a_\mu = 0 \,) \; ,
       \label{globG}
\end{equation}
one realizes that the first of its two terms corresponds to a coboundary 
with respect to the space-time exterior derivative while the second term 
is a BRS-coboundary with charges denoted by $G^a$ and $N^a$, respectively, 
\begin{equation} 
      Q^a =  \int d^3x \,  \partial_i F_{0 i}^a \,  +\,  \int d^3x \, 
             \{ Q_B , D_{0}^{ab} \bar c^b \} \, = \, G^a \, + \, N^a \; .
        \label{globC}
\end{equation}
For the first term herein there are only two options, it is either ill-defined
due to massless states in the spectrum of $\partial_\nu F_{\mu\nu}^a $, or else
it vanishes. 

In QED massless photon states contribute to the analogues of both currents
in~(\ref{globG}), and both charges on the r.h.s. in (\ref{globC}) are
separately ill-defined. One can employ an arbitrariness in the  definition of
the generator of the global gauge transformations (\ref{globC}), however, to
multiply the first term by a suitable constant so chosen that these massless
contributions cancel. This way one obtains a well-defined and unbroken global
gauge charge which replaces the naive definition in (\ref{globC}) above
\cite{Kug95}. Roughly speaking, there are two independent structures in the
globally conserved gauge currents in QED which both contain massless photon
contributions. These can be combined  to yield one well-defined charge as the
generator of global gauge transformations leaving the other independent
combination (the displacement symmetry) spontaneously broken which lead to the
identification  of photons with massless Goldstone bosons \cite{Nak90,Fer71}. 

If $\partial_\nu F_{\mu\nu}^a $ contains no massless
discrete spectrum on the other hand, {\it i.e.}, if there is no massless
particle pole in the Fourier transform of transverse gluon correlations, then
$G^a \equiv 0$.
In particular, this is the case for channels containing massive vector fields
in theories with Higgs mechanism, and it is expected to be also the case in
any color channel for QCD with confinement for which it actually represents one
of the two conditions formulated by Kugo and Ojima. 
In both these situations one has  
\begin{equation}
                       Q^a \, = \, N^a \, = \, \Big\{   Q_B \, , 
       \int d^3x \,   D_{0}^{ab} \bar c^b \Big\} \; ,
\end{equation}
which is BRS-exact. The second of the two conditions for confinement
is that this charge be well-defined in the whole of the indefinite metric space
$\mathcal{V}$. Together these conditions 
are sufficient to establish that all BRS-singlet physical
states in $\mathcal{H}$ are also color singlets, and that all colored states
are thus subject to the quartet mechanism. The 
second condition thereby provides the essential 
difference between Higgs mechanism and confinement. 
The operator $D_\mu^{ab}\bar c^b$ determining the charge $N^a$ will in
general contain a  {\em massless} contribution from the elementary
quartet due to the asymptotic field $\bar\gamma^a(x)$ in the  
antighost field,  $\bar c^a\, \stackrel{x_0 \to \pm\infty}{\longrightarrow}
\, \bar\gamma^a + \cdots $ (in the weak asymptotic limit), 
\begin{equation}
          D_\mu^{ab}\bar c^b \; \stackrel{x_0 \to \pm\infty}{\longrightarrow}
              \;   ( \delta^{ab} + u^{ab} )\,   \partial_\mu \bar\gamma^b(x) +
                 \cdots  \;  .
\end{equation}
Here, the dynamical parameters $ u^{ab} $ determine the contribution 
of the massless asymptotic state to the composite field $g f^{abc} A^c_\mu
\bar c^b  \, \stackrel{x_0 \to \pm\infty}{\longrightarrow}  \,
u^{ab} \partial_\mu \bar\gamma^b + \cdots $. These parameters can be obtained
in the limit $p^2\to 0$ from the Euclidean correlation functions of this
composite field, {\it e.g.},
\vspace{-.2cm}
\begin{equation}
\int d^4x \; e^{ip(x-y)} \,
\langle  \; D^{ae}_\mu c^e(x) \; gf^{bcd}A_\nu^d(y) \bar c^c (y) \; \rangle
\; =: \; \Big(\delta_{\mu \nu} -{p_\mu p_\nu \over p^2} \Big) \, u^{ab}(p^2)
\; .  \label{Corru}
\end{equation}
The theorem by Kugo and Ojima asserts that all $Q^a = N^a$ do not suffer from
spontaneous breakdown (and are thus well-defined), if and only if
\begin{eqnarray}
                 u^{ab} \equiv u^{ab}(0)  \stackrel{!}{=} - \delta^{ab} \; .
\label{KO1}
\end{eqnarray}
Then the massless states from the elementary quartet do not contribute to 
the spectrum of the current in $N^a$, and the equivalence between physical
BRS-singlet states and color singlets is established.\cite{Kug79,Nak90,Kug95}

In contrast, if $\mbox{det}(  \ID + u ) \not=0$, the global
gauge symmetry generated by the charges $Q^a$ in eq.~(\ref{globC}) is
spontaneuosly broken in each channel in which the gauge potential 
contains an asymptotic massive vector field \cite{Kug79,Nak90}.
While this massive vector state 
results to be a BRS-singlet, the massless Goldstone boson states which 
usually occur in some components of the Higgs field, replace the 
third component of the vector field in the elementary
quartet and are thus unphysical. 
Since the broken charges
are BRS-exact, this
symmetry breaking is not directly observable in the Hilbert space of physical
states $\mathcal{H}$.  

The condition $u = -\ID$ %, leading to well-defined charges $N^a$, can in
Landau gauge be shown by standard arguments employing Dyson--Schwinger
equations and Slavnov--Taylor identities to be 
equivalent to an infrared enhanced ghost propagator \cite{Kug95}.
In momentum space the non-perturbative ghost propagator of Landau gauge QCD  
is related to the form factor occuring in the correlations of
eq.~(\ref{Corru}), 
\begin{equation}
    D_G(p) = \frac{-1}{p^2}      \, \frac{1}{ 1 + u(p^2) } \, , \;\;
                 \mbox{with}  \; \;   
                 u^{ab}(p^2)  = \delta^{ab}  u(p^2) \, . \label{DGdef}
\end{equation}
The Kugo--Ojima confinement criterion, $u(0) = -1$, thus entails that the
Landau gauge ghost propagator should be more singular than a massless particle
pole in the infrared. Indeed, we will present evidence for this exact infrared
enhancement of ghosts in Landau gauge. 

The necessity for the absence of the massless particle pole in $\partial_\nu
F^a_{\mu\nu} $ in the Kugo-Ojima criterion shows that the (unphysical)
massless correlations to avoid the cluster decomposition property are {\em
not} the transverse gluon correlations. An infrared suppressed propagator for
the transverse gluons in Landau gauge confirms this condition. This holds
equally well for the infrared vanishing propagator obtained from
Dyson--Schwinger Equations \cite{Sti96,Sme98} and conjectured in the
studies of the implications of the Gribov horizon \cite{Gri78,Zwa92}, 
as for the infrared suppressed but possibly finite ones extraced from
improved lattice actions for quite large volumes \cite{Bon00}.
The infrared enhanced correlations responsible for the failure of the cluster
decomposition can be identified with the ghost correlations which at
the same time provide for the realization of the Kugo--Ojima criterion in
Landau gauge.

\vskip 5mm

%\newpage

\noindent
{\bf Verifying the Kugo--Ojima Confinement Criterion from the
Dyson--Schwin\-ger Equation for the Ghost Propagator\\}

In Landau gauge the gluon and ghost propagators are parametrized by
the two invariant functions $Z(k^2)$ and $G(k^2)$, respectively
(with $G(k^2) = 1/(1+u(k^2))$, {\it c.f.}, eq.~(\ref{DGdef})). 
In Euclidean momentum space one has
\begin{eqnarray} 
        D_{\mu\nu}(k) = \frac{Z(k^2)}{k^2} \, \left( \delta_{\mu\nu} -
        \frac{k_\mu k_\nu}{k^2} \right)  \; ,\quad
        D_G(k)  &=& - \frac{G(k^2)}{k^2}          \;.
\end{eqnarray}
The non-perturbative infrared behaviour of these 
functions can be studied with employing their  
Dyson--Schwinger equations \cite{Alk00,Rob00}.

The equation for the ghost propagator is the simplest of all QCD
Dyson--Schwinger equations. Besides the ghost and gluon propagators
it contains the ghost-gluon vertex function.  
In Landau gauge this 3-point function needs not to be renormalized.
Furthermore, it becomes bare whenever the out-ghost momentum vanishes.
This has the important consequence that it cannot be singular for 
vanishing ghost momenta.

Furthermore assuming that the QCD Green's  functions can be expanded
in asymptotic series\footnote{Note that this is not possible if the 
infrared slavery picture is correct. An infinite $\beta$-function for vanishing
scales prohibits such an expansion.}, {\it e.g.},
\begin{eqnarray} 
G(p^2;\mu^2) = \sum _{n} d_n \left( \frac {p^2}{\mu^2} \right)^{\delta_n} ,
\end{eqnarray}
the integral in the ghost Dyson--Schwinger equation can be split up in three
pieces. The infrared integral is complicated, and we have not treated it
analytically yet (see, however, ref.\ \cite{Ler01}). The ultraviolet integral,
on the other hand, does not contribute to the infrared behaviour. As a matter
of fact, it is the resulting equation for the  ghost wave function
renormalization constant $\widetilde Z_3$  which allows one to extract definite
information  \cite{Wat01} without using any truncation or specific ansatz
beyond the underlying assumption for the existence of asymptotic infrared
series for QCD Green's functions.

The results are that the infrared behaviour of the gluon and ghost propagators
are uniquely related: The gluon propagator is infrared suppressed as compared
to the one for a free particle, the ghost propagator is infrared enhanced.
This implies that the Kugo--Ojima confinement criterion is satisfied.

\vskip 5mm

\noindent
{\bf A Truncation Scheme for Gluon and Ghost Propagators\\}

\begin{figure}[t]
 \centerline{\epsfxsize=0.7\linewidth \epsfbox{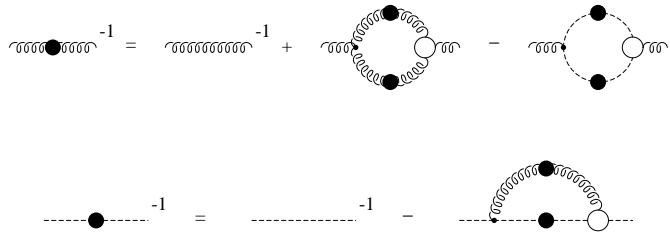}}
\caption{Diagrammatic representation of the truncated system of %coupled 
gluon and ghost DSEs.}
\label{GlGh}
\end{figure}

\begin{figure*}[t]

\parbox{.49\linewidth}{\hskip -.1cm\epsfxsize=0.98\linewidth
\epsfbox{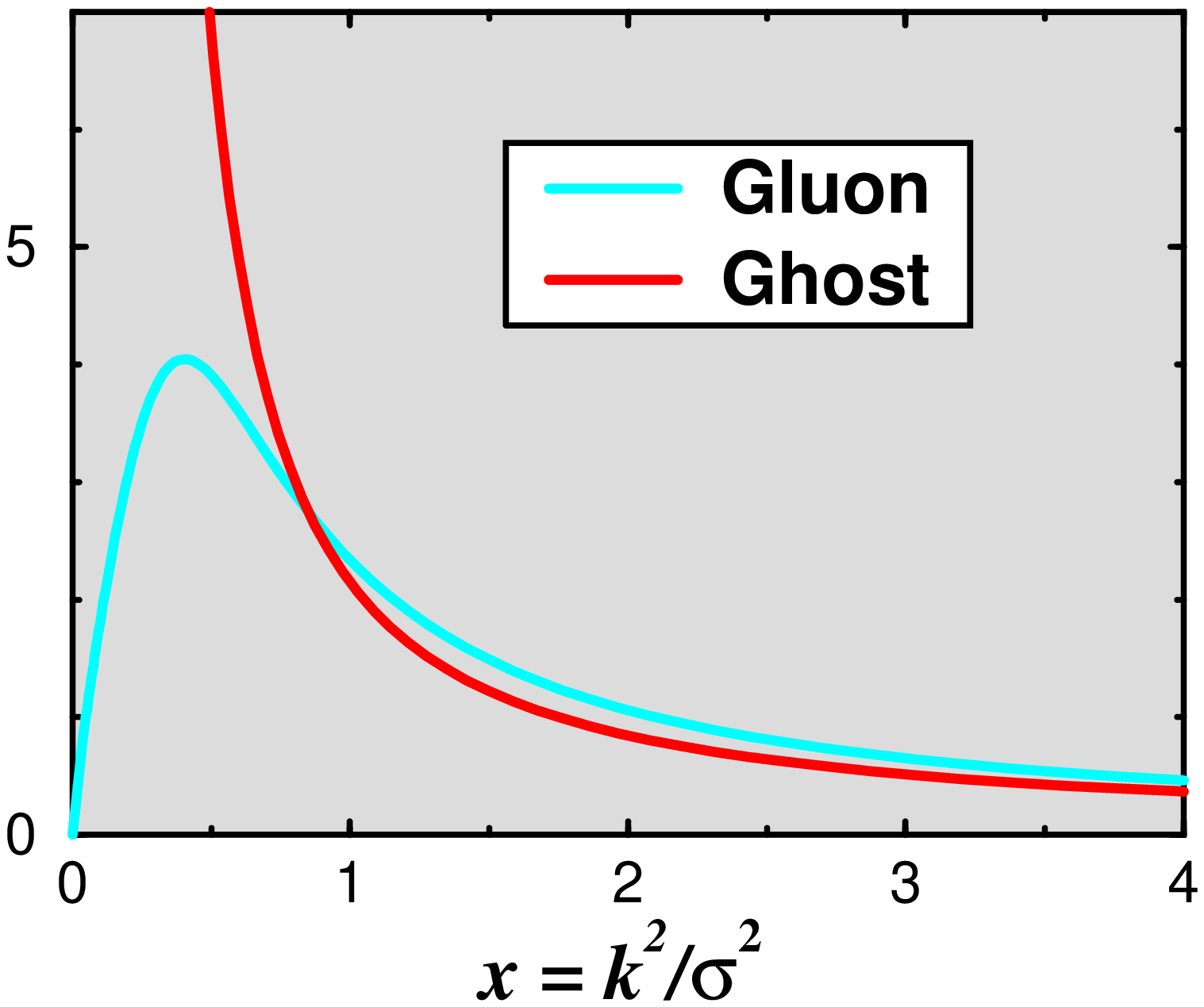}}
%\hskip .3cm
\hfill
\parbox{.46\linewidth}{\hskip .5cm\epsfxsize=0.83\linewidth
\epsfbox{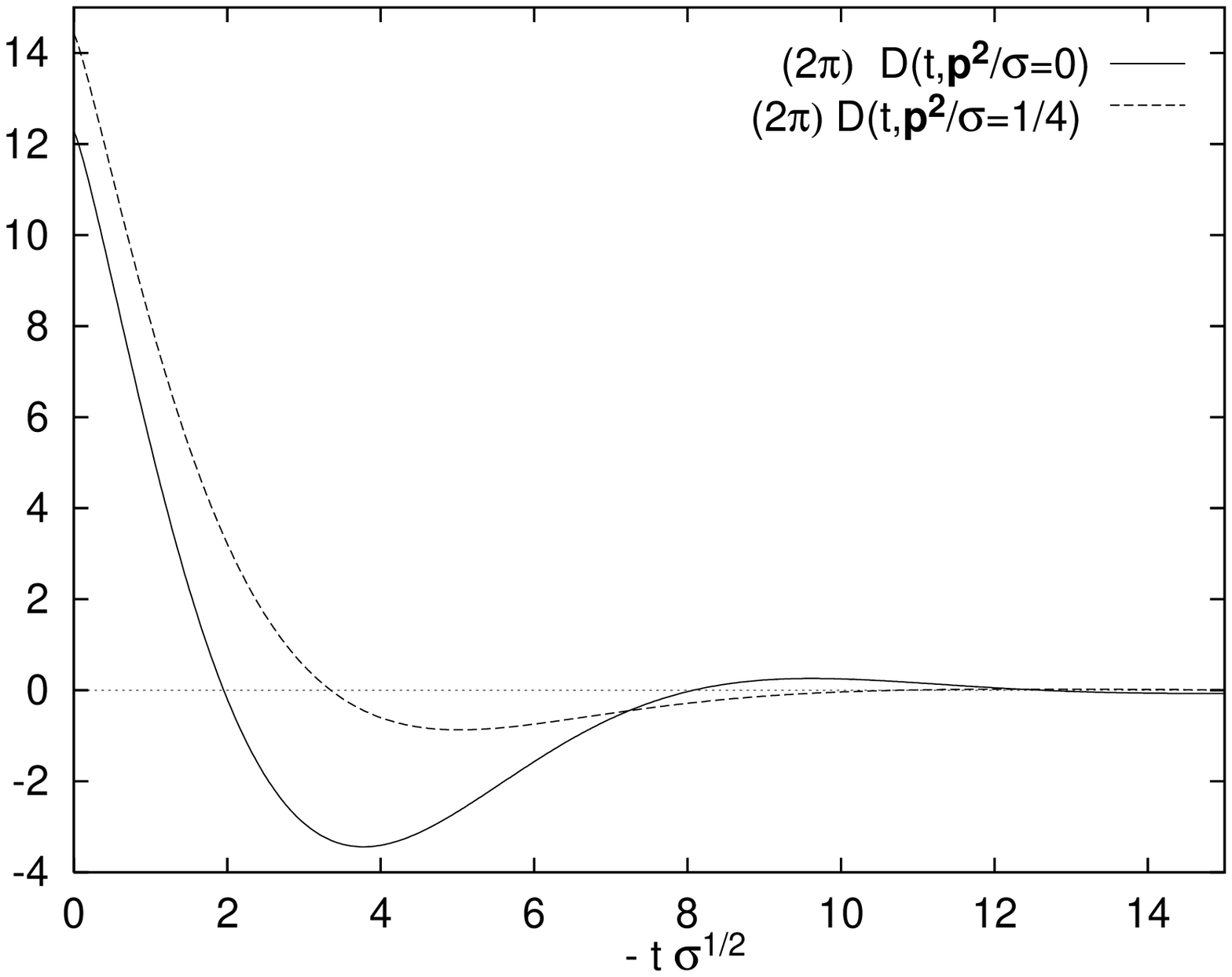}} 

\hskip .1cm\parbox{0.45\linewidth}{\refstepcounter{figure}  \label{ZG}
{\footnotesize Fig. \thefigure:  DSE solutions for $Z(x)$ and
$G(x)$\protect\cite{Sme98}.}}  
\hfill
\parbox{0.45\linewidth}{\refstepcounter{figure} \label{gluon_ft}
{\footnotesize Fig. \thefigure: $D(t, \hbox{\bf p}^2)$ from DSEs for
the gluon 
%\\[-3pt]
renormalization function $Z$ in Fig.~\ref{ZG}.}}
\vskip -.2cm
\end{figure*}

The known structures in the 3-point vertex functions, most importantly from
their Slavnov-Taylor identities and exchange symmtries, have been employed
to establish closed systems of non-linear integral equations that are
complete on the level of the gluon, ghost and quark propagators in Landau
gauge QCD. This is possible with systematically neglecting
contributions from explicit  4-point vertices to the propagator 
Dyson--Schwinger Equations
(DSEs) as well as non-trivial 4-point scattering kernels 
in the constructions of the 3-point vertices \cite{Sme98,Alk00}.
For the pure gauge theory this leads to the propagators DSEs 
diagrammatically represented in Fig.~\ref{GlGh} with
the 3-gluon and ghost-gluon vertices (the open circles) 
expressed in terms of the two functions $Z$ and $G$. 
Employing a one-dimensional approximation one obtains the numerical 
solutions sketched in Fig.~\ref{ZG} \cite{Sme98,Hau98}.

Asymptotic expansions of the solutions in the infrared yield the leading 
infrared behaviour analytically. It is thereby uniquely determined 
by one exponent $\kappa = (61 - \sqrt{1897})/19 \approx 0.92 $,
\begin{equation} 
   Z(k^2) \, \stackrel{k^2\to 0}{\sim}   
      \,  \left(\frac{k^2}{\sigma^2}\right)^{2\kappa}  \quad \mbox{and} \quad
          G(k^2) \, \stackrel{k^2\to 0}{\sim}    
         \, \left(\frac{\sigma^2}{k^2}\right)^{\kappa} \; , \label{IRB}
\end{equation}
for which the bounds $0 < \kappa < 1$ can be established requiring consistency
with Slavnov--Taylor identities \cite{Sme98}.
The renormalization group invariant momentum scale $\sigma $ represents a 
free parameter at this point which is later on fixed by choosing a definite
value for the strong coupling constant at some scale.  
The qualitative infrared behavior in eqs.~(\ref{IRB})
has been also found by studies of further truncated
DSEs \cite{Atk97}. Neither does it thus seem to depend on the particular
3-point vertices nor on employed approximations for angular integrals. 
All these solutions agree qualitatively and confirm the Kugo--Ojima
confinement criterion. 

There are also recent lattice simulations which test this criterion
directly\cite{Nak99}. Instead of $u^{ab} = -\delta^{ab}$ they obtain
numerical values of around $u = -0.7$  for the 
unrenormalized diagonal parts and zero (within  statistical errors) for the
off-diagonal parts. Taking into account the finite size effects on the
lattices employed in the simulations, these preliminary results
might still comply with the Kugo-Ojima confinement criterion.

\vspace{.2cm}
\vspace{.2cm}

Positivity violations of transverse gluon states are manifest in 
the spectral representation of 
the gluon propagator,
\begin{equation} 
       D(p^2) := \frac{Z(p^2)}{p^2}  = \int_0^\infty dm^2
         \, \frac{\rho(m^2)}{p^2 +  m^2} \; .
\end{equation}
From color antiscreening and unbroken global gauge symmetry
in QCD it follows that the spectral density  
asymptotically is negative and {\em
superconvergent} \cite{Oeh80,Nis94,Alk00}  
\begin{equation} 
  \rho(k^2) \stackrel{k^2\to\infty}{\sim}  - \frac{\gamma g^2}{k^2}
  \Big(g^2 \ln\frac{k^2}{\Lambda^2}\Big)^{-\gamma-1} \hskip -.3cm ,
  \quad\mbox{and}\;
  \int_0^\infty \!\! dm^2  \rho(m^2) = \left(
\frac{g_0^2}{g^2} \right)^\gamma  \to 0 \; , 
%\quad \mbox{since} \quad  \gamma > 0  \quad \mbox{for} \quad  N_f\le 9 
\end{equation}
since $ \gamma > 0  $ for $ N_f\le 9 $ in Landau gauge.
This implies that it contains contributions from quartet states 
(and does therefore not need to be gauge invariant unlike in QED).
Here, we consider the one-dimensional Fourier transform 
\begin{eqnarray}
  D(t,\hbox{\bf p}^2) =  \int \frac{dp_0}{2\pi}
    \frac{Z(p_0^2 + \hbox{\bf p}^2)}{p_0^2 + \hbox{\bf p}^2} \; e^{i p_0 t}
               \, = \, 
\int_{\sqrt{\mbox{\footnotesize\bf p}^2}}^\infty  d\omega \, \rho(\omega^2\!
-\! \mbox{\bf p}^2) \,  e^{-\omega t}\; , \label{eq:gluon_FT}
\end{eqnarray}
which for $\rho \ge 0$ had to be positive definite (and one had
$\frac{d^2}{dt^2} \ln D(t,\hbox{\bf p}) \ge 0$).
This is clearly not the case for the DSE solution shown in
Fig.~\ref{gluon_ft} which violates reflection positivity \cite{Alk00,Sme98}.
Even though no negative $D(t, \hbox{\bf p}^2)$ have been reported in 
lattice calculations yet, the available results \cite{Man99} 
agree in indicating that ln $D(t, \hbox{\bf p}^2)$ is
not the convex function of the Euclidean time it should be for positive
$\rho$ \cite{Man87,Nak95}. These are non-perturbative verifications of the
positivity violation for transverse gluon states which already occur in
perturbation theory. More significant for
confinement is the fact that no massless single transverse gluon 
contribution to $\rho$ exists for $Z(0) = 0$.

\vspace{.2cm}
\begin{figure*}[t]
\parbox{.49\linewidth}{\hskip -.5cm\epsfxsize=1.07\linewidth
\epsfbox{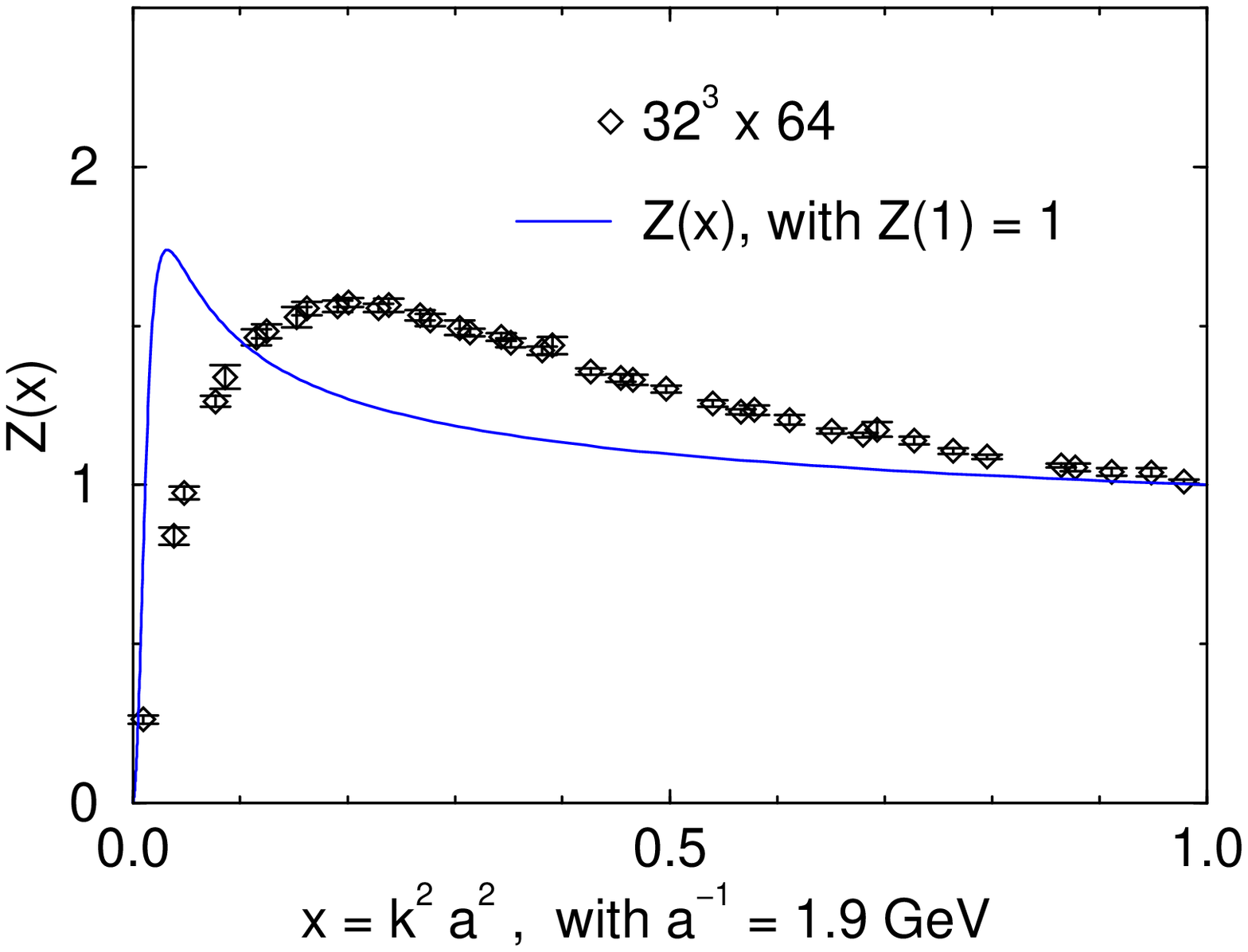}}
\hskip -.1cm
\parbox{.48\linewidth}{\hskip .2cm
\epsfxsize=0.97\linewidth\epsfbox{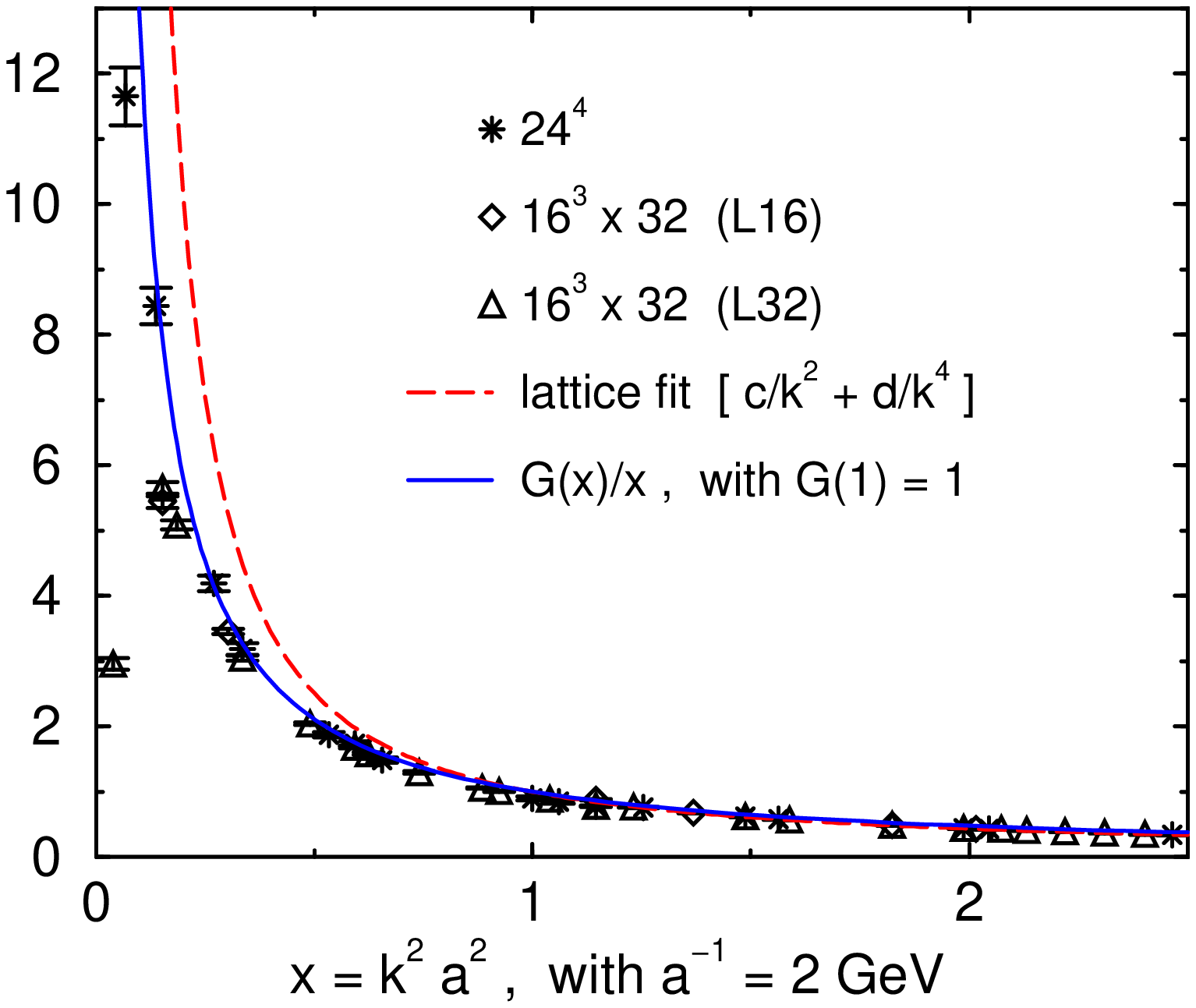}}

\hskip .2cm\parbox{0.42\linewidth}{
\refstepcounter{figure} \label{Der98} 
{\footnotesize Fig. \thefigure: The gluon renormalization function 
%\\[-3pt] 
from the DSE solutions of Ref.~\protect\cite{Sme98} (solid line) 
%\\[-3pt] 
and from the lattice data of Ref.~\protect\cite{Lei98}.}}  
\hskip .3cm \hfill
\parbox{0.48\linewidth}{\refstepcounter{figure}  \label{Sum95col} 
{\footnotesize Fig. \thefigure: The ghost propagator from
DSEs in Ref.~\protect\cite{Sme98} 
%\\[-3pt] 
(solid line) compared to data and fit (dashed with 
%\\[-3pt] 
$ca^2 =0.744 , \; da^4=0.256 $ for $x \ge 2$) from Ref.~\protect\cite{Sum96}.}}
\end{figure*}
\vspace{.2cm}

Confirmation of the important result that the gluon renormalization function
vanishes in the infrared and no massless asymptotic transverse gluon states
occur, {\it i.e.}, $Z(0) =0$, is seen in Fig.~\ref{Der98}, where 
the DSE solution of Fig.~\ref{ZG} is compared to lattice data \cite{Lei98} 
and it was further verified recently with improved lattice actions for large
volumes \cite{Bon00}. This infrared suppression as seen in lattice
calculations thereby seems to be weaker than the DSE result, 
apparently giving rise to an infrared finite gluon
propagator $D(k) \sim Z(k^2)/k^2 $ (corresponding to an exponent 
$\kappa = 1/2 $ in~(\ref{IRB})), but a vanishing one does not seem to be 
ruled out for the infinite volume limit \cite{Cuc97}.
Similar results with finite $D(0)$ are also 
reported from the Laplacian gauge which practically 
avoids Gribov copies \cite{Ale00}.

%\vspace{.2cm}
%\begin{figure*}[t]
%\parbox{.49\linewidth}{\hskip -.5cm\epsfxsize=1.07\linewidth
%\epsfbox{Der98.eps}}
%\hskip -.1cm
%\parbox{.48\linewidth}{\hskip .2cm
%\epsfxsize=0.97\linewidth\epsfbox{Sum95col.eps}}
%
%\hskip .2cm\parbox{0.42\linewidth}{
%\refstepcounter{figure} \label{Der98} 
%{\footnotesize Fig. \thefigure: The gluon renormalization function 
%%\\[-3pt] 
%from the DSE solutions of Ref.~\protect\cite{Sme98} (solid line) 
%%\\[-3pt] 
%and from the lattice data of Ref.~\protect\cite{Lei98}.}}  
%\hskip .3cm \hfill
%\parbox{0.48\linewidth}{\refstepcounter{figure}  \label{Sum95col} 
%{\footnotesize Fig. \thefigure: The ghost propagator from
%DSEs in Ref.~\protect\cite{Sme98} 
%%\\[-3pt] 
%(solid line) compared to data and fit (dashed with 
%%\\[-3pt] 
%$ca^2 =0.744 , \; da^4=0.256 $ for $x \ge 2$) from Ref.~\protect\cite{Sum96}.}}
%\end{figure*}
%\vspace{.2cm}

The infrared enhanced DSE solution for ghost propagator is compared
to lattice data in Fig.~\ref{Sum95col}. 
One observes quite compelling agreement, the numerical DSE solution fits the
lattice data at low momenta ($x \le 1$) significantly better than the fit to an
infrared singular form with integer exponents, $D_G(k^2) = c/k^2  + d/k^4$.
Though low momenta ($x<2$) were excluded in this fit, the authors concluded
that no reasonable fit of such a form was otherwise possible \cite{Sum96}.
Therefore, apart from the question about a possible maximum at the very
lowest momenta, the lattice calculation seems to confirm the infrared
enhanced ghost propagator 
with a non-integer exponent $0 < \kappa < 1$. 
The same qualitative conclusion has in fact been obtained in a more recent
lattice calculation of the ghost propagator in $SU(2)$ \cite{Cuc97}, where its
infrared dominant part was fitted best by $D_G \sim 1/(k^2)^{1+\kappa}$ for
an exponent $\kappa $ of roughly $ 0.3$ (for $\beta = 2.7$). 

To summarize, the qualitative infrared behavior in eqs.~(\ref{IRB}), an
infrared suppression of the gluon propagator together with an infrared
enhanced ghost propagator as predicted by the Kugo-Ojima criterion for the
Landau gauge, is confirmed by the presently availabe lattice
calculations. The exponents obtained therein ($0.3 < \kappa \le 0.5 $) both
seem to be consistently smaller than the one obtained in solving their DSEs.
Whether also the lattice data is thereby determined 
by one unique exponent $0<\kappa <1 $ for 
the infrared behavior of both propagators, has not yet been investigated
to our knowledge.
An independent confirmation of this combined infrared behavior which is 
indicative of an infrared fixed point would support the existence of the
unphysical massless states that are necessary to circumvent the decomposition
property for colored clusters.

\bigskip

\noindent
{\bf Acknowledgements}

\noindent
R.\ A.\  thanks Sebastian Schmidt and David Blaschke for organizing 
this stimulating workshop. 

\noindent
This work has been supported in part by the DFG under contract Al 279/3-3
and under contract We 1254/4-2.

\newpage

%% file: SOURCE/g1NEU.tex
\section*{\bf SU(2) gluon propagators from the lattice -- \\ a preview}
\addcontentsline{toc}{section}{\protect\numberline{}{SU(2) gluon propagators from the lattice -- a preview \\ \mbox{\it K. Langfeld}}}
\begin{center}
\vspace*{2mm}  
{Kurt Langfeld}\\[0.3cm]
{\small\it Institut f\"ur Theoretische Physik, Universit\"at T\"ubingen,\\
Auf der Morgenstelle 14, D--72076 T\"ubingen, Germany}
\end{center}
\newcounter{eqn18}[equation]
\setcounter{equation}{-1}
\stepcounter{equation}

\newcounter{bild18}[figure]
\setcounter{figure}{-1}
\stepcounter{figure}

\newcounter{tabelle18}[table]
\setcounter{table}{-1}
\stepcounter{table}

\newcounter{unterkapitel18}[subsection]
\setcounter{subsection}{-1}
\stepcounter{subsection}

\begin{abstract}
High accuracy numerical results for the SU(2) gluonic form factor 
are previewed for the case of Landau gauge. 
I focus on the information of quark confinement encoded in the 
gluon propagator. 
\end{abstract}

%\vfil
%\hrule width 5truecm
%\vskip .2truecm
\begin{quote} 

PACS:  11.15.Ha 12.38.Aw 

keywords: {\it Landau gauge, Gribov problem, gluon propagator, confinement, 
lattice gauge theory. }

\end{quote}
%\eject
%
\vspace{1cm}

{\bf Introduction.}
Two prominent methods to treat non-perturbative Yang-Mills theory will be 
addressed in this talk: the numerical simulation of lattice gauge theory 
(LGT) and the approach by means of the Dyson-Schwinger equations (DSE). 
While LGT covers all non-perturbative effects and, in particular, bears 
witness of quark confinement (see e.g.~\cite{bali95}), 
simulations including dynamical quarks are cumbersome despite the 
recent successes by improved algorithms~\cite{kap92} and the 
increase of computational power. 
At the present stage, systems at finite baryon densities are hardly 
accessible in the realistic case of an SU(3) gauge group~\cite{bar98}
(for recent successes see~\cite{eng99}). By contrast, the DSE approach 
can be easily extended for an investigation of quark 
physics~\cite{rob94,alk00} 
even at finite baryon densities~\cite{rob00}. Unfortunately,
the DSE approach requires a truncation of the infinite 
tower of equations, and this approximation is difficult to improve 
systematically. 
In addition, the DSE approach needs gauge fixing which is obscured by 
Gribov copies. Whether the standard Faddeev-Popov method of gauge fixing 
is appropriate in non-perturbative studies, is still under 
debate~\cite{bau98}. 

\vskip 0.3cm 
In order to merge the advantages of both approaches to low energy 
Yang-Mills theory, I will firstly address the gluon propagator 
of pure SU(2) lattice gauge theory in Landau gauge. The result 
can then be compared with the one provided by the solution of the 
coupled ghost-gluon Dyson equation~\cite{sme97,atk98,atk98b}. This will allow 
us to estimate the soundness of the truncations introduced 
to solve the equations (e.g.~vertex ansatz, angular approximation). 
Secondly, the gluon propagator is an 
one essential ingredient of the quark DSE. Two options are obvious: 
a parameterization of the lattice result for the gluon propagator 
enters the quark DSE. The corresponding solution of this equation 
provides informations on hadronic observables in quenched approximation 
i.e.~the backreaction of the quarks on the gluonic Greenfunctions 
is neglected. Once the reliability of the DSE approach to the ghost 
gluon system has been tested, the second option is to solve 
a truncated set of coupled ghost-gluon-quark DSEs, thereby, challenging 
the quenched approximation. 

\vskip 0.3cm 
In my talk, I will focus on the gluon propagator as inferred from 
the lattice calculation, and I will concentrate on the information on 
quark confinement which might be encoded in the gluon propagator. 
High accuracy data for the latter 
are obtained by a numerical method superior to 
existing techniques. Further informations and details of the numerical 
method will be presented in a forthcoming publication.

\vskip 0.3cm 
{\bf Lattice definition of the gluon field.} 
Before identifying the gluonic degrees of freedom in the lattice 
formulation, I briefly recall the definition of the gluon field 
in continuum Yang-Mills theory. 

\vskip 0.3cm 
The starting point for constructing Yang-Mills theories is the 
transformation property of the matter fields. In the case of an 
SU(2) gauge theory, we demand invariance under local SU(2) (say color) 
transformations of the quark fields 
\be 
q^\prime (x) \; = \; \Omega \, (x) \; q(x) \; , \hbo 
\Omega (x) \in {\mathrm SU(2) } \; . 
\label{eq:1} 
\en 
In order to construct a gauge invariant kinetic term for the 
quark fields, one defines the gauge covariant derivative $D_\mu 
:= \partial _\mu + i A_\mu^a t^a$, where $t^a$ are the generators 
of the SU(2) gauge group. Per definitionem, this covariant 
derivative homogeneously transforms under gauge transformations, 
\be 
D^\prime _\mu q^\prime (x) \; = \; \Omega (x) \, D_\mu (x) \, q(x) 
\label{eq:2} 
\en 
if the gluon fields transforms as 
\bea 
A^{a \, \prime }_{\mu }(x) &=& O^{ab}(x) \,  A^b _\mu (x) 
\; + \; f^{aef} \, O^{ec}(x) \, \partial _\mu O^{fc} \; , 
\label{eq:3a} \\ 
O^{ab}(x) &:=& 2 \, \tr \bigl\{ \Omega (x) \, t^a \, 
\Omega ^\dagger (x) \, t^b \bigr\} \; , \hbo 
O^{ab}(x) \, \in \, {\mathrm SO(3) } \; . 
\label{eq:3} 
\ena 
The crucial observation is that the gluon fields transform according 
to the {\it adjoint} representation while the matter fields are 
defined in the fundamental representation. 

\vskip 0.3cm 
Let us compare these definitions of fields with the ones in LGT. 
In LGT, a discretization of space-time with a lattice spacing $a$ 
is instrumental. The 'actors' of the theory are SU(2) matrices 
$U_\mu (x)$ which are associated with the links of the lattice. 
These links transform under gauge transformations as 
\be 
U^\prime _\mu (x) \; = \; \Omega (x) \, U_\mu (x) \, \Omega ^\dagger  
(x+\mu)  \; \hbo \Omega (x) \in {\mathrm SU(2) } \; . 
\label{eq:4} 
\en 

For comparison with the ab initio continuum formulation, I also 
introduce the adjoint links 
\bea 
\widetilde{U}_{\mu }^{ab} (x) &:=&  2 \, \tr \bigl\{ U_\mu (x) \, t^a \, 
U^\dagger _\mu (x) \, t^b \bigr\} \; , 
\label{eq:5} \\ 
\widetilde{U}_{ \mu }^{\prime }(x) &:=&  O(x) \, \widetilde{U}_{\mu } (x) \, 
O^T(x) \; , \hbo O(x) \in {\mathrm SO(3) } \; , 
\ena 
where $O(x)$ was defined in (\ref{eq:3}). 

\vskip 0.3cm 
In order to define the gluonic fields from lattice configurations, 
I exploit the behavior of the (continuum) gluon fields under 
gauge transformations (see~(\ref{eq:3a})), and identify the lattice 
gluon fields $\hat{A}_\mu ^a (x)$ as algebra fields of the adjoint 
representation, i.e. 
\be 
\widetilde{U}_{\mu }^{ab} (x) \; =: \; \biggl[ \exp \bigl\{ \epsilon ^f  
\, \hat{A}^f_\mu (x) \, a \bigr\} \, \biggr] ^{ab} \; , 
\hbo \big(\epsilon ^f \bigr)^{ac} := \epsilon ^{afc} \; , 
\label{eq:6} 
\en 
where the total anti-symmetric tensor $\epsilon ^{abc}$ acts as generator 
for the SU(2) adjoint representation, and where $a$ denotes the lattice 
spacing.

\vskip 0.3cm 
For later use, it is convenient to have an explicit formula for the 
(lattice) gluon fields $\hat{A}_\mu ^a (x)$ in terms of the 
SU(2) link variables $U_\mu (x)$. Usually, these links are given  in terms 
of four vectors of unit length 
\be 
U_\mu (x) \; = \; u_\mu ^0(x) \, + \, i \, \vec{u}_\mu (x) \, 
\vec{\tau} \; , \hbo \bigl[ u^0_\mu (x)\bigr]^2 \, + \, 
\bigl[ \vec{u}_\mu (x)\bigr]^2 \; = \; 1 \; , 
\label{eq:7} 
\en 
where $\tau ^b$ are the Pauli matrices. 
Using these variables, a straightforward calculation yields 
\be 
\hat{A}^b_\mu (x) \, a \; + \; {\cal O}(a^2) \; = \; 2 \, u^0_\mu (x) 
\, u^b_\mu (x) \; , \hbox to 4cm{ without summation over  } \, \mu \; .
\label{eq:8} 
\en 
I point out that (\ref{eq:8}) is a novel definition of the lattice 
gluon field. 

\vskip 0.3cm 
Finally, I illustrate the definition (\ref{eq:8}). Let us assume that 
we have exploited the gauge degrees of freedom (see (\ref{eq:4})) 
to bring the SU(2) link elements $U_\mu (x)$ as close as possible to 
the unit matrix, 
\be 
\Omega (x): \; \sum _{\{x\}, \mu } \tr \, U^\prime _\mu (x) 
\rightarrow \, \mathrm{max} \; .
\label{eq:9} 
\en 
In this gauge, I decompose the link variables by 
\be 
U^\prime _\mu (x) \; = \; Z_\mu (x) \, \exp \biggl\{ i \hat{A}^b_\mu (x) 
\, t^b \, a \biggr\} \; , 
\label{eq:10} 
\en 
where $\hat{A}^b_\mu (x) $ is implicitly defined by (\ref{eq:6}) and 
$Z_\mu (x) \in \{-1,+1\}$. Indeed, the lattice gluon fields (\ref{eq:8}) 
do not change when $U_\mu (x) \rightarrow (-1) U_\mu(x)$. Hence, the 
fields $\hat{A}^b_\mu (x) $ are relegated to the SO(3) coset space. 
I here propose 
to disentangle the information carried by the center and coset 
parts of the link variables by studying the $Z_\mu(x)$ and $\hat{A}^b_\mu (x)$ 
correlation functions independently. I stress, however, that in Landau 
gauge (\ref{eq:9}) the role of the $Z_\mu (x)$ is de-emphasized 
($Z_\mu (x) \rightarrow 1$). 
In particular, I do not expect a vastly different gluon propagator 
when other (more standard) definitions of the lattice gluon fields 
are used~\cite{cuc97,bon01}. 

\vskip 0.3cm 
{\bf Gauge fixing.} 
In the continuum formulation, calculations employing gauge fixed 
Yang-Mills theory use only gauged gluon fields which satisfies the 
gauge condition, e.g. 
\be 
\partial _\mu A^{\prime \, a }_\mu (x) \; = 0 \; \hbox to 3cm 
{\hfill (Landau gauge) \hfill } \; , 
\label{eq:11} 
\en 
and rely on the assumption that the Faddeev-Popov determinant corrects 
the probabilistic weight in an appropriate way. This assumption is true 
if the gauge condition picks a unique solution $\Omega (x)$ of (\ref{eq:11}) 
for a given field $A^a_\mu (x)$. Unfortunately, the Landau gauge 
condition generically admits several solutions depending on the 
''background field'' $A_\mu^b(x)$ which is the subject of gauge fixing 
(Gribov problem). Thus, the above assumption 
seems not always be justified~\cite{bau98}. Further restrictions 
on the space of possible solutions are required~\cite{zwa94}.

\vskip 0.3cm 
Let us contrast the continuum gauge fixing with its lattice analog. 
In a first step, link configurations $U_\mu(x)$ are generated by means 
of the gauge invariant action without any bias to a gauge condition. 
In a second step, the gauge-fixed ensemble is obtained by 
adjusting the gauge matrices $\Omega (x)$ (see~(\ref{eq:4})) 
until the gauged link ensemble satisfies the gauge condition. 
This procedure obviously guarantees the correct probabilistic measure 
for the gauged configurations, and gauge invariant quantities 
which are calculated with the gauged configurations evidently 
coincide with the ones obtained from un-fixed configurations. 
However, the Gribov problem re-appears as the problem of finding ''the 
name of the gauge''. Let me illustrate this last point: 
The naive Landau gauge condition for the lattice gluon field, i.e. 
\be 
\hat{\partial }_\mu \hat{A}^{\prime \, b} _\mu (x) \; = \; 0 
\label{eq:12} 
\en 
is satisfied if we seek an {\it extremum} of the variational condition 
(\ref{eq:9}). If we restrict the variety of solutions $\Omega (x)$ 
which extremize (\ref{eq:9}) to those solutions which {\it maximize } 
the functional (\ref{eq:9}), we confine ourselves to the case where 
the Faddeev-Popov matrix is positive semi-definite. The fraction of 
the configuration space of gauge fixed fields $\hat{A}^{\prime \, b} _\mu $ 
is said to lie within the first Gribov horizon. A numerical algorithm 
which obtains the gauge transformation matrices $\Omega (x)$ from 
the condition (\ref{eq:9}) still samples a particular set of local 
maxima. Different algorithms might differ in the subset of chosen 
local maxima, and, hence, implement different gauges. A conceptual 
solution to the problem is to restrict the configuration space 
of gauge fixed fields $\hat{A}^{\prime \, b} _\mu $ to the 
so-called {\it fundamental modular region}. In the lattice simulation, 
this amounts to picking the {\it global maximum } of the 
variational condition (\ref{eq:9}). In practice, finding the 
global maximum is a highly non-trivial task. Here, I adopt two extreme 
cases of gauge fixing: firstly, I will study the gluon propagator of 
the gauge where an iteration over-relaxation algorithm almost 
randomly averages over the local maxima of the variational condition 
(\ref{eq:9}). This result will then be compared with the gluon propagator 
of a gauge where a simulated annealing algorithm searches for the 
global maximum. It will turn out that the gluon propagators of both 
gauges agree within statistical error bars. 

\vskip 0.3cm
{\bf The numerical simulation:} 
The link configurations are generated using the Wilson action. 
I refrain from using a ''perfect action'' since I am interested in 
the gluon propagator in the full momentum range; simulations 
using perfect actions recover a good deal of continuum physics 
at finite values of the lattice spacing at the cost of a non-local action. 
For practical simulations, perfect actions are truncated which is 
poor approximation at high energies where the full non-locality of 
the action must come into play. 

\vskip 0.3cm
Here, calculations were performed using a $16^3 \times 32$ lattice. 
The dependence 
of the lattice spacing on $\beta $ (renormalization), i.e. 
\be 
\sigma a^2 (\beta ) \; = \; 0.12 \, \exp \biggl\{ - \frac{ 6 \pi^2}{11} 
\, \bigl( \beta - 2.3 \bigr) \biggl \} \, , \hbo 
\sigma := (440 \,{\mathrm MeV}) ^2 \; , 
\label{eq:20} 
\en 
is appropriate for $\beta \in [2.1, 2.6]$ for the achieved numerical
accuracy. 

\vskip 0.3cm
Once gauge-fixed ensembles are obtained by implementing a variational 
gauge condition (see discussion of previous section), the gluon 
propagator is calculated using 
\be 
D^{ab}_{\mu \nu } (x-y) \; = \; \bigl\langle \hat{A}^a_\mu (x) \, 
\hat{A}^b_\nu (y) \, \bigr\rangle _{MC} \; , 
\label{eq:21} 
\en 
where the Monte-Carlo average is taken over 200 properly thermalized 
gauge configurations. Of particular interest is the gluonic form factor 
$F(p^2)$ which is defined by 
\be 
D(p^2) \; = \; \int D^{aa}_{\mu \mu } (x) \; \exp \bigl\{ ipx \bigr\} 
\; d^4x \; , \hbo 
D(p^2) = \frac{F(p^2)}{p^2} \; . 
\label{eq:22} 
\en 
Since in Landau gauge the propagator is diagonal in color and transversal 
in Lorentz space, the form factor $F(p^2)$ contains the full information. 
\begin{table}
\caption{ Simulation parameters: $L=32 a$: lattice extension, 
$\Lambda $: UV cutoff } 
\begin{tabular}{cccccc}
$\beta $ & 2.1 & 2.2 & 2.3 & 2.4 & 2.5  \\ 
L [fm] & 8.6 & 6.6 & 5.0 & 3.8 & 2.9 \\ 
$\Lambda $ [GeV] & 2.3 & 3.0 & 4.0 & 5.2 & 6.8 \\
\end{tabular}
\end{table}

\vskip 0.3cm 
{\bf Results I: Landau gauge } 
The lattice momentum in units of the lattice spacing is given by 
\be 
p_x \, a \; = \; 2 \, \sin \left( \frac{\pi}{N} n_x \right) \; , 
\hbo x = 1 \ldots 4 
\label{eq:23} 
\en 
where $n_x$ is an integer which numbers the Matsubara frequency and 
$N$ is the number of lattice points (in $x$-direction). 
\begin{figure}[t]
\centerline{
\epsfxsize=0.5\linewidth 
\epsfbox{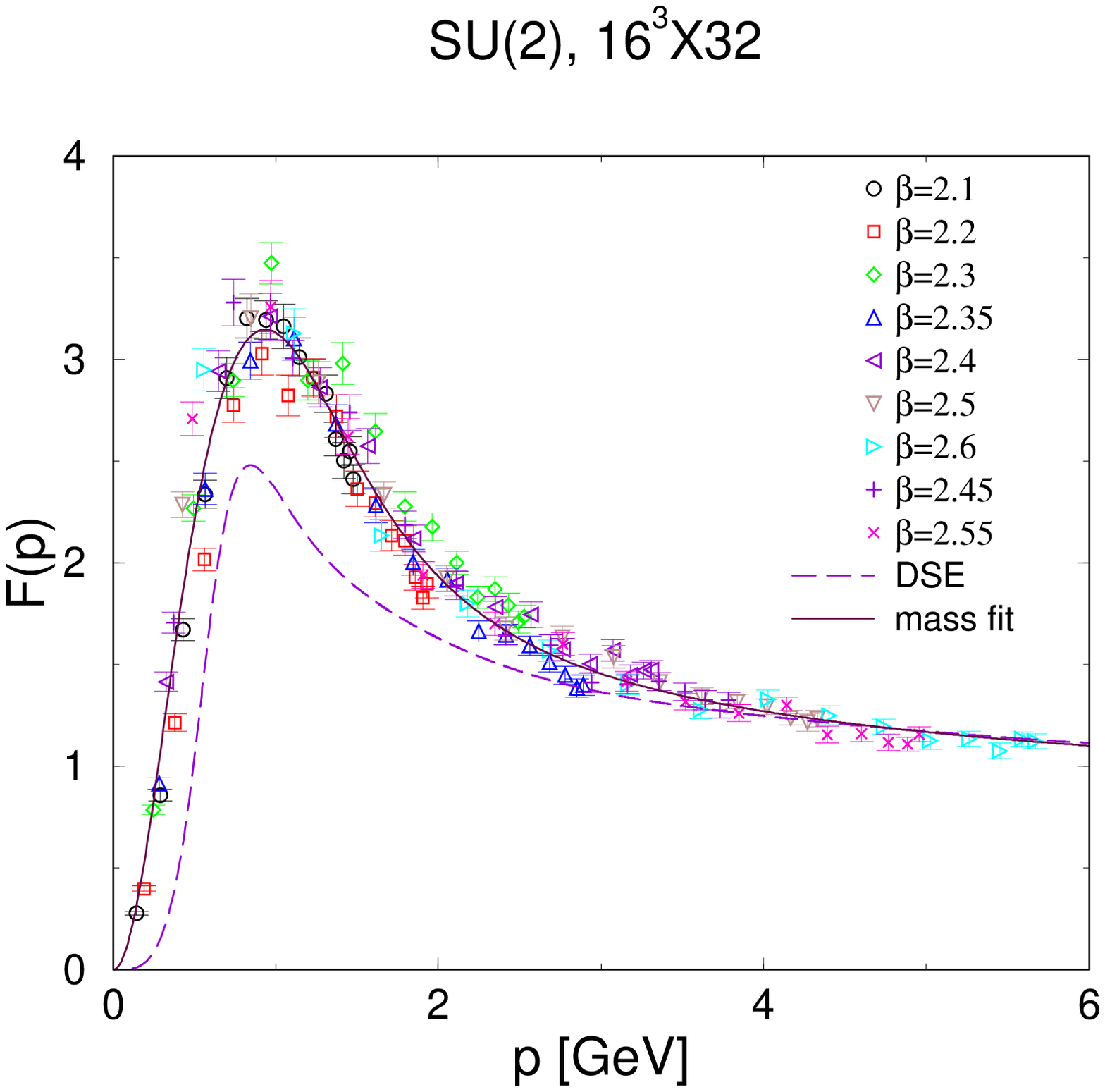}
\epsfxsize=0.55\linewidth 
\epsfbox{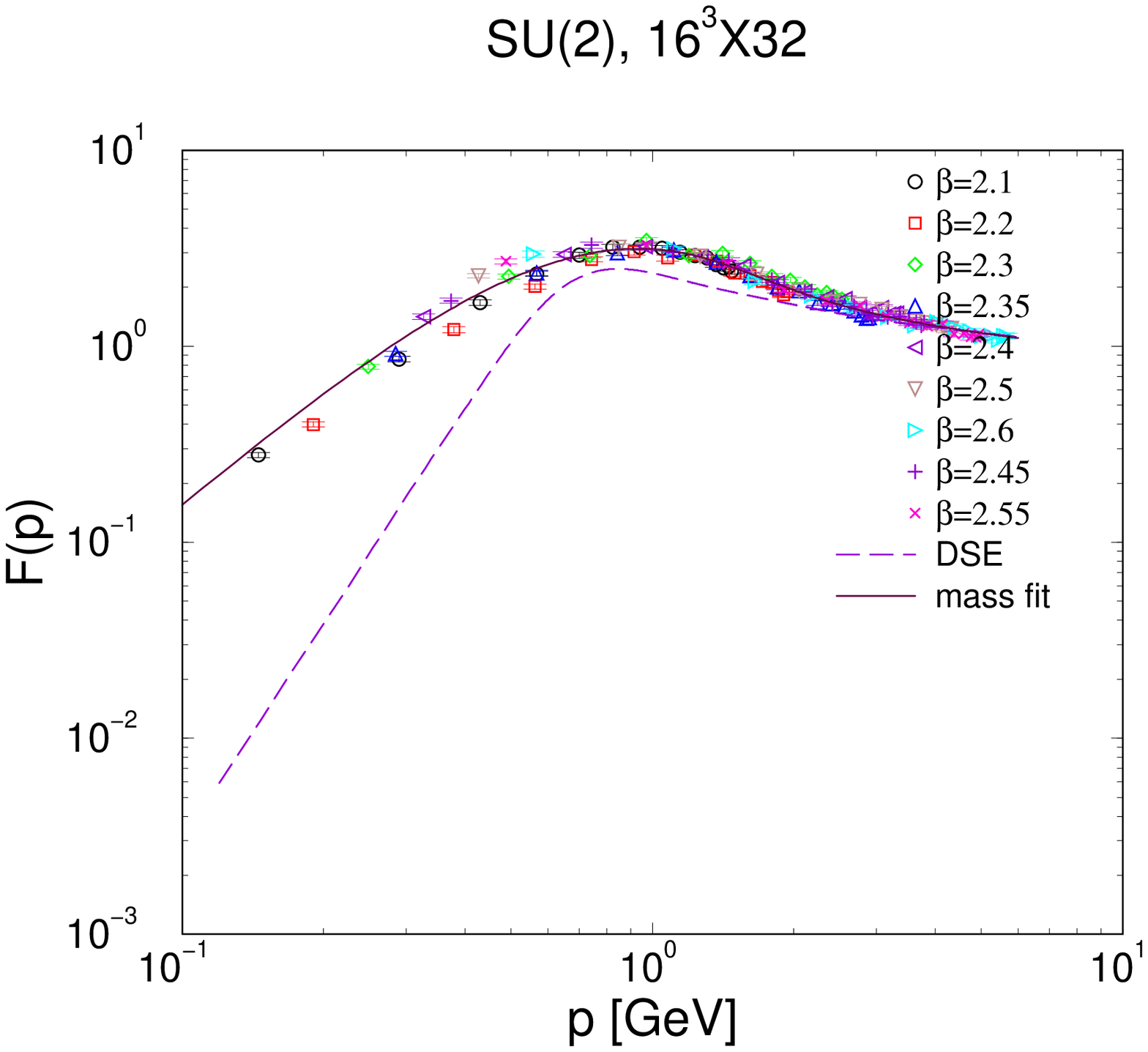}
}
\caption{The gluonic form factor $F(p^2)$ as function of the momentum 
   transfer (left panel: linear scale; right panel: log-log scale). 
   Also shown is the solution of the set of DSEs proposed 
   in [9] which have been solved for the case of $SU(2)$ [10].
}
\label{fig:1}
\end{figure}
Physical 
units for the momentum can be obtained by using (\ref{eq:20}). 
Calculations 
with different $\beta $-values correspond to simulations with a  
different UV-cutoff $\Lambda := \pi / a(\beta )$. In order to 
obtain the {\it renormalized} gluon propagator, the gluonic wave 
function renormalization is chosen to yield a finite (given) value 
for the form factor at fixed momentum transfer. 

\vskip 0.3cm
Figure \ref{fig:1} shows my result for the form factor $F(p^2)$ where 
the condition (\ref{eq:9}) was implemented with an iteration over-relaxation 
method. The data obtained with different $\beta $-values are identical 
within numerical accuracy, thus signaling a proper extrapolation 
to the continuum limit.

At high momentum the lattice data are consistent with the behavior 
known from perturbation theory, 
\be 
F(p^2) \; \propto \; 1/ \left[ log \frac{p^2}{\mu ^2 } \right]^{13/22} \; , 
\hbo p^2 \gg \mu ^2 \approx (1 \, \mathrm{GeV})^2 \; . 
\label{eq:24} 
\en 
The lattice data are compared with the solution of the gluon-ghost 
DSE~\cite{sme97}\footnote{ I thank Chr.~Fischer for communicating his 
DSE solution for the SU(2) case prior to publication.}. 
From the DSE studies one expects a scaling law at small momentum
\be 
F(p^2) \; \propto \left[ p^2 \right]^{2 \kappa } \; , \hbo 
p^2 \ll \mu ^2 \approx (1 \, \mathrm{GeV})^2 \; . 
\label{eq:25}  
\en 
Depending on the truncation of the Dyson tower of equations and on 
the angular approximation of the momentum loop integral, one finds 
$\kappa = 0.92$~\cite{sme97} or $\kappa = 0.77$~\cite{atk98} 
or $\kappa \rightarrow 1$~\cite{atk98b}. The 
lattice data are consistent with $\kappa = 0.5$ corresponding to 
an infrared screening by a gluonic mass. Also shown is the coarse grained 
''mass fit'' ($\mu = 1\, $GeV) 
\be 
F(p^2) \; = \; \frac{ p^2}{p^2+m_1^2} \biggl[ \frac{ \mu ^4 }{p^4+m_2^4} 
\; + \; \frac{s}{\left[ log \left(\frac{m_l^2}{\mu ^2 } + \frac{p^2}{\mu ^2 } 
\right) \right]^{13/22} } \biggr] 
\label{eq:26} 
\en 
which nicely reproduces the lattice data within the statistical error bars. 

\vskip 0.3cm 
{\bf Results II: gluon propagator and confinement } 
In order to get a handle on the information of quark confinement 
encoded in the gluon propagator in Landau gauge, I now {\it change by hand} 
the SU(2) Yang-Mills theory to a theory which does not confine quarks. 
It is instructive to compare the gluon propagator of the modified 
theory with the SU(2) result (see figure \ref{fig:1}). 

\vskip 0.3cm 
In the Maximal Center gauge~\cite{deb97}, the mechanism of quark confinement 
can be understood by a percolation of vortices which acquire physical 
relevance in the continuum limit~\cite{la98}. An intuitive picture 
in terms of vortex physics is also available for the deconfinement phase 
transition at finite temperatures~\cite{la99}. Reducing the full Yang-Mills 
configurations to their vortex content still yields the full string 
tension~\cite{deb97}. Vice versa, removing these vortices from the 
Yang-Mills ensemble results in a vanishing string tension. 
This observation was used in~\cite{for99} to verify the impact of the 
vortices on chiral symmetry breaking. 

\begin{figure}[t]
\centerline{
\epsfxsize=0.53\linewidth 
\epsfbox{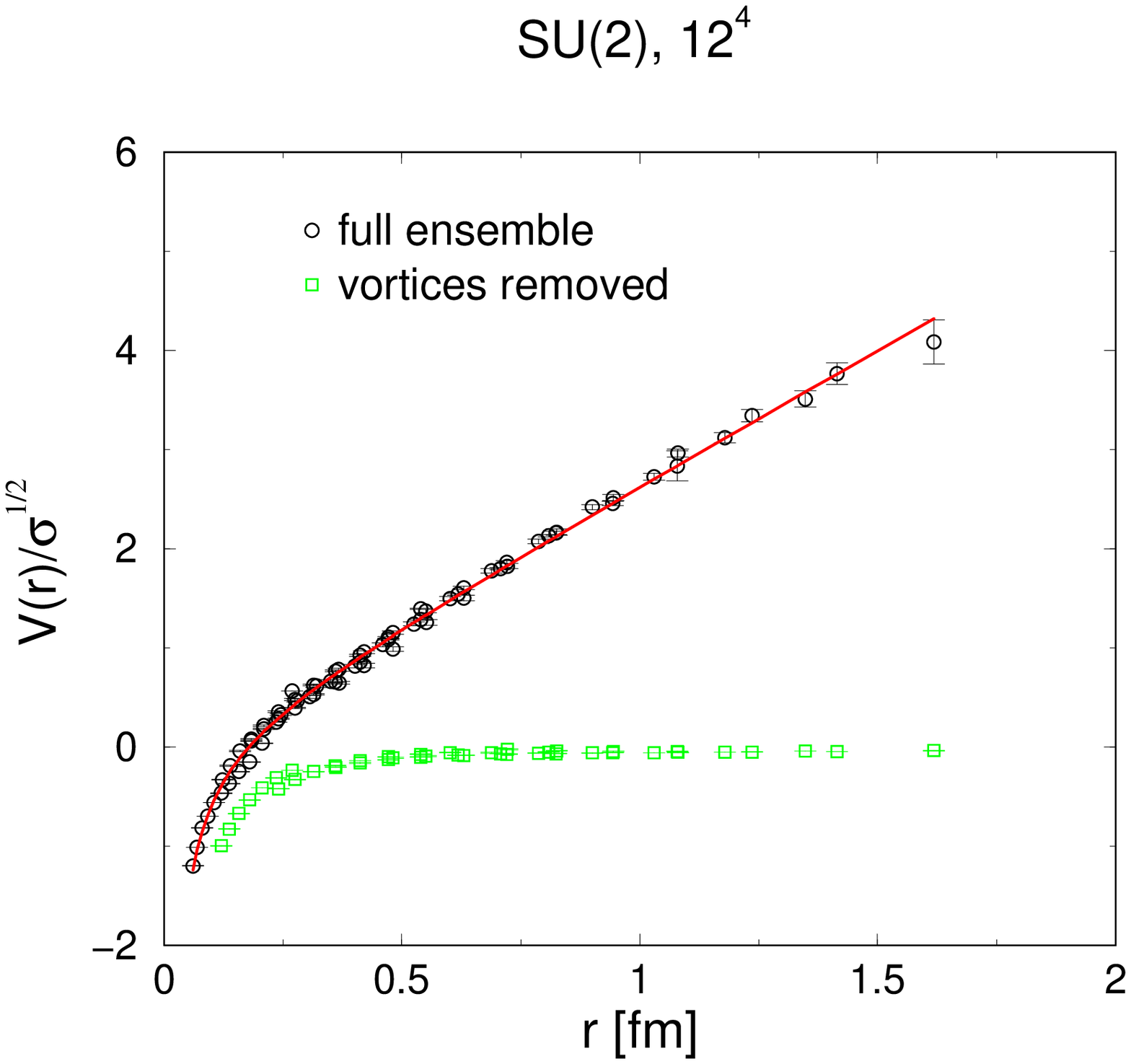}
\epsfxsize=0.5\linewidth 
\epsfbox{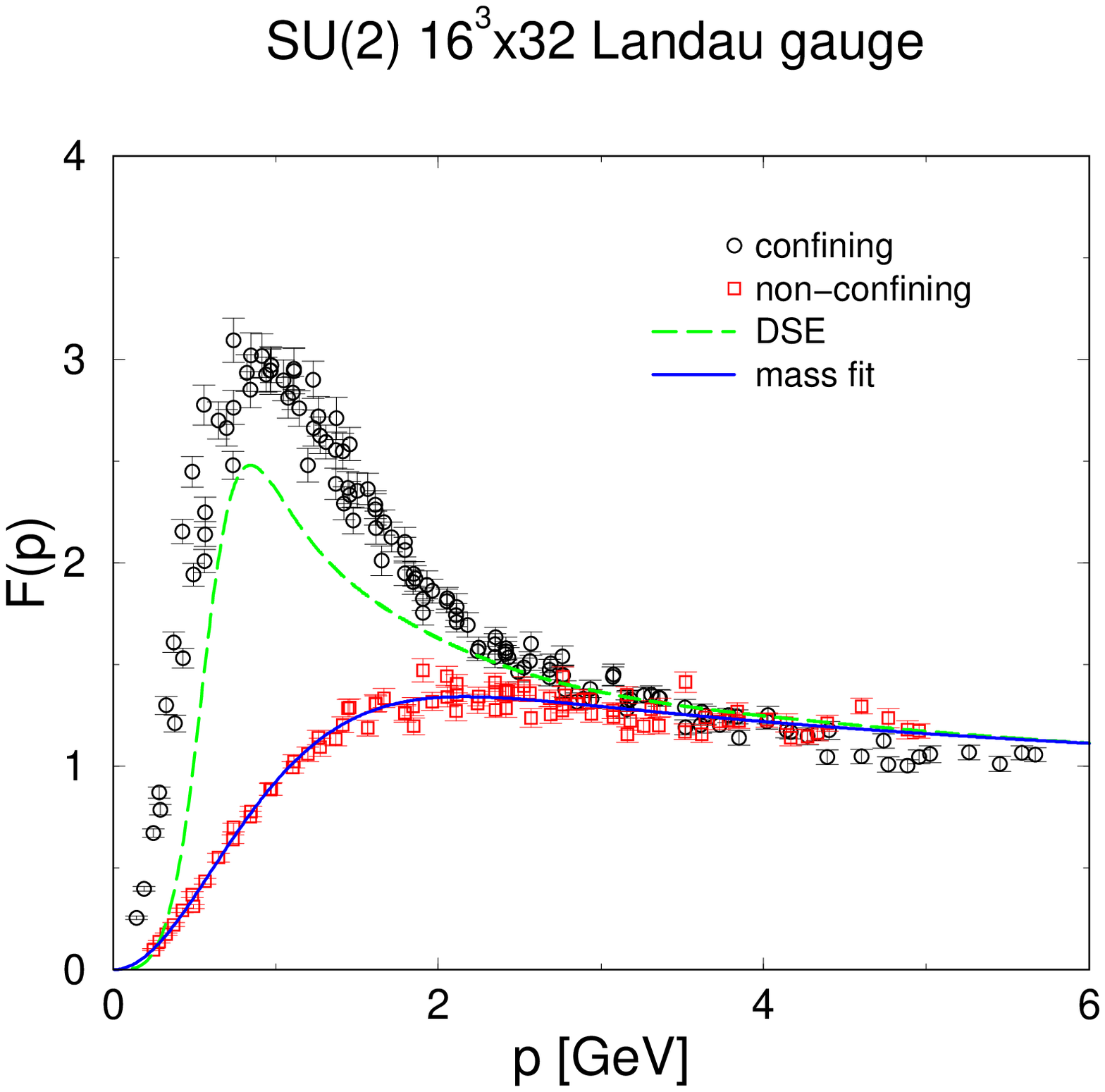}
}
\caption{ The static quark anti-quark potential (left panel) and 
   the corresponding gluonic form factors (right panel); DSE solution 
   from [10]. } 
\label{fig:2}
\end{figure}
\vskip 0.3cm 
The static quark anti-quark potential in figure \ref{fig:2} demonstrates 
that a removal of the center vortices produces a 
non-confining theory. Figure \ref{fig:2} also shows the gluonic form 
factor obtained from the modified ensemble. The striking feature is 
that the strength of the form factor in the infra-red momentum range 
is drastically reduced. 

\begin{figure}[t]
\centerline{
\epsfxsize=0.5\linewidth 
\epsfbox{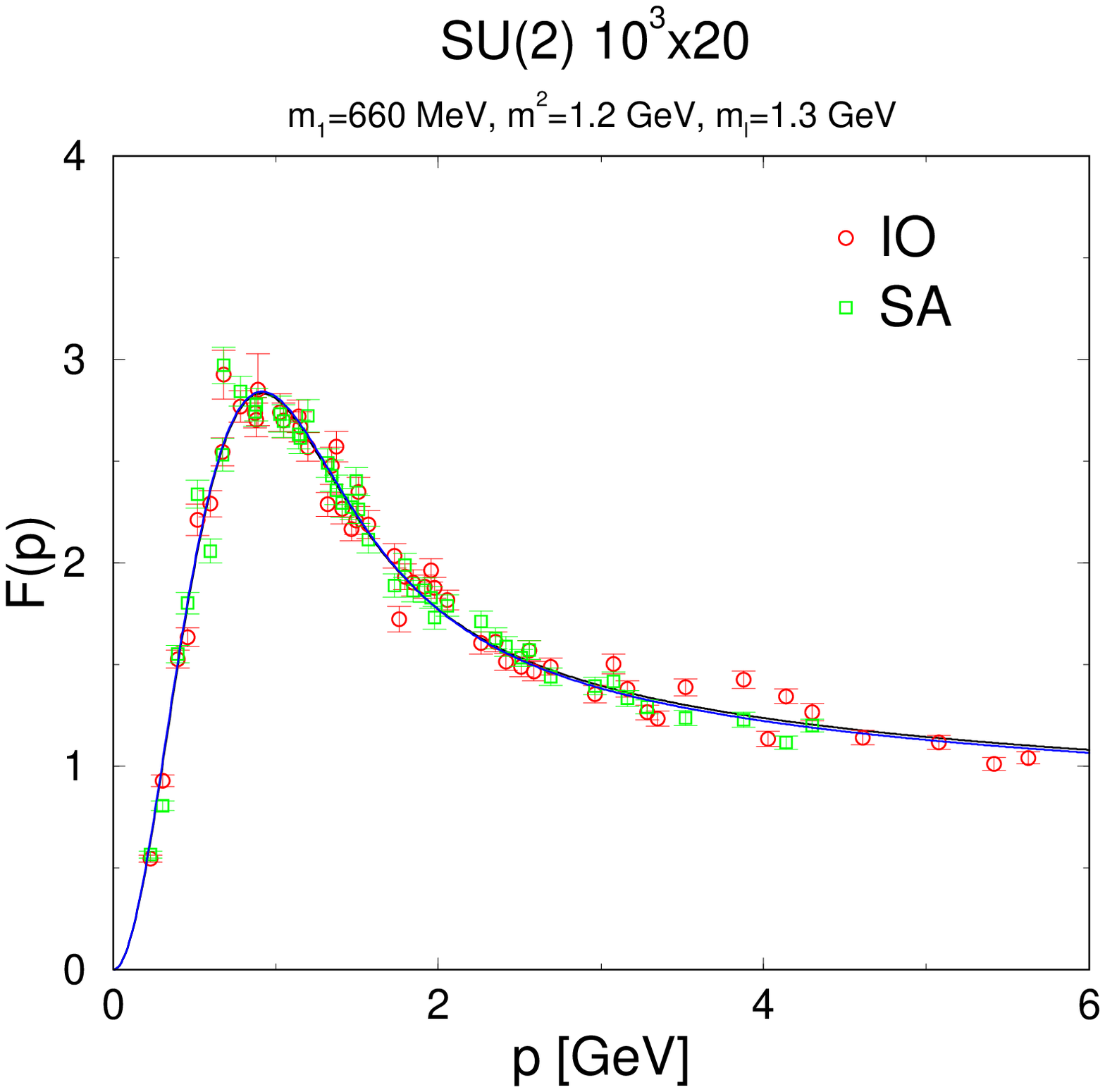}
\epsfxsize=0.5\linewidth 
\epsfbox{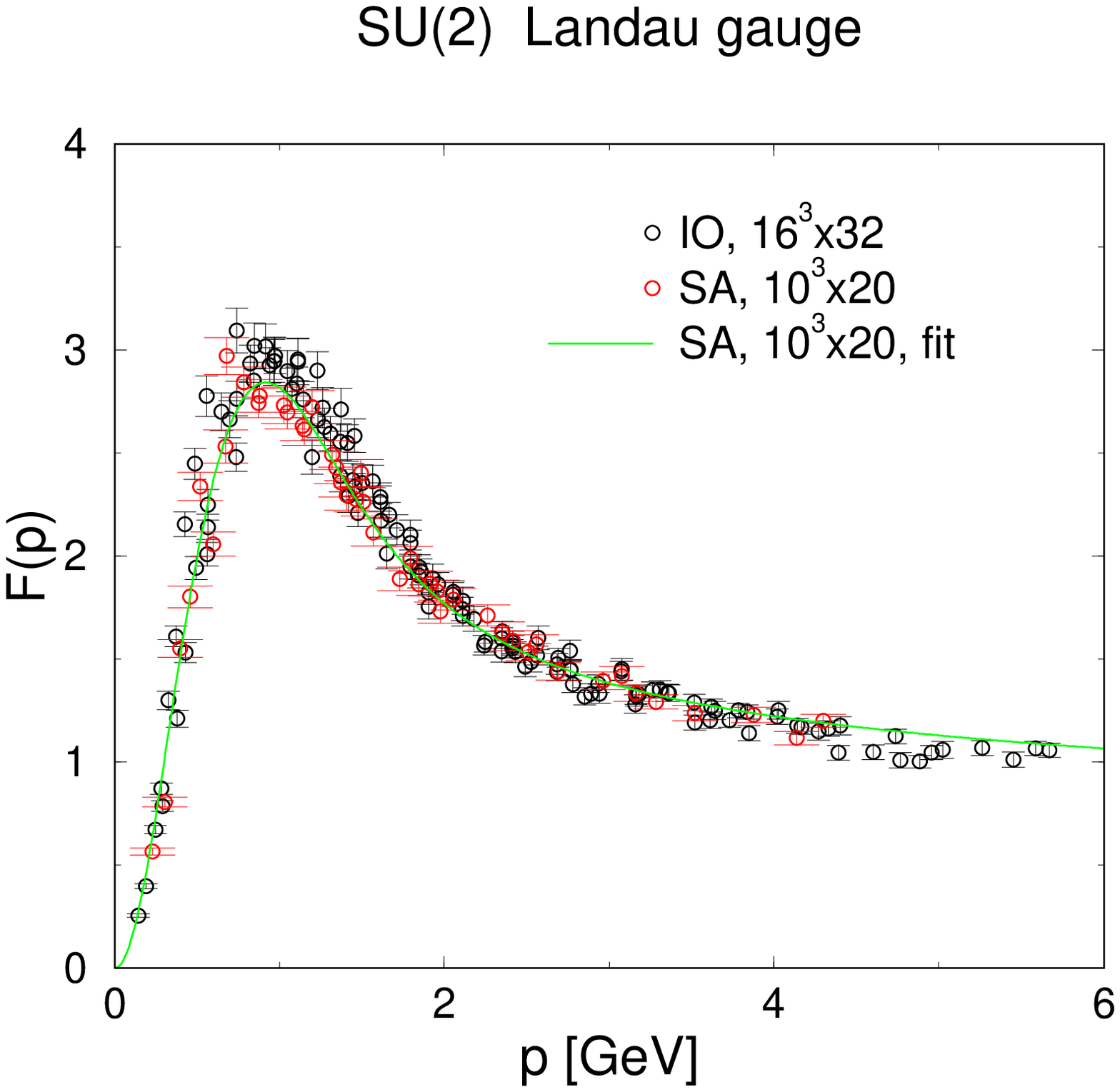}
}
\caption{The gluonic form factor $F(p^2)$ for a $10^3\times 20$ lattice 
   in the gauge IA and SA, respectively (left panel) and compared with 
   previous results (IA, $16^3\times 32$) (right panel). } 
\label{fig:3}
\end{figure}

\vskip 0.3cm 
{\bf Results III: the Gribov noise } 
Finally, let us check how strongly the gluonic form factor $F(p^2)$ 
depends on the choice of gauge, i.e. on the sample of maxima of the 
variational condition (\ref{eq:9}) selected by the algorithm. 
For this purpose, I adopt an extreme point of view by comparing the gauge 
implemented by the iteration over-relaxation (IA) algorithm with the 
gauge obtained by simulated annealing (SA). The results of the form factor 
in both cases are shown in figure~\ref{fig:3}. I find, in agreement 
with~\cite{cuc97}, that, in the case of the gluonic form factor, 
the Gribov noise is comparable with statistical noise for data generated 
with $\beta \in [2.1,2.5]$ (scaling window). 

\vspace{1cm}
{\bf Acknowledgements.} 
I thank my coworkers J.~Gattnar and H.~Reinhardt. I am indebted to 
J.~Bloch, R.~Alkofer and Chr.~Fischer for helpful discussions.

\newpage

%% file: SOURCE/topqcd.tex
\section*{\bf Gaugeless Unconstrained QCD and Monopoles}
\addcontentsline{toc}{section}{\protect\numberline{}{Gaugeless Unconstrained QCD and Monopoles \\ \mbox{\it V. Pervushin}}}
\begin{center}
\vspace*{2mm}
{Victor Pervushin}\\[0.3cm]
{\small\it BLTP, Joint Institute for Nuclear Research, 141980 Dubna, Russia} 
\end{center}

\medskip 
%PACS number(s):04.60.-m, 04.20.Cv, 98.80.Hw (QCD) 
\medskip

\begin {abstract} 
 
"Equivalent unconstrained systems" for QCD obtained 
by resolving the Gauss law are discussed. 
We show that the effects of hadronization, confinement, 
spontaneous chiral symmetry breaking and $\eta_0$-meson mass 
can be hidden 
in solutions of the non-Abelian Gauss constraint in the class of 
functions of topological gauge transformations, in the form of 
a monopole, a zero mode of the Gauss law, and a rising potential. 
 
\end{abstract} 

\newcounter{eqn11}[equation]
\setcounter{equation}{-1}
\stepcounter{equation}

\newcounter{bild11}[figure]
\setcounter{figure}{-1}
\stepcounter{figure}

\newcounter{tabelle11}[table]
\setcounter{table}{-1}
\stepcounter{table}

%\newcounter{kapitel11}[section]
%\setcounter{section}{-1}
%\stepcounter{section}

\newcounter{unterkapitel11}[subsection]
\setcounter{subsection}{-1}
\stepcounter{subsection}

%\vspace*{1cm} 
 
%\newpage 
%(Key-words: QCD, Gauss law, topology, monopole, zero mode, 
%hadronization, confinement, U(1)-problem) 
 
%--------------------------------------------------------------%%%% 
\subsection{Introduction} 
%--------------------------------------------------------------%%%%(page 1) 
 
The consistent dynamic description of gauge constrained systems 
was one of the most fundamental problems of theoretical physics 
in the 20th century. There is an opinion that 
the highest level of solving the problem of quantum description of 
gauge relativistic constrained systems is the Faddeev-Popov (FP) integral 
for unitary perturbation theory~\cite{fp1}. In any case, just this FP 
integral was the basis to prove renormalizability of the unified theory 
of electroweak interactions in papers by 't Hooft and Veltman marked by 
the 1999 Nobel prize. 
 
Another opinion is that the FP integral has only the intuitive status. 
The most fundamental level of the description of gauge constrained 
systems is the derivation of "equivalent unconstrained systems" (EUSs)
compatible with the simplest variation methods of the Newton mechanics 
and with the simplest quantization by the Feynman path integral. 
It was the topic of Faddeev's paper~\cite{f} " Feynman integral 
for singular Lagrangians" where 
the non-Abelian EUS was obtained (by 
explicit resolving the Gauss law), in order to justify the intuitive 
FP path integral~\cite{fp1} by its equivalence to the Feynman path 
integral. Faddeev managed to prove the equivalence of the Feynman integral 
to the FP one only for the scattering amplitudes~\cite{f} where 
all particle-like excitations of the fields are on their mass-shell. 
However, this equivalence is not proved and becomes doubtful for the cases of 
bound states, zero modes and other collective phenomena 
where these fields are off their mass-shell. It is just the case of QCD. 
In this case, the FP integral in an arbitrary relativistic gauge can lose 
most interesting physical phenomena 
hidden in the explicit solutions of constraints~\cite{n,rev}. 
 
The present paper is devoted to the derivation an EUS for QCD 
in the class of functions of topologically nontrivial 
transformations, in order to present here a set of arguments in favor of 
that physical reasons of hadronization and confinement in QCD 
can be hidden in the explicit solutions of the non-Abelian constraints. 
 
\subsection{Equivalent Unconstrained Systems in QED} 
 
The Gauss law constraint is the equation for 
the time component of a gauge field 
\be  \label{glqed} 
\frac{\delta W}{\delta A_0}=0~\Rightarrow~~~\partial_j^2A_0= 
\partial_k \dot A_k +J_0 
\ee 
in the frame of reference with an axis of time 
$l^{(0)}_{\mu}=(1,0,0,0)$. 
Heisenberg and Pauli~\cite{hp} noted that the gauge 
($\partial_k A^*_k \equiv 0$) is distinguished, 
and Dirac~\cite{cj} constructed the corresponding ("dressed") variables 
$A^*$ in the explicit form 
\be \label{df} 
ieA^*_k=U(A)(ieA_k+\partial_k )U(A)^{-1}~,~~~~~~~~ 
U(A)=\exp[ie\frac{1}{\partial_j^2 }\partial_kA_k ]~, 
\ee 
using for the phase the time integral of the spatial part of the 
Gauss law $\partial_k\dot A_k $. The action for an EUS for QED is derived by 
the substitution of the solution of the Gauss law in terms of the "dressed" 
variables into the initial action 
\be\label{qed} 
W_{\rm Gauss-shell}= W^*_{l^{(0)}}(A^*,E^*)~. 
\ee 
The peculiarity of the EUS for QED is 
the electrostatic phenomena described by the monopole class of functions 
($f(\vec x)= O(1/r), |\vec x|=r \rightarrow \infty$). 
 
The EUS  can be quantized by 
the Feynman path integral in the form 
\bea \label{qedf} 
Z_F[l^{(0)},J^*] &=&\int\limits_{ }^{ }d^2A^* d^2 E^* \nonumber\\
&&\exp\bigg\{iW^*_{l^{(0)}} [A^*,E^*]+i\int\limits_{ }^{ }d^4x 
[J^*_k \cdot A^*_k-J_0^*\cdot A_0^*]\bigg\} 
\eea 
where $J^*$ are physical sources. 
This path integral depends on the axis of time $l^{(0)}_{\mu}=(1,0,0,0)$. 
 
One supposes that the dependence on the frame ($l^{(0)}$) can be removed 
by the transition from the Feynman integral of the EUS ~(\ref{qedf}) 
to perturbation theory in any relativistic-invariant 
gauge $f(A)=0$ with the FP determinant 
\be \label{qedr} 
Z_{FP}[J] =\int\limits_{ }^{ }d^4A \delta[f(A)]\Delta_{FP} 
\exp\bigg\{iW[A]-i\int\limits_{ }^{ }d^4xJ_{\mu} \cdot A^{\mu}\bigg\}~. 
\ee 
This transition is well-known as a "change of gauge", and it 
is fulfilled in two steps \\ 
I) the change of the variables $A^*$~(\ref{df}), and \\ 
II) the change of the physical sources $J^*$ of the type of 
\be \label{ps} 
A^*_k(A)J_k^*=U(A)\left(A_k-\frac{i}{e}\partial_k \right)U^{-1}(A) J^*_k 
~\Rightarrow~A_{k}J^{k}~. 
\ee 
At the first step, all electrostatic monopole physical phenomena 
are concentrated in the Dirac gauge factor $U(A)$~(\ref{df}) 
that accompanies the physical sources $J^*$. 
 
At the second step, changing the sources~(\ref{ps}) we lose the Dirac factor 
together with the whole class 
of electrostatic phenomena including the Coulomb-like instantaneous bound 
state formed by the electrostatic interaction. 
 
Really, the FP perturbation theory in the relativistic gauge contains only 
photon propagators with the light-cone singularities forming 
the Wick-Cutkosky bound states with the spectrum  differing 
\footnote{The author thanks W. Kummer who pointed out that in 
Ref. \cite{kum} the difference between the Coulomb atom and 
the Wick-Cutkosky bound states in QED has been demonstrated.} 
from the observed one which corresponds to the instantaneous Coulomb 
interaction. 
Thus, the restoration of the explicit relativistic form of EUS($l^{(0)}$) 
by the transition to a relativistic gauge 
loses all electrostatic "monopole physics" with the Coulomb bound states. 
 
In fact, a moving relativistic atom in QED is described by the usual boost 
procedure for the wave function, which corresponds to a change of the 
time axis $l^{(0)}\Rightarrow l$, i.e., 
motion of the Coulomb potential~\cite{yaf} itself 
\begin{eqnarray} 
{ W}_{C} 
= \int d^4 x d^4 y \frac{1}{2} 
J_{l}(x) 
V_C(z^{\perp}) 
J_{l}(y) \delta(l \cdot z) \,\,\, , 
\end{eqnarray} 
where 
$ 
J_{l} = l_{\mu} J^{\mu} \,\, , \,\, 
z_{\mu}^{\perp} = 
z_{\mu} - l_{\mu}(z \cdot l) \,\, , \,\, 
z_\mu = (x - y)_\mu  \,\, . \,\, 
$ 
This transformation law and the relativistic covariance of EUS 
in QED has been predicted by von Neumann~\cite{hp} and 
proven by Zumino \cite{z} on the level of the algebra of 
generators of the Poincare group. 
Thus, on the level of EUS, the choice of a gauge is reduced to the choice of 
a time axis (i.e., the reference frame). A time axis is chosen 
to be parallel to the total momentum of a bound state, so that the 
coordinate of the potential always coincides with the space of the relative 
coordinates of the bound state wave function to satisfy the 
Yukawa-Markov principle~\cite{ym} and the Eddington concept of 
simultaneity ("yesterday's electron and today's proton do not make 
an atom")~\cite{ed}. 
 
In other words, each instantaneous bound state in QED has a proper EUS, 
and the relativistic generalization of the potential model is not only 
the change of the form of the potential, but sooner the change of a 
direction of its motion in four-dimensional space to lie along the 
total momentum of the 
bound state. The relativistic covariant unitary perturbation theory 
in terms of instantaneous bound states has been constructed in~\cite{yaf}. 
 
\subsection{Unconstrained QCD} 
 
\subsubsection{Topological degeneration and class of functions} 
 
We consider the non-Abelian $SU_c(3)$ theory with the action functional 
\be \label{u} 
W=\int d^4x 
\left\{\frac{1}{2}({G^a_{0i}}^2- {B_i^a}^2) 
+ \bar\psi[i\gamma^\mu(\partial _\mu+{\hat 
A_\mu}) 
-m]\psi\right\}~, 
\ee 
where $\psi$ and $\bar \psi$ are the fermionic quark fields. 
We use the conventional notation for the non-Abelian electric 
and magnetic fields 
\be \label{v} 
G_{0i}^a = \partial_0 A^a_i - D_i^{ab}(A)A_0^b~,~~~~~~ 
B_i^a=\epsilon_{ijk}\left(\partial_jA_k^a+ 
\frac g 2f^{abc}A^b_jA_k^c\right)~, 
\ee 
as well as the covariant derivative 
$D^{ab}_i(A):=\delta^{ab}\partial_i + gf^{acb} A_i^c$. 
 
The action (\ref{u}) is invariant with respect to gauge transformations 
$u(t,\vec x)$ 
\be \label{gauge1} 
{\hat A}_{i}^u := u(t,\vec x)\left({\hat A}_{i} + \partial_i 
\right)u^{-1}(t,\vec x),~~~~~~ 
\psi^u := u(t,\vec x)\psi~, 
\ee 
where ${\hat A_\mu}=g\frac{\lambda^a }{2i} A_\mu^a~$. 
 
It is well-known~\cite{fs} that the initial data of all fields are 
degenerated with respect to the stationary gauge transformations 
$u(t,\vec{x})=v(\vec{x})$. 
The group of these transformations represents 
the group of three-dimensional paths lying  on the three-dimensional 
space of the $SU_c(3)$-manifold with the homotopy group 
$\pi_{(3)}(SU_c(3))=Z$. 
The whole group of stationary gauge transformations is split into 
topological classes marked by the integer number $n$ (the degree of the map) 
which counts how many times a three-dimensional path turns around the 
$SU(3)$-manifold when the coordinate $x_i$ runs over the space where it is 
defined. 
The stationary transformations $v^n(\vec{x})$ with $n=0$ are called the small 
ones; and those with $n \neq 0$ 
\be \label{gnl} 
{\hat A}_i^{(n)}:=v^{(n)}(\vec{x}){\hat A}_i(\vec{x}) 
{v^{(n)}(\vec{x})}^{-1} 
+L^n_i~,~~~~L^n_i=v^{(n)}(\vec{x})\partial_i{v^{(n)}(\vec{x})}^{-1}~, 
\ee 
the large ones. 
 
The degree of a map 
\be \label{gn2} 
{\cal N}[n] 
=-\frac {1}{24\pi^2}\int d^3x ~\epsilon^{ijk}~ Tr[L^n_iL^n_jL^n_k]=n~. 
\ee 
as the condition for 
normalization  means that the large transformations 
are given in the  class of functions with the spatial asymptotics 
${\cal O} (1/r)$. 
Such a function $L^n_i$~(\ref{gnl}) is given by 
\be \label{class0} 
v^{(n)}(\vec{x})=\exp(n \hat \Phi_0(\vec{x})),~~~~~ 
\hat \Phi_0=- i \pi\frac{\lambda_A^a x^a}{r} f_0(r)~, 
\ee 
where the antisymmetric SU(3) matrices are denoted by 
$$\lambda_A^1:=\lambda^2,~\lambda_A^2:=\lambda^5,~\lambda_A^3:=\lambda^7~,$$ 
and $r=|\vec x|$. 
The function $f_0(r)$ satisfies the boundary conditions 
\be \label{bcf0} 
f_0(0)=0,~~~~~~~~~~~~~~ 
f_0(\infty)=1~, 
\ee 
so that the functions $L_i^n$ disappear at spatial infinity 
$\sim {\cal O} (1/r)$. 
The functions $L_i^n$ belong to monopole-type class of 
functions. It is evident that the transformed physical fields $A_i$ 
in~(\ref{gnl}) should be given in the same class of functions. 
 
The statement of the problem is {\bf to construct an 
equivalent unconstrained system (EUS) for the non-Abelian fields 
in this monopole-type class of functions}. 
 
\subsubsection{The Gauss Law Constraint} 
 
So, to construct EUS, one should solve the non-Abelian Gauss law constraint 
\cite{n,vp1} 
\be \label{gaussd} 
\frac{\delta W}{\delta A_0}=0~~~~~ \Rightarrow 
(D^2(A))^{ac} { A_0}^c = D_i^{ac}(A)\partial_0 A_i^c+ j_0^a~, 
\ee 
where $j_\mu^a=g\bar \psi \frac{\lambda^a}{2} \gamma_\mu\psi$ 
is the quark current. 
 
As dynamical gluon fields $A_i$ belong to a class of 
monopole-type functions, we restrict ouselves to 
ordinary perturbation theory around a static monopole $\Phi_i(\vec x) $ 
\be \label{bar} 
A^c_i(t,\vec{x}) = {\Phi}^c_i(\vec{x}) + \bar A^c_i(t,\vec{x})~, 
\ee 
where $\bar A_i$ is a weak perturbative part 
with the asymptotics at the spatial infinity 
\be \label{ass1} 
\hat {\Phi}_i(\vec{x})= O(\frac{1}{r}),~~~~~~~~ 
\bar A_i(t,\vec{x})|_{\rm asymptotics} = O(\frac{1}{r^{1+l}})~~~~(l > 1)~. 
\ee 
We use, as an example, the Wu-Yang monopole~\cite{wy,fn} 
\be \label{wy} 
\Phi_i^{WY}= 
- i \frac{\lambda_A^a}{2}\epsilon_{iak}\frac{x^k}{r^2} f_1^{WY},~~~~~~~ 
f^{WY}_{1}=1 
\ee 
which is a solution of classical equations everywhere besides 
the origin of coordinates. 
To remove a sigularity at the origin of coordinates and regularize 
its energy, the Wu-Yang monopole is changed by the 
Bogomol'nyi-Prasad-Sommerfield (BPS) monopole~\cite{bps} 
\be 
f^{WY}_{1}~\Rightarrow~ 
f^{BPS}_{1}= 
\left[1 - \frac{r}{\epsilon \sinh(r/\epsilon)}\right]~,~~~~~ 
\int\limits_{ }^{ }d^3x [B^a_i(\Phi_k)]^2 =\frac{4\pi}{g^2 \epsilon}~, 
\ee 
to take the limit of zero size $\epsilon~\rightarrow~ 0$ at the 
end of the calculation of spectra and matrix elements. 
This case gives us the possibility to obtain the phase of the 
topological transformations~(\ref{class0}) in the form of 
the zero mode of the covariant Laplace operator in the monopole field 
\be \label{lap} 
(D^2)^{ab}({\Phi_k^{BPS}})({\Phi}_0^{BPS})^b(\vec{x})=0~. 
\ee 
The nontrivial solution of this equation is well-known~\cite{bps}; 
it is given by equation~(\ref{class0}) where 
\be \label{lapb} 
f_0^{BPS}=\left[ \frac{1}{\tanh(r/\epsilon)}-\frac{\epsilon}{r}\right] 
\ee 
with the boundary conditions~(\ref{bcf0}). 
This zero mode signals about a topological excitation of 
the gluon system as a whole in the form of the solution ${\cal Z}^a$  of 
the homogeneous equation 
\be\label{zm} 
(D^2(A))^{ab}{\cal Z}^b=0~, 
\ee 
i.e., a zero mode of the Gauss law constraint~(\ref{gaussd})~\cite{vp1,p2} 
with the asymptotics at the space infinity 
\be \label{ass} 
\hat {\cal Z}(t,\vec{x})|_{\rm asymptotics}=\dot N(t)\hat \Phi_0(\vec{x})~, 
\ee 
where $\dot N(t)$ is the global variable of this topological excitation 
of the gluon system as a whole. 
From the mathematical point of view, this means that 
the general solution of the inhomogeneous equation~(\ref{gaussd}) 
for the time-like component $A_0$ 
is a sum of the homogeneous equation~(\ref{zm}) 
and a particular solution 
${\tilde A}_0^a$ of the inhomogeneous one~(\ref{gaussd}): 
$A_0^a = {\cal Z}^a + {\tilde A}^a_0$~. 
 
The zero mode of the Gauss constraint and the 
topological variable $N(t)$ allow us to remove the topological 
degeneration of all fields by the non-Abelian generalization of 
the Dirac dressed variables~(\ref{df}) 
\be \label{gt1} 
0=U_{\cal Z}(\hat {\cal Z}+\partial_0)U_{\cal Z}^{-1}~,~~~~ 
{\hat A}^*_i=U_{\cal Z}({\hat A}^I+\partial_i)U_{\cal Z}^{-1},~~~ 
\psi^*=U_{\cal Z}\psi^I~, 
\ee 
where the spatial asymptotics of $U_{\cal Z}$ is 
\be \label{UZ} 
U_{\cal Z}=T\exp[\int\limits^{t} dt' 
\hat {\cal Z}(t',\vec{x})]|_{\rm asymptotics} 
=\exp[N(t)\hat \Phi_0(\vec{x})]=U_{as}^{(N)}~, 
\ee 
and $A^I=\Phi+\bar A,\psi^I$ are the degeneration free variables 
with the Coulomb-type gauge in the monopole field 
\be \label{qcdg} 
D_k^{ac}(\Phi)\bar A^c_k=0~. 
\ee 
In this case, the topological degeneration of all color fields 
converts into the degeneration of only one global topological 
variable $N(t)$ with respect to a shift of this variable on integers: 
$(N~\Rightarrow~ N+n,~ n=\pm 1,\pm 2,...)$. 
One can check~\cite{bpr} that the Pontryagin index for 
the Dirac variables~(\ref{gt1}) with the 
assymptotics~(\ref{ass1}),~(\ref{ass}),~(\ref{UZ}) is determined 
by only the diference of the final and initial values of 
the topological variable 
\be \label{pont} 
\nu[A^*]=\frac{g^2}{16\pi^2}\int\limits_{t_{in} }^{t_{out} }dt 
\int\limits_{ }^{ }d^3x G^a_{\mu\nu} {}^*G^{a\mu\nu}=N(t_{out}) -N(t_{in}) 
\ee 
The considered case corresponds 
to the factorization of the phase factors of the topological 
degeneration, so that 
the physical consequences of the degeneration with respect to the 
topological nontrivial initial data are determined by the gauge of 
the sources of the Dirac dressed fields $A^*,\psi^*$ 
\be \label{tcsa} 
W^*_{l^{(0)}}(A^*) + \int\limits_{ }^{ }d^4x J^{c*}A^{c*}= 
W^*_{l^{(0)}}(A^I) +\int\limits_{ }^{ } d^4x J^{c*}A^{c*}(A^I)~. 
\ee 
The nonperturbative 
phase factors of the topological degeneration can lead to 
a complete destructive interference of color amplitudes~\cite{n,p2,pn} 
due to averaging over all parameters of the degenerations, in particular 
\be \label{conf} 
<1|\psi^*|0>=<1|\psi^I|0> \lim\limits_{L \to \infty} 
\frac{1}{2L} \sum\limits_{n=-L }^{n=+L } U_{as}^{(n)}(x)=0~. 
\ee 
This mechanism of confinement due to the interference 
of phase factors (revealed by the explicit 
resolving the Gauss law constraint~\cite{n}) disappears 
after the change of "physical" sources $A^*J^*~\Rightarrow~A J$ (that 
is called the transition to another gauge). Another gauge of the sources 
loses the phenomenon of confinement, like 
a relativistic gauge of sources 
in QED~(\ref{ps}) loses the phenomenon of electrostatics in QED. 
 
\subsubsection{Physical Consequences} 
 
The dynamics of physical variables including the topological one 
is determined by the constraint-shell action of an equivalent unconstrained 
system (EUS) as a sum 
of the zero mode part, and the monopole and perturbative ones 
\be \label{csa} 
W^*_{l^{(0)}}=W_{Gauss-shell}=W_{\cal Z}[N]+W_{mon}[\Phi_i]+W_{loc}[\bar A]~. 
\ee 
The action for an equivalent unconstrained system~(\ref{csa}) in the 
gauge~(\ref{qcdg}) with a monopole and a zero mode has been 
obtained  in the paper~\cite{bpr} following the paper~\cite{f}. 
This action contains the dynamics of the topological variable in the 
form of a free rotator 
\be \label{ktg} 
W_{\cal Z}=\int\limits_{ }^{ }dt\frac{{\dot N}^2 I}{2};~~~ 
I=\int\limits_{V}d^3x(D^{ac}_i(\Phi_k)\Phi^c_0)^2= 
\frac{4\pi}{g^2}(2\pi)^2\epsilon~, 
\ee 
where $\epsilon$ is a size of the BPS monopole considered 
as a parameter of the infrared regularization which disappears 
in the infinite volume limit. The dependence of $\epsilon$ on 
volume can be chosen so that the density of energy was finite. 
In this case, the U(1) anomaly can lead to additional mass of the 
isoscalar meson due to its mixing with the topological variable~\cite{bpr}. 
The vacuum wave function of the topological free motion 
in terms of the Ponryagin index~(\ref{pont}) takes 
the form of a plane wave $\exp(i P_N \nu[A^*])$. 
The well-known instanton wave 
function~\cite{hooft} appears for nonphysical values of the topological 
momentum $P_N=\pm i 8 \pi^2/g^2$ that points out the possible status 
of instantons as nonphysical solutions with the zero energy 
in Euclidean space-time 
\footnote{ The author is grateful to V.N. Gribov for the discussion of 
the problem of instantons in May of 1996, in Budapest.}. 
In any case, such the Euclidean solutions 
cannot describe the phenomena of the type of the complete destructive 
interference~(\ref{conf}). 
 
The Feynman path integral for the obtained unconstrained system in 
the class of functions of the topological transformations 
takes the form (see~\cite{bpr}) 
\bea \label{qcdf} 
Z_F[l^{(0)},J^{a*}]&=&\int\limits_{ }^{ } DN(t) 
\int\limits_{ }^{ }\prod\limits_{c=1 }^{c=8 } 
[d^2A^{c*} d^2 E^{c*}]\nonumber\\ 
&&\times\exp\left\{iW^*_{l^{(0)}} [A^*,E^*]+i\int\limits_{ }^{ }d^4x 
[J^{c*}_{\mu} \cdot A^{c*}_{\mu}]\right\}~, 
\eea 
where $J^{c*}$ are physical sources. 
 
The perturbation theory in the sector of local excitations 
is based  on the Green function $1/D^2(\Phi)$ as the inverse 
differential operator of the Gauss law which is 
the non-Abelian generalization of the Coulomb potential. 
As it has been shown in~\cite{bpr}, the non-Abelian Green function 
in the field of the Wu-Yang monopole 
is the sum of a Coulomb-type potential and a rising one. 
This means that the instantaneous quark-quark interaction 
leads to spontaneous chiral symmetry breaking~\cite{yaf,fb}, 
goldstone mesonic bound states~\cite{yaf}, glueballs~\cite{fb,ac}, and 
the Gribov modification of the asymptotic freedom formula~\cite{ac}. 
If we choose a time-axis $l^{(0)}$ 
along the total momentum of bound states~\cite{yaf} 
(this choice is compatible with the experience of QED in the description 
of instantaneous bound states), we get the bilocal generalization of 
the chiral Lagrangian-type mesonic interactions~\cite{yaf}. 
 
The change of variables $A^*$ of the type of~(\ref{df}) 
with the non-Abelian Dirac factor 
\be \label{dirqcd} 
U(A)=U_{\cal Z}\exp\left\{\frac{1}{D^2(\Phi)} D_j(\Phi)\hat A_j\right\} 
\ee 
and the change of the Dirac dressed sources $J^*$ can remove all 
monopole physics, including confinement and hadronization, 
like similar changes~(\ref{df}),~(\ref{ps}) in QED (to get a 
relativistic form of the Feynman path integral) 
remove all electrostatic phenomena in the relativistic gauges. 
 
The transition to another gauge faces the problem of zero 
of the FP determinant $det D^2(\Phi)$ (i.e. the Gribov ambiguity~\cite{g} of 
the gauge~(\ref{qcdg})). It is the zero mode of the second class 
constraint. The considered example~(\ref{qcdf}) shows that 
the Gribov ambiguity (being simultaneously the zero mode of the first 
class constraint) cannot be removed by the change of gauge 
as the zero mode is the inexorable consequence of internal dynamics, like the 
Coulomb field in QED. Both the zero mode, in QCD, and the Coulomb field, 
in QED, have nontrivial physical consequences discussed above, 
which can be lost by the standard gauge-fixing scheme.

\subsection{Instead of Conclusion} 
 
The variational methods of describing dynamic systems were created 
for the Newton mechanics. All their peculiarities (including 
time initial data, spatial boundary conditions $O(1/r)$, time evolution, 
spatial localization, the classification 
of constraints, and equations of motion in the Hamiltonian approach) reflect 
the choice of a definite frame of reference distinguished by the axis of time 
$l^{(0)}_{\mu}=(1,0,0,0)$. This frame determines also the 
EUS for the relativistic gauge theory. 
This equivalent system is compatible with the simplest variational 
methods of the Newton mechanics. 
The manifold of frames corresponds to 
the manifold of "equivalent unconstrained systems". 
{\bf The relativistic invariance means that a complete set of physical states 
for any equivalent system coincides with the one for another 
equivalent system}~\cite{bww}. 
 
This Schwinger's treatment of the relativistic invariance is confused 
with the naive  understanding of the relativistic invariance as 
{\bf independence on the time-axis of each physical state}. 
The latter is not obliged, and it can be possible only for the QFT 
description of local elementary excitations on their mass-shell. 
 
For a bound state, even in QED, the dependence on the time-axis exists. 
In this case, the time-axis is chosen to lie along the total momentum of 
the bound state in order to get the relativistic covariant dispersion law and 
invariant mass spectrum. 
This means that for the description of the processes 
with some bound states with different total momenta we are forced 
to use also some corresponding EUSs. 
Thus, a gauge constrained system can be completely covered by a set of EUSs. 
This is not the defect of the theory, but the method developed 
for the Newton mechanics. 
 
What should we choose to prove confinement and compute the 
hadronic spectrum in QCD: equivalent unconstrained systems
obtained by the honest and direct resolving of constraints, or 
relativistic gauges with the lattice calculations in the Euclidean 
space with the honest summing of all diagrams that lose from the 
very beginning all constraint effects? 
 
\subsection*{Acknowledgments} 
 
\medskip 
 
I thank  Profs. D. Blaschke, J. Polonyi, and G. R\"opke 
for critical discussions. This research and the participation at the 
workshop ``Quark matter in Astro- and Particle Physics'' was supported
in part by funds from the Deutsche Forschungsgemeinschaft and the Ministery 
for Education, Science and Culture in Mecklenburg- Western Pommerania.

%%\end{document} 
\newpage

%% file: SOURCE/polnoiNEU.tex
\pagestyle{plain}
\section*{\bf Confinement as crossover}
\addcontentsline{toc}{section}{\protect\numberline{}{Confinement as Crossover \\ \mbox{\it J. Polonyi}}}
\begin{center}
\vspace*{2mm}
{Janos Polonyi}\\[0.3cm]
{\small\it polonyi@fresnel.u-strasbg.fr\\
\vspace{0.2cm}
Laboratory of Theoretical Physics, Louis Pasteur University\\
3 rue de l'Universit\'e 67084 Strasbourg, Cedex, France\\
Department of Atomic Physics, L. E\"otv\"os University\\
P\'azm\'any P. S\'et\'any 1/A 1117 Budapest, Hungary}
\end{center}

\begin{abstract}
The order parameter of confinement together with
the haaron model of the QCD vacuum is reviewed and it is
pointed out that the confining forces are generated by the
non-renormalizable, invariant Haar-measure vertices of the
path integral. A hybrid model is proposed for the description of the
crossover leading to the confining vacuum. This scenario suggests
that the differences between the low and the high temperature
phases of QCD should be looked for in the quark channels instead of the
hadronic sector.
\end{abstract}

\newcounter{eqn9}[equation]
\setcounter{equation}{-1}
\stepcounter{equation}

\newcounter{bild9}[figure]
\setcounter{figure}{-1}
\stepcounter{figure}

\newcounter{tabelle9}[table]
\setcounter{table}{-1}
\stepcounter{table}

%\newcounter{kapitel9}[section]
%\setcounter{section}{-1}
%\stepcounter{section}

\newcounter{unterkapitel9}[subsection]
\setcounter{subsection}{-1}
\stepcounter{subsection}

\subsection{Introduction}
Having no systematic derivation of the confining
forces in pure Yang-Mills theories the studies of the long
range properties of the hadronic vacuum are usually based on
model computations \cite{bag}-\cite{haaron}.
In order to get closer
to the understanding of the problem it is obviously better to
use models which contain at least partially the original
gluonic degrees of freedom.

The bag model \cite{bag} is based on
weakly interacting quarks and gluons, the confinement
is realized by the difference of the energy density of
two different vacua: inside and outside of the bag. No colored
degrees of freedom are supposed to exist outside.
The boundary of the bag, considered as a dynamical degree
of freedom is obviously non-renormalizable. In the Abelian dual
superconductor model \cite{dual} the Schwinger relation which asserts
that the electric and the magnetic charges are inversely proportional
to each other indicates that the magnetic condensate is due to a
non-renormalizable coupling constant. The stochastic confining
model \cite{stoch} does not shed light on this question since it is based
on the cluster expansion leaving the origin of the correlations
an open question.
The invariant Haar-measure terms of the functional
integral which yield the string tension within the
haaron model \cite{haaron} are non-renormalizable, as well.

It is worthwhile noting that haaron model is the only one
where the non-renormalizable term which is responsible for the
string tension is already present in the original asymptotically free
Yang-Mills Lagrangian. In fact, the gauge invariant, non-perturbative
lattice regularization is based on the invariant Haar measure for
the gauge group valued link variables.
The logarithm of the Haar measure, treated as a
local interaction potential in the action is non-polynomial
and thereby non-renormalizable. Nevertheless the cut-off
can be removed by suppressing these vertices
sufficiently fast in the continuum limit \cite{riesz}.

We are confronted with an interesting possibility:
How can it happen that the leading infrared force of the Yang-Mills
theory comes from non-renormalizable vertices? The string tension,
being a dimensionful parameter, can be generated by a relevant
operator only. Since the non-renormalizable terms are irrelevant
these vertices influence the interaction at the cut-off scale
only and could have been left out from the theory according to the
universality.
The solution of this apparent paradox is rather simple \cite{cren}:
Any theory with internal scale has at least two scaling regimes,
an UV and an IR one, separated by a crossover at the internal scale.
The non-renormalizable vertices are indeed irrelevant in the
UV scaling regime but they might become relevant at the
IR side of the crossover, in the IR scaling regime, where
the IR forces are generated. 

Section B overviews briefly the symmetry and the order parameter
related to the confinement. An effective theory, the haaron model,
to describe confinement as a destructive interference is mentioned
in Section C. The lesson of the manner the confining force
is generated in this model is discussed in Section D.
The finite temperature aspects of confinement as a crossover phenomenon
are touched upon in Section E. Finally, Section F is for the
conclusion.

\subsection{Order parameter and destructive interfence}
The order parameter for confinement is given in terms
of the analytically continued massive quark propagator,
\be\label{orderp}
\Omega(x,t)=tr\la\psi(x,t)\bar\psi(x,t+i\beta)\ra,
\ee
where the trace is over the color and the spin indices and $\beta=1/T$. 
Note that for infinitely heavy quarks in a time independent environment
this expression reduces to the Polyakov line
\be
\omega(x)=trPe^{i\int_0^\beta d\tau A_0(x,\tau)}
\ee
up to a constant multiplicative factor, where $A_0(x,\tau)$ is the
temporal component of the Euclidean gauge field.

$\Omega$ displays the status of the symmetry with respect to the
center of the gauge group which is the group $Z_n$ for the gauge group
$SU(n)$. The order parameter can be easily introduced even for
Yang-Mills models in continuous Minkowski space-time \cite{mech}
and it remains a manifestly gauge invariant, well defined observable.

The dynamical quarks make the picture more complicated and we should
distinguish two competing confinement mechanisms, a hard and a soft
one. Both are driven by the increase of the effective coupling strength
as the color charges are separated.
The hard confining mechanism of the Yang-Mills models is responsible
for the flux tube formation and the linearly rising potential between
a static quark-anti quark pair. The soft mechanism is due to the
Dirac-sea polarization and, similarly to the supercritical vacuum
of QED \cite{supcrit}, shields the isolated quarks \cite{gribov}.
The soft mechanism cuts short the hard one and saturates
the linearly rising potential when the flux tube
between a static quark-anti quark pair is broken by the polarization
of the Dirac-sea.

This "deconfining " vacuum-polarization effect appears in the
dynamics of our order parameter, as well: The formal center
symmetry is broken by the fermion determinant in the
grand canonical ensemble. But it is easy to see that the Legendre
transformation of the baryon number between the canonical and
the grand canonical ensemble is ill defined in the thermodynamical
limit due to the confinement mechanism. In fact, the free energy
is infinitely large for states with non-vanishing triality\footnote{
The $n$-ality of a multi-quark state with $N$ quarks and $\bar N$
anti-quarks in an $SU(n)$ gauge model is defined as $t=N-\bar N({\rm mod} n)$.}
which turns the free energy into a non-differentiable function of
the baryon number density in the thermodynamical limit,
and the control of the baryon number by a chemical
potential into a highly non-trivial problem \cite{canonical}. It is the
canonical ensemble for the triality \cite{canonical}, \cite{canonicao}
which should rather be used
in this case and this ensemble is formally center symmetrical.
But this formal symmetry is broken spontaneously at low temperatures
\cite{canonical}. At high temperature the center symmetry is broken
dynamically by the gluon kinetic energy. There is
no reason to expect that the two unrelated symmetry breaking mechanisms
would generate the same expectation value for $\Omega$ thereby
\eq{orderp} remains to be an order parameter which
experiences a non-analytic dependence on the environmental variables
at the deconfinement transition \cite{canonical}.

The dynamical picture of confinement with $SU(n)$ as color gauge group
is the following \cite{mech}: The configuration space for global gauge
rotations, $SU(n)/Z_n$, is multiply connected. Its fundamental group,
the center $Z_n$ can be used to lump the time-dependent gluon field
configurations into $n$-tuples in such a manner that the
trajectories of an $n$-tuple correspond to the same initial or end
points in the multiply connected space $SU(n)/Z_n$ but differ on the
covering space, $SU(n)$. The center symmetry of the pure gluonic system
makes the action $S$ of the trajectories of an $n$-tuple degenerate
in the absence of quarks. Thus the contribution of $n$ trajectories
of an $n$-tuple to the transition amplitude is
\be
{\cal A}_E=ne^{-S},~~~{\cal A}_M=ne^{iS}
\ee
in Euclidean and Minkowski space-time.
When a spectator quark is propagating along with the gluons
then it picks up the $Z_n$ phase of the center transformation
and the contribution of an $n$-tuple is vanishing
due to the destructive interference between the homotopy classes,
\be\label{interf}
{\cal A}_E=\sum_{\ell=1}^ne^{i{2\pi\over n}\ell-S},~~~
{\cal A}_M=\sum_{\ell=1}^ne^{i{2\pi\over n}\ell+iS}.
\ee
Notice that the non-positive definite phase factor comes
from the projection operator which is supposed to install Gauss' law
and may lead to a destructive interference even for imaginary time.
The semiclassical expansion, saturated by Wu-Yang monopoles
in the Prasad-Sommerfeld limit \cite{perv} supports the confinement as a
destructive interference phenomenon, as well.

For an $SU(2)$ gauge model the center symmetry expresses the
invariance of a three vector under rotation by angle $2\pi$
and the destructive interference is due to the factor $-1$
the spinors of the fundamental representation collect during
a rotation by $2\pi$.

The high temperature deconfining transition is due to the too
high kinetic energy barrier for gluons
to follow the trajectories in the whole homotopy class \cite{mech}. In the
high temperature phase "there is no time" to realize all homotopy
class, the destructive interference is prohibited and quarks can
propagate. Note that the kinetic
energy driven dynamical symmetry breaking occurs at high energy
or in short time processes contrary to the potential energy
governed spontaneous symmetry breaking which is observed at low energy
or long time. When dynamical quarks are present then the
spontaneous breakdown of the center symmetry selects a
homotopy class which dominates the sum \eq{interf} and leads
to the screening of the isolated triality charge.

\subsection{Haaron model}
We start this brief summary of the haaron model with a remark about
the importance of keeping the exact gauge invariance in a computation
to extract the string tension in Yang-Mills theory. Suppose
that gauge invariance is implemented in an approximate manner and
the state with a static quark-anti quark pair is
$|q\bar q\ra_0+|q\bar q\ra_1$ where $|q\bar q\ra_0$ has the proper
transformation rule under gauge transformations and $|q\bar q\ra_1$
not. The non-covariant component $|q\bar q\ra_1$ appears
to contain uncontrollable charge distribution. The expression
\be
\left(\la q\bar q|_0+\la q\bar q|_1\right)H
\left(|q\bar q\ra_0+|q\bar q\ra_1\right)
=\la q\bar q|_0H|q\bar q\ra_0+\la q\bar q|_1H|q\bar q\ra_1
\ee
for the static potential shows that the charges in the component $|q\bar q\ra_1$
will break the flux tube for a sufficiently large separation
of the test charges and saturate the potential. The gauge non-covariant
components shield off the string tension when they are present with any
small amplitude.\footnote{This does not present serious
problem in QED where the gauge non-covariant contributions are
not gaining importance by non-perturbative effects in the absence
of the confining forces.}

There is another indication of the strong relation between gauge
invariance and the confining forces. The effective theory for the
Polyakov line obtained in the strong coupling expansion
shows that the minimum of the effective potential
is at a vanishing value of the order parameter at low temperature
due to the presence of the invariant Haar-measure for the gauge field
\cite{effth}. The replacement of the Haar-measure
with a non-compact measure may keep the gauge invariance intact at any
finite order of the loop expansion but certainly would break it
at a non-perturbative level.

Guided by these remarks we wish to base our effective theory
for the confining forces on the invariant Haar-measure
in the path integral. Consider $SU(2)$ Yang-Mills theory for
simplicity where the center transformation amounts to
$\Omega\to-\Omega$. The lattice regularized path integral
is given as
\be
\int D_H[aA_\mu(x)]e^{-S_{YM}[aA_\mu(x)]},
\ee
where $a$ stands for the lattice spacing. The invariant integration
measure for $A_0^j(x)=u(x)\omega^j(x)$, $j=1,2,3$, $(\omega^j(x))^2=1$
can be written as
\be\label{hmeas}
D_H[aA_0(x)]=D[\omega(x)]D[u(x)]e^{\sum_x\log\sin^2au(x)}
\ee
in terms of the flat integration measure $D[u(x)]$ for $u(x)$.
The center transformation, $u(x)\to u(x)+\pi/a$, performed
at a given equal-time hypersurface is a symmetry of the periodic
potential $a^{-4}\log\sin^2au$ appearing\footnote{The factor
$a^{-4}$ is to compensate the space-time integration volume $d^4x$
of the potential in the action. It makes the measure term
vanishing in dimensional regularization.} in \eq{hmeas}.
Let us integrate out the UV modes from the path integral and lower the
cut-off. This step makes the dimensional transmutation explicit and
induces an effective
gauge theory model with dimensional parameters. First we choose a
gauge in this theory where the temporal component $A_0(x)$ is diagonal,
$\omega^a(x)=\delta^{a,3}$ and then we set the non-diagonal
components of the gauge field to zero,
leaving behind a compact $U(1)$ gauge model. The Feynman gauge is
chosen for this model. As the final step, we set the spatial
component of the Abelian gauge field to zero. What we find at the end
is an effective theory for the diagonal, temporal component of the
original gauge field, $u(x)$. The corresponding effective
action will be approximated in the framework of the gradient
expansion by the form
\be
S_{eff}[u]=\int d^4x\left[\hf(\partial_\mu u(x))^2-V(u(x))\right].
\ee
The potential, the remnant of the invariant Haar-measure
is periodic $V(u+2\pi/\ell)=V(u)$. The periodicity reflects
the discrete center symmetry of the original theory and allows
the Fourier representation $V(u)=\sum_nv_m\cos m\ell u$.
The center symmetry requires
that the symmetry $u\to u+2\pi/\ell$ is respected by the vacuum.
The obvious consequence of this symmetry is the absence of any barrier
in the effective potential between the periodic minima. This leads
to the flattening of the effective potential for $u$ and the masslessness
of the field $u(x)$.

It is well known that the sine-Gordon model is equivalent with a
Coulomb gas. The simplest way to see this is to expand the generating
functional in the coupling constant $v$,
\bea\label{resum}
Z[\rho]&=&\int D[u]e^{-\hf u\cdot G^{-1}\cdot u+iu\cdot\rho+\hf v_m\int dx
(e^{im\ell u(x)}+e^{-im\ell u(x)})}\nonu
&=&\sum_n{(v_m/2)^n\over n!}\prod_{j=1}^n\int dx_j\sum_{\sigma_j=\pm1}
\int D[u]e^{-\hf u\cdot G^{-1}\cdot u+iu\cdot(\rho+\sigma)}\nonu
&=&e^{-\hf\rho\cdot G\cdot\rho}\sum_n{(v_m/2)^n\over n!}
\prod_{j=1}^n\int dx_j\sum_{\sigma_j=\pm1}
e^{-\hf\sigma\cdot G\cdot\sigma}e^{-\rho\cdot G\sigma},
\eea
where $G$ is the massless propagator and
$\sigma(x)=m\ell\sum_{j=1}^n\sigma_j\delta(x-x_j)$.
This is the grand canonical partition function of a four dimensional
gas of particles interacting with the inverse of the massless propagator,
the Coulomb potential. The first exponential in the last line represents
the perturbative self-interaction of the external source $\rho$,
the second stands for the self-interaction of the particles and
finally the third one describes the interaction between the source
and the particles. Let us ignore the inter-particle forces and the
partition function for non-interacting particles can be resummed,
\be
Z[\rho]\approx e^{\hf\rho\cdot G\cdot\rho+\hf v_m\int dx
\cos(im\ell\int dyG(x-y)\rho(y))}.
\ee
These steps, repeated for each Fourier modes give
\bea\label{freeh}
Z[\rho] &=& \int D[u]e^{-\int d^4x[\hf(\partial_\mu u(x))^2-V(u(x))
-i\rho(x)u(x)]}\nonu
&\approx& e^{-\hf\rho\cdot G\cdot\rho+\int dxV(i\int dyG(x-y)\rho(y))}.
\eea
The particles representing the vertices of the perturbation
expansion are called haarons since their contributions
come from the invariant measure of the original path integral.

It is worth mentioning three applications of the partial resummation
\eq{freeh}. The leading long range part of the static potential
between a quark-anti quark pair separated by $x$ is
\be
-\int d^3yV\left({i\over|y|}-{i\over|y-x|}\right)\approx-2V''(0)|x|,
\ee
giving the string tension
\be\label{strt}
\sigma=-2V''(0).
\ee
Notice that in the
center symmetry broken phase there is no protection against mass generation
and the massive propagator does not give linearly rising potential.
Thus the measure term which gives vanishing contribution in the
UV regime and is included to assure the full gauge invariance only
actually generates the leading long range force. Since it generates
a new dimensional parameter, the string tension, the measure
term must be relevant in the IR regime.

The second application of the resummation gives a confining version
of the NJL model. We start with the Lagrangian
\be
L=\hf(\partial_\mu u)^2-V(u)+i\bar\psi\partial_\mu\gamma^\mu\psi-igj~,
\ee
where $j=u\bar\psi\gamma^0\sigma_z\psi$ and perform the free haaron gas
resummation yielding
\bea
L_{eff}&=&i\bar\psi(x)\partial_\mu\gamma^\mu\psi(x)
-V\left(ig\int d^4yG(x-y)j(y)\right)\nonu
&\approx&i\bar\psi(x)\partial_\mu\gamma^\mu\psi(x)
-\hf g^2V''(0)\int d^4yj(x)G_2(x-y)j(y)\nonu
&=&i\bar\psi(x)\partial_\mu\gamma^\mu\psi(x)+g\sqrt{-V''(0)}j(x)\phi(x)
+\hf\phi(x)\Box^2\phi(x)~,
\eea
where
\be
G_2(x-y)=\int d^4zG(x-z)G(z-y)=\int{d^4p\over(2\pi)^4}
{1\over p^4}e^{-ip(x-y)}~,
\ee
and the auxiliary field $\phi(x)$ was introduced in order to
render the Lagrangian local. Notice that the $1/p^4$ propagator
of the auxiliary field, coupled to the quarks in the same manner as $u(x)$,
confines the color charges with a linearly rising potential and the
string tension is \eq{strt}.

The third application of the resummation is the computation
of the quenched quark propagator. The grand canonical partition function,
the last line of Eq. \eq{resum} when $\rho$ is replaced by the quark current
$j(x)$ shows that the quarks are propagating in the imaginary long range field 
\be
u_{y,n}(x)={in\ell\over4\pi^2(x-y)^2}
\ee
of the haarons. After performing the
Wick rotation into Minkowski space-time this external field becomes
real. The destructive interference between the homotopy classes
appears in this effective model as the destructive interference
between the scattering processes of a quark off the gas of haarons.
The long range haaron field makes the phase shift diverging and
the fast rotating phase of the scattered state
cancels the quark propagator when the averaging over the haaron
distributions is performed. In order to understand the propagation
of a meson qualitatively let us assume that the haaron field
at $x$ and $y$ is identical or completely uncorrelated when $|x-y|<\xi$
or $|x-y|>\xi$, respectively where $\xi\approx\Lambda^{-1}_{QCD}$
is the correlation length
of the haaron gas. As long as the quark and the anti-quark of the meson
propagates within the distance $\xi$ the phase shift suffered by them
is canceled and the haarons do not influence much the propagation.
When the color charges are separate from each other more than
$\xi$ then the statistically independent phase shift suppresses
the amplitude. The result is that the world lines can not separate
more then the distance $\xi$, the confinement radius.

\subsection{Crossover in the vacuum}
Let us consider the thought-experiment when the hadronic matter
is viewed by a microscope of adjustable space-resolution. When
details below the distance scale $\Lambda^{-1}_{QCD}$ are considered
we find partons, i.e. quarks and gluons. As the resolution becomes
worse and details on the scale well above $\Lambda^{-1}_{QCD}$ are
seen only then hadrons and glueballs are found. The interactions
between quarks and gluons on the one hand, and between hadrons on the
other hand, are very different. This difference can simply be
recorded by following the scale dependence generated by them.

There are at least two different scaling regimes in any non-scale
invariant theory, an UV and an IR one separated by a crossover
at the internal scale of the theory, $\Lambda^{-1}_{QCD}$ in our case,
c.f. Fig. \ref{ev}. The
UV scaling reflects asymptotically free forces between quarks and
gluons in the UV regime and short ranged Yukawa interactions
among the asymptotic states, hadrons, on the IR side. In pure
Yang-Mills theory glueballs are the asymptotic states in the IR
and color charges remain strongly bound by the linearly rising
potential. What was surprising in the haaron model picture is
that the measure vertices of the action which are non-renormalizable,
i.e. irrelevant in UV scaling regime can generate the leading
long range force. There must be a change in the behavior of the
measure vertices as we move towards the IR directions which explains
their increased importance in the confining forces. The most
natural scenario is that these operators, being irrelevant in the
UV scaling regime become relevant in the IR side of the crossover.

This scenario raises a more general question, the possibility that
non-renormalizable operators might play an important role in low
energy physics. It is easy to see that this surprising phenomenon
does not take place in models with mass gap $m\not=0$. These models
display a correlation length $\xi\approx1/m$ and the evolution of the
running coupling constant slows down at distance $x \gg \xi$.
In fact, the evolution of the coupling constants is driven by the
contribution of the modes around the running cut-off and the
fluctuations at the scale $x \gg \xi$ are suppressed
by $\exp(-x/\xi)$. The absence of runaway trajectories of the
renormalization group flow indicates that all non-Gaussian operators
are irrelevant in the IR scaling regime\footnote{An irrelevant coupling
constant may naturally be important if its fixed point value is
not small.}.

Theories without mass gap may develop new relevant operators
by the help of collinear or simple IR divergences which may drive
the run-away trajectories. The $\phi^4$ model in the mixed
phase possesses a non-renormalizable operator which is relevant
at low energies \cite{grg}. The condensation mechanism
in general can easily generate radically new scaling laws
\cite{tree}. When gauge symmetry is protecting against mass generation
then the four fermion interaction, the effective vertex
responsible for the emergence of the BCS phase, turns out to be relevant
at low energies \cite{shankar}. The interaction vertices between
hadronic states are irrelevant in QCD because the colorless sector
is massive. The lesson of the haaron model is that the integral measure
vertices become relevant in the IR scaling regime of the colored
channels.

There are two ways to deal with changing scaling laws. The
phenomenological approach is the matching where one introduces different
models for the UV and the IR scaling regimes and tries to match them
at the crossover. A more instructive procedure can be constructed by recalling
one of the rules of the renormalization group studies: All coupling
constants which are generated by the blocking and might turn out
to be important should be present in the action from the very
beginning. In fact, the renormalization group flow is a reliable
source of information about the interactions only if the
truncation of the space of Hamiltonians does not remove
important pieces. This principle, applied to QCD suggests the
introduction of colorless composite operators which control
the hadronic states,
\newpage
\bea
Z = &\int& D[\bar\psi]D[\psi]D_H[A_\mu]D[\bar\chi]D[\chi]D[\bar\Psi]D[\Psi]\times\nonu
&&e^{iS_{QCD}[\bar\psi,\psi,A_\mu]+iS_H[\bar\chi,\chi,\bar\Psi,\Psi]
+i\int dx(G_\chi\bar\chi\psi\psi\psi+G_\Psi\bar\Psi\bar\psi\psi+h.c.)},
\eea
the fields $\chi$ and $\Psi$ correspond to baryon and meson
states and $S_H[\bar\chi,\chi,\bar\Psi,\Psi]$ is the action for
a hadronic field theory.
Since we are interested in the low energy phenomena we
can fix the original cut-off at a sufficiently high but finite energy
scale $\Lambda_0$. The coupling constants $G_\chi$ and $G_\Psi$
govern the strength of interactions between the
hadronic and the colored states and their initial value is
$G_\chi(\Lambda_0)=G_\Psi(\Lambda_0)=0$, together with the hadronic coupling constants in
$S_H$. This scheme is not a double counting since it is cast
in the path integral formalism, it is a
possible parameterization of the effective action.

Such a hybrid model should hold the key to the understanding of
the confinement phenomenon because it offers a singularity-free
description of the crossover. As the cutoff is lowered the
non-renormalizable coupling strengths remain small and the
asymptotically free coupling $g$ grows. When the crossover is reached
then $g$ explodes in perturbative QCD and an IR Landau-pole arises because
the long range correlations of the ground state are supposed to be
generated by the asymptotically free vertices. But such a
hybrid model offers the following alternative. In the presence
of operators which are important in the IR scaling regime
there is a chance that $g$ stays finite because the desired
long range correlations can be established first in the colored
and after that in the neutral sector by the renormalization of the measure term
as in the haaron model and the hadronic coupling constants,
respectively.

\subsection{Crossover at high temperature}
\begin{figure}
\begin{minipage}{5cm}
\epsfxsize=5cm
\epsfysize=5cm
\centerline{\epsfbox{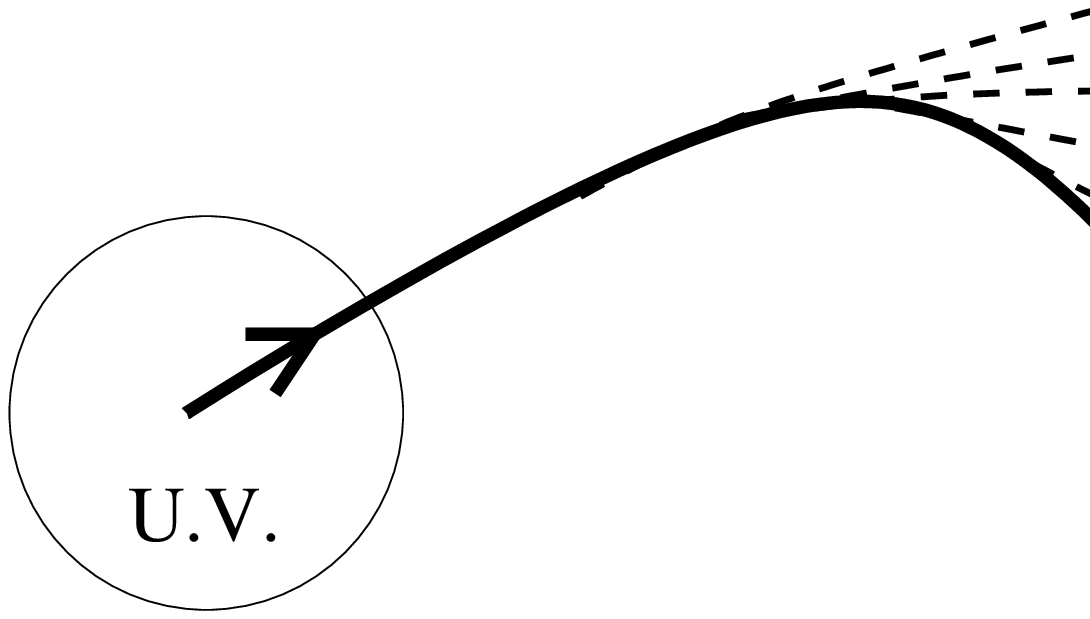}}
\end{minipage}
\caption{\label{ev}The renormalization group flow of QCD shown
for different temperatures. The solid line starting at the
UV and ending at the IR fixed point, encircled by the limit of
asymptotic scaling regimes, corresponds to $T=0$. The crossover
separating the UV and IR scaling regimes is at $Cr$,
$x\approx\Lambda^{-1}_{QCD}$. The flow at finite temperature
displays temperature dependence when $x>1/T$ and gives
two qualitatively different IR fixed point manifolds
for $T<T_{dec}$ and $T>T_{dec}$.}
\end{figure}

The real RHIC experiment is different, we are interested
in the long distance correlations and quasiparticle structure
at high temperature assuming that thermal equilibrium is
an acceptable approximation. How does the temperature
modify the scaling laws and the renormalized trajectory?
It is obvious that the renormalized trajectory is
in good approximation temperature independent at the
observational length scale $x \ll 1/T$ and the temperature
induced effects show up around $x\approx1/T$, as shown qualitatively in
Fig. \ref{ev}. For $T<T_{dec}$ the clusterization of the
color charges can be best understood as the impossibility
of screening the $1/3$ color charge of a quark by
multi-gluon states whose color charge is sum of integers,
$\sum\pm1\not=1/3$. The absence of screening mechanism
leads to confining forces. The deconfining phase transition
can be characterized in the Hamiltonian description
by the improper implementation of the Gauss' law
projection operator which does not exclude certain states with
infinitely many gluons. These states carry the color charge
of a quark or anti-quark \cite{mech}. The result is
the possibility of screening a quark color charge by a gluon cloud
whose wave functional is multi-valued,
the rearrangement of the infinite sum $\sum\pm1$ in such an
order that it converges to $1/3$. The effect of the temperature at the
deconfining transition is the removal of the linear
potential between triality charges by vacuum-polarization,
a mechanism similar to the polarization of the Dirac-sea
when dynamical quarks are present. A sort of soft confinement
mechanism is operating in the high temperature phase of the
pure glue system. The deconfined quarks are rendered colorless
and only their flavor quantum numbers reveal their quark content.

%\begin{figure}
%\begin{minipage}{5cm}
%\epsfxsize=5cm
%\epsfysize=5cm
%\centerline{\epsfbox{decev.eps}}
%\end{minipage}
%\caption{\label{ev}The renormalization group flow of QCD shown
%for different temperatures. The solid line starting at the
%UV and ending at the IR fixed point, encircled by the limit of
%asymptotic scaling regimes, corresponds to $T=0$. The crossover
%separating the UV and IR scaling regimes is at $Cr$,
%$x\approx\Lambda^{-1}_{QCD}$. The flow at finite temperature
%displays temperature dependence when $x>1/T$ and gives
%two qualitatively different IR fixed point manifolds
%for $T<T_{dec}$ and $T>T_{dec}$.}
%\end{figure}

The renormalization group flow of the pure glue system should have
two different manifolds of IR fixed points, for $T>T_{dec}$
and $T<T_{dec}$, as shown in Fig. \ref{ev}. The high temperature
fixed points should be qualitatively similar to those of full
QCD at low temperature, the role of quarks are being played
by gluon states with multi-valued wave functionals.

For full QCD the difference between the low and the high
temperature fixed point manifolds is more subtle since the
soft confinement mechanism is operating in both phases, by
means of quark or gluon states with multi-valued wave functionals.
I believe that the quasiparticles of the high temperature fixed point
are similar to those of the low temperature phase except that
a new quark "flavor" appears in the form of gluonic states
with multi-valued wave functionals. The difference between the
real and this fake quark can be detected by electro-weak
currents only, the color charges being equivalent.
The main features of this scenario are the screening of the
color charge of the deconfined quark and the presence of the
usual hadronic bound states. As of the former, a careful
numerical study should be carried out measuring the gluonic
color charge polarization around a deconfined quark.
There are two indirect numerical evidences supporting
the latter conjecture, the presence of the usual hadronic
bound states. The first is the observation that the
spacelike string tension is to a large extent temperature independent
and remains non-vanishing even at high temperature \cite{spacel},
indicating that the equal time long-range correlations of the
multi-quark states do not change at the deconfinement phase transition.
Another result, indicating the unimportance of the string tension from the
point of view of the structure of the hadronic states at $T=0$ is that
the hadronic structure functions are qualitatively reproduced
after cooling, a modification of the gluonic configurations
which removes the string tension.

\subsection{Conclusion}
It was argued in the framework of the haaron model that the leading
long range forces between a quark-anti quark pair are generated by
non-renormalizable vertices. This phenomenon motivates a look into
the confinement problem following the strategy of the renormalization
group and suggests that the confinement characterizes the IR scaling
regime.

A lesson learned in dealing with the renormalization group is that
all important operators should be present in the initial
Hamiltonian even with vanishing coupling strength in order to
understand the appearance of the dynamically generated, new
kind of forces. Such a point of view motivates a hybrid
model which contains both the quark-gluon and the hadronic fields.
Their difference is set by the initial condition
for the renormalization group flow only: a finite value for the
asymptotically free coupling constant and a vanishing strength for the
hadrons. The non-renormalizable measure term is
supposed to generate a crossover in this model where the
hadronic coupling constants turn on and induce the interactions
of nuclear physics.

Such a description of the long range structure, together with the
screening mechanism of the quark color charges by gluons
available at high temperature suggests that the main
difference between the high and the low temperature phases
is not in the hadronic but rather in the quark sector of QCD.
Possible examples are the following: The chromomagnetic monopoles,
being "hedgehog" configurations relate color and spin. The gluon
polarization cloud around a deconfined quark with even or odd number
of monopoles possesses integer or half-integer spin, respectively. In
this manner the violation of the superselection rule for the
charge induces a similar violation for the spin and the
deconfined quark state is actually the sum of components with
Bose and Fermi statistics \cite{mech}. Another triality-related
effect is the deviation of the temperature of the quark and gluonic
degrees of freedom at the deconfinement phase transition point,
$T_q=T_{gl}/3$ \cite{thqua}.

\newpage

%$\left. \right.$

%%\end{document}

%% file: SOURCE/rostock00NEU.tex
\section*{\bf Lattice simulations of QCD-like theories at non-zero density}
\addcontentsline{toc}{section}{\protect\numberline{}{Lattice Simulations of QCD-like Theories at Non-Zero Density
\\ \mbox{\it J. Skullerud}}}
\begin{center}
\vspace*{2mm}
{Jonivar Skullerud}\\[0.3cm]
%%\email{jonivar@mail.desy.de}
%%\homepage{http://www.bigfoot.com/~jonivar/}
{\small\it DESY Theory Group, Notkestra{\ss}e 85, D--22603 Hamburg, Germany\\
Current address: ITF, Universiteit van Amsterdam,
Valckenierstraat 65, 1018 XE Amsterdam, The Netherlands}
\end{center}
%%\maketitle

\begin{abstract}
One way of avoiding the complex action problem in lattice QCD at
non-zero density is to simulate QCD-like theories with a real action,
such as two-colour QCD.  The symmetries of two-colour QCD with quarks
in the fundamental and in the adjoint representation are described,
and the status of lattice simulations is reviewed, with particular
emphasis on comparison with predictions from chiral perturbation
theory.  Finally, we discuss how the lessons from two-colour QCD may
be carried over to physical QCD.
\end{abstract}

\newcounter{eqn6}[equation]
\setcounter{equation}{-1}
\stepcounter{equation}

\newcounter{bild6}[figure]
\setcounter{figure}{-1}
\stepcounter{figure}

\newcounter{tabelle6}[table]
\setcounter{table}{-1}
\stepcounter{table}

%\newcounter{kapitel6}[section]
%\setcounter{section}{-1}
%\stepcounter{section}

\newcounter{unterkapitel6}[subsection]
\setcounter{subsection}{-1}
\stepcounter{subsection}

\subsection{Introduction}
\label{sec:intro}

Recently, there has been a considerable interest in QCD at non-zero
chemical potential, after a number of model studies have indicated a
rich phase structure \cite{Alford:1997zt,Rapp:1998zu,Alford:1998mk}
(for a review, see \cite{Rajagopal:2000wf}).  Clearly, it would be
desirable if these predictions could be tested by first-principles,
non-perturbative studies, e.g.\ lattice QCD.  Unfortunately, lattice
simulations of QCD at non-zero baryon density using standard methods
are in practice impossible because the action (and the fermion
determinant) becomes complex once the chemical potential is
introduced, causing importance sampling to fail.  It is possible to
split the determinant into a modulus and a phase, simulating with the
modulus of the determinant as the measure and reweighting the
observables with the phase,
\begin{equation}
\bra O\ket = \frac{\bra\bra O\,\mbox{arg(det} M)\ket\ket}
                    {\bra\bra \mbox{arg(det} M)\ket\ket},
\label{eq:obs}
\end{equation}
where $\bra\bra\ldots\ket\ket$ denotes the expectation value with
respect to the positive real measure.  However, the denominator in
(\ref{eq:obs}) is effectively the ratio of the partition functions of
two different theories: the true theory and one with a positive real
measure.  This should scale as $\exp(-\Delta F)$, where $\Delta F$ is
the difference in free energy between the two theories.  Since the
free energy is an extensive quantity, the computational effort
required to obtain a reliable sample rises exponentially with the
volume.

A number of approaches have been tried to overcome this problem (see
\cite{Hands:2001jn} for a recent review).  In
the Hamiltonian formalism, the problem does not arise.  Analytical
results have been obtained in the strong coupling limit
\cite{Gregory:1999pm,Luo:2000xi}, but so far no method for numerical
simulations exists.

With an imaginary chemical potential \cite{Alford:1998sd}, the action
becomes real and positive, so simulations are straightforward.  The
problem is whether an analytical continuation to real $\mu$ is
possible.  It works at high temperature
\cite{Lombardo:1999cz,Hart:2000ef}, where also other approaches may be
successfully employed \cite{Hart:2000ef}, but does not seem to be
possible at zero or low temperatures.  Imaginary chemical potential
can also be used to formulate a quenched limit of QCD in the
background of a non-zero number of static quarks \cite{Engels:1999tz}.

Cluster algorithms \cite{Chandrasekharan:2000ew} may provide a way of
eliminating the sign problem by summing analytically over
configurations in a cluster in such a way that the contribution to the
partition function from each cluster is always positive definite.  So
far, these methods have been applied to a number of spin models;
however, an application to QCD has yet to be found.

Finally, the sign problem may be avoided by simulating theories which
resemble QCD, but have a real action even at non-zero chemical
potential.  One such theory is QCD at non-zero isospin density
\cite{Son:2000xc,Hands:2000hi}, which is of intrinsic interest because
it corresponds to part of the phase diagram for asymmetric nuclear
matter.  Another class of theories encompasses two-colour QCD with
fermions in the fundamental representation, as well as QCD with
adjoint fermions, for any number of colours.  The remainder of this
review will focus on what can be learnt from lattice simulations of
these theories.

\subsection{Theories with real action}
\label{sec:realaction}

The chemical potential $\mu$ is introduced on the lattice by
multiplying the forward timelike links by $e^\mu$ and the backward
timelike links by $e^{-\mu}$ \cite{Hasenfratz:1983ba}.  It can be
shown \cite{Kogut:2000ek,Hands:2000ei} that the determinant $\det{M}$
of the fermion matrix $M$ in two-colour QCD is real, even for non-zero
$\mu$, both in the continuum and on the lattice.  However, it is only
possible to demonstrate that it is positive \cite{Hands:2000ei} in the
cases of continuum or Wilson adjoint fermions and staggered
fundamental fermions.  Indeed, we will see in section
\ref{sec:adjoint} that in the case of staggered adjoint fermions there
are configurations with a negative determinant, leading to a sign
problem at large $\mu$.

\subsubsection{Symmetry breaking pattern}
\label{sec:symmbreaking}

In the chiral limit, the action for two-colour QCD with $N$ flavours
has a $\U{N}_L\otimes\U{N}_R$ symmetry, which for staggered fermions
is manifest as independent $\U{N}$ symmetries for the even and odd
sites.  At $\mu=0$ this enlarges to a $\U{2N}$ symmetry.  This can be
seen most easily by introducing new fields,
\begin{equation}
\bar{X}_e = (\chibar_e,-\chi^T_e\tau_2) \qquad X_o = 
\left(\begin{array}{c}
\chi_o\\-\tau_2\bar\chi_o^T
      \end{array} \right) \\
\end{equation}
for (staggered) fundamental quarks, and
\begin{equation}
\bar{X}_e = (\chibar_e,\chi^T_e) \qquad X_o = 
\left(\begin{array}{c}
\chi_o\\\bar\chi_o^T
      \end{array}\right)
\end{equation}
for adjoint quarks.  The action can then be written as
\begin{eqnarray}
S={1\over2}\sum_{x_e, \nu}\eta_\nu(x)
\biggl[\bar X_e(x)&\left(\matrix{e^{\mu\delta_{\nu,0}}&0\cr
             0 &e^{-\mu\delta_{\nu,0}}\cr}\right)&
                           U_\nu(x)X_o(x+\hat\nu) \, -  \\
\bar X_e(x)&\left(\matrix{e^{-\mu\delta_{\nu,0}}&0\cr
             0 &e^{\mu\delta_{\nu,0}}\cr}\right)&
U_\nu^\dagger(x-\hat\nu)X_o(x-\hat\nu)\biggr] \nonumber
\end{eqnarray}
where $x_e$ denotes the even sites. In the continuum, the equivalent
fields are
\begin{equation}
\text{fundamental:} \quad 
\Psi = \left(\begin{array}{c}\psi_L\\
         \sigma_2\tau_2\psi_R^{*}\end{array}\right)
\qquad
\text{adjoint:} \quad
\Psi = \left(\begin{array}{c}\psi_L\\ \sigma_2\psi_R^{*}\end{array}\right)
\end{equation}
which gives the continuum lagrangian
\begin{equation}
{\mathcal{L}} = i \Psi^\dagger \sigma_\nu(D_\nu-\mu B_\nu)\Psi \,
\qquad B_\nu = \left(\begin{array}{cc} 1 & 0 \\ 0 & -1
\end{array}\right) \delta_{\nu0} \, .
\end{equation}
The explicity chiral symmetry breaking term in the Wilson fermion
action means that there is no equivalent enlarged symmetry for Wilson
fermions; however, new fields may be introduced analogously to the
continuum case, and the enlarged symmetries will be broken by 
${\cal O}(a)$ terms in the action.

The chiral condensate can be written in terms of the new fields,
\begin{equation}
\bar\chi\chi=
\bar X_e\!\left(\begin{array}{cc}0&\One\\
                \pm\One&0\end{array}\right)\!{T\over2}\bar X_e^{tr}+
X_o^{tr}\!\left(\begin{array}{cc}0&\One\\
                \pm\One&0\end{array}\right)\!{T\over2}X_o
\label{eq:condf}
\end{equation}
for staggered fermions, and
\begin{equation}
\psibar\psi = \Psi^T\sigma_2\frac{T}{2}
\left(\begin{array}{cc}0&-\One\\\pm\One&0\end{array}\right)
\Psi + {\rm h.c.}
\label{eq:cond-cont}
\end{equation}
in the continuum and for Wilson fermions.  In both cases, the $+$ sign
is for fundamental fermions and the $-$ sign for adjoint, while $T$ is
$\tau_2$ for fundamental fermions and 1 for adjoint.  A nonzero chiral
condensate thereby breaks down the $\U{2N}$ symmetry to O($2N$) for
fundamental fermions and Sp($2N$) for adjoint fermions, giving rise to
$N(2N+1)$ and $N(2N-1)$ Goldstone modes respectively.  Of these, there
will be $N^2$ mesonic states, while the remaining $N(N\pm1)$ will be
diquarks.  In the continuum, and for Wilson fermions, the pattern will
be the opposite (modulo the 1 mode destroyed by the axial anomaly in
the continuum), but for 1 flavour of fundamental quarks, there is no
chiral symmetry in the first place.  From this we see that in the case
of $N=1$ adjoint staggered fermions, and only in this case, are there
no diquark Goldstone modes.

For $m\neq0$, all the pseudo-Goldstone modes remain degenerate, with
masses $m_\pi\propto\sqrt{m}$.  As the chemical potential $\mu$
increases, the ground state will begin to be populated with baryonic
matter.  The transition to a ground state containing matter occurs
when $\mu=\mu_o\simeq m_b/n_q$, where $m_b$ is the mass of the
lightest baryon, and this baryon contains $n_q$ quarks.  At this
point, the baryon number density $n$ becomes non-zero, where $n$ is
given by
\begin{eqnarray}
\label{eq:n-wilso}
n & =  \half\big\bra&\psibar(x)e^\mu(\gamma_0-1)U_0(x)\psi(x+\hat0) \\
 & & +
 \psibar(x+\hat0)e^{-\mu}(\gamma_0+1)U_0^\dagger(x)\psi(x)\big\ket
\nonumber
\end{eqnarray}
for Wilson fermions, and
\begin{equation}
n=\half\big\bra\bar\chi(x)\eta_0(x)[e^\mu U_0(x)\chi(x+\hat0)
 +  e^{-\mu}U_0^\dagger(x-\hat0)\chi(x-\hat0)]\big\ket. 
\label{eq:n}
\end{equation}
for staggered fermions.
Where there are diquark Goldstone modes, those states will be the
lightest baryons in the spectrum. This means that for most variants of
two-colour QCD we expect $\mu_o\simeq m_\pi/2$, in contrast to the
much larger value $m_N/3$ expected in real (three-colour) QCD.  The
exception is two-colour QCD with one flavour of adjoint staggered
quarks.

In the limit of small $m$ and $\mu$, the behaviour of $\pbp$, the
diquark condensate $\bra\psi\psi\ket$, and $n$
as functions of $m$ and $\mu$ can be calculated in chiral perturbation
theory.  If we define the rescaled variables
\begin{equation}
x=\frac{2\mu}{m_{\pi0}} \,, \quad 
y=\frac{\pbp}{\pbp_0} \,, \quad z=\frac{\bra\psi\psi\ket}{\pbp_0}\,, \quad 
\tilde n=\frac{m_{\pi0}n}{8m\pbp_0} \, ,
\label{eq:rescale}
\end{equation}
where the 0 subscript denotes values at $\mu=0$, the prediction from
$\chi$PT for the models with diquark Goldstone modes is
\cite{Kogut:2000ek}
\begin{equation}
y=\left\{\begin{array}{c}
        1\\
        1\over x^2
         \end{array} \right.
\quad
z=\left\{\begin{array}{c}
        0\\\sqrt{1-\frac{1}{x^4}}
         \end{array}\right.
\quad
\tilde n=\left\{\begin{array}{cl}
        0&;x\!<\!1\\
        {x\over4}\left(1-{1\over x^4}\right)&;x\!>\!1
               \end{array}\right.
\label{eq:chipt}
\end{equation}

\subsubsection{Diquark condensation}
\label{sec:diquark}

At large chemical potential, the relevant degrees of freedom will be
quarks with momenta near the Fermi surface.  The attractive
quark--quark interaction will give rise to instability with respect to
condensation of diquark pairs at opposite sides of the Fermi surface.
In physical QCD, the diquark condensate cannot be a colour singlet, so
the gauge symmetry is spontaneously broken, giving rise to the
phenomenon of colour superconductivity.

In two-colour QCD, on the other hand, there may be the possibility of
gauge singlet diquarks, which will be energetically favoured compared
to non-singlet states.  Indeed, in the previous section we saw that
most variants of two-colour QCD have diquark Goldstone modes, which
will be the preferred channel for diquark condensation.  In the $N=1$
staggered adjoint model, this is not the case, and we do not know {\em
a priori} in which channel the condensation will occur.  We must
proceed by constructing possible operators which obey the Pauli
principle and making additional assumptions about locality, Lorentz
structure and gauge invariance \cite{Hands:2000ei}.  At least one of
the possible condensates constructed this way gives rise to a colour
superconding ground state.

The standard way of computing the diquark condensate on the lattice is
to introduce a diquark source term into the action
\cite{Morrison:1998ud}.  For two-colour QCD with fundamental staggered
quarks the action then becomes
\begin{eqnarray}
S_F & = & \sum_{x,y}\bar\chi(x)M_{xy}\chi(y) 
+ \sum_x\frac{j}{2}\left[\chi^T(x)\tau_2\chi(x)
  +\bar\chi(x)\tau_2\bar\chi^T(x)\right] \\
 & = & (\bar\chi,\chi^T)
 \left(\begin{array}{cc} j\tau_2 & \half M \\
  \half M & j\tau_2 \end{array}\right)
 \left(\begin{array}{c} \!\bar\chi^T\\ \!\!\chi\end{array}\right)
\equiv X^T{\mathcal{A}}[j]X \, .
\label{eq:action-diquark}
\end{eqnarray}
In this case, the partition function becomes proportional to the
Pfaffian $\text{Pf}{\mathcal{A}}[j]$.  The diquark
condensate $\bra\chi^T\tau_2\chi\ket$ may be evaluated by taking
\begin{equation}
\bra\chi^T\tau_2\chi\ket = \lim_{j\to0}\frac{1}{2V}
 \left\bra\Tr\left\{{\mathcal{A}}^{-1}
\left(\begin{array}{cc}\tau_2&0\\0&\tau_2\end{array}\right)\right\}\right\ket
\label{eq:diquark-j}
\end{equation}
An alternative approach \cite{Aloisio:2000nr} is to rewrite the
Pfaffian as
\begin{equation}
\mathrm{Pf}{\mathcal{A}}[j] = \mathrm{Pf}(B+j) = \pm\sqrt{\det(B+j)}
\label{eq:pfaff-expand}
\end{equation}
where
\begin{equation}
B = \left(\begin{array}{cc} 0 & \half M\tau_2\\ -\half M\tau_2 &
0\end{array}\right) \, .
\end{equation}
This can be expanded as a polynomial in $j$ by diagonalising $B^2$,
obviating the need to simulate at non-zero diquark source.  Since this
gives the Pfaffian at any $j$, it can also be used to determine the
diquark condensate using the probability distribution function
\cite{Azcoiti:1995dq}.

\subsection{Simulations with fundamental quarks}
\label{sec:fund}

In the past year and a half, a number of groups have been performing
lattice simulations of two-colour QCD with fundamental staggered
fermions both at zero
\cite{Hands:1999zv,Aloisio:2000if,Aloisio:2000rb,Bittner:2000rf,Hands:2000hi}
and non-zero \cite{Alles:2000qi,Liu:2000in} temperature.  Also, one
group is performing simulations with Wilson fermions
\cite{Muroya:2000qp}.

\begin{figure}[p]
\begin{center}
%\begin{sideways}
\epsfig{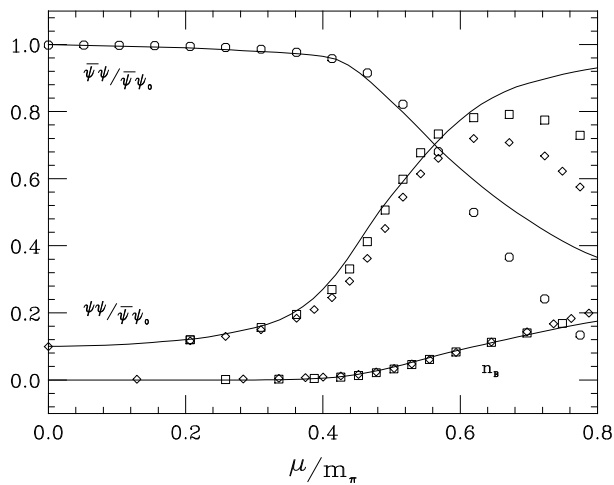}
%\end{sideways}
\end{center}
%\vspace{-0.8cm}
\caption{Baryon density, chiral condensate and diquark condensate
vs. chemical potential, from \protect\cite{Aloisio:2000if}.  The solid
lines are the predictions of (\protect\ref{eq:chipt}).}
\label{fig:aloisio-eos}
\end{figure}

\begin{figure}[p]
\begin{center}
%\begin{sideways}
\epsfig{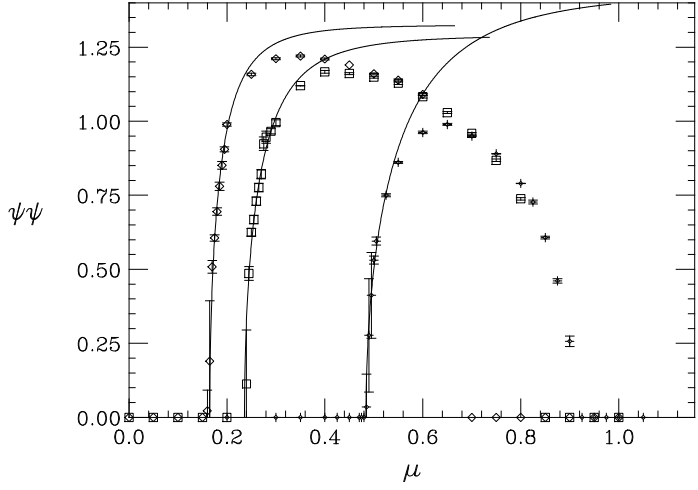}
%\end{sideways}
\end{center}
\caption{Diquark condensate for a $6^4$ lattice, $N_f=1$, for
$m=0.025$ (diamonds), 0.05 (squares) and 0.2 (stars) at strong
coupling, from \protect\cite{Aloisio:2000rb}, with the predictions of
(\protect\ref{eq:chipt}).} 
\label{fig:aloisio-diq}
\end{figure}

Aloisio {\em et al.}\ \cite{Aloisio:2000if,Aloisio:2000rb} have
performed simulations in the strong coupling limit for a number of
quark masses, flavours and lattice volumes.
Fig.~\ref{fig:aloisio-eos} shows results for the chiral condensate,
the diquark condensate and the baryon number density, at $m=0.2$ and a
non-zero source $j=0.02$, for two different lattice sizes.  Also shown
are the $\chi$PT predictions from (\ref{eq:chipt}).
The agreement between the prediction and the numerical results is
quite striking, considering that this is far from the continuum
limit.  This suggests a weak $\beta$ dependence. 
Fig.~\ref{fig:aloisio-diq} shows the diquark condensate at zero
diquark source, for $N_f=1$ and a range of quark masses.  Again, we
see a very good agreement with the prediction (\ref{eq:chipt}).
At larger $\mu$, we see that the value of $\bra\psi\psi\ket$ drops,
going to zero at high $\mu$.  This second transition is due to lattice
artefacts connected with the saturation of lattice sites with
fermions.  In the infinite volume limit it is expected to disappear.

Simulations at non-zero $\mu$ with a diquark source have been
performed by Kogut and Sinclair \cite{Hands:2000hi}.  Results for the
diquark condensate and chiral condensate are shown in
fig.~\ref{fig:sinclair-cond}.  Again, we see the chiral condensate
dropping and the diquark condensate rising for $\mu\gtrsim m_\pi/2$,
in agreement with $\chi$PT.  We also see the same large-$\mu$
saturation behaviour for the diquark condensate as in
\cite{Aloisio:2000rb}. 
Fig.~\ref{fig:sinclair-mass} shows results for pion and scalar diquark
masses.  The scalar diquark mass falls roughly as $m_\pi-2\mu$ as
$\mu$ approaches $m_\pi/2$.  The pion mass remains constant up to
$\mu\approx m_\pi/2$, after which it falls to zero.  This is again in
accordance with the expectation from $\chi$PT.

The spectrum of the Dirac operator has been studied in some detail by
the Vienna group \cite{Bittner:2000rf} for staggered fermions and by
the Hiroshima group \cite{Muroya:2000qp} for Wilson fermions.  A
preliminary study of topology at non-zero temperature has also been
performed \cite{Alles:2000qi}.

\subsection{Simulations with adjoint quarks}
\label{sec:adjoint}

As indicated in section \ref{sec:symmbreaking}, two-colour QCD with
one flavour of adjoint staggered fermions has features which makes it
in some senses more `QCD-like' than other variants of two-colour QCD.
In particular, it has no diquark Goldstone modes, so we expect an
onset transition at a value of the chemical potential different from
$m_\pi/2$ --- possibly at $\mu=m_N/3$ where $m_N$ denotes the mass of
the lightest three-quark baryon (the `nucleon').  It also has a sign
problem.

\begin{figure}[tb]
\begin{center}
\epsfig{width=0.70\colw,file=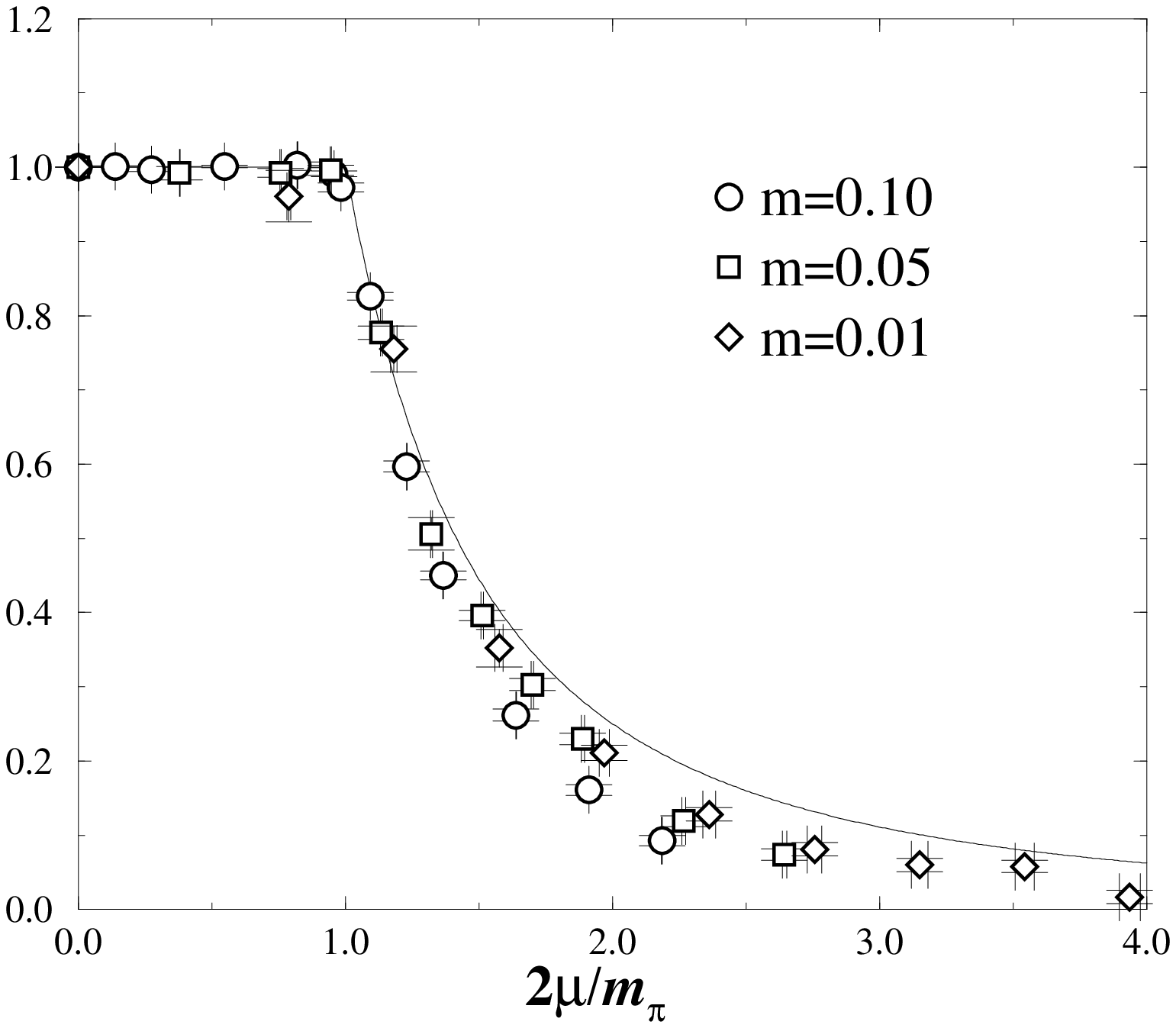}
\end{center}
%\vspace{-0.8cm}
\caption{Chiral condensate vs. chemical potential for one flavour of
adjoint staggered quarks,
using the rescaled variables of eq. (\ref{eq:rescale}); from
\protect\cite{Hands:2000yh}. 
\label{fig:unipbp}}
\end{figure}

\begin{figure}[tb]
\begin{center}
\epsfig{width=0.70\colw,file=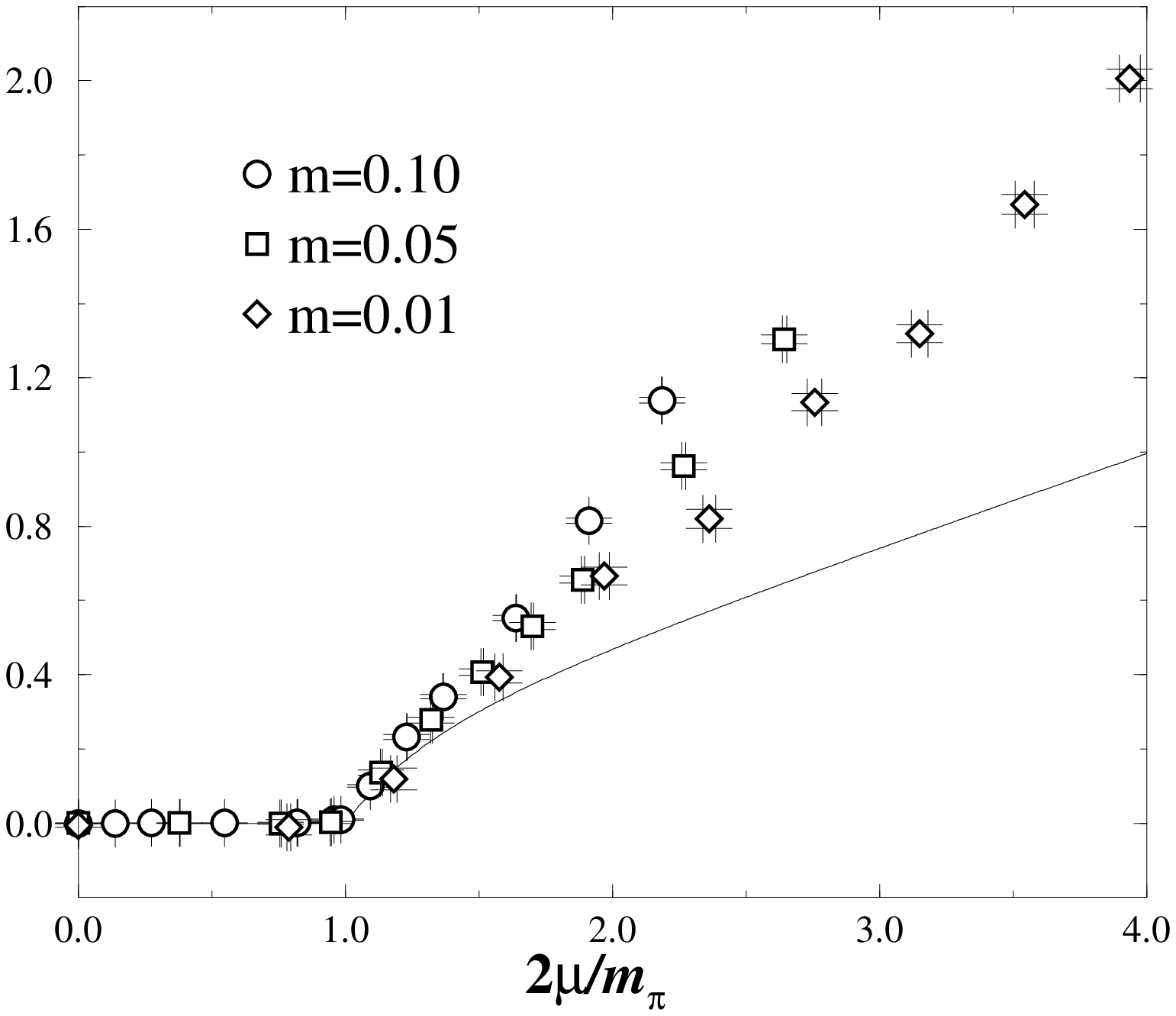}
\caption{Baryon density vs. chemical potential for one flavour of
adjoint staggered quarks, using the rescaled variables
of eq. (\ref{eq:rescale}); from \protect\cite{Hands:2000yh}.
\label{fig:uniden}}
\end{center}
\end{figure}

The theory has been simulated \cite{Hands:2000ei,Hands:2000yh} using
two different algorithms: Hybrid Monte Carlo, which is not able to
change the sign of the determinant, and therefore only simulates the
positive determinant sector of the theory, and the two-step
multibosonic algorithm \cite{Montvay:1996ea}, which is able to take the sign
properly into account.  Figs~\ref{fig:unipbp} and \ref{fig:uniden}
show the results from HMC simulations for $y$ and $\tilde n$ of
(\ref{eq:rescale}) respectively, for a range of values for the quark
mass $m$ and chemical potential $\mu$.  Up to $x\sim1.5$, the data
collapse onto a universal curve, which agrees well with the
predictions of (\ref{eq:chipt}).  Even at larger $x$, the data for the
chiral condensate lie close to the $\chi$PT prediction.  At
$x\gtrsim2$, however, the data for different $m$ diverge, indicating
that $\chi$PT may be breaking down.  Results for the plaquette
\cite{Hands:2000yh} show a drop in its value for $\mu\geq m_\pi/2$,
presumably due to Pauli blocking, while the pion mass appears to agree
with the $\chi$PT prediction $m_\pi=2\mu$ at $x>2$.

The agreement between the predictions of $\chi$PT and these results
paradoxically enough presents a problem, since this model is not
supposed to contain any diquark Goldstone modes, and thus the $\chi$PT
predictions of (\ref{eq:chipt}) are not valid in this case.  In
particular, there should not be any onset transition at $\mu=m_\pi/2$.
The suspicion must be that this contradiction is due to the fact that
HMC does not change the sign of the determinant, and that it therefore
simulates the wrong theory --- a theory with conjugate quarks.

The simulation points for the TSMB algorithm were selected to focus on
the effect of the sign, with one point at $\mu=0$, one just past the
HMC onset transition, and one deeper into the dense region.  With this
algorithm, a reweighting factor $r$ and the sign of $\det{M}$ must be
determined for each configuration \cite{Montvay:1999ty}.  The
expectation value of an observable $O$ is then determined by the ratio
\begin{equation}
\bra O\ket={{\bra O\times r\times sign\ket}\over
                  {\bra r\times sign\ket}}.
\label{eq:Osign}
\end{equation}
The results for $\pbp$ and $n$, together 
with the corresponding HMC results, are summarised in Table
\ref{tab:tsmb}. For TSMB at $\mu\not=0$ we also include 
observables determined separately in each sign sector, defined by
$\bra O\ket_\pm=\bra O\times r\ket_\pm/\bra r\ket_\pm$.
At $\mu=0.0$ the two algorithms agree, as they should.  Also, the
results in the positive determinant sector for TSMB at larger $\mu$
agree with the HMC results.
However, the results for the negative determinant sector are
significantly different.  This difference has the effect of bringing
the average both for $\pbp$ and for $n$ back to values consistent with
the $\mu=0$ values.  This is an indication that at $\mu=0.36$ and
quite possibly also at $\mu=0.4$, the system is still in the vacuum
phase, which means that the onset transition in this model occurs at a
larger $\mu$ than for other variants of two-colour QCD.  This is
consistent with the symmetry-based
arguments of section \ref{sec:symmbreaking} that this model has no
baryonic Goldstone modes.

\subsection{Conclusions}

Substantial progress has been made recently in lattice simulations of
two-colour QCD at non-zero density, with both fundamental and adjoint
quarks.  The simulations with fundamental quarks nicely reproduce the
predictions from chiral perturbation theory for the chiral condensate,
diquark condensate, and baryon density, except at very large chemical
potentials.  Here, saturation effects are observed, especially for the
diquark condensate.  The meson and diquark spectrum is also being
analysed, with preliminary results for the pion and scalar diquark
masses again in rough agreement with chiral perturbation theory.

Two-colour QCD with one flavour of adjoint staggered quark is not
expected to have any diquark Goldstone modes, unlike all other
variants of two-colour QCD.  It also has a sign problem.  Simulations
of this model restricted to the sector with a positive fermion
determinant reproduce the predictions of chiral perturbation theory
for theories with diquark Goldstone modes, including an early onset
transition at $\mu\approx m_\pi/2$ --- although the breakdown of
$\chi$PT may be observed at larger $\mu$.  When configurations with
negative determinants are included, we find a strong correlation
between the sign and the value of observables.  This effect appears to
lead to a cancellation of the early onset transition.  This
observation may hold the key to understanding the problem of the
premature onset which has bedevilled previous attempts to simulate
physical QCD at non-zero density.

The presence of diquark modes which are degenerate with the pion means
that the study two-colour QCD is of limited usefulness when it comes
to studying directly the onset transition, hadron spectrum and diquark
condensation in physical QCD at non-zero density.  However, useful
experience may be gained by comparing the results of lattice
simulations with those of other methods which are also applicable to
physical QCD.  Of particular interest would be the study of
gluodynamics, where SU(2) and SU(3) are expected to exhibit similar
behaviour, even at non-zero $\mu$.  Thus it might be possible to cast
light on the deconfinement transition at high $\mu$ and low $T$ by
studying two-colour QCD.

\subsection*{Acknowledgments}
This work is supported by  the TMR network ``Finite temperature phase 
transitions in particle physics'', EU contract ERBFMRX--CT97--0122.

\setlength{\unitlength}{0.9cm} %% for a5 size paper
\begin{figure}
\begin{picture}(11.5,6)
\put(-0.5,0){\vbox{\epsfig{width=0.5\colw,file=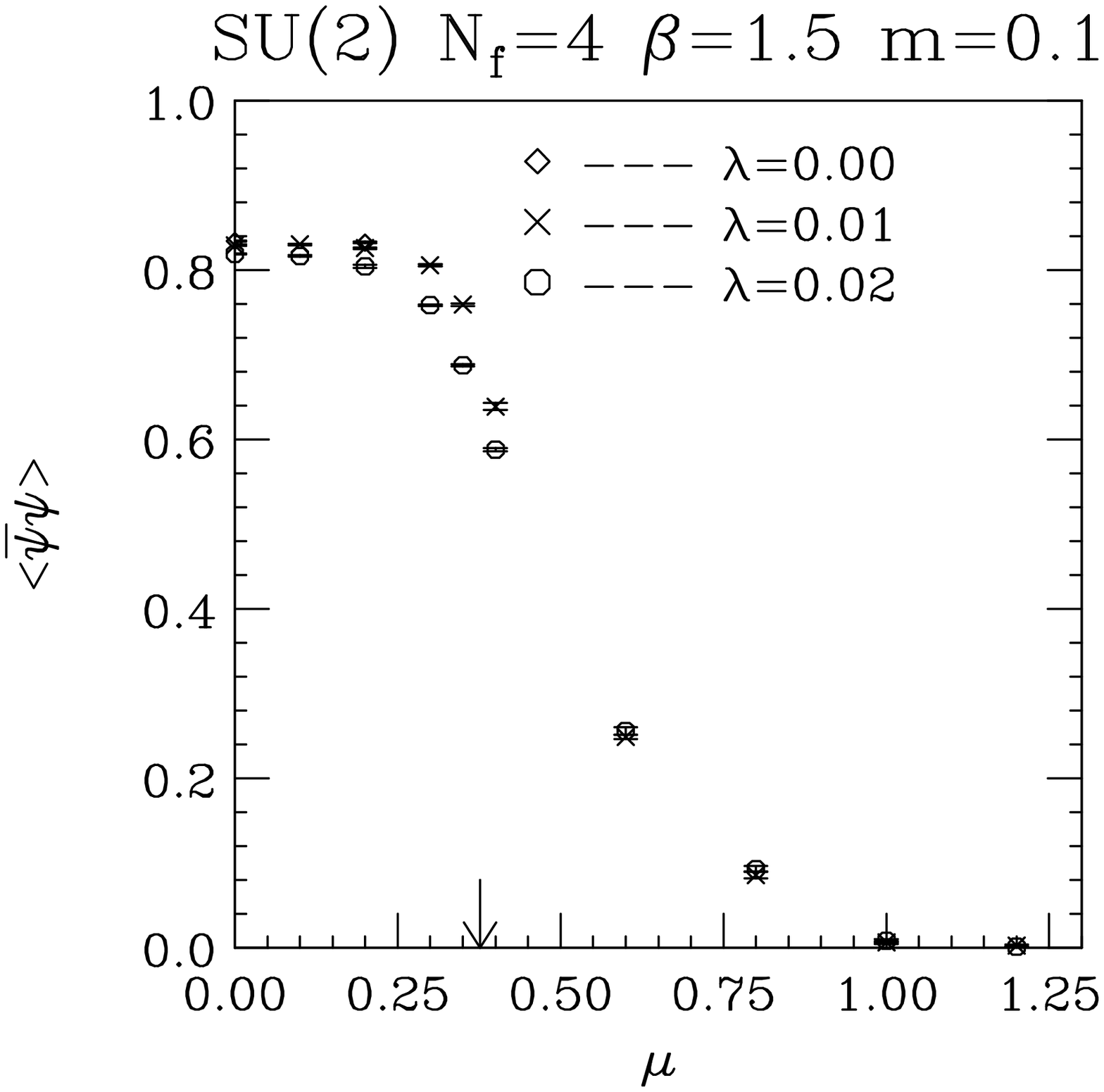}}}
\put(5.5,0){\vbox{\epsfig{width=0.5\colw,file=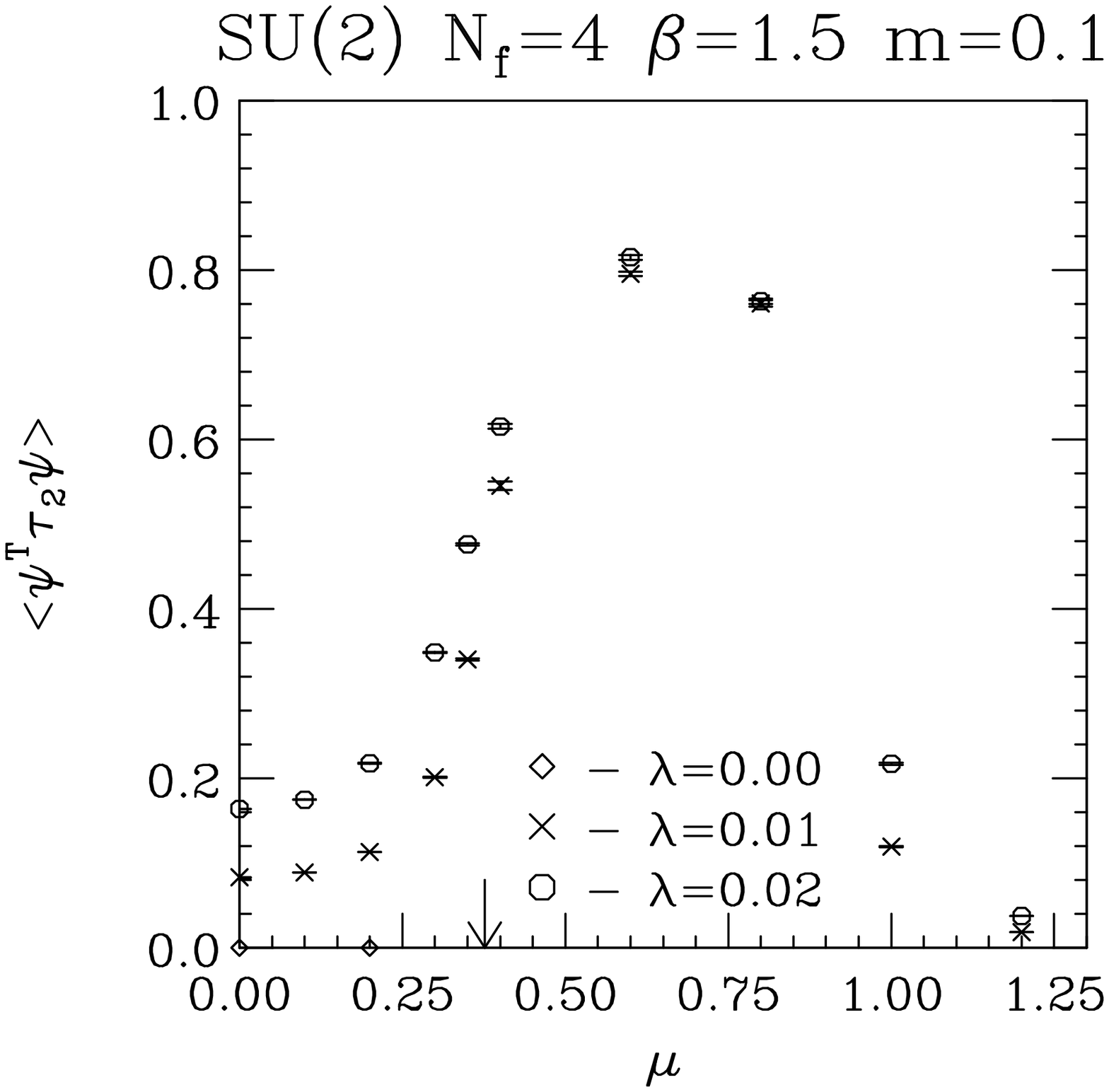}}}
\end{picture}
\caption{Chiral condensate (left) and diquark condensate (right) vs.\
chemical potential, for one flavour fundamental staggered quarks, on
an $8^4$ lattice; from
\protect\cite{Hands:2000hi}.  The arrow indicates $\mu\approx m_\pi/2$.}
\label{fig:sinclair-cond}
\end{figure}
\newpage
\begin{figure}
\begin{picture}(11.5,6)
\put(2.5,0){\vbox{\epsfig{width=0.5\colw,file=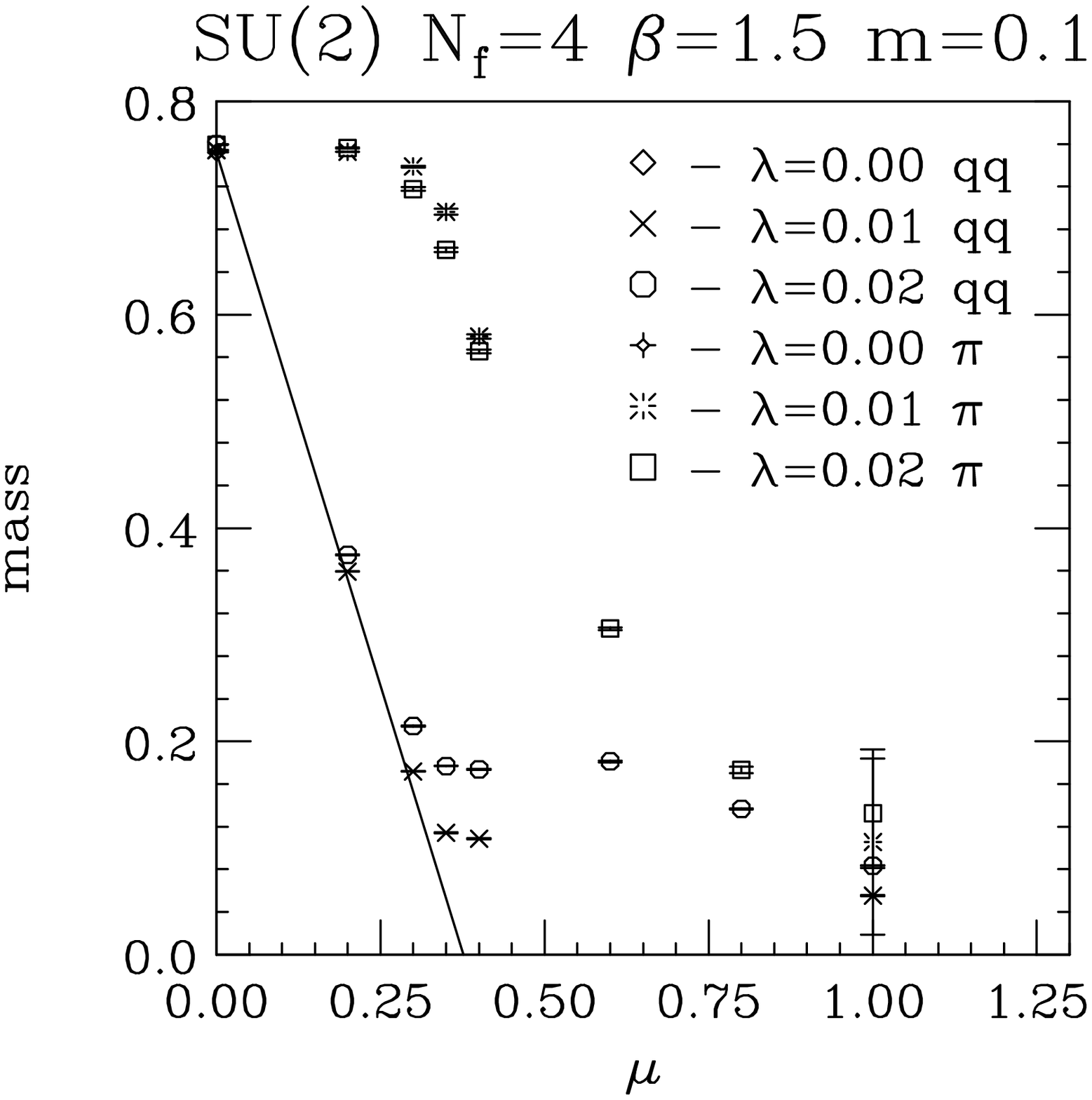}}}
\end{picture}
\caption{Pion and scalar diquark masses as functions of chemical
potential, for one flavour of fundamental staggered quarks, on an $8^4$
lattice; from \protect\cite{Hands:2000hi}.  The straight line is
$m=m_\pi-2\mu$.}
\label{fig:sinclair-mass}
\end{figure}


\begin{table*}[tbh]
\begin{tabular}{llllll}
 & \cc{$\mu$} & \multicolumn{3}{c}{TSMB} & \cc{HMC} \\ \cline{3-5}
 & & \cc{$\bra O\ket$} & \cc{$\bra O\ket_+$} & \cc{$\bra O\ket_-$} & 
\\ \hline
$\cbc$ & 0.0 & \m1.526(2) & & & \m1.526(1) \\
 & 0.36 & \m1.528(8) & 1.507(7) & 1.108(55) & \m1.497(5) \\
 & 0.4  & \m1.59(13) & 1.275(13) & 1.177(14) & \m1.261(12) \\ \cline{2-6}

$n$ & 0.0 & $-0.0004(8)$ & & & $-0.0002(3)$ \\
 & 0.36 & $-0.0053(66)$ & 0.0086(57) & 0.279(35) & \m0.0203(41) \\
 & 0.4  & $-0.041(84)$ & 0.138(11) & 0.193(12) & \m0.1462(100) \\
\end{tabular}
\vspace*{0.3cm}
\caption{A comparison of results between TSMB and HMC for one flavour
of adjoint staggered quarks.  Due to long autocorrelation times, the
errors in the HMC results may be underestimated.\label{tab:tsmb}}
\end{table*}

\newpage

%$\left. \right.$

%%\end{document}

%% file: SOURCE/HIC.tex
\addcontentsline{toc}{section}{$\left. \right.$\hspace{0.3cm} \bf Kinetic Theory and Heavy Ion Collisions}
$\left. \right.$
\vspace*{5cm}
\begin{center}
{\Large Contributions on}
\end{center}
\begin{center}
{\Large {\bf Kinetic Theory and Heavy Ion Collisions}}
\end{center}
\newpage

%% file: SOURCE/arne.tex
\newcounter{eqn26}[equation]
\setcounter{equation}{-1}
\stepcounter{equation}

\newcounter{bild26}[figure]
\setcounter{figure}{-1}
\stepcounter{figure}

\newcounter{tabelle26}[table]
\setcounter{table}{-1}
\stepcounter{table}

\newcounter{unterkapitel26}[subsection]
\setcounter{subsection}{-1}
\stepcounter{subsection}
 
\section*{\bf Kinetic Equations of QED Plasmas in Strong External Fields}
\addcontentsline{toc}{section}{\protect\numberline{}{Kinetic Equations of QED Plasmas in Strong External Fields \\ \mbox{\it A. H\"oll, V. Morozov, G. R\"opke}}}
\begin{center}
\vspace*{2mm}
A.~H\"oll$^{\dagger}$, V.~Morozov$^{*}$, 
G.~R\"opke$^{\dagger}$\\[0.3cm]  
{\small\it $^\dagger$Rostock University, 18051 Rostock, Germany\\ 
$^{*}$Moscow State Institute of Radioengineering, Electronics and Automation, Moscow, Russia} 
\end{center} 

\begin{abstract} 
A covariant density matrix approach to kinetic theory of QED plasmas
in strong external fields is discussed. 
Applying Fleming's hyperplane formalism, Schr\"odinger picture
correlation functions are expressed on spacelike hyperplanes in
Minkowski space and their corresponding equations of motion are
derived.
Additionally the nonequilibrium evolution of the statistical operator
must be treated, leading to the problem of including initial
correlations. 
A spinor decomposition of the Wigner matrix in spinor space is
performed and the classical limit of these equations discussed.
Finally it is shown how to write the kinetic equation in a coavariant
form. 
\end{abstract} 
%%%%%%%%%%%%%%%%%%%%%%%%%%%%%%%%%%%%%%%%%%%%%%%%%%%%%%%%%%%%%%%%%%%%%%%%%%%%%%

%\setcounter{equation}{0}
%1
\subsection{Introduction}
 
In recent years the theoretical study of high-temperature, dense
relativistic plasmas occurring in astrophysics as well as in
high-intense short-pulse lasers~\cite{Sprangle90,Gibbon1_96} is of
increasing interest. The nonequilibrium state of such systems
consisting of fermions and photons should be described by relativistic
covariant equations~\cite{Bezzerides,DeGroot80,OchsHeinz98}.

A particular point is the occurrence of strong fields which needs a
special treatment when introducing perturbation theory. We will give a 
systematic approach within a Schr\"odinger picture. As an example, the
kinetic equations in mean-field approximation are considered. 

We use the system of units with $c=\hbar=1$. The signature of the metric
tensor is $(+, -, -, -)$.
%%%%%%%%%%%%%%%%%%%%%%%%%%%%%%%%%%%%%%%%%%%%%%%%%%%%%%%%%%%%%%%%%%%%%%%%%%%%%%
%\setcounter{equation}{0}
%2
\subsection{Covariant Density Matrix Approach on Hyperplanes}
A quantum system can be characterized by the knowledge of the state
vector of the system $|\Psi\rangle$. 
It is determined by a complete set of commuting
observables. In order to formulate a relativistic invariant theory we
are lead to the question how measurements of observers in different
frames of reference with respect to the system must be related with each 
other. Here we follow the idea of
Bogoliubov, Fleming and others~\cite{Bogoliubov51,Fleming65,Fleming66}.
We define the operators and the state vector on spacelike hyperplanes
$\sigma^{}_{n,\tau}$ in Minkowski space. 
These planes are defined as
\begin{equation}
\label{EqPlane}
x\cdot n=\tau, \qquad
n^2=n^{\mu} n^{}_{\mu}=1,
\end{equation}
with $n_{\mu}^{}$ the unit normal vector of the plane and $\tau$ a
scalar parameter.
A special frame is the ``laboratory'' frame, were 
$n_{\mu}^{} = (1,0,0,0)$ is a system with the observers being at rest
with respect to the system.

Thus we write the state vector as a functional of the hyperplane 
$|\Psi[\sigma^{}_{n,\tau}]\rangle$ and consider the operators
$\hat{O}(n,\tau)$ as frame-dependent quantities.
In what follows we will consider a Schr\"odinger picture
formulation. That means that the operators are fixed to the initial
plane chosen, whereas the state vector will evolve in timelike
direction, i.e. depends on $\tau$. 
The space-time four-vector can always be decomposed as
\begin{equation}
\label{DecompX}
x^{\mu}_{}=n^{\mu}\tau + x^{\mu}_{\trans},
\qquad
\tau=n\cdot x,
\end{equation}
where
\begin{equation}
\label{XTrans}
x^{\mu}_{\trans}=\Delta^{\mu}_{\ \nu}\, x^{\nu},
\quad
\Delta^{\mu}_{\ \nu}=\delta^{\mu}_{\ \nu} -n^{\mu} n^{}_{\nu}.
\end{equation}
Now we
must answer the question how state vectors located on  different planes 
(i.e. state vectors measured by observers in different frames of
references with respect to the system) are related with each other.
In order to obtain a relativistic invariant formulation these state
vectors must be related by Lorentz transformations (boosts) 
$\sigma'=\Lambda\sigma$
\begin{equation}
\label{LorTrf}
\sigma\to
\sigma'=L\sigma: \quad x\to x'=\Lambda x.
\end{equation}
With $U(\Lambda)$ a unitary representation of the Lorentz
transformation $\Lambda$ we can write
\begin{equation}
\label{TrfStVec}
U(L)\left|\Psi[L\sigma]\right\rangle=
\left|\Psi[\sigma]\right\rangle.
\end{equation}

If we consider an infinitesimal timelike translation (i.e. in the
direction of $n_{\mu}^{}$) we can derive a 
relativistic Schr\"odinger equation
\begin{equation}
\label{SchrEq}
i\frac{\partial}{\partial\tau}
\left|\Psi(n,\tau)\right\rangle= \hat H(n) \left|\Psi(n,\tau)\right\rangle
\end{equation}
with the Hamiltonian on the hyperplane given by
\begin{equation}
\label{HamOnPlane}
\hat H(n)=\hat P^{}_{\mu} n^{\mu}.
\end{equation}
$P^{}_{\mu}$ in Eq.~(\ref{HamOnPlane}) is the energy-momentum four
vector.

Going now to a mixed quantum ensemble we can find the dynamical
equation for the frame dependent statistical operator $\varrho(n,\tau)$
\begin{equation}
\label{VonNEq}
\frac{\partial\varrho(n,\tau)}{\partial\tau}
- i\left[\varrho(n,\tau),\hat H^{\tau}(n)\right]=0.
\end{equation}
In the ``laboratory'' frame Eq.~(\ref{VonNEq}) reduces to the von
Neumann equation.

In order to proceed and to derive kinetic equtions for Schr\"odinger
picture correlation functions, we needto define the quantum
Hamiltonian, defined on the plane. A detailed discussion is presented
in~\cite{Hoell1_01}, which lead to 
\begin{equation}
\label{TotalHam:Q} 
\hat H^{\tau}(n)=
\hat H^{}_{D}(n)+\hat H^{}_{EM}(n)+ \hat H^{}_{\rm int}(n) +
\hat H^{\tau}_{\rm ext}(n),
\end{equation}
where $\hat H^{}_{D}(n)$
and $\hat H^{}_{EM}(n)$ are the Hamiltonians for free fermions and the
electro-magnetic (EM) field (due to 
polarization) respectively, $\hat H^{}_{\rm int}(n)$ is the
interaction term, and $\hat H^{\tau}_{\rm ext}(n)$
describes the external EM
field effects. In the Schr\"odinger picture the explicit
expressions for these terms are
\begin{eqnarray}
\label{DirHam:Q}
& & \hat{H}^{}_{D}(n)= \int_{\sigma^{}_{n}} d\sigma\,
\,\hat{\!\bar \psi} \left( -\frac{i}{2} \gamma^{\mu}_{\trans}(n)
\nablaleftright^{}_{\mu} +
\, m \right)\hat{\psi}
\\
\label{EMHam:Q}
& &
\hat{H}^{}_{EM}(n)=
\int_{\sigma^{}_{n}}  d\sigma\,
\left(
\frac{1}{4} \hat{F}_{\trans \mu\nu}^{} \hat{F}_{\trans}^{\mu\nu}
- \frac{1}{2} \hat{\Pi}_{\trans \mu}^{}\hat{\Pi}_{\trans}^{\mu}
\right)
\\
\label{IntHam:Q}
& &
\hat{H}^{}_{\rm int}(n)=
\int_{\sigma^{}_{n}}  d\sigma\,
\,\hat{\!j}^{}_{\!\trans \mu} \hat{A}_{\trans}^{\mu}
+ \frac{1}{2}
\int_{\sigma^{}_{n}} d \sigma \int_{\sigma^{}_{n}}
d \sigma_{}^{\prime} \,
\,\hat{\!j}_{\!\longi}^{}(x_{\trans}^{})
G(x_{\trans}^{} - x_{\trans}^{\prime} )
\,\hat{\!j}_{\!\longi}^{}(x_{\trans}^{\prime})
\\
\label{ExtHam:Q}
& &
\hat{H}^{\tau}_{\rm ext}(n)=\int_{\sigma^{}_{n}} d\sigma\,
\,\hat{\!j}^{}_{\!\mu}(x^{}_{\trans}) {A}^{\mu}_{\rm ext}(\tau, x^{}_{\trans}).
\end{eqnarray}
In these equations we apply the Coulomb gauge condition on the plane 
$\nabla^{}_{\mu} A^{\mu}_{\trans}=0$.
Further $\hat{\Pi}_{\trans}^{\mu}$ is the canonical momentum to 
$\hat{A}^{\mu}_{\trans}$ and $G$
the Green function of the Poisson equation 
\begin{equation}
\label{GrFunc:Eq}
\nabla^{}_{\mu}\nabla^{\mu} G(x^{}_{\trans})=\delta^{3}(x^{}_{\trans}).
\end{equation}
In the EM field Hamiltonian~(\ref{EMHam:Q}) the transverse field
strength tensor was defined
\begin{equation}
\label{Ftens:trans}
\hat{F}^{\mu\nu}_{\trans}=
\nabla^{\mu} \hat{A}^{\nu}_{\trans} -
\nabla^{\nu} \hat{A}^{\mu}_{\trans},
\qquad
\nabla^{\nu} = \Delta_{\mu}^{\nu} \partial_{}^{\mu}.
\end{equation}

We can now observe the averaged values of the dynamical operators 
$\hat{A}^{\nu}_{\trans}$ and $\hat{\Pi}^{\nu}_{\trans}$ are non-zero.
In particular in the case of strong external fields these values can
be large due to large polarization effects in the plasma.
As shown in~\cite{Hoell1_01} we apply a unitary transformation of
these operators and can show, that
the condensate values are eliminated
\begin{equation}
\label{ZeroCond}
\left\langle
\hat{A}^{\mu}_{\trans}(x^{}_{\trans})
\right\rangle^{\tau}_{\varrho^{}_{C}}=
\left\langle\hat{\Pi}^{\mu}_{\trans}(x^{}_{\trans})
\right\rangle^{\tau}_{\varrho^{}_{C}}=0.
\end{equation}
with 
\begin{equation}
\label{StatOp:Transf}
\varrho^{}_{C}(n,\tau)=
{\rm e}^{i\hat{C}(n,\tau)}\,\varrho(n,\tau)\,
{\rm e}^{-i\hat{C}(n,\tau)},
\end{equation}
and the operator $\hat C(n,\tau)$ given by
\begin{equation}
\label{OperC}
\hat{C}(n,\tau)= \int\limits_{\sigma^{}_{n}} d\sigma\,
\left\{
A^{\mu}_{\trans}(x) \hat{\Pi}^{}_{\trans\mu}(x^{}_{\trans})
- {\Pi}^{}_{\trans\mu}(x) \hat{A}^{\mu}_{\trans}(x^{}_{\trans})
\right\}.
\end{equation}
The redefinition of the statistical operator leads to a modification
of the relativistic von Neumann equation~(\ref{VonNEq})
\begin{equation}
\label{VonNEq:Trnsf}
\frac{\partial\varrho^{}_{C}(n,\tau)}{\partial\tau} -
i\left[\varrho^{}_{C}(n,\tau),\hat{\mathcal H}^{\tau}(n)\right]=0
\end{equation}
with the effective Hamiltonian $\hat{\mathcal H}^{\tau}(n)$. This
effective Hamiltonian has the advantage to separate out the mean-field
part from the quantum fluctuations. It is therefore suitable for
perturbative expansions with respect to quantum effects.
The effective Hamiltonian has now the form
\begin{equation}
\label{EffHam:2}
\hat{\mathcal H}^{\tau}(n)= \hat{\mathcal H}^{\tau}_{0}(n)
+ \hat{\mathcal H}^{\tau}_{\rm int}(n),
\end{equation}
with the mean-field part
\begin{equation}
\label{ZeroHam}
\hat{\mathcal H}^{\tau}_{0}(n)=
\hat{H}^{}_{D}(n) + \hat{H}^{}_{EM} +
\int\limits_{\sigma^{}_{n}} d\sigma\,
\,\hat{\!j}^{}_{\mu}(x^{}_{\trans})\,{\mathcal A}^{\mu}(x)
\end{equation}
describing  free photons and fermions interacting with the total
electro-magnetic field
\begin{equation}
\label{EM:totA}
{\mathcal A}^{\mu}(x) = A^{\mu}_{\rm ext}(x) +
A^{\mu}(x)
\end{equation}
and the interaction term containing only quantum fluctuations
\begin{eqnarray}
\label{EffInt}
& &
\hat{\mathcal H}^{\tau}_{\rm int}(n)=
\int\limits^{}_{\sigma^{}_{n}} d\sigma\,
\Delta\,\hat{\!j}^{\,\mu}_{\trans}(x^{}_{\trans};\tau)\,
\hat{A}^{\mu}_{\trans}(x^{}_{\trans})
\nonumber\\[6pt]
& &
\hspace*{60pt}
{}+{1\over2} \int\limits_{\sigma^{}_{n}} d\sigma
\int\limits_{\sigma^{}_{n}} d\sigma' \,
\Delta\,\hat{\!j}^{}_{\longi}(x^{}_{\trans};\tau)\,
G(x^{}_{\trans} - x^{\prime}_{\trans})\,
\Delta\,\hat{\!j}^{}_{\longi}(x^{\prime}_{\trans};\tau).
\end{eqnarray}
The operator $\Delta \hat{j}_{}^{\mu}(x^{}_{\trans};\tau)$
represents the quantum deviation from the averaged fermion current
operator 
\begin{equation}
\label{DeltaJ}
\Delta\,\hat{\!j}^{\,\mu}(x^{}_{\trans};\tau)=
\hat{\!j}^{\,\mu}(x^{}_{\trans})-
\langle \,\,\hat{\!j}^{\,\mu}(x^{}_{\trans}) \rangle^{\tau}
\end{equation}

%2
\subsection{The Kinetic Equation on the one-particle Level}
Defining the fermionic one-particle density operator 
\begin{equation}
\label{DensOper:Com}
\hat f^{}_{aa'}(x^{}_{\trans},x^{\prime}_{\trans})=
-{1\over2}\big[
\hat\psi^{}_{a}(x^{}_{\trans}),
\,\hat{\!\bar \psi}^{}_{a'}(x^{\prime}_{\trans})
\big],
\end{equation}
we derive a kinetic equation of the form (see~\cite{Hoell1_01})
\begin{eqnarray}
\label{KinEq:F}
\hspace*{-10pt}
\frac{\partial}{\partial\tau}\,
f^{}_{aa'}(x^{}_{\trans},x^{\prime}_{\trans};\tau)
&=&
-i\left\langle \big[
\hat f^{}_{aa'}(x^{}_{\trans},x^{\prime}_{\trans}),
\hat{\mathcal H}^{\tau}_{0}(n)\big]
\right\rangle^{\tau}_{\varrho^{}_{\rm rel}} 
\nonumber \\
& &
+ I^{(f)}_{aa'}(x^{}_{\trans},x^{\prime}_{\trans};\tau).
\end{eqnarray}
This equation has to be completed by the solution of
Eq.~(\ref{VonNEq:Trnsf}), written in an approximate form as
\begin{equation}
\label{StOp2}
\varrho^{}_{C}(n,\tau)=
\varrho^{}_{\rm rel}(n,\tau) +
\Delta{\varrho}(n,\tau).
\end{equation}
In Eq.~(\ref{StOp2}) the nonequilibrium statistical operator
$\varrho^{}_{C}(n,\tau)$ is decomposed into a relevant part,
describing the initial (equilibrium or nonequilibrium) distribution
and the nonrelevant part $\Delta{\varrho}(n,\tau)$.
Since we want to work consistently on the one-particle level, we
assume the one-particle density operator 
$\hat f^{}_{aa'}(x^{}_{\trans},x^{\prime}_{\trans})$ to be the
relevant operator. This method is known as
Zubarev's method~\cite{ZubMorRoep1}.
The collision term
$I^{(f)}_{aa'}(x^{}_{\trans},x^{\prime}_{\trans};\tau)$ in
Eq.~(\ref{KinEq:F}) was derived as
\begin{eqnarray}
\label{ColInt:F}
& &
I^{(f)}_{aa'}(x^{}_{\trans},x^{\prime}_{\trans};\tau)=
-i\left\langle \big[
\hat f^{}_{aa'}(x^{}_{\trans},x^{\prime}_{\trans}),
\hat{\mathcal H}^{\tau}_{\rm int}(n)\big]
\right\rangle^{\tau}_{\varrho^{}_{\rm rel}}
\nonumber\\[6pt]
& &
\hspace*{85pt}
{}-i\,{\rm Tr}
\left\{
 \big[
\hat f^{}_{aa'}(x^{}_{\trans},x^{\prime}_{\trans}).
\hat{\mathcal H}^{\tau}_{\rm int}(n)
\big]\,\Delta{\varrho}(n,\tau)\right\}.
\end{eqnarray}
%%%%%%%%%%%%%%%%%%%%%%%%%%%%%%%%%%%%%%%%%%%%%%%%%%%%%%%%%%%%%%%%%%%%%%%%%%%%%%

%\setcounter{equation}{0}
%4
\subsection{The Mean-Field Kinetic Equation} 
%%%%%%%%%%%%%%%%%%%%%%%%%%%%%%%%%%%%%%%%%%%%%%%%%%%%%%%%%%%%%%%%%%%%%%%%%%%%%%
In this section we consider the mean-field part of
Eq.~(\ref{KinEq:F}), i.e. neglecting the collision term
$I^{(f)}_{aa'}(x^{}_{\trans},x^{\prime}_{\trans};\tau)$.
This frequently used approximation is especially well justified for
strong external fields, were collisions become less important.

In view of discussing the classical limit of the kinetic equation it
is advantageous to express $f$ in a mixed representation in coordinate and
momentum space. This well known Wigner representation~\cite{Wigner1_32}
can be defined as 
\begin{eqnarray}
\label{Wigner:Def}
& &
%\hspace*{-5pt}
W^{}_{aa'}(x^{}_{\trans},p^{}_{\trans};\tau)=
\int d^4 y\, {\rm e}^{ip\cdot y}\,\delta(y\cdot n)\,
\nonumber\\[6pt]
& &
%\hspace*{30pt}
{}\times
\exp\left\{
ie\Lambda(x^{}_{\trans} +\mbox{$1\over2 $}y^{}_{\trans},
x^{}_{\trans}-\mbox{$1\over2 $}y^{}_{\trans};\tau)
\right\}
f^{}_{aa'}
\left(x^{}_{\trans}+\mbox{$1\over2 $}y^{}_{\trans},
x^{}_{\trans}-\mbox{$1\over2 $}y^{}_{\trans};\tau\right)
\end{eqnarray}
with the gauge function
\begin{eqnarray}
\label{GaugeFunc}
\Lambda(x^{}_{\trans},x^{\prime}_{\trans};\tau)
&=&
\int\limits_{x^{\prime}_{\trans}}^{x^{}_{\trans}}
{\mathcal A}^{}_{\trans\mu}(\tau,R^{}_{\trans})\,
dR^{\mu}_{\trans}
\nonumber\\[6pt]
{}&\equiv&
\int\limits_{0}^{1}
ds \left(x^{\mu}_{\trans} - x^{\prime\mu}_{\trans}\right)
{\mathcal A}^{}_{\trans\mu}\big(
\tau, x^{\prime}_{\trans} +s(x^{}_{\trans} -
x^{\prime}_{\trans}) \big).
\end{eqnarray}
The phase factor $\exp\{ie\Lambda\}$ appearing in
Eq.~(\ref{Wigner:Def}) guarantees the gauge-invariance of the Wigner
function (see for example~\cite{VasakGyulassyElze87}).

With the Wigner transformation we finally can derive a mean-field
kinetic equation, written in matrix notation 
($W \equiv \left[ W_{aa'}^{}\right]$) as
\begin{equation}
\label{Wigner:Eq:D}
{\sf D}^{}_{\tau}W=-{imc\over\hbar}\,\big[\gamma^{}_{\longi},W\big]
-{i\over2}\,{\sf D}^{}_{\trans\mu}
\left[S^{\mu},W\right] - {1\over\hbar}\,{\sf P}^{}_{\mu}
\left\{S^{\mu},W\right\},
\end{equation}
where we have introduced the operators
\begin{eqnarray}
& &
\label{DTau:D}
{\sf D}^{}_{\tau}= \frac{\partial}{\partial\tau}
-{e\over c}\int\limits^{1/2}_{-1/2} ds\,
n^{\mu} {\mathcal F}^{}_{\mu\nu}
\left(\tau,x^{}_{\trans} -is\hbar\nabla^{}_{p}\right)
\nabla^{\nu}_{p},
\\[6pt]
& &
\label{DTrans:D}
{\sf D}^{}_{\trans\mu}=
\nabla^{}_{\mu}
- {e\over c}\int\limits_{-1/2}^{1/2} ds\,
{\mathcal F}^{}_{\trans\mu\nu}
\left(\tau,x^{}_{\trans} -is\hbar\nabla^{}_{p}\right)
\nabla^{\nu}_{p},
\\[6pt]
& &
\label{PTrans:D}
{\sf P}^{}_{\mu}=
p^{}_{\trans\mu}
-{ie\hbar\over c} \int\limits_{-1/2}^{1/2} s\,ds\,
{\mathcal F}^{}_{\trans\mu\nu}
\left(\tau,x^{}_{\trans} -is\hbar\nabla^{}_{p}\right)
\nabla^{\nu}_{p}.
\end{eqnarray}
Note that we recovered the factors of $\hbar$ and $c$ in these
equations, since this will be important for the discussion of the
classical form of these equations.
Further we would like to mention that these equations coincide with
the kinetic equation derived in~\cite{BGR91} if we restrict ourselves to 
the ``laboratory'' frame.
%%%%%%%%%%%%%%%%%%%%%%%%%%%%%%%%%%%%%%%%%%%%%%%%%%%%%%%%%%%%%%%%%%%%%%%%%%%%%%

%\setcounter{equation}{0}
%5
\subsection{Spinor Decomposition and the Classical Limit}
In order to obtain a deeper physical inside into the rather
complicated matrix structure of Eq.~(\ref{Wigner:Eq:D}) we expand the
Wigner function in a complete basis in spinor space
\begin{equation}
\label{Wigner:Decomp}
W={1\over4}\left( I\mathcal{W} + \gamma^{}_{\mu}
\mathcal{W}^{\mu}_{} + \gamma_{\mbox{\footnotesize 5}}^{} \mathcal{W}^{}_{(P)} +
\gamma^{}_{\mbox{\footnotesize 5}}\gamma^{}_{\mu} \mathcal{W}^{\mu}_{(A)} + \sigma^{}_{\mu\nu}
\mathcal{W}^{\mu\nu}_{} \right) .
\end{equation}
In Eq.~(\ref{Wigner:Decomp}) we use the matrix notation with $\mathcal{W}$,
$\mathcal{W}^{\mu}_{}$, $\mathcal{W}^{}_{(P)}$, $\mathcal{W}^{\mu}_{(A)}$ and
$\mathcal{W}^{\mu\nu}_{}$, the scalar, vector, pseudo-scalar, axial-vector and
tensor coefficient function of the Wigner matrix $W$ respectively. 
Using this decomposition we derive a coupled set of equations for
these functions from the kinetic equation~(\ref{Wigner:Eq:D}).

In the clasical limit $\hbar\rightarrow 0$ 
the operators~(\ref{DTau:D}) $\ldots$ (\ref{PTrans:D}) reduce to the
local expressions
\begin{eqnarray}
\label{P:Class}
& &
{\sf P}^{}_{\mu}=p^{}_{\trans\mu},
\\[8pt]
\label{Dtau:Class}
& &
{\sf D}^{}_{\tau}=
\frac{\partial}{\partial\tau} -\frac{e}{c}\,
n^{\mu}{\mathcal F}^{}_{\mu\nu}\,\nabla^{\nu}_{p},
\\[8pt]
& &
\label{Dtr:Class}
{\sf D}^{}_{\trans\mu\nu}=
\nabla^{}_{\mu} -\frac{e}{c}\,{\mathcal F}^{}_{\trans\mu\nu}\nabla^{\nu}_{p}.
\end{eqnarray}
In that limit analytic solutions from the derived set of equations can 
only be found for the scalar part $\mathcal{W}$ and the 
longitudinal projection $\mathcal{W}_{\longi}^{}$. The corresponding
solution is given by
\begin{equation}
\label{WW:Class2}
\begin{array}{l}
\displaystyle
{\sf D}^{}_{\tau}
\left(\frac{\epsilon^{}_{p}}{mc^2}\, {\mathcal W} \right) +
\frac{v^{\mu}_{\trans}}{c}\,
{\sf D}^{}_{\trans\mu} {\mathcal W}^{}_{\longi}=0,
\\[10pt]
\displaystyle
{\sf D}^{}_{\tau} {\mathcal W}^{}_{\longi} +
\frac{v^{\mu}_{\trans}}{c}\,
{\sf D}^{}_{\trans\mu}\left(
\frac{\epsilon^{}_{p}}{mc^2}\, {\mathcal W}
\right)=0.
\end{array}
\end{equation}
In Eqs.~(\ref{WW:Class2}) we have introduced the transverse velocity
on the hyperplane
\begin{equation}
\label{Veloc:Tr}
v^{\mu}_{\trans}= \frac{c^2}{\epsilon^{}_{p}}\,p^{\mu}_{\trans}
\end{equation}
and the dispersion relation for fermions  on the hyperplane
\begin{equation}
\label{Disper:Plane}
\epsilon^{}_{p}=c\sqrt{m^2c^2 -p^{2}_{\trans}}.
\end{equation}
Finally we define the classical distribution functions
for the particles $w$ and for the anti-particles $\bar{w}$
\begin{eqnarray}
\label{DistrSig:Part}
& &
w(x^{}_{\trans},p^{}_{\trans};\tau)=
\frac{1}{2}\left\{
\frac{\epsilon^{}_{p}}{mc^2}\,
{\mathcal W}(x^{}_{\trans},p^{}_{\trans};\tau)+
{\mathcal W}^{}_{\longi}(x^{}_{\trans},p^{}_{\trans};\tau)
\right\},
\\[10pt]
\label{DistrSig:Anti}
& &
\bar{w}(x^{}_{\trans},p^{}_{\trans};\tau)=
\frac{1}{2}\left\{
\frac{\epsilon^{}_{p}}{mc^2}\,
{\mathcal W}(x^{}_{\trans},-p^{}_{\trans};\tau)-
{\mathcal W}^{}_{\longi}(x^{}_{\trans},-p^{}_{\trans};\tau)
\right\}
\end{eqnarray}
and we obtain the classical kinetic equations on the plane
\begin{eqnarray}
\label{PartKE:Plane}
& &
\left(\frac{\partial}{\partial\tau} + \frac{v^{\mu}_{\trans}}{c}\,
\nabla^{}_{\mu}\right)w
-\frac{e}{c}\left(
n^{\mu}{\mathcal F}^{}_{\mu\nu} +
\frac{v^{\mu}_{\trans}}{c}\,{\mathcal F}^{}_{\trans\mu\nu}\right)
\nabla^{\nu}_{p}w=0,
\\[10pt]
\label{AntiKE:Plane}
& &
\left(\frac{\partial}{\partial\tau} + \frac{v^{\mu}_{\trans}}{c}\,
\nabla^{}_{\mu}\right)\bar{w}
+\frac{e}{c}\left(
n^{\mu}{\mathcal F}^{}_{\mu\nu} +
\frac{v^{\mu}_{\trans}}{c}\,{\mathcal F}^{}_{\trans\mu\nu}\right)
\nabla^{\nu}_{p} \bar{w}=0.
\end{eqnarray}
We observe, that the equations for the particles and anti-particles
are decoupled.
The equations~(\ref{PartKE:Plane}) and (\ref{AntiKE:Plane}) are well
adopted for specific calculations, since we easily can change the
frame of reference by changing the hyperplane.
However, due to this plane dependence the equations are not manifest 
covariant.
Eliminating the normal vector $n_{\mu}^{}$ from the equations, we can
write these equations in a covariant fashion.
To do this we introduce the invariant distribution functions for the
particles and anti-particles
\begin{eqnarray}
\label{Inv:Part}
& &
f(x,p)= \delta\left(p^{}_{\longi} - \epsilon^{}_{p}/c\right)
w(x^{}_{\trans},p^{}_{\trans};\tau ),
\\[6pt]
\label{Inv:Anti}
& &
\bar{f}(x,p)= \delta\left(p^{}_{\longi} - \epsilon^{}_{p}/c\right)
\bar{w}(x^{}_{\trans},p^{}_{\trans};\tau ).
\end{eqnarray}
Further we introduce the unit-four vector
\begin{equation}
\label{FVeloc}
u^{\mu}=\frac{\epsilon^{}_{p}}{mc^2}\,n^{\mu}
+ \frac{p^{\mu}_{\trans}}{mc}=
\frac{1}{mc}\left[
p^{\mu} - n^{\mu}\left( p^{}_{\longi} -\epsilon^{}_{p}/c\right)
\right].
\end{equation}
In terms of the invariant distributions~(\ref{Inv:Part}),
(\ref{Inv:Anti}) and the unit four-vector~(\ref{FVeloc}) we find the
covariant kinetic equations 
\begin{eqnarray}
\label{InvKE:Part}
& &
u^{\mu}\left(
\partial^{}_{\mu} -
\frac{e}{c}\,{\mathcal F}^{}_{\mu\nu}(x)\partial^{\nu}_{p}
\right)f(x,p)=0,
\\[10pt]
\label{InvKE:Anti}
& &
u^{\mu}\left(
\partial^{}_{\mu} +
\frac{e}{c}\,{\mathcal F}^{}_{\mu\nu}(x)\partial^{\nu}_{p}
\right)\bar{f}(x,p)=0.
\end{eqnarray}

%4
\subsection{Conclusions} 

A general approach to relativistic covariant kinetic equations within
a density matrix approach was given. Strong classical fields are
treated by a canonical transformation so that perturbation expansions
with respect to the quantum fluctuations are applicable. Neglecting the
collision term, a relativistic mean-field equation was obtained. After
spinor decomposition, the classical limit has been shown to be in
agreement with known results.

The mean-field approximation will be improved systematically in future
works considering collision processes.  The method presented here gives
the possibility to include also higher order correlations in
non-equilibrium. Interesting applications are, e.g., bremsstrahlung
and pair creation in strong fields.

\newpage

%% file: SOURCE/trentoNEU.tex
\section*{\bf Gluon Pair Production from a Space-Time Dependent Classical Chromofield via Vacuum Polarization}
\addcontentsline{toc}{section}{\protect\numberline{}{Gluon Pair Production from a Space-Time Dependent Classical Chromofield via Vacuum Polarization \\ \mbox{\it G.C. Nayak, D.D. Dietrich, W. Greiner}}}
\begin{center}
\vspace*{2mm}
{Gouranga~C.~Nayak, Dennis~D.~Dietrich, and Walter~Greiner}\\[0.3cm]
{\small\it J. W. Goethe- Universit\"at, Institut f\"ur Theoretische Physik,\\
60054 Frankfurt am Main, Germany}
\end{center}

\newcounter{eqn19}[equation]
\setcounter{equation}{-1}
\stepcounter{equation}

\newcounter{bild19}[figure]
\setcounter{figure}{-1}
\stepcounter{figure}

\newcounter{tabelle19}[table]
\setcounter{table}{-1}
\stepcounter{table}

\newcounter{unterkapitel19}[subsection]
\setcounter{subsection}{-1}
\stepcounter{subsection} 

%%%%%%%%%%%%%%%%%%%%%%%%%%%%%%%%%%%%%%%%%%%%%%%%%%%%%%%%%%%%%%%%%%%%%%%%%%%%%%%
\begin{abstract}
We investigate the production of gluon pairs from a space-time dependent
classical chromofield via vacuum polarization within the framework of the
background field method of QCD. The investigation of the production of gluon 
pairs is important in the study of the evolution of the quark-gluon plasma in 
ultra-relativistic heavy-ion collisions at RHIC and LHC.
\end{abstract}
%\bigskip
%%%%%%%%%%%%%%%%%%%%%%%%%%%%%%%%%%%%%%%%%%%%%%%%%%%%%%%%%%%%%%%%%%%%%%%%%%%%%%%
\subsection{Introduction}

Ultra-relativistic heavy-ion collisions at RHIC and LHC will provide the best 
opportunity to study the color deconfined state of matter, namely the 
quark-gluon plasma (QGP).
The space-time evolution of the QGP can be split into different stages:
1. the pre-equilibrium,
2. the equilibrium, and 
3. the hadronization stage.
One of the central problems in these experiments is to study how partons are 
formed and how their distribution function evolves in space-time to form an 
equilibrated quark-gluon plasma (if at all).
High momentum partons ($p_T\ge1GeV$), {\it i.e.} minijets are calculated using 
pQCD.
Soft Parton Production is treated differently. There exist various model 
approaches:
1)
In the HIJING model soft parton production is treated via string formations.
2) 
In the color flux-tube model, an extension of the model named before, they 
are treated via the creation of a classical chromofield.
When partons and a classical chromofield are simultanously present, a 
relativistic non-abelian transport equation has got to be solved.

% % % % % % % % % % % % % % % % % % % % % % % % % % % % % % % % % % % % % % % %
\subsection{Field and Particle Dynamics}

The space-time evolution of the partons can be studied by solving 
relativistic non-abelian transport equations for quarks and gluons 
\cite{nayak,heinz}. 
As the chromofield exchanges color with quarks and gluons, color
is a dynamical quantity. 
The time evolution of the classical color charge 
follows Wong's equations \cite{wong}:

\be
{{{dQ^a} \over d{\tau}}=gf^{abc}u_\mu Q^bA^{c\mu}}.
\ee

There is also a non-abelian version of the Lorentz-force equation:

\be
{{dp^{\mu}} \over d{\tau}}=gQ^aF^{a\mu\nu}u_{\nu}
\ee

Taking the above equations into account, one finds the relativistic 
non-abelian transport equation \cite{nayak,heinz}:

\be
[ p_{\mu} \partial^\mu + g Q^a F_{\mu\nu}^a p^\nu 
\partial^\mu_p
+g f^{abc} Q^a A^b_\mu
p^{\mu} \partial_Q^c ]  f(x,p,Q)=C+S.
\label{trans}
\ee

Note that there are seperate transport equations for quarks, anti-quarks, and
gluons. 
The single-particle distribution function $f(x,p,Q)$ is defined in the 
14-dimensional extended phase space of co-ordinate, momentum, and 
SU(3)-color.
The first term on the LHS of Eq.(\ref{trans}) corresponds to convective flow,
the second to the non-abelian generalization of the effect of the Lorentz 
force, and
the third term describes the precession of the color charge in the presence of 
a classical field.
On the RHS there is the collision term $C$ and the source term for particle
production $S$.
For any system containing field and particles one has the following 
conservation equation:

\be
\partial_\mu T^{\mu\nu}_{mat}+\partial_\mu T^{\mu\nu}_{f}=0
\ee

which is coupled with the above transport equation for the description of the 
QGP. The evolution of the plasma depends crucially on the source term $S$ 
which contains all the information about how partons are produced from the 
classical chromofield.

%%%%%%%%%%%%%%%%%%%%%%%%%%%%%%%%%%%%%%%%%%%%%%%%%%%%%%%%%%%%%%%%%%%%%%%%%%%%%%%
\subsection{Parton Production from a Space-Time Dependent Chromofield}

The background field method of QCD is a suitable method to describe the 
production of partons from the QCD vacuum via vacuum polarization in the 
presence of a classical chromofield. 
Let us apply the background field method of QCD in order to describe the
production of $q\bar q$-pairs. 

The situation is 
similar to that of $e^+e^-$-pair production described by Schwinger 
\cite{sch51} in QED.
For a space-time dependent classical field $A_{cl}$ 
the amplitude for $e^+e^-$-pair production (see Fig.(\ref{nayak1})) from the 
vacuum is given by:

\be
M=<k_1,k_2|S^{(1)}|0>~ =~\\
-ie\bar{u}(k_1)\gamma_{\mu}A_{cl}^{\mu}(K=k_1+k_2)
v(k_2)
\ee 

% % % % % % % % % % % % % % % % % % % % % % % % % % % % % % % % % % % % % % % %
\begin{figure}[thb]
\begin{center}
\pfig{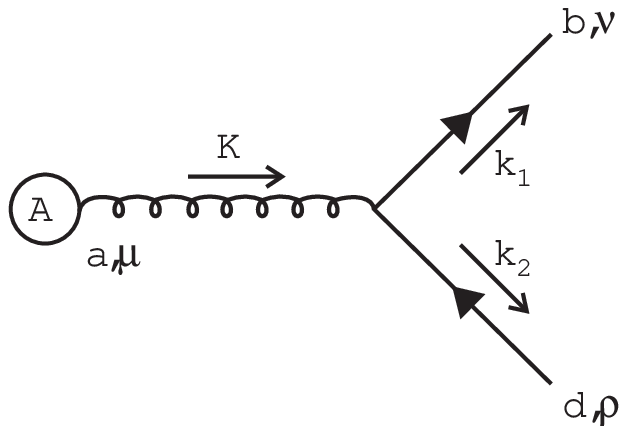}{4.75cm}
{Vacuum polarization diagram for the production of fermions in lowest order}
\end{center}
\end{figure}
% % % % % % % % % % % % % % % % % % % % % % % % % % % % % % % % % % % % % % % %

What, by the general formula:

\be
W^{(1)}=  
\int\frac{d^3k_1}{(2\pi)^3 2k_1^0}\frac{d^3k_2}{{(2\pi)}^3 2k_2^0}
\int d^4K(2\pi)^4\delta^{(4)}(K-k_1-k_2)\sum_{spin}|M|^2
\ee

leads to the pair-production probability \cite{izju}:

\bea
W_{e^+e^-}^{(1)}=
\frac{\alpha}{3}\int d^4K ~ 
{(1-\frac{4m_e^2}{K^2})}^{\frac{1}{2}}(1+\frac{2m_e^2}{K^2})\times\\
{[|K\cdot A_{cl}(K)|^2-K^2|A_{cl}(K)|^2]}
\eea

where the $d^4K$-integral is defined for $K^2>4m_e^2$.
Simillarly, carrying out the same procedure in the non-abelian theory
one finds for the amplitude for $q\bar{q}$-pair production: 

\be
M=ig\bar{u}^i(k_1)\gamma_{\mu}T^a_{ij}A_{cl}^{a\mu}(k_1+k_2)v^j(k_2)
\ee

and for the corresponding probability \cite{nay99}:

\bea
W_{q\bar{q}}^{(1)}=
\sum_{f} \frac{\alpha_s}{6} 
\int d^4K ~
{(1-\frac{4m_f^2}{K^2})}^{\frac{1}{2}}(1+\frac{2m_f^2}{K^2})\times\\
~[|K\cdot A_{cl}^a(K)|^2-K^2|A_{cl}^a(K)|^2].
\eea

% % % % % % % % % % % % % % % % % % % % % % % % % % % % % % % % % % % % % % % %
\subsubsection{Gluon-Pair Production from a Space-Time Dependent Chromofield}

As conventional QCD cannot describe the interaction between a classical 
chromofield and a quantum gluon, one has to fall back on the background field
method of QCD. 
This problem did not arise in QED, as there is no direct interaction between 
the classical field and the photon.
That method was first introduced by DeWitt \cite{dew} and further developped 
by 't Hooft \cite{tho}. In the background field method of QCD, one defines:

\be
A^{a \mu}=
A_{cl}^{a \mu}+A_q^{a \mu},
\ee

where $A_{cl}$ will not be quantized. So the generating functional excluding 
quarks is:

\be
Z[J,A_{cl}]=
\int [dA_q] \det M_G 
\exp(i [S[A_q+A_{cl}]-\frac{1}{2\alpha}G\cdot G + J \cdot A_q]),
\ee

with the classical action:

\be
S[A_q+A_{cl}]=-\frac{1}{4}~\int d^4x ~{(F^{a\mu \nu})}^2,
\ee

where the field-tensor is defined as:

\bea
F^{a\mu\nu}=
\partial^{\mu}{(A_q^{a\nu}+A_{cl}^{a\nu})}-
\partial^{\nu}{(A_q^{a\mu}+A_{cl}^{a\mu})}+\\
g~f^{abc}~{(A_q^{b\mu}+A_{cl}^{b\mu})}{(A_q^{c\nu}+A_{cl}^{c\nu})}.
\eea

The gauge fixing term $G^a$ is chosen  following 't Hooft:

\be
G^a=\partial^{\mu}A_q^{a\mu}+g~f^{abc}~A_{cl}^{b\mu}A_q^{c\mu}.
\ee

The matrix element of $M_G$ is given by:
 
\be
{(M_G(x,y))}^{ab}=\frac{\delta(G^a(x))}{\delta \theta^b (y)}
\ee

which is the functional derivative 
of the gauge fixing term with respect to the infinitesimal change of the 
gauge parameter $\theta$ of the gauge transformation

\be
\delta A_q^{a\mu}= -f^{abc}~\theta^b {(A_q^{c\mu}+A_{cl}^{c\mu})}
~+~\frac{1}{g}~\partial^{\mu}\theta^a.
\ee

Writing $\det M_G$ as functional integral over the ghost field, one obtains 
for the generating functional:

\bea
Z[J,A_{cl},\xi,\xi^{*}]=
\int[dA_q][d\chi][d\chi^{*}]
~\nonumber \\
\times
\exp(i[S[A_q+A_{cl}]+
       S_{ghost}-
       \frac{1}{2\alpha}G\cdot G+J \cdot A_q+\chi^{*} \xi + \xi^{*} \chi]),
\eea

where $\xi$ and $\xi^*$ are source functions for the ghosts and the 
ghost-part of the action is given by:

\bea
S_{ghost}=
-\int d^4x
\chi^{\dagger}_a[\Box^2 \delta^{ab}+
                 g{\overleftarrow{\partial}}_{\mu}f^{abc}(A_{cl}^{c\mu}+
                 A_q^{c\mu})
~\nonumber \\
                 -gf^{abc}A^{c\mu}\partial_{\mu}+
                 g^2f^{ace}f^{edb}A^c_{cl\mu}(A_{cl}^{d\mu}+A_q^{d\mu})]\chi_b.
\eea

Feynman rules involving a classical chromofield, gluons and ghosts
can now be constructed from the above generating functional \cite{abb}.
The vertices involving the coupling of two gluons to the classical field are 
given by:

\be
(V_{1A})^{abd}_{\mu\nu\rho}=
gf^{abd}[-2g_{\mu\rho}K_{\nu}+
g_{\nu\rho}(k_1-k_2)_{\mu}+
2g_{\mu\nu}K_{\rho}]
\ee

for coupling to the classical field once and by

\bea
(V_{2A})^{abcd}_{\mu\nu\lambda\rho}=-ig^2
[f^{abx}f^{xcd}
(g_{\mu\lambda}g_{\nu\rho}-g_{\mu\rho}g_{\nu\lambda}+g_{\mu\nu}g_{\lambda\rho})
~\nonumber \\
+f^{adx}f^{xbc}
(g_{\mu\nu}g_{\lambda\rho}-g_{\mu\lambda}g_{\nu\rho}-g_{\mu\rho}g_{\nu\lambda})
~\nonumber \\
+f^{acx}f^{xbd}
(g_{\mu\nu}g_{\lambda\rho}-g_{\mu\rho}g_{\nu\lambda})]
\eea

for coupling to the classical field twice.

% % % % % % % % % % % % % % % % % % % % % % % % % % % % % % % % % % % % % % % %
\begin{figure}[thb]
\begin{center}
\pfig{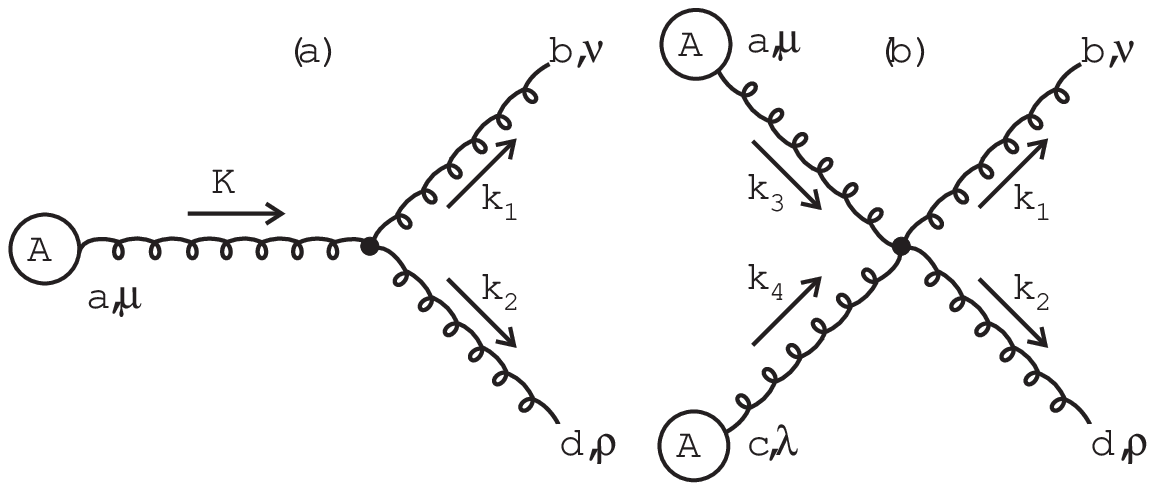}{10cm}
{Vacuum polarization diagrams for the production of gluons in lowest order.}
\end{center}
\end{figure}
% % % % % % % % % % % % % % % % % % % % % % % % % % % % % % % % % % % % % % % %

Note that the above vertices are different from the three and four gluon 
vertices used in conventional QCD and that from hereon the classical field is 
denoted only by $A$ not $A_{cl}$.
The gluon production amplitude $M=<k_1k_2|S^{(1)}|0>$ is defined in a way so 
that $S^{(1)}$ contains all interaction terms of the Lagrangian density 
involving two $Q$-fields, {\it i.e.}:

\newpage

\bea
S^{(1)}=S^{(1)}_G+S^{(1)}_{GF}
~\nonumber \\
=i\int d^4x(
-\frac{1}{2}F^a_{\mu\nu}[A]gf^{abc}Q^{b\mu}Q^{c\nu}
~\nonumber \\
-\frac{1}{2}(\partial_{\mu}Q^a_{\nu}-\partial_{\nu}Q^a_{\mu})
    gf^{abc}(A^{b\mu}Q^{c\nu}+Q^{b\mu}A^{c\nu})
~\nonumber \\
-\frac{1}{4}g^2f^{abc}f^{ab'c'}(A^b_{\mu}Q^c_{\nu}+Q^b_{\mu}A^c_{\nu})
                               (A^{b'\mu}Q^{c'\nu}+Q^{b'\mu}A^{c'\nu})
)
~\nonumber \\
+i \int d^4x(
-\partial_{\lambda}Q^{a\lambda}gf^{abc}A^b_{\kappa}Q^{c\kappa}
~\nonumber \\
-\frac{1}{2}g^2f^{abc}f^{ab'c'}A^b_{\lambda}Q^{c\lambda}
                               A^{b'}_{\kappa}Q^{c'\kappa}
).
\eea

The total amplitude $M=M_{1A}+M_{2A}$ consists of a contribution by the 
three-vertex (see Fig.(\ref{nayak2})(a)):

\bea
M_{1A}=
\frac{(2\pi)^2}{2}\int d^4K\delta^{(4)}(K-k_1-k_2)
~\nonumber \\
A^{a\mu}(K)\epsilon^{b\nu}(k_1)\epsilon^{d\rho}(k_2)(V_{1A})^{abd}_{\mu\nu\rho}
\eea

and one by the four-vertex (see Fig.(\ref{nayak2})(b)):

\bea
M_{2A}=
\frac{1}{4}\int d^4k_3 d^4k_4\delta^{(4)}(k_1+k_2-k_3-k_4)
~\nonumber \\
A^{a\mu}(k_3)A^{c\lambda}(k_4)\epsilon^{b\nu}(k_1)\epsilon^{d\rho}(k_2)
(V_{2A})^{abcd}_{\mu\nu\lambda\rho}.
\eea

The above amplitudes include all the weight factors needed in order 
to retrieve the corresponding Lagrangian density. Now, we again calculate 
the pair production probability:

\be
W=\sum_{spin}
\int\frac{d^3k_1}{(2\pi)^32k_1^0}\frac{d^3k_2}{(2\pi)^32k_2^0}|M|^2.
\ee

To obtain the correct physical gluon polarizations in the final state we use:

\be
\sum_{spin}\epsilon^{\nu}(k_1)\epsilon^{*\nu'}(k_1)=
\sum_{spin}\epsilon^{\nu}(k_2)\epsilon^{*\nu'}(k_2)=-g^{\nu\nu'}
\ee

for the spin-sum and afterwards deduct the corresponding 
ghost contributions.

% % % % % % % % % % % % % % % % % % % % % % % % % % % % % % % % % % % % % % % %
\begin{figure}[thb]
\begin{center}
\pfig{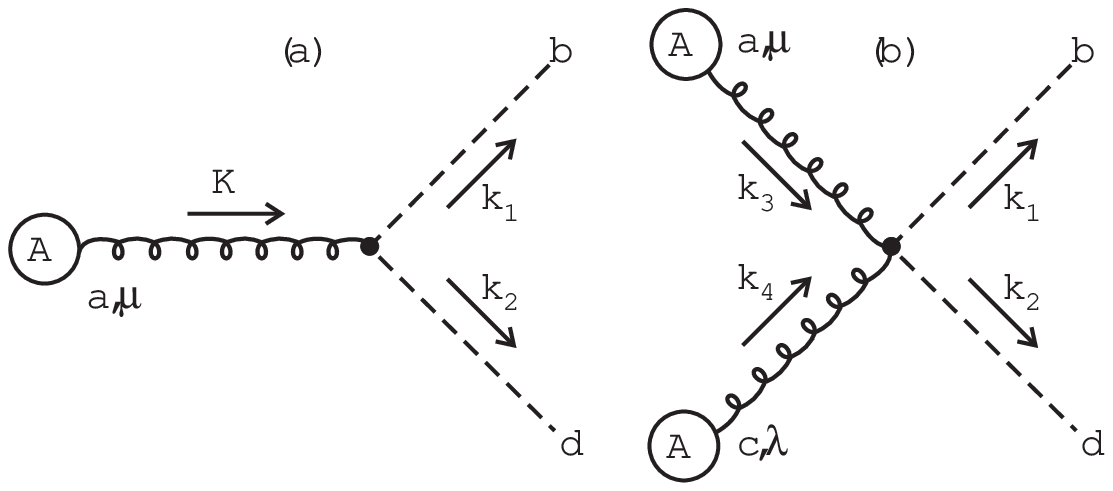}{10cm}
{Vacuum polarization diagram for the production of ghosts in lowest order}
\end{center}
\end{figure}
% % % % % % % % % % % % % % % % % % % % % % % % % % % % % % % % % % % % % % % %

The probability for the gluon part becomes:

\bea
W^{g}
=\frac{10}{8}\alpha_S\int d^4K
~\nonumber \\
((A^a(K)\cdot A^{*a}(K))K^2-(A^a(K)\cdot K)(A^{*a}(K)\cdot K))
~\nonumber \\
+\frac{3ig\alpha_S}{4}\int d^4Kd^4k_3
~\nonumber \\
f^{aa'c'}[(A^a(K)\cdot A^{*a'}(-k_3))(A^{*c'}(K-k_3)\cdot K)]
~\nonumber \\
+\frac{\alpha_Sg^2}{16}\int d^4k_3d^4k'_3d^4K
~\nonumber \\
((A^a(k_3)\cdot A^c(K-k_3))(A^{*a'}(k'_3)\cdot A^{*c'}(K-k'_3))
~\nonumber \\
\times
(f^{abx}f^{xcd}+f^{adx}f^{xcb})(f^{a'bx'}f^{x'c'd}+f^{a'dx'}f^{x'c'b})
~\nonumber \\
+12f^{acx}f^{a'c'x}\times
~\nonumber \\
(A^a(k_3)\cdot A^{*a'}(k'_3))(A^c(K-k_3)\cdot A^{*c'}(K-k'_3))).
\eea

Now, we calculate the ghost part. The vertices involving two ghosts 
and one classical field and two ghosts and two classical fields 
respectively are given by:

\be
(V^{FP}_{1A})^{abd}_{\mu}=+gf^{abd}(k_1-k_2)_{\mu}
\ee

and:

\be
(V^{FP}_{2A})^{abcd}_{\mu\lambda}=
-ig^2g_{\mu\lambda}(f^{abx}f^{xcd}+f^{adx}f^{xcb}).
\ee

The corresponding amplitude for the ghosts reads:

\be
(M^{FP})^{bd}=(M^{FP}_{1A})^{bd}+(M^{FP}_{2A})^{bd}
\ee

with (see Fig.(\ref{nayak3})(a)):

\be
(M_{1A}^{FP})^{bd}=
\frac{(2\pi)^2}{2}\int d^4K(2\pi)^4\delta^{(4)}(k_1+k_2-K)
~\nonumber \\
A^{a\mu}(K)(V^{FP})^{abd}_{\mu}
\ee

and (see Fig.(\ref{nayak3})(b)):

\bea
(M_{2A}^{FP})^{bd}=
\frac{1}{4}\int d^4k_3 d^4k_4\delta^{(4)}(k_1+k_2-k_3-k_4)
~\nonumber \\
A^{a\mu}(k_3)A^{c\lambda}(k_4)(W^{FP})^{abcd}_{\mu\lambda}.
\eea

The probability in this case is simply: 

\be
W^{FP}=
\int\frac{d^3k_1}{(2\pi)^32k_1^0}\frac{d^3k_2}{(2\pi)^32k_2^0}
(M^{FP})^{bd}(M^{FP})^{*bd},
\ee

which becomes:

\bea
W^{FP}=
-\frac{\alpha_S}{8}\int d^4K\delta^{(4)}(K-k_1-k_2)
~\nonumber \\
((A^a(K)\cdot A^a(K))K^2-(A^a(K)\cdot K)(A^a(K)\cdot K))
~\nonumber \\
-\frac{\alpha_Sg^2}{32}\int d^4Kd^4k_3d^4k'_3
~\nonumber \\
(A^a(k_3)\cdot A^c(K-k_3))(A^{a'}(k'_3)\cdot A^{c'}(K-k'_3))\times
~\nonumber \\
(f^{abx}f^{xcd}+f^{adx}f^{xcb})(f^{a'bx'}f^{x'c'd}+f^{a'dx'}f^{x'c'b}).
\eea

The real gluon-pair production probability is given by $W_{gg}=W^{g}-W^{FP}$.

Instead of the probabilities for pair production, one can also consider the 
corresponding source terms which then ultimatively enter the transport 
equation. The source terms are equal to the probability per unit of time and
per unit volume of the phase space. Some calculations yield \cite{ddd}:

\bea
\frac{dW_{q\bar{q}}^{(1)}}{d^4x d^3k}=
\frac{g^2m}{(2\pi)^5\omega}~A^a_{\mu}(x)
~e^{i k \cdot x}
\int d^4x_2 ~A^a_{\nu}(x_2)~e^{-ik\cdot x_2}
~\nonumber \\
(i[k^{\mu} (x-x_2)^{\nu}
  +(x-x_2)^{\mu} k^{\nu}
  +k \cdot (x-x_2)g^{\mu\nu}]
~\nonumber \\ 
(\frac{K_0(m \sqrt{-(x-x_2)^2})m \sqrt{-(x-x_2)^2}+2K_1(m \sqrt{-(x-x_2)^2})}{[\sqrt{-(x-x_2)^2}]^3})
~\nonumber \\
 -m^2 g^{\mu\nu}
 \frac{K_1(m \sqrt{-(x-x_2)^2})}{\sqrt{-(x-x_2)^2}}).
\label{dwqq3}
\eea

for the quarks and:

\bea
\frac{dW_{gg}}{d^4xd^3k}&=&
\frac{1}{(2\pi)^5 k^0}\int d^4x' e^{ik\cdot(x-x')}\frac{1}{(x-x')^2}
~\nonumber \\
&\times&
\{
\frac{3}{4}g^2
A^{a\mu}(x)A^{a\mu'}(x')
[3k_{\mu}k_{\mu'}
-8g_{\mu\mu'}k^{\nu}i\frac{(x-x')_{\nu}}{(x-x')^2}
~\nonumber \\
&+& 5(k_{\mu}i\frac{(x-x')_{\mu'}}{(x-x')^2}
  +k_{\mu'}i\frac{(x-x')_{\mu}}{(x-x')^2})
+\frac{6g_{\mu\mu'}}{(x-x')^2} \nonumber \\
&-&12\frac{(x-x')_{\mu}(x-x')_{\mu'}}{(x-x')^4}] \nonumber \\
&-&3ig^3 A^{a\mu}(x')A^{c\lambda}(x')A^{a'\mu'}(x)
f^{a'ac}K_{\lambda}g_{\mu\mu'}
~\nonumber \\
&-& \frac{1}{16}g^4
A^{a\mu}(x)A^{c\lambda}(x)A^{a'\mu'}(x')A^{c'\lambda'}(x')
~\nonumber \\
&\times&~[g_{\mu\lambda}g_{\mu'\lambda'}(f^{abx}f^{xcd}+f^{adx}f^{xcb})
                                 (f^{a'bx'}f^{x'c'd}+f^{a'dx'}f^{x'c'b})
~\nonumber \\
 &+& 24g_{\mu\mu'}g_{\lambda\lambda'}f^{acx}f^{a'c'x}]
\}.
\label{dwgg}
\eea

for the gluons.
It can be checked that the above results are gauge invariant with respect to 
type-(I)-gauge transformations \cite{ddd}. 

%  %  %  %  %  %  %  %  %  %  %  %  %  %  %  %  %  %  %  %  %  %  %  %  %  %  %
\subsection{Discussion}

The above results are still to complicated in order to directly get an idea 
about their content, so we look at them for a special, purely time dependent 
model field.

\be
A^{a3}(t)=A_{in}e^{-|t|/t_0},~t_0>0,~a=1,...,8
\label{as},
\ee

and all other components are equal to zero. 
Many other forms could have been taken. We have 
chosen this option just to get a feeling for how the source term in the 
phase-space behaves.
The actual form of the decay of the classical field can only be 
determined from a self consistent solution of the relativistic non-abelian 
transport equations.
The above choice yields:

\bea
\frac{dW_{q\bar{q}}}{d^4x d^3k}=
16\frac{\alpha_S}{(2\pi)^2}
(A_{in})^2
e^{2 i \omega t}
e^{-|t|/t_0}
\frac{t_0}{1+4\omega^2t_0^2}
\frac{m_T^2}{\omega^2},
\label{dWqqs}
\eea

with $m_T^2=m^2+k_T^2$ where $k_T$ is the transverse momentum and:

\bea
\frac{dW_{gg}}{d^4xd^3k}
=
\frac{24\alpha_S}{(2\pi)^2}(A_{in})^2e^{2ik^0t}e^{-|t|/t_0}
\frac{t_0}{1+4(k^0)^2t_0^2}(-3-\frac{k_T^2}{(k^0)^2})
~\nonumber \\
+
\frac{36\alpha_S^2}{2\pi}(A_{in})^4e^{2ik^0t}e^{-2|t|/t_0}
\frac{t_0}{1+(k^0)^2t_0^2}\frac{1}{(k^0)^2}.
\label{dWtots}
\eea

\begin{figure}
\begin{center}
\refstepcounter{dafigcounter}
\begin{minipage}[h]{5cm}
\epsfig{figure=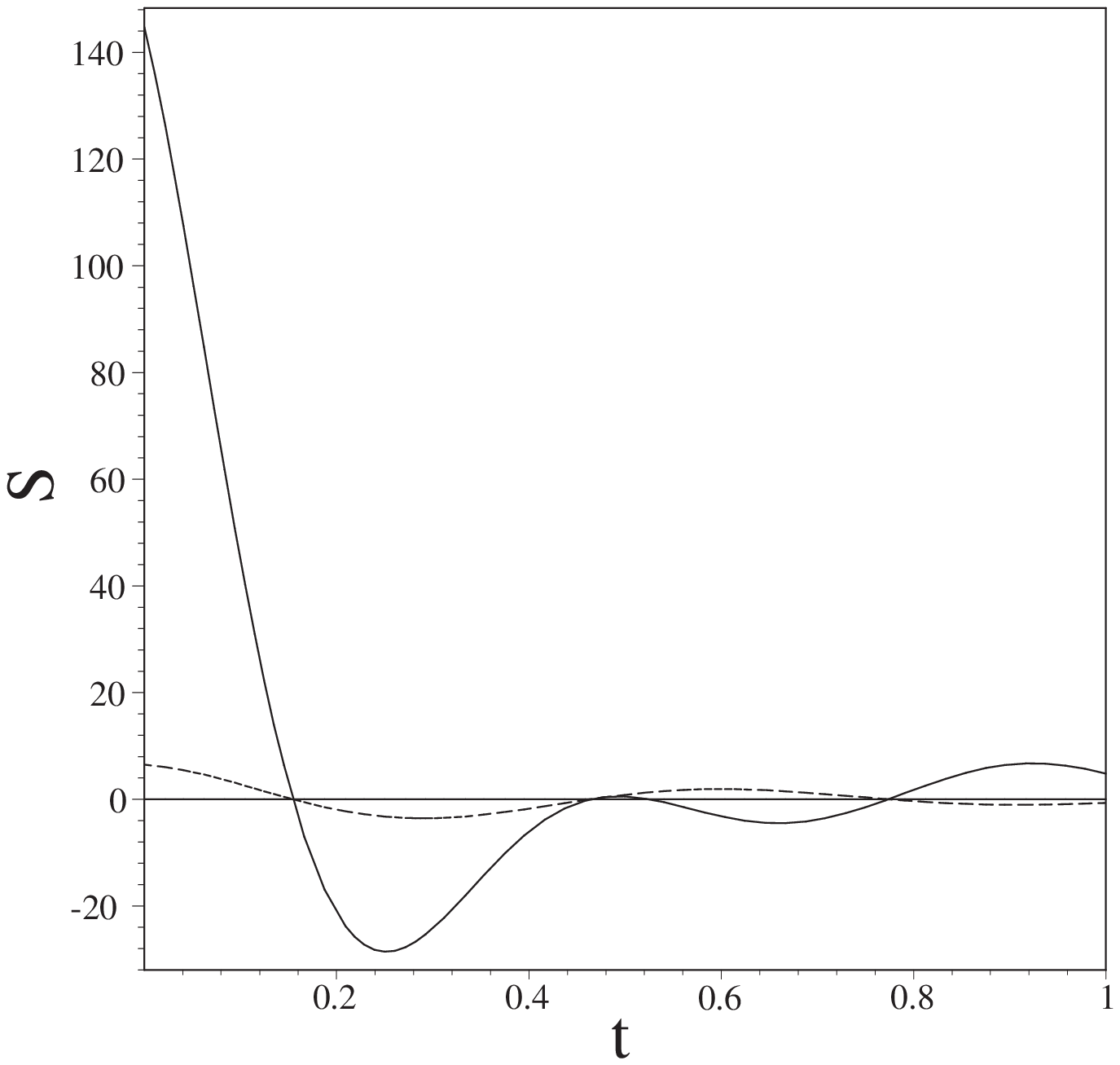, width=4.75cm}
\end{minipage}
\begin{minipage}[h]{5cm}
\epsfig{figure=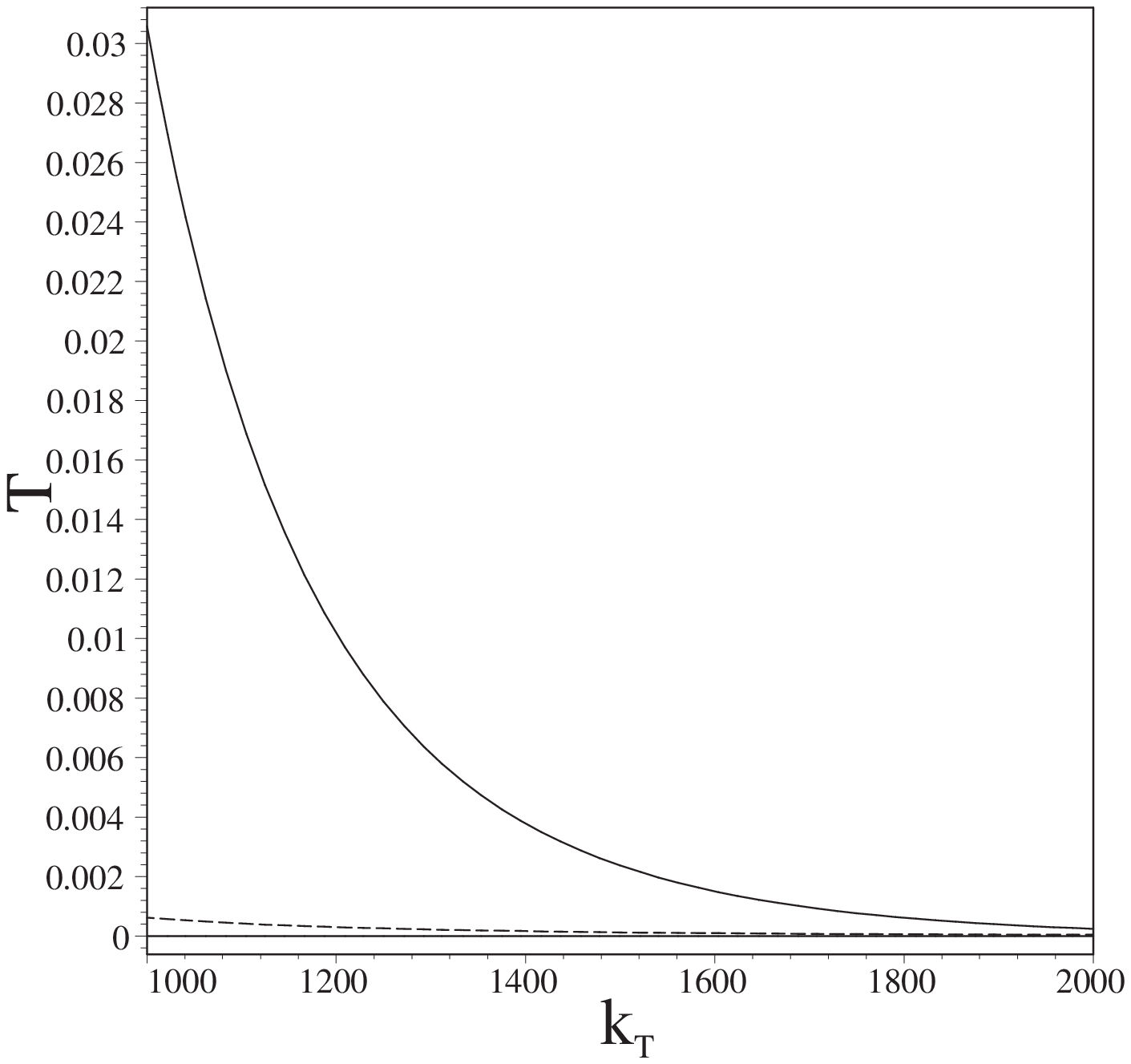, width=5cm}
\end{minipage}
\label{graph}
\end{center} 
\small {\bf Fig.~\thedafigcounter}
\
Source term $S$ [MeV] for quarks (dashed) and gluons (solid) production 
respectively versus 
time $t$ [fm/c] and time-integrated source-tem $T$ for quarks and gluons 
versus $k_T$ [MeV] for the above choice of the model field. 
\end{figure}

We choose the following parameters: $\alpha_S=0.15$, 
$A_{in}=1.5GeV$, $k_T=1.5GeV$, $y=0$, and $t_0=0.5fm$. Additionally, the 
quarks are considered to be massless.
On the LHS of Fig.(\ref{graph}), the oscillatory behavior of the source terms 
$S$ seems to indicate that there exist periods of particle creation 
and particle annihilation which follow each other periodically.
This oscillatory behavior of the source term will play a crucial role once 
it is included in a self consistent transport calculation. It can also be 
seen in the Figure that there are considerably more gluons produced than 
quarks.
On the RHS of Fig.(\ref{graph}), the time-integrated source terms $T$ can be 
regarded as a measure 
for the net-production of particles in an infinitesimal volume around any 
given point in the phase-space. It does not show the oscillatory behavior 
which gives a totally different picture for different times.
In future, we will include these source terms in the transport equation in 
order to study the production and equilibration of the QGP at RHIC and LHC.

% % % % % % % % % % % % % % % % % % % % % % % % % % % % % % % % % % % % % % %  
% % % % % % % % % % % % % % % % % % % % % % % % % % % % % % % % % % % % % % %

\noindent
{\small \bf Acknowledgements}

\noindent
G.C.N. would like to thank Alexander von Humboldt foundation for financial
support.

%%%%%%%%%%%%%%%%%%%%%%%%%%%%%%%%%%%%%%%%%%%%%%%%%%%%%%%%%%%%%%%%%%%%%%%%%%%%%%%

\newpage

%% file: SOURCE/002eproc.tex
\section*{\bf Pair Creation and Plasma Oscillations}
\addcontentsline{toc}{section}{\protect\numberline{}{Pair Creation and Plasma Oscillations \\ {\it S.M. Schmidt, A.V. Prozorkevich, D.V. Vinnik,} \\ {\it M.B. Hecht, C.D. Roberts}}}
\begin{center}
\vspace*{2mm}
{
{A.V. Prozorkevich and D.V. Vinnik}~$^{\dagger}$\\
{S.M. Schmidt}~$^{\ddagger}$\\
{M.B. Hecht and C.D. Roberts}~$^{*}$
}\\[.3cm]
{\small\it $\dagger$~Physics Department, Saratov State University,\\
410071 Saratov, Russian Federation\\
$\ddagger$~Fachbereich Physik, Universit\"at Rostock, D-18051 Rostock, Germany\\
$*$~Physics Division, Building 203, Argonne National Laboratory,\\
Argonne, IL 60439-4843, USA}
\end{center}
\begin{abstract}
We describe aspects of particle creation in strong
fields using a quantum kinetic equation with a relaxation-time approximation
to the collision term.  The strong electric background field is determined by
solving Maxwell's equation in tandem with the Vlasov equation.  Plasma
oscillations appear as a result of feedback between the background field and
the field generated by the particles produced.  The plasma frequency depends
on the strength of the initial background field and the collision frequency,
and is sensitive to the necessary momentum-dependence of dressed-parton
masses.
\end{abstract}

\newcounter{eqn8}[equation]
\setcounter{equation}{-1}
\stepcounter{equation}

\newcounter{bild8}[figure]
\setcounter{figure}{-1}
\stepcounter{figure}

\newcounter{tabelle8}[table]
\setcounter{table}{-1}
\stepcounter{table}

%\newcounter{kapitel8}[section]
%\setcounter{section}{-1}
%\stepcounter{section}

\newcounter{unterkapitel8}[subsection]
\setcounter{subsection}{-1}
\stepcounter{subsection}

\begin{center}
{{\it\small Pacs Numbers}\small : 05.20.Dd, 25.75.Dw, 05.60.Gg,
12.38.Mh}
\end{center}
%
%%\maketitle
%

%--------------------start--------------------
%
\vspace{.5cm}
Ultra-relativistic heavy-ion collisions are complicated processes and their
understanding requires a microscopic modelling of all stages: the formation,
evolution and hadronisation of a strongly coupled plasma. If the energy
density produced in the interaction volume is large enough, then the relevant
degrees of freedom are quarks and gluons. The terrestrial recreation of this
quark gluon plasma (QGP) will aid in understanding phenomena such as the big
bang and compact stars.

Construction of the Relativistic Heavy Ion Collider at the Brookhaven
National Laboratory is complete and the initial energy density: $\varepsilon
\sim 10-100$ GeV/fm$^3$, expected to be produced in the collisions at this
facility is certainly sufficient for QGP formation.  Experimentally there are
two parameters that control the conditions produced: the beam/target
properties and the impact parameter.  Varying these parameters changes the
nature of the debris measured in the detectors.  Signals of QGP formation and
information about its detailed properties are buried in that
debris~\cite{signals}.  Predicting the signals and properties requires a
microscopic understanding of the collisions, including their non-equilibrium
aspects.

In the space-time evolution of a relativistic heavy ion collision the initial
state is a system far from equilibrium.  This system then evolves to form an
equilibrated QGP, and the investigation of that evolution and the signals
that characterise the process are an important contemporary aspect of QGP
research.

The formation of a QGP is commonly described by two distinct mechanisms: the
perturbative parton picture~\cite{Gribovpartons} and the string
picture~\cite{anders83}.  In the parton picture the colliding nuclei are
visualised as clouds of partons and the plasma properties are generated by
rapid, multiple, short-range parton-parton interactions.  In the string
picture the nuclei are imagined to pass through one another and stretch a
flux tube between them as they separate, which decays via a nonperturbative
particle-antiparticle production process.  These approaches are complimentary
and both have merits and limitations.  Once the particles are produced the
subsequent analysis proceeds using Monte-Carlo event generators
\cite{partonMC,Geiger95,stringMC}.

\begin{figure}[t]
\centering{\ \epsfig{figure=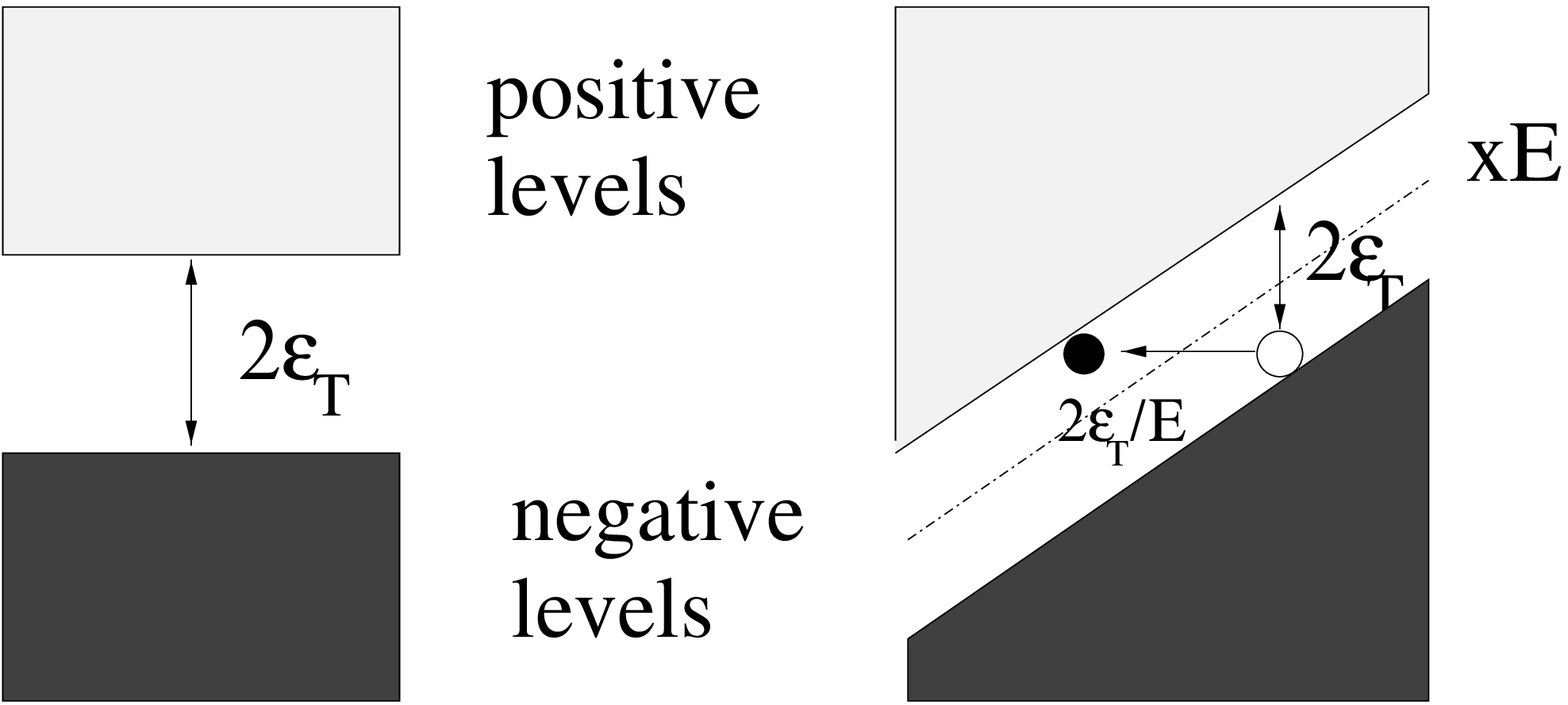,height=4cm}}\vspace*{2ex}

\caption{\label{fig1} {\em Left}: For $E=0$ the vacuum is characterised by a
completely filled negative-energy Dirac sea and an unoccupied positive-energy
continuum, separated by a gap: $2\, \varepsilon_{\rm T}= 2\,(m^2 +
p_\perp^2)^{1/2}$. {\em Right}: Introducing a constant external field:
$\vec{E} = \hat{e}_x E$, which is produced by a potential: $A^0= -E \vec{x}$,
tilts the energy levels.  In this case a particle in the negative-energy sea
will tunnel through the gap with a probability $\sim
\exp(-\pi\varepsilon_T^2/eE)$.  Succeeding, it will be accelerated by the
field in the $-x$-direction, while the hole it leaves behind will be
accelerated in the opposite direction.  The energy-level distortion is
increased with increasing $E$ and hence so is the tunneling probability.
$eE$ can be related to the flux-tube string-tension.}
\end{figure}

Herein we employ the nonperturbative flux tube picture \cite{nuss}.  A flux
tube is characterised by a linearly rising, confining quark-antiquark
potential: $V_{q{\bar q}}(r)=\sigma\,r$.  The string tension can be estimated
in lattice simulations using static quark sources, which yields $\sigma \sim
4 \Lambda_{\rm QCD}^2\sim 1\,$GeV$/$fm.  This string tension can be viewed as
a strong background field that destabilises the vacuum and the instability is
corrected through particle-antiparticle production via a process akin to the
Schwinger mechanism~\cite{Sch}.  Figure~\ref{fig1} is an artist's impression
of this process.

Assuming a constant, uniform, Abelian field, $E$, one is able to derive an
expression for the rate of particle production via this nonperturbative
mechanism
\begin{equation}
S(p_\perp)=\frac{dN}{dtdVd^2p_\perp}=|eE|\ln
\bigg[1+\exp\bigg(-\frac{2\pi(m^2+p_\perp^2)}{|eE|}\bigg)\bigg].
\end{equation}
where $m$ is the mass and $e$ the charge of the particles produced.  It is
plain from this equation that the production rate is enhanced with increasing
electric field and suppressed for a large mass and/or transverse momentum.

The production of charged particles leads naturally to an internal
current.  
That current produces an electric field
that increasingly screens and finally completely neutralises the background
field so that particle creation stops.  However, the current persists and the
field associated with that internal current restarts the production process
but now the field produced acts to retard and finally eliminate the
current. \ldots This is the back-reaction phenomenon and the natural
consequences are time dependent fields and currents~\cite{Back}.  One
observable and necessary consequence is plasma oscillations.  Their
properties, such as frequency and amplitude, depend on the initial strength
of the background field and the frequency of interactions between the
partons.  It is clear that very frequent collisions will rapidly damp plasma
oscillations and equilibrate the system.

The process we have described can be characterised by four distinct
time-scales:\\[-4.0ex]
\begin{enumerate}
\item the quantum time, $\tau_{qu}$, which is set by the Compton wavelength
of the particles produced and identifies the time-domain over which they can
be localised/identified as ``particles;''\vspace*{-1.5ex}
\item the tunneling time, $\tau_{tu}$, which is inversely proportional to the
flux tube field strength and describes the time between successive tunnelling
events;\vspace*{-1.5ex}
\item the plasma oscillation period, $\tau_{pl}$, which is also inversely
proportional to field strength in the flux tube but is affected by other
mechanisms as well; \vspace*{-1.5ex}
\item and the collision period, $\tau_{coll}$, which is the mean time between
two partonic collision events.\vspace*{-1.5ex}
\end{enumerate}
Each of the time-scales is important and the behaviour of the plasma depends
on their relation to each other~\cite{review,kme}; e.g., non-Markovian
(time-nonlocality) effects are dramatic if $\tau_{qu}\sim\tau_{tu}$.  Herein
we focus on the interplay of the larger time scales and consider a system for
which $\tau_{pl}\sim\tau_{coll}$.

Particle creation is a natural outcome when solving QED in the presence of a
strong external field and a formulation of this problem in terms of a kinetic
equation is useful since, e.g., transport properties are easy to explore.
The precise connection between this quantum kinetic equation and the mean
field approximation in non-equilibrium quantum field theory \cite{neq} is not
trivial and the derivation yields a kinetic equation that, importantly, is
non-Markovian in character~\cite{Rau,basti,gsi,prd}.

The single particle distribution function is defined as the vacuum
expectation value, in a time-dependent basis, of creation and annihilation
operators for single particle states at time $t$ with three-momentum
$\vec{p}$: $a^\dagger_{\vec{p}}(t)$, $a_{\vec{p}}(t)$; i.e., 
\begin{equation}
f(\vec{p},t):= \langle 0 | a^\dagger_{\vec{p}}(t)\,a_{\vec{p}}(t)|0\rangle\,.
\end{equation}
The evolution of this distribution function is described by the following
quantum Vlasov equation:
\begin{eqnarray}
\label{10}
\lefteqn{
\frac{df_\pm(\wvp,t)}{dt}=\frac{\partial f_\pm(\wvp,t) }{\partial
t}+eE(t)\frac{\partial f_\pm(\wvp,t)}{\partial P_\parallel(t)} }&&\\
\nonumber &=&\frac{1}{2}{\cal W}_\pm(t)\int_{-\infty}^t dt'{\cal
W}_\pm(t')\times [1 \pm 2 f_\pm(\wvp,t')]\cos[x(t',t)]+C_\pm(\wvp,t)\,,
\end{eqnarray}
where the lower [upper] sign corresponds to fermion [boson] pair creation,
$C$ is a collision term and ${\cal W}_\pm$ are the transition amplitudes.
The momentum is defined as $\wvp=(p_1,p_2,P_\parallel(t))$, with the
longitudinal [kinetic] momentum $P_\parallel(t)=p_3-eA(t)$.

We approximate the collision-induced background field by an external,
time-dependent, spatially homogeneous vector potential: $A_\mu$, in Coulomb
gauge: $A_0=0$, taken to define the $z$-axis: ${\vec A} = (0,0,A(t))$.  The
corresponding electric field, also along the $z$-axis, is
\begin{equation}
E(t) = -{\dot A}(t)=-\frac{dA(t)}{dt}\,.
\end{equation}

For fermions~\cite{Rau,gsi} and bosons~\cite{kme,gsi} the transition
amplitudes are
\begin{equation}
\label{12}
{\cal W}_-(t)=\frac{eE(t)\varepsilon_\perp}{\omega^2(t)}\,,\;\;
{\cal W}_+(t)=\frac{eE(t)P_\parallel(t)}{\omega^2(t)}\,,
\end{equation}
where the transverse energy
$\varepsilon_\perp=\sqrt{m^2+\vec{p}_\perp^{\,2}}$, $\vec{p}_\perp = (p_1,p_2)$,
and $\omega(t)=\sqrt{\varepsilon_\perp^2+P_\parallel^2(t)}$ is the total
energy.  In Eq.~(\ref{10}),
\begin{equation}
\label{30} 
x(t^\prime,t) = 2[\Theta(t)-\Theta(t^\prime)]\,,\;\;
\Theta(t) = \int^t_{-\infty}dt^\prime \omega(t^\prime)\,,
\end{equation}
is the dynamical phase difference.

The time dependence of the electric field is obtained by solving the Maxwell
equation
\begin{equation}
\label{Edot}
-{\ddot A}^\pm(t)= {\dot E}^\pm(t)= - j^{ex}(t) - j_{cond}(t) - j_{pol}(t)
\end{equation}
where the three components of the current are: the external current
generated, obviously, by the external field; the conduction current
\begin{equation}
\label{jcond}
j_{cond}(t)= g_\pm
e\int\!\frac{d^3p}{(2\pi)^3}\,\frac{P_\parallel(t)}{\omega(\wvp,t)}
f_\pm(\wvp,t) \,,
\end{equation}
associated with the collective motion of the charged particles; and the
polarisation current
\begin{equation}
\label{jpol}
j_{pol}(t) = g_\pm e \int \!\!  \frac{d^3P}{(2\pi)^3}
\frac{P_\parallel(t)}{\omega(\wvp,t)} \left[ \frac{S(\wvp,t)}{{\cal
W}_\pm(\wvp,t)}-\frac{e\, \dot E^\pm(t)\,P_\parallel(t) }
{8\,\omega^4(\vec{P},t)} \right]
\bigg(\frac{\epsilon_\perp}{P_\parallel(t)}\bigg)^{g_\pm-1}\!\!\!,
\end{equation}
which is proportional to the production rate.  Here $g_-=2$, $g_+=1$ and all
fields and charges are understood to be fully renormalised, which has a
particular impact on the polarisation current~\cite{bloch}.

\begin{figure}[t]
\centerline{\epsfig{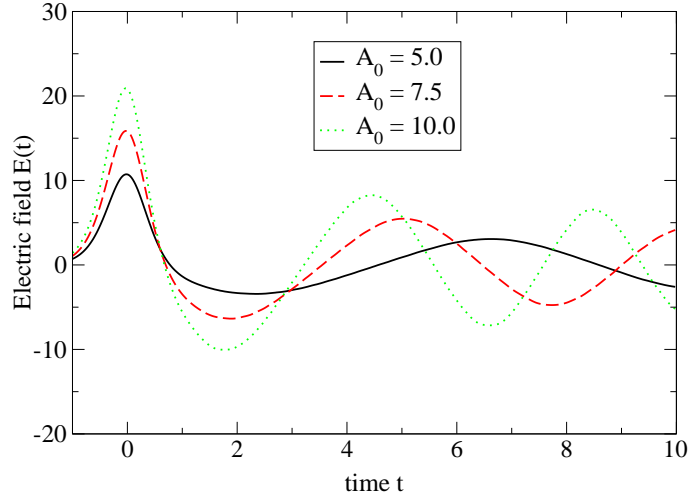}}
\vspace*{0.2cm}
\caption{Time evolution of the fermion's electric field obtained with three
different initial field strengths.  The amplitude and frequency of the plasma
oscillations increases with an increase in the strength of the external
impulse field, Eq.~(\protect\ref{Eext}).  The reference time-scale is the
lifetime $b$ of the impulse current.  (The field and time are given in
arbitrary units, and $b=0.5$.)\label{fig2}}
\end{figure}

We mimic a relativistic heavy ion collision by using an impulse profile for
the external field
\begin{equation}
\label{Eext}
E_{ex}=-\frac{A_0}{b} {\rm sech}^2(t/b)\,,
\end{equation}
which is our two-parameter model input: the width $b$ and the amplitude $A_0$
are chosen so that the initial conditions are comparable to
typical/anticipated experimental values.  This is the seed in a solution of
the coupled system of Eqs.~(\ref{10}) and (\ref{Edot})--(\ref{jpol}), and in
Fig.~\ref{fig2} we illustrate the result for the electric field {\it in the
absence of collisions}.

%%
%\begin{figure}[t]
%\centerline{\epsfig{figure=elect.eps,width=9.0cm}}
%%
%\caption{Time evolution of the fermion's electric field obtained with three
%different initial field strengths.  The amplitude and frequency of the plasma
%oscillations increases with an increase in the strength of the external
%impulse field, Eq.~(\protect\ref{Eext}).  The reference time-scale is the
%lifetime $b$ of the impulse current.  (The field and time are given in
%arbitrary units, and $b=0.5$.)\label{fig2}}
%\end{figure}

The qualitative features are obvious and easy to understand.  The external
impulse (the collision) is evident: the portion of the curve roughly
symmetric about $t=0$, where the field assumes its maximum value $A_0/b$.  It
produces charged particles and accelerates them, producing a positive current
and an associated field that continues to oppose the external field until the
net field vanishes.  At that time particle production ceases and the current
reaches a maximum value.  However, the external field is dying away: it's
lifetime is $t\sim 0.5$, so that the part of the electric field due to the
particles' own motion quickly finds itself too strong.  The excess of field
strength begins to produce particles.  It accelerates these in the opposite
direction to the particles generating the existing current whilst
simultaneously decelerating the particles in that current.  That continues
until the particle current vanishes, at which point the net field has
acquired its largest negative value.  Particle production continues and with
that a negative net current appears and grows. \ldots\ Now a pattern akin to
that of an undamped harmonic oscillator has appeared.  In the absence of
other effects, such as collisions, it continues in a steady state with the
magnitude and period of the plasma oscillations determined by the two model
parameters that characterise the collision.

The quantum kinetic equation, Eq.~(\ref{10}), describes a system far from
equilibrium and therefore any realistic collision term $C$ valid shortly
after the impact must be expected to have a very complex form.  However, as
Fig.~\ref{fig2} illustrates, the evolution at not-so-much later times is
determined by the properties of the particles produced and not by the violent
nonequilibrium effects of the collision.  This observation suggests that the
parton plasma can be treated as a quasi-equilibrium system and that the
effects of collisions can be represented via a relaxation time
approximation~\cite{bloch,memory,eis,hydro,nayak2}:
\begin{equation}
C(\vec{p},t)=\frac{1}{\tau(t)}[f_-^{eq}(\vec{p},t)-f(\vec{p},t)]\,,
\end{equation}
with $\tau(t)$ an in general time-dependent ``relaxation time'' (although we
use a constant value $\tau(t)=\tau_r$ in the calculations reported herein)
where
\begin{equation}
\label{eq}
f_-^{eq}(\vec{p},T(t))= \left[ \exp\left( \frac{p_{\nu}u^{\nu}(t)}{T(t)}\right) + 1
\right]^{-1}
\end{equation}
is the quasi-equilibrium distribution function for fermions.  Here $T(t)$ is
a local-temperature and $u^\nu(t)$, $u^2=1$, is a hydrodynamical
velocity~\cite{hydro}.  (With our geometry, $u^{\nu}(t) =
(1,0,0,v(t))/[1-v^2(t)]^{1/2}$.)

Our definition of quasi-equilibrium is to require that at each $t$ the energy
and momentum density in the evolving plasma are the same as those in an
equilibrated plasma; i.e., we require that
\begin{equation}
\label{eneneq}
\epsilon_f(t) = \epsilon^{eq}(t)\,,\; \vec{p}_f(t) = \vec{p}^{\,eq}(t)
\end{equation}
where, as one would expect, 
\begin{equation}
\epsilon^{eq}(t)= \int\frac{d^3p}{(2\pi)^3}\,\omega(\vec{p},t)\,
f_-^{eq}(\vec{p},t)\,,\; 
\vec{p}^{\,eq}(t)= \int\frac{d^3p}{(2\pi)^3}\, \vec{p}(t)\,f_-^{eq}(\vec{p},t)\,,
\end{equation}
and 
\begin{eqnarray}
\epsilon_f(t) & = & \int\frac{d^3p}{(2\pi)^3}\,\omega(\vec{p},t)\left[f_-(\vec{p},t)-
z_{f_-}(\vec{p},t)\right] \,, \\
\vec{p}_f(t) & = &\int\frac{d^3p}{(2\pi)^3}\,\vec{p}(t)\left[f_-(\vec{p},t)-
z_{f_-}(\vec{p},t)\right]\,,
\end{eqnarray}
where 
\begin{equation}
z_{f_-}(\vec{p},t) = \bigg(\frac{e\varepsilon_{\perp}}{4 \omega^3}\bigg)^2\bigg[
E(t)^2 - \frac{e^{-2t/\tau_r}}{\tau_r}\int \limits_{-\infty}^{t}dt' E(t')^2
e^{2t'/\tau_r} \bigg]
\end{equation}
is a regularising counterterm.  Adding Eqs.~(\ref{eneneq}) to the system of
coupled equations embeds implicit equations for the temperature profile,
$T(t)$, and collective velocity, $v(t)$.\footnote{A shortcoming of the
approach presented thus far is that it neglects the possibility of
dissipative inelastic scattering events transforming electric field energy
directly into temperature.  However, we have almost completed an extension
valid in that case.}

Solving the complete set of coupled equations is a straightforward but time
consuming exercise.  For the present illustration we employ the minor
simplification of assuming that the equilibrium energy density is that of a
two-flavour, massless, free-quark gas; i.e.,
\begin{equation}
\epsilon^{eq}(T(t)) = \frac{7\pi^2}{10} T^4(t)
\end{equation}
and then proceed.  

\begin{figure}[t]
\centerline{\epsfig{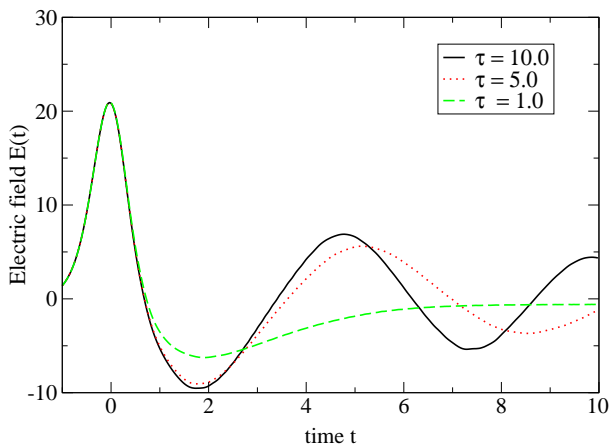}}
\vspace*{0.2cm}
\caption{Time evolution of the electric field for three different relaxation
times.  As one would anticipate, collisions damp the plasma oscillations:
their frequency and amplitude decreases with decreasing relaxation time.
Solution obtained using the impulse profile, Eq.~(\protect\ref{Eext}), with
$A_0=10$, $b=0.5$.  The field and the time are given in arbitrary
units.\label{fig3} }
\end{figure}

The solution obtained for the electric field using the impulse profile,
Eq.~(\ref{Eext}), is depicted Fig.~\ref{fig3}.  Here the behaviour is
analogous to that of a damped oscillator: the frequency and amplitude of the
plasma oscillations diminishes with increasing collision frequency,
$1/\tau_r$ (friction).

%%
%\begin{figure}[t]
%\centerline{\epsfig{figure=elecoll.eps,width=8.0cm}}
%%
%\caption{Time evolution of the electric field for three different relaxation
%times.  As one would anticipate, collisions damp the plasma oscillations:
%their frequency and amplitude decreases with decreasing relaxation time.
%Solution obtained using the impulse profile, Eq.~(\protect\ref{Eext}), with
%$A_0=10$, $b=0.5$.  The field and the time are given in arbitrary
%units.\label{fig3} }
%\end{figure}
%%

In the examples presented hitherto we have employed a constant fermion mass.
However, in QCD the dressed-quark mass is momentum-dependent~\cite{cdragw,jonivar}
and that momentum-dependence is significant when $m_0\lsim \Lambda_{\rm
QCD}$, where $m_0$ is the current-quark mass.  That can be illustrated using
a simple, instantaneous Dyson-Schwinger equation model of QCD, introduced in
Ref.~\cite{hirsch}, which yields the following pair of coupled equations for
the scalar functions in the dressed-quark propagator: $S(p)=1/[i\gamma\cdot p
\,A(\vec{p}^{\,2}) + B(\vec{p}^{\,2})]$,
\begin{eqnarray}
\label{dseB}
B(\vec{p},t)&=&m_0 +
\eta\frac{B(\vec{p},t)}{\sqrt{\vec{p}^2\,A^2(\vec{p},t)+B^2(\vec{p},t)}}\,(1-2f_-(\vec{p},t))\,,\\
\label{dseA}
A(\vec{p},t)&=&\frac{2B(\vec{p},t)}{m_0+B(\vec{p},t)}\,,
\end{eqnarray}
where $\eta$ is the model's mass scale.  In this preliminary, illustrative
calculation we discard the distribution function in Eq.~(\ref{dseB}).

%\begin{figure}[t]
%\centerline{\epsfig{figure=mass.eps,width=7.0cm}}
%%
%\caption{The scalar functions characterising a dressed-$u$-quark propagator,
%as function of momentum, obtained using the simple DSE model introduced in
%Ref.~\protect\cite{hirsch}, see Eqs.~(\protect\ref{dseB}),
%(\protect\ref{dseA}).  We plot the functions as obtained at
%$T=T_c=0.17\,$GeV, which is the critical temperature for deconfinement in the
%model~\protect\cite{bastiscm}.
%\label{fig4} }
%\end{figure}

\begin{figure}[t]
\centerline{\epsfig{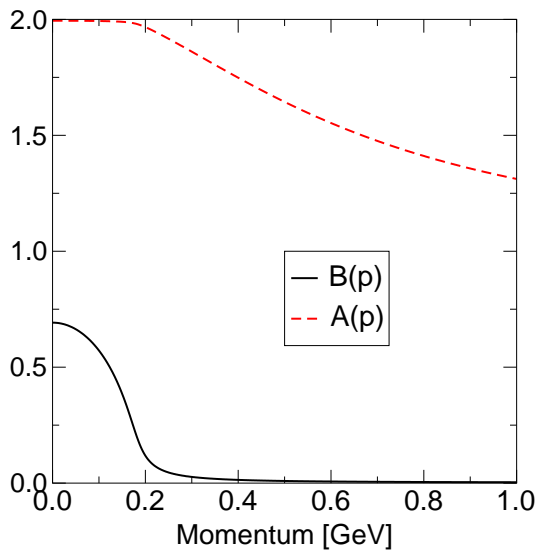}}
\vspace*{0.2cm}
\caption{The scalar functions characterising a dressed-$u$-quark propagator,
as function of momentum, obtained using the simple DSE model introduced in
Ref.~\protect\cite{hirsch}, see Eqs.~(\protect\ref{dseB}),
(\protect\ref{dseA}).  We plot the functions as obtained at
$T=T_c=0.17\,$GeV, which is the critical temperature for deconfinement in the
model~\protect\cite{bastiscm}.
\label{fig4} }
\end{figure}

The solution obtained using a current-quark mass $m_0=5\,$MeV and with $\eta
= 1.33\,$GeV is depicted in Fig.~\ref{fig4}.  This value of the mass-scale
parameter can be compared with the potential energy in a QCD string at the
confinement distance, $V_{q\bar q}(r=1\,\mbox{fm}) = \sigma r \simeq (2
\Lambda_{\rm QCD})^2 (1/ \Lambda_{\rm QCD}) = 4 \Lambda_{\rm QCD}\sim
1.0\,$GeV.  The dressed-quark mass function is
$m(\vec{p}^{\,2},T)=B(\vec{p}^{\,2},T)/A(\vec{p}^{\,2},T)$, and $m(0,T)$
provides a practical estimate of the $T$-dependent constituent-quark
mass~\cite{mr97}: in this example $m(0,T_c)\approx 0.35\,$GeV.

\begin{figure}[t]
\centerline{\epsfig{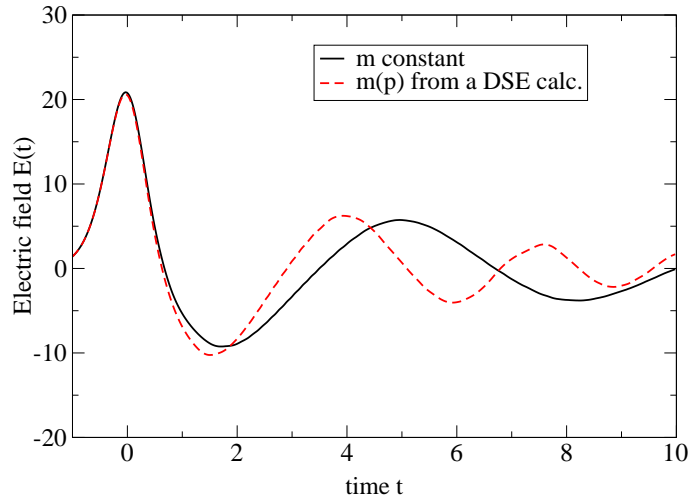}}
\vspace*{0.2cm}
\caption{Electric field as function of time for $A_0=10.0$, $\tau=5.0$,
$b=0.5$.  We compare the results obtained using particles produced with a
constant mass, $m= 1$, with those obtained when the mass is
momentum-dependent.  For illustrative simplicity, we use the profile in
Fig.~\protect\ref{fig4} but normalised such that, $m(0)=m$.  The mass-scale
is arbitrary.\label{fig5} }
\end{figure}

In our flux tube model for particle production we produce fermions with
different momenta and following this discussion it is clear that the
effective mass of the particles produced must be different for each momentum.
That will affect the production and evolution of the plasma.  To illustrate
that we have repeated the last calculation using $m(p)$ wherever the particle
mass appears in the system of coupled equations for the single particle
distribution function.  The effect on the electric field is depicted in
Fig.~\ref{fig5}, wherein it is evident that the plasma oscillation frequency
is increased when the momentum-dependent mass is used because many of the
particles produced are now lighter than the reference mass and hence respond
more quickly to changes in the electric field.

We have sketched a quantum Vlasov equation approach to the study of particle
creation in strong fields.  Using a simple model to mimic a relativistic
heavy ion collision, we solved for the single particle distribution function
and the associated particle currents and electric fields.  Plasma
oscillations are a necessary feature of all such studies.  We illustrated
that the oscillation frequency depends on: the initial energy density reached
in the collision, it increases with increasing energy density; and the
particle-particle collision probability, it decreases as this probability
increases; and that it is sensitive to the necessary momentum-dependence of
dressed-particle mass functions.  The response in each case is easy to
understand intuitively and that is a strength of the Vlasov equation
approach.

\bigskip

\hspace*{-\parindent}{{\large\bf Acknowledgments.}~A.V.P.\ is grateful for
financial support provided by the Deutsche Forschungsgemeinschaft under
project no.\ 436 RUS $17/102/00$.  This work was supported by the US
Department of Energy, Nuclear Physics Division, under contract no.\
W-31-109-ENG-38; the US National Science Foundation under grant no.\
INT-9603385; and benefited from the resources of the National Energy Research
Scientific Computing Center.

%______________________________ References ______________________________
\begin{small}

\end{small}
%%----------------------------------------
%%\end{document}
\newpage

%% file: SOURCE/paper.tex
\section*{\bf Kinetic Equilibration of Gluons in High-Energy Heavy Ion Collisions}
\addcontentsline{toc}{section}{\protect\numberline{}{Kinetic Equilibration of Gluons in High-Energy Heavy Ion Collisions \\ \mbox{\it J. Serreau}}}
\begin{center}
\vspace*{2mm}
{J. Serreau}\\[0.3cm]
{\small\it LPT, B\^{a}timent 210, Universit\'e Paris-Sud,\\ 
91405 Orsay, France\footnote{Laboratoire associ\'e au 
Centre National de la Recherche Scientifique - URA00063.}}
\end{center}

\newcounter{eqn20}[equation]
\setcounter{equation}{-1}
\stepcounter{equation}

\newcounter{bild20}[figure]
\setcounter{figure}{-1}
\stepcounter{figure}

\newcounter{tabelle20}[table]
\setcounter{table}{-1}
\stepcounter{table}

\newcounter{unterkapitel20}[subsection]
\setcounter{subsection}{-1}
\stepcounter{subsection} 

\begin{abstract}
 We study the kinetic equilibration of gluons produced in the 
 very early stages of a high energy heavy ion collision. We
 include only $gg \rightarrow gg$ elastic processes, which we
 treat in a ``self-consistent'' relaxation time approximation.
 We compare two scenarios describing the initial state of the 
 gluon system, namely the saturation and the minijet scenarios, 
 both at RHIC and LHC energies. We find that elastic collisions
 alone are not sufficient to rapidly achieve kinetic equilibrium
 in the longitudinally expanding fireball. This contradicts a 
 widely used assumption.
\end{abstract}

\subsection{Introduction}

A central question in the study of high-energy heavy ion collision
is that of the possible formation and thermalization of a deconfined
state of matter, the quark-gluon plasma (QGP). It is widely believed 
that a large amount of gluons with transverse momentum $p_t \sim 1-2$~GeV 
are produced in the very early stages of such collisions~\cite{minijet,Mueller}. 
Wether this dense gas of partons thermalizes before hadronization is 
a question of great interest for interpreting the data at RHIC and LHC. 
So far, most of the predictions concerning QGP signatures rely on the 
assumption that the system is at least in kinetic equilibrium. 
\par
It is widely assumed in the literature (see {\it e.g.}~\cite{Biro}) that 
elastic collisions between partons, which randomize their momenta, rapidly 
drive the system toward kinetic equilibrium, on time scales $\lesssim 1$~fm
(see {\it e.g.}~\cite{BMPR}). It was argued however
that, due to the effect of longitudinal expansion at early times, at least
elastic collisions may not be effective enough~\cite{Wong}. Also, the
emphasis was put on the importance of the initial condition which
characterizes the partonic system just after the collision~\cite{Mueller,HeisWang}.
The purpose of the work\footnote{This work has been done in collaboration with
   D.~Schiff (LPT, Orsay). An extended version will soon be available~\cite{JSDS}.}
reported here is to examine this question again.
We consider only gluons and assume that already at early times a
local Boltzmann equation can be written for the partonic phase space 
distribution. We include only $2 \rightarrow 2$ elastic processes in the
small-scattering angle limit and work in the relaxation time
approximation. We compare the case where initial conditions are given by
the saturation scenario (see~\cite{Mueller}), to the case where the initial
gluons are produced incoherently in semi-hard (perturbative) processes,
the so-called minijet scenario~\cite{minijet}. By consistently including the
effect of collisions in the calculation of the relaxation time, we 
reproduce semi-qualitative features of the exact solution of the Boltzmann
equation, recently obtained numerically in~\cite{Raju}. We follow the 
approach toward kinetic equilibrium by testing the isotropy of different
observables. This way of characterizing equilibration is more satisfying than 
that used in~\cite{Raju}, and leads to different conclusions. 
In particular, we find that, in both scenarios, the system does not equilibrate
at RHIC energies. More generally, we show that the assumption that elastic
collisions are efficient enough to rapidly achieve kinetic equilibrium is
not reliable. Due to the longitudinal expansion, the actual kinetic 
equilibration time is an order of magnitude bigger than the
typical $1$~fm estimate usually assumed.
\par
In Section~\ref{model} we present the model used to describe the time evolution
of the local partonic distribution in the central region of the collision.
We describe in particular the method we use to compute ``self-consistently''
the relaxation time. The results of our analysis of kinetic equilibration
in the saturation and minijet scenarios are presented in Section~\ref{result}.
More details about the work presented here can be found in~\cite{JSDS}.

\subsection{The model}
\label{model}

Shortly after being produced, because of the expansion, the parton 
system is rapidly diluted enough so one can reliably describe 
it as a gas of classical particles. One can then use a classical Boltzmann 
equation to describe the time evolution of the local phase-space distribution
$f(\vec p,\vec x,t)$: $df/dt={\mathcal C}$, where ${\mathcal C}$ is the 
collision integral. Here we consider only elastic $gg \rightarrow gg$ 
scatterings in the small-angle limit. Assuming one-dimensional expansion 
at early times and longitudinal boost invariance in the central region 
of the collision ($z\simeq 0$) in the center of mass frame, one obtains
a simple equation at constant $p_z t$. It reads, in the leading logarithmic
approximation for the collision term~\cite{Landau,Mueller},
\begin{equation}
\label{BE}
 \left. \pr_t f(\vec p,t) \right|_{p_z t} = 
 {\mathcal L} \, N_0 \nabla_p^2 f(\vec p,t) + 
 2 {\mathcal L} \, N_{-1} \vec\nabla_p \left[\vec v \, f(\vec p,t)\right]  \, ,
\end{equation}\noindent
where $\vec v = \vec p/p$, and where ${\mathcal L} = 2 \pi \alpha_S^2 \frac{N_c^2}{N_c^2-1} \, \int d\chi/\chi$ is a logarithmically divergent
integral which is physically regulated by medium effects (see for 
example~\cite{Raju} and references therein).
We have defined the moments $N_s=\langle p^s \rangle = 
\int_{\vec p} \, p^s \, f(\vec p,t)$, with the notation $\int_{\vec p} \, 
\equiv \, 2(N_c^2-1) \int d^3p/(2\pi)^3$. 
\par
The Boltzmann equation~(\ref{BE}) has been numerically solved in the
saturation scenario in Ref.~\cite{Raju}. Here we solve the equations
for moments of the distribution in a relaxation time approximation
(RTA), taking into account ``self-consistently'' the dynamical information
contained in~(\ref{BE}). The RTA consists in replacing the collision term
of the Boltzmann equation by an exponential relaxation term~\cite{Baym}:
\begin{equation}
\label{RTA} 
 \left. \pr_t f \right|_{p_z t = \mbox{cte}} \equiv -(f-f_{eq})/\theta \, ,
\end{equation} 
with $f_{eq}(\vec p,t) = \lambda (t) \, \exp (-p/T(t))$.
The parameters $\lambda$ and $T$ are determined at each time by using the
conservation of energy and of particle number in elastic collisions. In the
RTA, these leads to the following equations for the energy and particle number
densities per unit volume $\epsilon = N_1$ and $n=N_0$:
\begin{eqnarray}
\label{engy}
 \epsilon (t) & = & \epsilon_{eq} (t) = 
 6 \frac{N_c^2-1}{\pi^2} \, \lambda(t) \, T^4(t) \, , \\
\label{numb}
 n(t) & = & n_{eq} (t) = 
 2 \frac{N_c^2-1}{\pi^2} \, \lambda(t) \, T^3(t) \, .
\end{eqnarray}
We stress here that the time-dependent parameter $\lambda(t)$ is needed in 
order to enforce particle number conservation (which gives the exact 
equation $t \, n(t) =$~cte) and energy conservation independently. 
Also it is important to note that the parameter $T(t)$ does not have
the physical meaning of a local temperature unless the system has
reached the hydrodynamic regime (where $\lambda=$cte and 
$t^{1/3} T=$cte~\cite{Bjorken}).
\par
The third parameter of the RTA ansatz, namely the relaxation time $\theta$,
contains the dynamical information about the collisions. For simplicity, 
we assume it to be momentum-independent. To compute $\theta$, we shall 
identify the r.h.s of Eqs.~(\ref{BE}) and (\ref{RTA}). This cannot be done
directly and one has instead to identify some moment of these r.h.s:
\begin{equation}
\label{momeq}
 \int_{\vec p} \, m(\vec p) \, {\mathcal C} (\vec p,t) =  
 - \frac{\langle m \rangle (t) - \langle m \rangle_{eq} (t)}{\theta_m (t)} \, ,
\end{equation}\noindent
where $m(\vec p)$ is an arbitrary function, $\langle m \rangle_{(eq)} = 
\int_{\vec p} \, m(\vec p) \, f_{(eq)}(\vec p,t)$, and $\theta_m (t)$ is
the associated relaxation time. Being interested in kinetic equilibration, 
that is in the relaxation of the momentum distribution of the typical 
particles of the system toward isotropy, we choose $m$ so that it picks
up the corresponding momentum scale in the distribution, and then the 
relevant relaxation time. The momentum scale we are interested in is that
of the particles which build the ``thermodynamic'' quantities, like the energy
density, or the longitudinal and transverse pressure densities 
$P_L = \langle p_z^2/p \rangle$ and $P_T = \langle p_\perp^2/p \rangle$,
where $p_\perp^2=(p_x^2 + p_y^2)/2$. A pertinent choice leading to simple
equations is\footnote{The choices $\langle m \rangle = P_L$ 
   or $\langle m \rangle = P_T$ give the same results a those presented below.
   The choice $\langle m \rangle = \epsilon$ is equivalent to the equation of 
   energy conservation (see Eq.~(\ref{engy})).}
$\langle m \rangle = P_L - P_T$. Using Eqs.~(\ref{BE}) and~(\ref{momeq}),
one obtains the following equation for $\theta(t)$:
\begin{equation}
\label{momeq1}
  \frac{N_1^z - N_1^\perp}{\theta} = 
  4 \, {\mathcal L} \, N_0 \, ( N_{-1}^z - N_{-1}^\perp ) +
  2 \, {\mathcal L} \, N_{-1} \, ( N_0^z - N_0^\perp ) \, ,
\end{equation}\noindent
where we defined $N_s^z = \langle p_z^2 \, p^{s-2} \rangle$ and 
$N_s^\perp = \langle p_\perp^2 \, p^{s-2} \rangle$. For a particular
choice of initial distribution, we solve Eqs.~(\ref{RTA}), (\ref{engy}), 
(\ref{numb}) and (\ref{momeq}) numerically (for details see~\cite{JSDS}).
We can then follow the system toward kinetic equilibrium by measuring
the isotropy of different observables. 

\subsection{Kinetic equilibration}
\label{result}

Here we present results for two different initial conditions relevant to
high energy nuclear collisions at RHIC and LHC: the saturation and minijet 
scenarios. In both cases we use the distributions proposed in the literature,
which we recall below (see the corresponding references for details).

\begin{description}
\item[- Saturation scenario:] the initial distribution has 
the form~\cite{Mueller,Raju}
$$
 f_0 (\vec p) = f_{sat} (\vec p,t_0) = 
 \frac{c}{\alpha_S N_c} \, \delta(p_z) \, \Theta (Q_s^2 - p_t^2) \,
$$
where $t_0$ is the initial time, $Q_s$ is the saturation momentum and
$c \sim 1$ is a numerical constant. The values of these parameters corresponding 
to RHIC and LHC energies are taken from~\cite{Raju}: $c \simeq 1.3$, 
$t_0=0.4 \, (0.18)$~fm and $Q_s=1 \, (2)$~GeV at RHIC (LHC).

\item[- Minijet scenario:] the initial distribution has the form~\cite{Dumitru}
$$
 f_0 (\vec p) = f_{jet} (\vec p,t_0) = \exp (-p/T_{jet}) \, ,
$$
the parameter $T_{jet}$ being determined from the initial energy and particle
number densities. One has $t_0 = 0.18 \, (0.09)$~fm and $T_{jet}=0.535 \, 
(1.13)$~GeV at RHIC (LHC).
\end{description}
\par
\begin{figure}[htbp]
\epsfxsize=4.25in \centerline{ \epsfbox{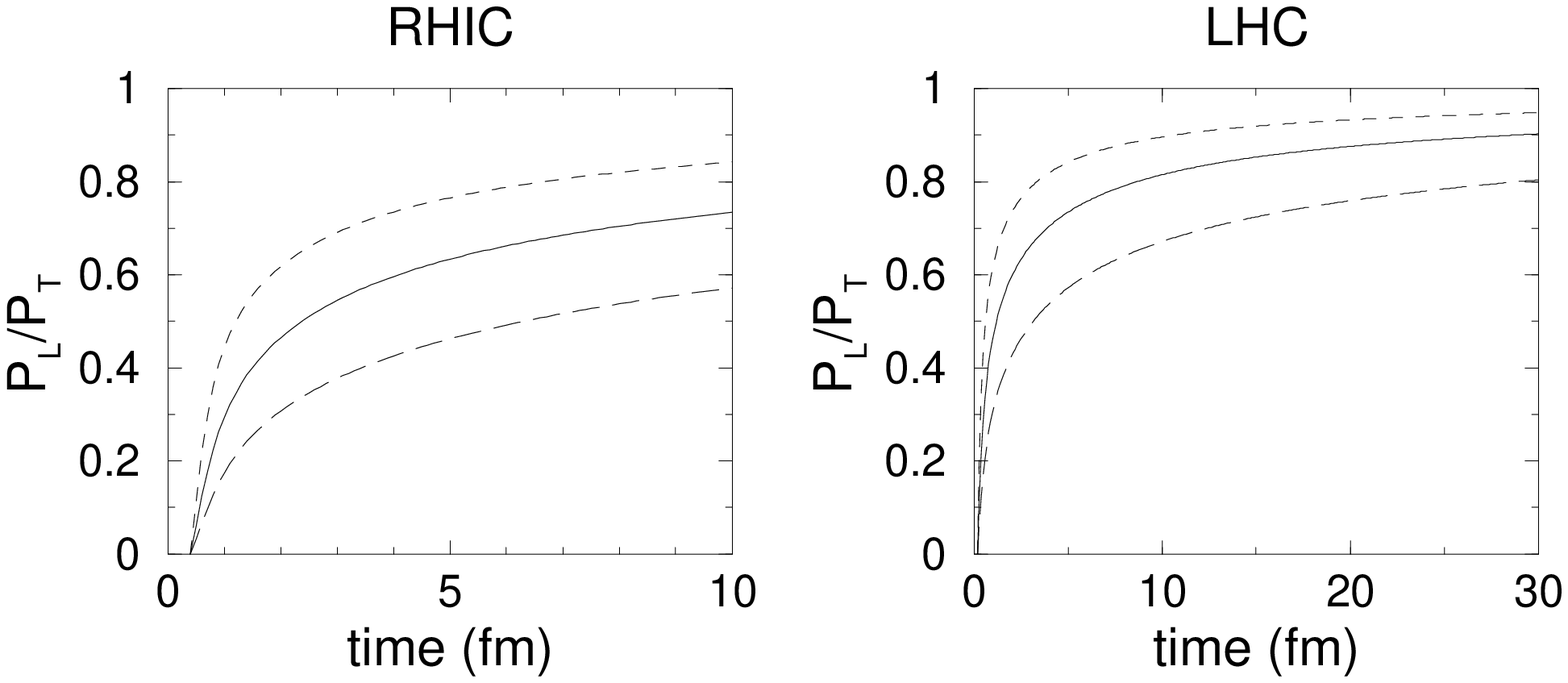}}
\caption{\small Time evolution of the ratio of longitudinal and transverse 
pressures in the saturation scenario at RHIC and LHC. The solid curve corresponds
to the choice $\alpha_S=0.3$. The upper (lower) curves are obtained by 
multiplying (dividing) the collision integral $\mathcal C$ by a factor $2$. 
This gives a rough estimate of the uncertainties of our simple description.} 
\label{fig_sat}
\end{figure}
\begin{figure}[htbp]
\epsfxsize=4.25in \centerline{ \epsfbox{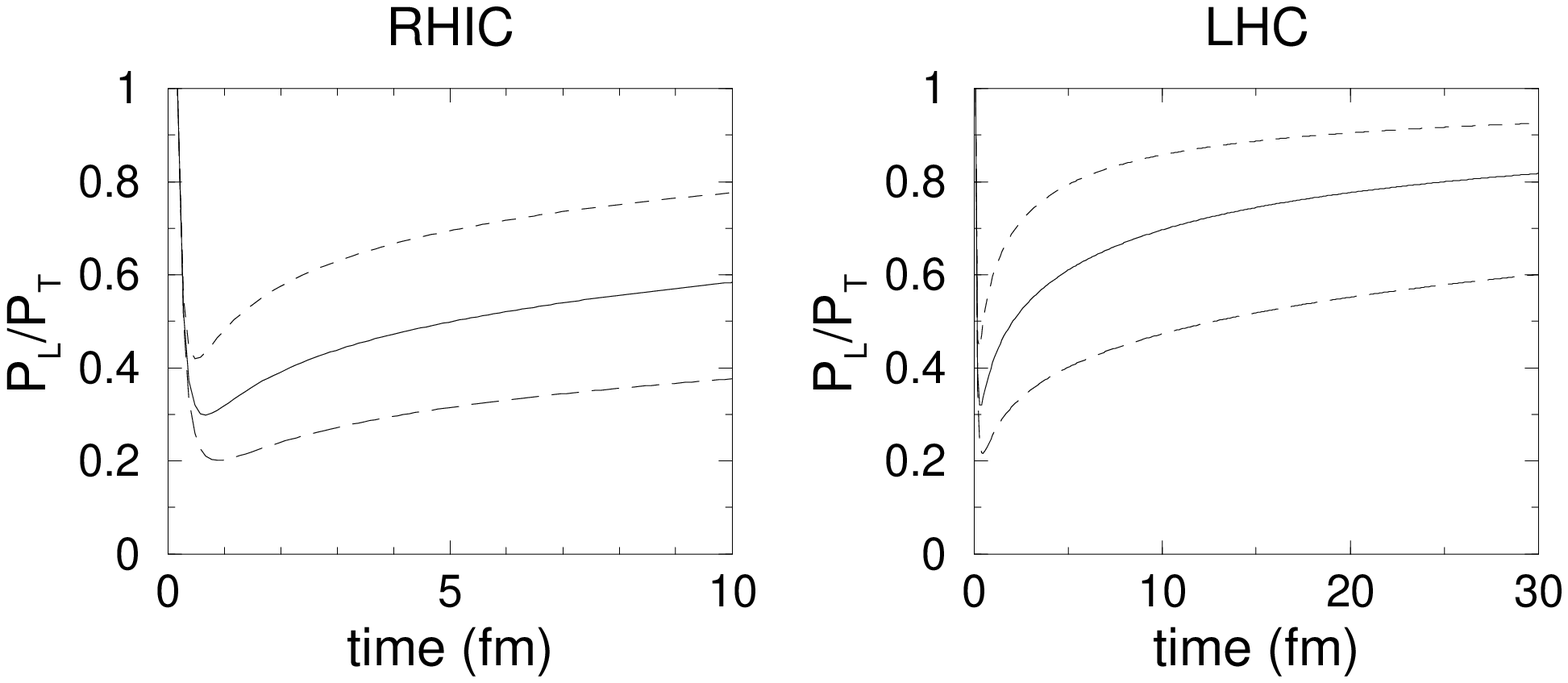}}
\caption{\small The same as Fig.~\ref{fig_sat} for the minijet scenario.} 
\label{fig_jet}
\end{figure}

In each case we measure the anisotropy of the distribution by means of the 
ratios $R_k=N_k^z/N_k^\perp$ of longitudinal and transverse moments, which 
should approach $1$ as the gas equilibrates. We stop the time evolution when
the particle number density becomes less than $n_c = 1$/fm$^3$, after which
the partonic description becomes meaningless. In both 
scenarios, the corresponding time is $t_{max}\approx 10$~fm at RHIC and 
$t_{max} \approx 30$~fm at LHC. Although our approximation of longitudinally
boost-invariant geometry can only hold for times $\lesssim R$, where 
$R \sim 5$~fm is the transverse size of the system (the radius of the incident
nuclei), we present the results for $t\le t_{max}$ in order to show how the
system behaves. The solid curves in Figs.~\ref{fig_sat} and \ref{fig_jet} 
show the time evolution of the ratio $R_1=P_L/P_T$ at RHIC and LHC energies 
for the saturation and minijet scenario respectively, where we took 
$\alpha_S=0.3$. We stress here that the results obtained in the RTA 
are in good agreement with the exact solution obtained in~\cite{Raju} in
the saturation scenario.


In order to estimate roughly the uncertainties of our description, in particular
those related to the choice of $\alpha_S$ and to the details of the screening
of the logarithmically divergent integral $\mathcal L$ (see above), 
we simply multiply and divide the collision integral by $2$. These 
respectively result in the dotted and dashed curves of Figs.~\ref{fig_sat} 
and \ref{fig_jet}. We see in both scenarios that, at RHIC energy, the 
distribution is still far from being isotropic ($P_L/P_T \lesssim 0.8$), 
even for $t\sim 10$~fm. At LHC, for $t\sim 10$~fm, although the saturation 
scenario seems more favorable, no conclusion can be made because of the
uncertainties of our description. What can be said however 
is that the typical equilibration time is at least of the order of a few fermis. 
This contradicts the usual assumption that elastic collisions are sufficient 
to achieve kinetic equilibrium on very short ($\lesssim 1$~fm) time-scales.

\subsection{Conclusion}

We studied the kinetic equilibration of the gluon gas produced in the very 
early times of a very high energy heavy ion collision. We studied in particular 
two different types of initial state relevant to these collisions: the saturation
and minijet scenarios. Assuming that, already at early times, the system can be 
described by a classical Boltzmann equation for the local partonic phase-space
distribution, we investigate the role of $2 \rightarrow 2$ processes by using a
relaxation time approximation. By measuring the anisotropy of different 
observables, we can follow the system towards kinetic equilibrium. Our 
results show that elastic collisions are not as efficient as usually
believed to achieve kinetic equilibration: because of the effect of longitudinal
expansion in the early stages, the typical equilibration time is of the order 
of a few fermis, which is comparable to the typical life-time of the partonic 
system.

\acknowledgments

I wish to thank D.~Blaschke and S.~Schmidt for the nice meeting 
in Trento and for giving me the opportunity to present this work.
I acknowledge enlighting discussions with R.~Baier, A.~Krzywicki and 
A.H.~Mueller.

\newpage

%% file: SOURCE/psi-A5.tex
\section*{\bf Photoproduction of Charmonia and Total Charmonium-Proton Cross 
Sections}
\addcontentsline{toc}{section}{\protect\numberline{}{Photoproduction of Charmonia and Total\\ Charmonium-Proton Cross Sections \\ \mbox{\it J. H\"ufner, Yu.P. Ivanov, B.Z. Kopeliovich, A.V. Tarasov}}}
\begin{center}
\vspace*{2mm}
{J.~H\"ufner$^{a,b}$, Yu.P.~Ivanov$^{a,b,c}$, B.Z.~Kopeliovich$^{b,c}$ 
  and A.V.~Tarasov$^{a,b,c}$}\\[0.3cm]
{\small\it $^a$ Institut f\"ur Theoretische Physik der Universit\"at,
  Germany\\
$^b$ Max-Planck Institut f\"ur Kernphysik, Germany\\
$^c$ Joint Institute for Nuclear Research, Dubna, Russia}
\end{center}
%%\maketitle

% ---- Abstract --------------------------------------------
\begin{abstract}

Elastic virtual photoproduction cross sections $\gamma^*p\to\Jpsi(\psi')\,p$
and total charmonium-nucleon cross sections for $\Jpsi,\ \psi'$ and $\chi$
states are calculated in a parameter free way with the light-cone dipole
formalism and the same input: factorization in impact parameters, light-cone
wave functions for the $\gamma^*$ and the charmonia, and the universal
phenomenological dipole cross section which is fitted to other data.
Very good agreement with data for the cross section of charmonium 
photoproduction is found in a wide range of $s$ and $Q^2$. We also
calculate the charmonium-proton cross sections whose absolute values
and energy dependences are found to correlate strongly with the sizes
of the states.

\end{abstract}

\newcounter{eqn5}[equation]
\setcounter{equation}{-1}
\stepcounter{equation}

\newcounter{bild5}[figure]
\setcounter{figure}{-1}
\stepcounter{figure}

\newcounter{tabelle5}[table]
\setcounter{table}{-1}
\stepcounter{table}

%\newcounter{kapitel5}[section]
%\setcounter{section}{-1}
%\stepcounter{section}

\newcounter{unterkapitel5}[subsection]
\setcounter{subsection}{-1}
\stepcounter{subsection}

%%\newpage

% ------ Introduction --------------------------------------
\subsection{Introduction}\label{intro}

The dynamics of production and interaction of charmonia has drawn attention
since their discovery back in 1973. As these heavy mesons have a small size
it has been expected that hadronic cross sections may be calculated relying
on perturbative QCD. The study of charmonium production became even more
intense after charmonium suppression had been suggested as a probe for the
creation and interaction of quark-gluon plasma in relativistic heavy ion
collisions \cite{Satz}. Since we will never have direct experimental 
information on charmonium-nucleon total cross sections one has to
extract it from other data for example from elastic photoproduction
of charmonia $\gamma p \to \Jpsi(\psi')\ p\,$. The widespread believe
that one can rely on the vector dominance model (VDM) is based on
previous experience the with photoproduction of $\rho$ mesons but
fails badly for charmonia.

Instead, one may switch to the quark basis, which should be equivalent to
the hadronic basis because of completeness. In this representation the
procedure of extracting $\sigma^{\Jpsi\,p}_{tot}$ from photoproduction data
cannot be realized directly, but has to be replaced by a different strategy.
Namely, as soon as one has expressions for the wave functions of charmonia and
the universal dipole cross section $\sigma_{q\bar q}(\rT,s)$, one can predict
both, the experimentally known charmonium photoproduction cross sections and
the unknown $\sigma^{\Jpsi(\psi')\,p}_{tot}$. If the photoproduction data
are well described one may have some confidence in the predictions for the
$\sigma^{\Jpsi(\psi') p}_{tot}$.

In the light-cone dipole approach the two processes, photoproduction
and charmonium-nucleon elastic scattering look as shown in Fig.~\ref{Fig-D}
\cite{KZ}.
\newpage
%
% --- Figure: Diagrams -------------------------------------
\BF
\centerline{
  \scalebox{0.3}{\includegraphics{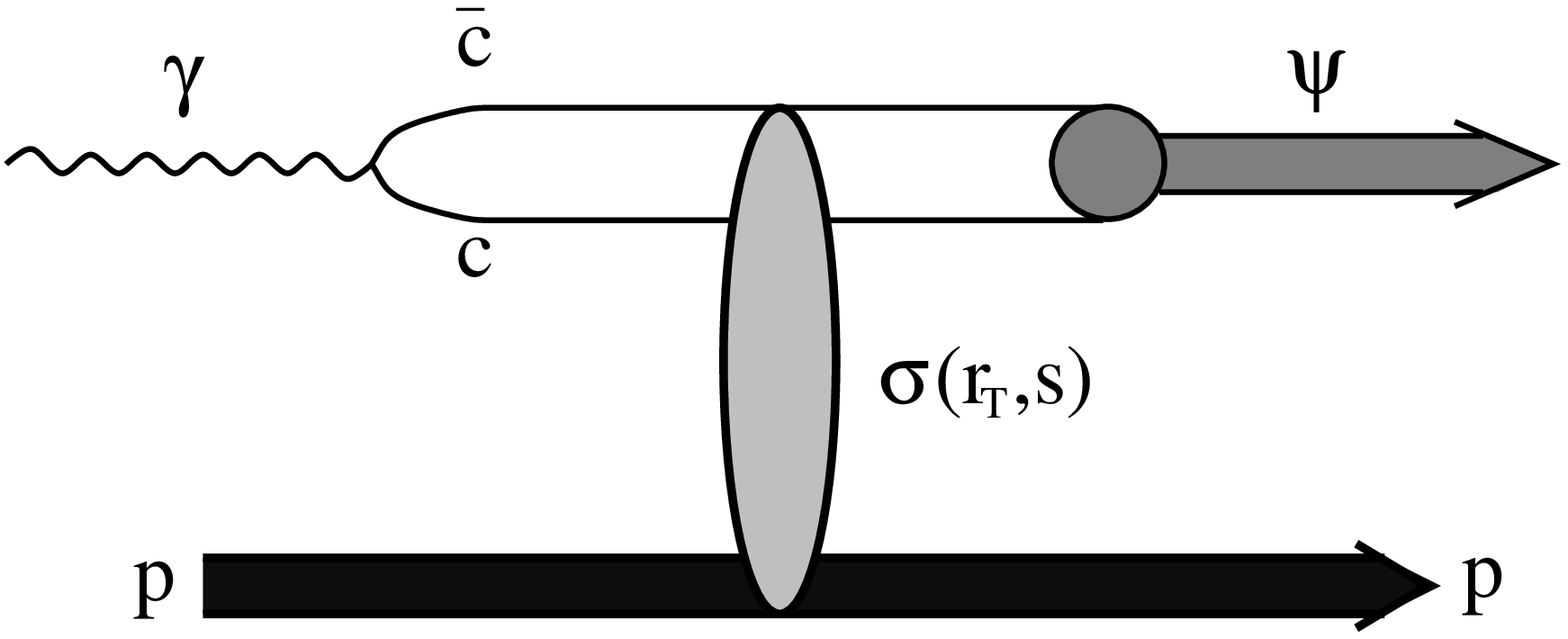}}~~
  \scalebox{0.3}{\includegraphics{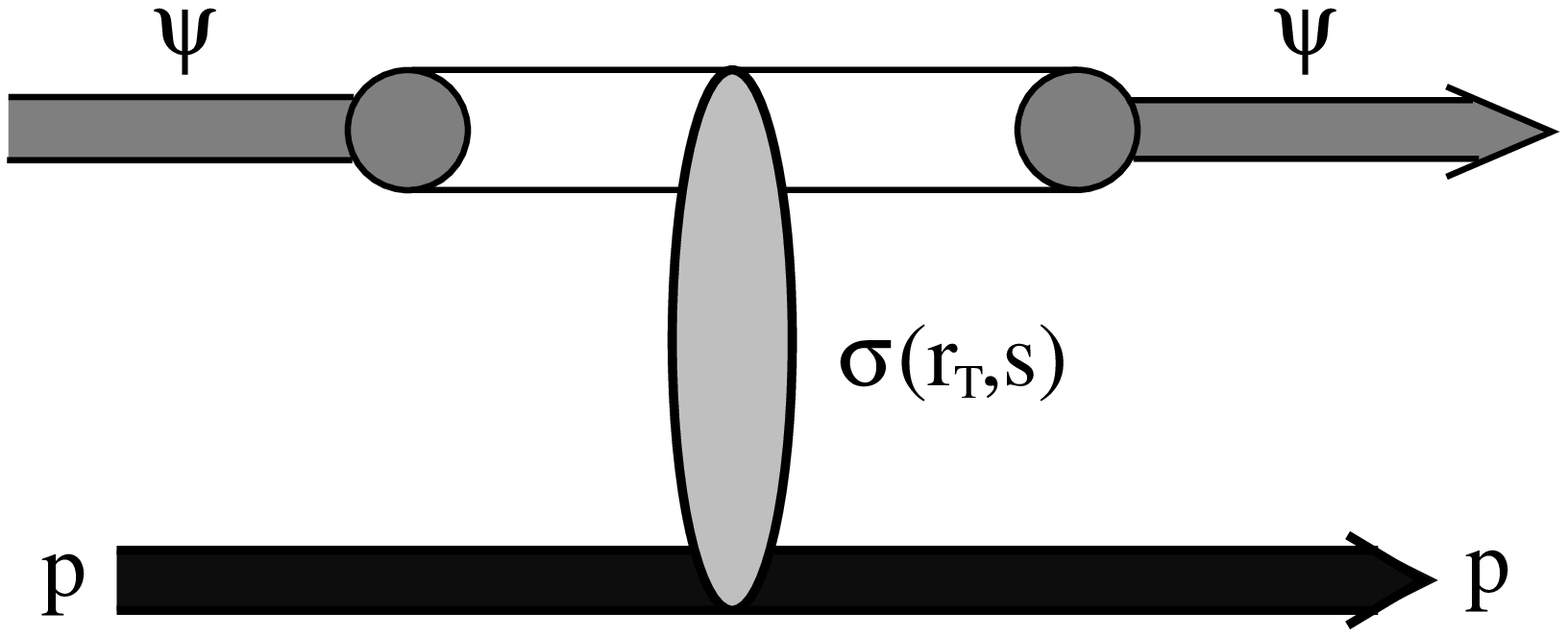}}
}
\Caption{
  \label{Fig-D}
  Schematic representation of the amplitudes for the reactions $\gamma^*p%
  \to \psi p$ (left) and $\psi\,p$ elastic scattering (right) in the rest
  frame of the proton. The $c\bar c$ fluctuation of the photon and the $\psi$
  with transverse separation $\rT$ and c.m. energy $\sqrt{s}$ interact with
  the target proton via the cross section $\sigma(\rT,s)$ and produce
  a $\Jpsi$ or $\psi'$.
}
\EF
\vspace*{-0.5cm}
The corresponding expressions for the forward amplitudes read 
%\BA
%  \label{Mgam}
%  {\cal M}_{\gamma^* p}(s,Q^2) &=& \sum_{\mu,\bar\mu}
%   \,\int\limits_0^1 \!\!d\alpha \!\int\!\! d^2\brT
%   \,\Phi^{\!*(\mu,\bar\mu)}_{\psi}(\alpha,\brT) \,\sigma_{q\bar q}(\rT,s) 
%   \,\Phi^{(\mu,\bar\mu)}_{\gamma^*}(\alpha,\brT,Q^2) \ ,\\
%  \label{Mpsi}
%  {\cal M}_{\psi\,p}(s) &=& \sum_{\mu,\bar\mu}
%    \,\int\limits_0^1 \!\!d\alpha \!\int\!\! d^2\brT 
%    \,\Phi^{\!*(\mu,\bar\mu)}_{\psi}(\alpha,\brT) \,\sigma_{q\bar q}(\rT,s)
%    \,\Phi^{(\mu,\bar\mu)}_{\psi}(\alpha,\brT) \ .
%\EA
\BA
  \label{Mgam}
  {\cal M}_{\gamma^* p}(s,Q^2) &=& \sum_{\mu,\bar\mu}
   \,\int\limits_0^1 \!\!d\alpha \!\int\!\! d^2\brT
   \,\Phi^{\!*(\mu,\bar\mu)}_{\psi}(\alpha,\brT) \,\sigma_{q\bar q}(\rT,s) 
   \,\Phi^{(\mu,\bar\mu)}_{\gamma^*}(\alpha,\brT,Q^2) \nonumber \\
   &&
\EA
\BA
  \label{Mpsi}
  {\cal M}_{\psi\,p}(s) &=& \sum_{\mu,\bar\mu}
    \,\int\limits_0^1 \!\!d\alpha \!\int\!\! d^2\brT 
    \,\Phi^{\!*(\mu,\bar\mu)}_{\psi}(\alpha,\brT) \,\sigma_{q\bar q}(\rT,s)
    \,\Phi^{(\mu,\bar\mu)}_{\psi}(\alpha,\brT) \ . \left.~~\right.
\EA
Here the summation runs over spin indexes $\mu$, $\bar\mu$ of the $c$ and
$\bar c$ quarks, $Q^2$ is the photon virtuality, $\Phi_{\gamma^*}(\alpha,%
\rT,Q^2)$ is the light-cone distribution function of the photon for a
$c\bar c$ fluctuation of separation $\rT$ and relative fraction $\alpha$ of
the photon light-cone momentum carried by $c$ or $\bar c$. Correspondingly,
$\Phi_{\psi}(\alpha,\brT)$ is the light-cone wave function of $\Jpsi$,
$\psi'$ and $\chi$ (only in Eq.~\ref{Mpsi}). The dipole cross section
$\sigma_{q\bar q}(\rT,s)$ mediates the transition ({\it cf\/}
Fig.~\ref{Fig-D}). We discuss the various ingredients.

% ------ Photoproduction -----------------------------------
\subsection{Light-cone dipole formalism}\label{lc-formalism}

The light cone variable describing longitudinal motion which is invariant
to Lorentz boosts is the fraction $\alpha=p_c^+/p_{\gamma^*}^+$ of the
photon light-cone momentum $p_{\gamma^*}^+ = E_{\gamma^*}+p_{\gamma^*}$
carried by the quark or antiquark. In the nonrelativistic approximation
(assuming no relative motion of $c$ and $\bar c$) $\alpha=1/2$ (e.g.
\cite{KZ}), otherwise one should integrate over $\alpha$ (see Eq.~\Ref{Mgam}).
For transversely ($T$) and longitudinally ($L$) polarized photons  
the perturbative photon-quark distribution function in Eq.~\Ref{Mgam} 
reads \cite{ks,bks},
\BE
  \label{psi-g}
  \Phi_{T,L}^{(\mu,\bar\mu)}(\alpha,\brT,Q^2) =
    \frac{\sqrt{N_c\,\alpha_{em}}}{2\,\pi}\,Z_c
    \,\chi_c^{\mu\dagger}\,\widehat O_{T,L}
    \,\widetilde\chi_{\bar c}^{\bar\mu}\,K_0(\epsilon\rT) \ ,
\EE
where 
\BE
  \label{tildechi}
  \widetilde\chi_{\bar c} = i\,\sigma_y\,\chi^*_{\bar c}\ ;
\EE
$\chi$ and $\bar\chi$ are the spinors of the $c$-quark and antiquark
respectively, $\widehat O_{T,L}$ are operators acting on the spin,
$Z_c=2/3$. $K_0(\epsilon\rT)$ is the modified Bessel function with
\BE
  \label{eps-Q}
  \epsilon^2 = \alpha(1-\alpha)Q^2 + m_c^2\ .
\EE

% ......... Dipole cross sections ..........................
\subsubsection{Phenomenological dipole cross section}
\label{sec-cross}

The dipole formalism for hadronic interactions introduced in \cite{ZKL}
expands the hadronic cross section over the eigen states of the interaction
which in QCD are the dipoles with a definite transverse separation (see
\Ref{Mgam}). Correspondingly, the values of the dipole cross section
$\sigma_{q\bar q}(\rT)$ for different $\rT$ are the eigenvalues of the
elastic amplitude operator. This cross section is flavor invariant, due
to universality of the QCD coupling, and vanishes like $\sigma_{q\bar q}%
(\rT) \propto r^2_T$ for $\rT\!\!\to\!0$. The latter property
is sometimes referred to as color transparency.

The total cross sections for all hadrons and (virtual) photons are known
to rise with energy. Apparently, the energy dependence cannot originate 
from the hadronic wave functions in Eqs.~(\ref{Mgam}, \ref{Mpsi}), but
only from the dipole cross section. In the approximation of two-gluon
exchange used in \cite{ZKL} the dipole cross section is constant, the
energy dependence originates from higher order corrections related to
gluon radiation. On the other way, one can stay with two-gluon exchange,
but involve higher Fock states which contain gluons in addition to the
$q\bar q$. Both approaches correspond to the same set of Feynman graphs.
We prefer to introduce energy dependence into $\sigma_{q\bar q}(\rT,s)$
and not include higher Fock states into the wave functions.

Since no reliable way to sum up higher order corrections is known so
far, we use a phenomenological form which interpolates between the two
limiting cases of small and large separations. Few parameterizations
are available in the literature, we choose two of them which are simple,
but quite successful in describing data and denote them by the initials of
the authors as ``GBW'' \cite{GBW} and ``KST'' \cite{KST} and give the
explicit expression for KST:
\BA
  \label{KST}
  \mbox{``KST'':}~~~~~~~~~
  \sigma_{\bar qq}(\rT,s) &=& \sigma_0(s) \left[1 - e^{-\rT^2/r_0^2(s)}
  \right]\ .~~~~~~
\EA
The values and energy dependence of hadronic cross sections is guaranteed
by the choice of
\BA
  \sigma_0(s) &=& 23.6 \left(\frac{s}{s_0}\right)^{\!\!0.08} 
  \left(1+\frac38 \frac{r_0^2(s)}{\left<r^2_{ch}\right>}\right)\mb\ ,\\
  r_0(s)      &=& 0.88 \left(\frac{s}{s_0}\right)^{\!\!-0.14}  \fm\ .
\EA
The energy dependent radius $r_0(s)$ is fitted to data for the proton
structure function $F^p_2(x,Q^2)$, $s_0 = 1000\GeV^2$ and the mean square of
the pion charge radius $\left<r^2_{ch}\right>=0.44\fm^2$. The improvement at
large separations leads to a somewhat worse description of the proton structure
function at large $Q^2$. Apparently, the cross section dependent on energy,
rather than $x$, cannot provide Bjorken scaling. Indeed, parameterization
\Ref{KST} is successful only up to $Q^2\approx 10\GeV^2$. 

% ......... Wave functions .................................
\vspace*{-3mm}
\subsubsection{Charmonium wave functions}\label{sec-wave}

The spatial part of the $c\bar c$ pair wave function satisfying the 
Schr\"odinger equation
\vspace*{-3mm}
\BE
  \label{Schroed}
  \left(-\,\frac{\Delta}{m_c}+V(r)\right)
   \,\Psi_{nlm}(\vr)=E_{nl}
   \,\Psi_{nlm}(\vr)
\EE
is represented in the form
\BE
  \label{wf}
  \Psi(\vr) = \Psi_{nl}(r) \cdot Y_{lm}(\theta,\varphi) \ ,
\EE
where $\vr$ is 3-dimensional $c\bar c$ separation, $\Psi_{nl}(r)$ and
$Y_{lm}(\theta,\varphi)$ are the radial and orbital parts of the wave
function. The equation for radial $\Psi(r)$ is solved with the help
of the program \cite{Lucha}. Four different potentials $V(r)$ have been
used (``BT'' \cite{BT}, ``COR'' \cite{COR}, ``LOG'' \cite{LOG} and
``POW'' \cite{POW}), of which we give one example (``COR'' - Cornell
potential \cite{COR}):
\BE
  \label{COR}
  V(r) = -\frac{k}{r} + \frac{r}{a^2}
\EE
with $k=0.52$, $a=2.34\GeV^{-1}$ and $m_c=1.84\GeV$.

The results of calculations for the radial part $\Psi_{nl}(r)$ of the $1S$
and $2S$ states are depicted in Fig.~\ref{Fig-y}. For the ground state all
the potentials provide a very similar behavior for $r>0.3\fm$, while for 
small $r$ the predictions differ by up to $30\%$. The peculiar property
of the $2S$ state wave function is the node at $r\approx 0.4\fm$ which causes
strong cancelations in the matrix elements Eq.~\Ref{Mgam} and as a result,
a suppression of photoproduction of $\psi'$ relative to $\Jpsi$
\cite{KZ,Benhar}.
% --- Figure: Radial wave functions ------------------------
\BF
\centerline{
  \scalebox{0.3}[0.26]{\includegraphics{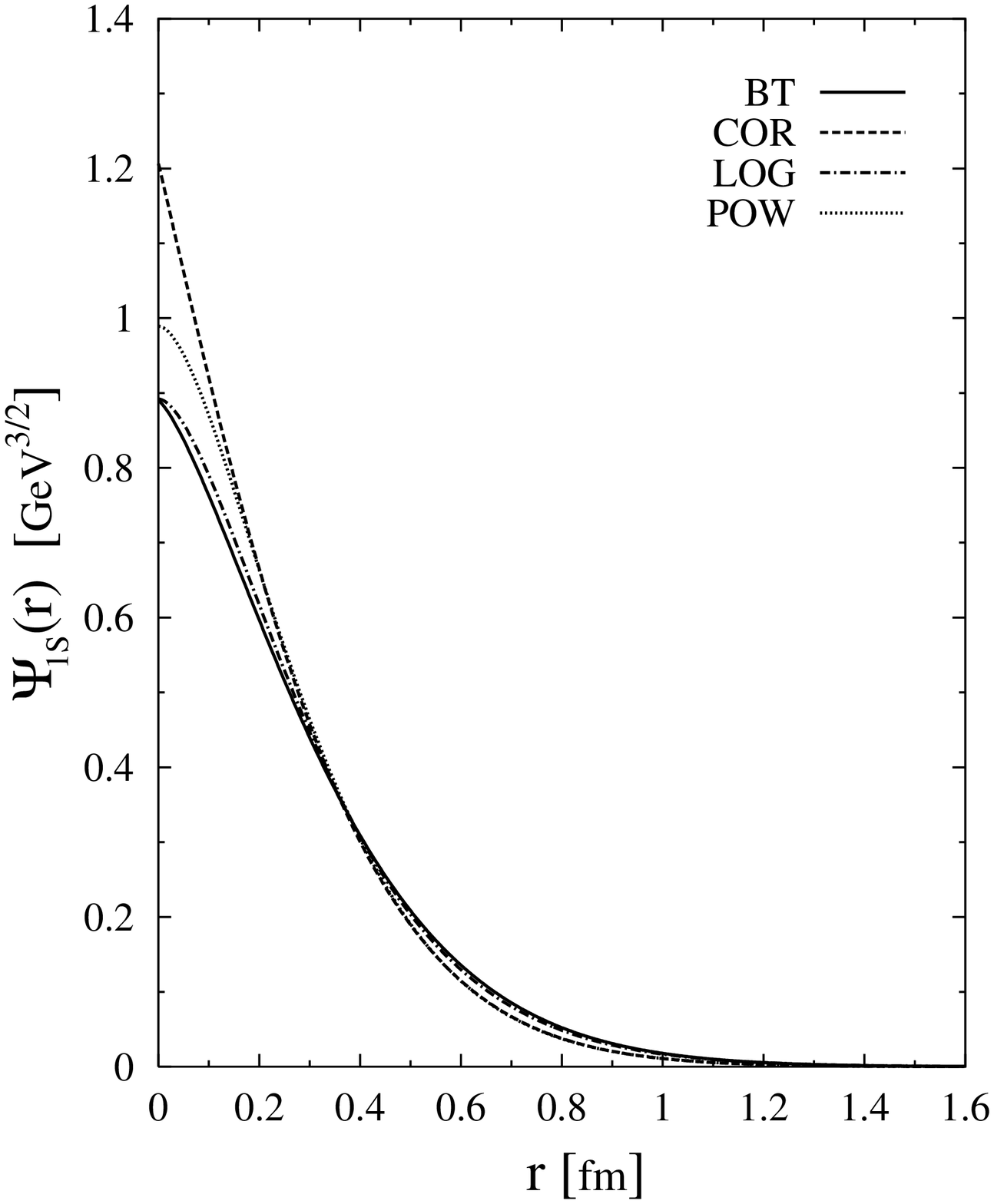}}~~
  \scalebox{0.3}[0.26]{\includegraphics{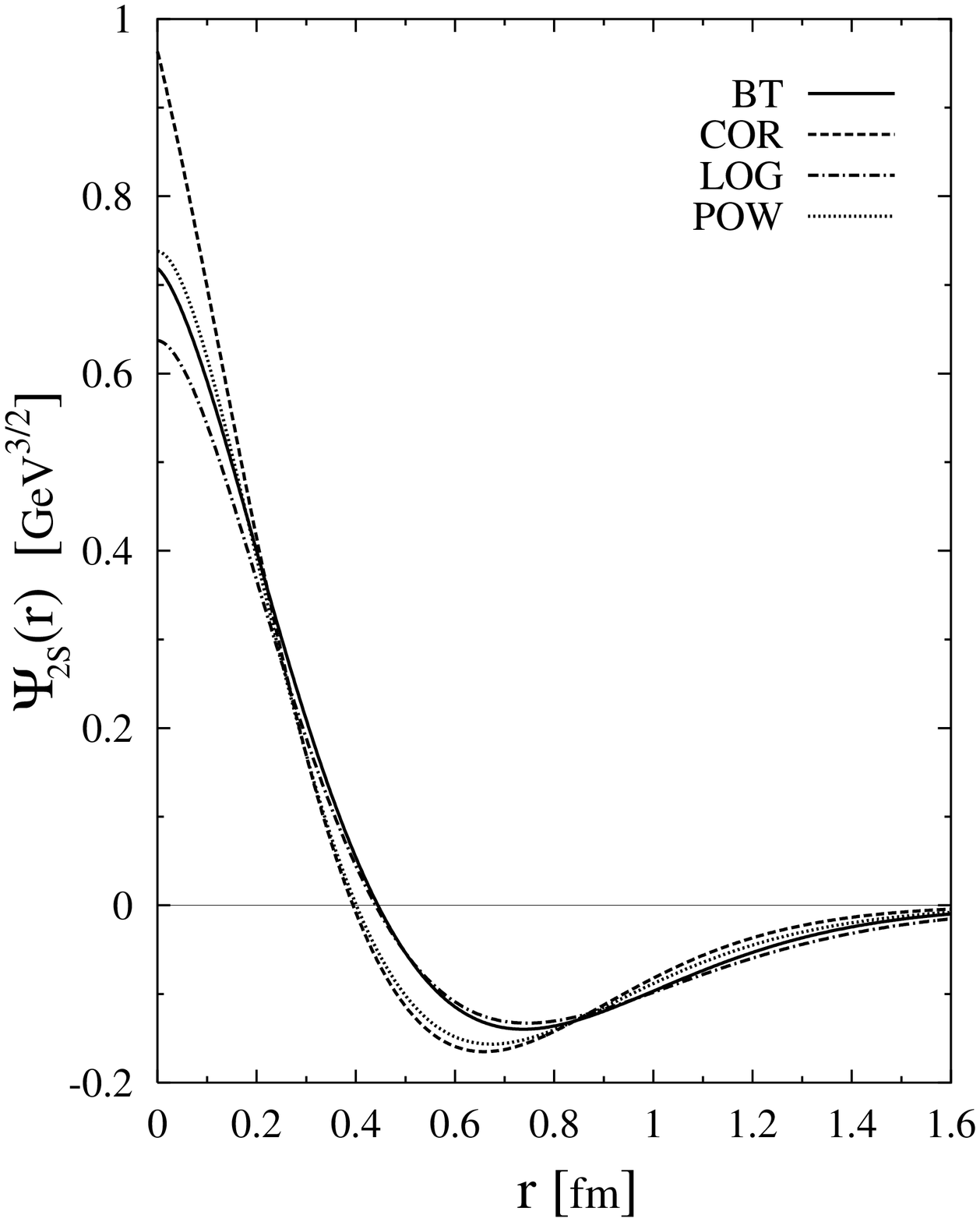}}
}
\Caption{
\label{Fig-y}
  The radial part of the wave function $\Psi_{nl}(r)$ for the $1S$ and
  $2S$ states calculated with four different potentials (see text).
}
\EF

% ......... Light cone wave functions ......................
\subsubsection{Light-cone wave functions for the bound states}\label{sec-spin}

As has been mentioned, the lowest Fock component $|c\bar c\ra$ in the
infinite momentum frame is not related by simple Lorentz boost to the
wave function of charmonium in the rest frame. This makes the problem
of building the light-cone wave function for the lowest $|c\bar c\ra$
component difficult, no unambiguous solution is yet known. There are
only recipes in the literature, a simple one widely used \cite{Terent'ev},
is the following. One applies a Fourier transformation from coordinate to
momentum space to the known spatial part of the non-relativistic wave
function \Ref{wf}, $\Psi(\vr)\Rightarrow\Psi(\vp)$, which can be written
as a function of the effective mass of the $c\bar c$, $M^2=4(p^2+m_c^2)$,
expressed in terms of light-cone variables
\vspace*{-2mm}
\BE
  M^2(\alpha,\pT) = \frac{\pT^2+m_c^2}{\alpha(1-\alpha)}\ .
\EE 
In order to change integration variable $p_L$ to the light-cone variable
$\alpha$ one relates them via $M$, namely $p_L=(\alpha-1/2)M(p_T,\alpha)$.
In this way the $c\bar c$ wave function acquires a kinematical factor
\BE
  \label{lc-wf-p}
  \Psi(\vp) \Rightarrow
  \sqrt{2}\,\frac{(p^2+m_c^2)^{3/4}}{(\pT^2+m_c^2)^{1/2}}
  \cdot \Psi(\alpha,\bpT)
  \equiv \Phi_\psi(\alpha,\bpT) \ .
\EE

This procedure is used in \cite{Hoyer} and the result is applied to 
calculation of the amplitudes \Ref{Mgam}. This leads to the enhancement
of $\Psi'$ photoproduction.

The 2-dimensional spinors $\chi_c$ and $\chi_{\bar c}$ describing $c$
and $\bar c$ respectively in the infinite momentum frame are known to be
related via the Melosh rotation \cite{Melosh,Terent'ev} to the spinors 
$\bar\chi_c$ and $\bar\chi_{\bar c}$ in the rest frame:
\BA
  \nonumber
  \bf\overline{\chi}_c        &=& \widehat R(  \alpha, \bpT)\,\chi_c\ ,\\
  \bf\overline{\chi}_{\bar c} &=& \widehat R(1-\alpha,-\bpT)\,\chi_{\bar c}\ ,
  \label{Melosh}
\EA
where the matrix $R(\alpha,\bpT)$ has the form:
\BE
  \widehat R(\alpha,\bpT) = 
    \frac{  m_c+\alpha\,M - i\,[\vec\sigma \times \vec n]\,\bpT}
    {\sqrt{(m_c+\alpha\,M)^2+\pT^2}} \ .
\label{matrix}
\EE

Since the potentials we use in section~\ref{sec-wave} contain no spin-orbit
term, the $c\bar c$ pair is in $S$-wave. In this case spatial and spin 
dependences in the wave function factorize and we arrive at the following
light cone wave function of the $c\bar c$ in the infinite momentum frame
\BE
  \label{lc-wf}
  \Phi^{(\mu,\bar\mu)}_\psi(\alpha,\bpT) =
     U^{(\mu,\bar\mu)}(\alpha,\bpT)\cdot\Phi_\psi(\alpha,\bpT)\ ,
\EE
where 
\BE
  U^{(\mu,\bar\mu)}(\alpha,\bpT) = 
    \chi_{c}^{\mu\dagger}\,\widehat R^{\dagger}(\alpha,\bpT)
    \,\vec\sigma\cdot\vec e_\psi\,\sigma_y
    \,\widehat R^*(1-\alpha,-\bpT)
    \,\sigma_y^{-1}\,\widetilde\chi_{\bar c}^{\bar\mu}
\EE
and $\widetilde\chi_{\bar c}$ is defined in \Ref{tildechi}. The Melosh
rotation is extremely important and changes the calculations for photo
production by up to a factor two.

% ------ Comparison with data ------------------------------
\vspace*{-2mm}
\subsection{Calculations and comparison with data}\label{data}

\vspace*{-1mm}
\subsubsection{$s$ and $Q^2$ dependence of $\sigma(\gamma^*p \to \Jpsi\,p$)}
\label{s-q2}

\vspace*{-1mm}
Now we are in a position to calculate the cross section of charmonium
photoproduction
\vspace*{-4mm}
\BE
  \label{sigma-gp}
  \sigma_{\gamma^*p\to\psi p}(s,Q^2) = 
  \frac{\vert\widetilde{\cal M}_T(s,Q^2)\vert^2 + 
  \eps\,\vert\widetilde{\cal M}_L(s,Q^2)\vert^2}{16\,\pi\,B}\ ,
\EE
where $\eps$ is the photon polarization (for H1 data $\la\eps\ra=0.99$) and
$B=4.73\,GeV^{-2}$ is the slope parameter in reaction $\gamma^* p\to\psi p$
\cite{H1-s}. $\widetilde{\cal M}_{T,L}$ includes also the correction for
the real part of the amplitude:
\BE
  \label{re/im}
  \widetilde{\cal M}_{T,L}(s,Q^2) = {\cal M}_{T,L}(s,Q^2)
   \,\left(1 - i\,\frac{\pi}{2}\,\frac{\partial
   \,\ln{\cal M}_{T,L}(s,Q^2)}{\partial\,\ln s} \right)\ .
\EE
The results for $J/\psi$ are compared with the data in Fig.~\ref{Fig-s}.
% --- Figure: s-dependence for J/psi -----------------------
\vspace*{-2mm}
\BF
\centerline{\scalebox{0.6}[0.5]{\includegraphics{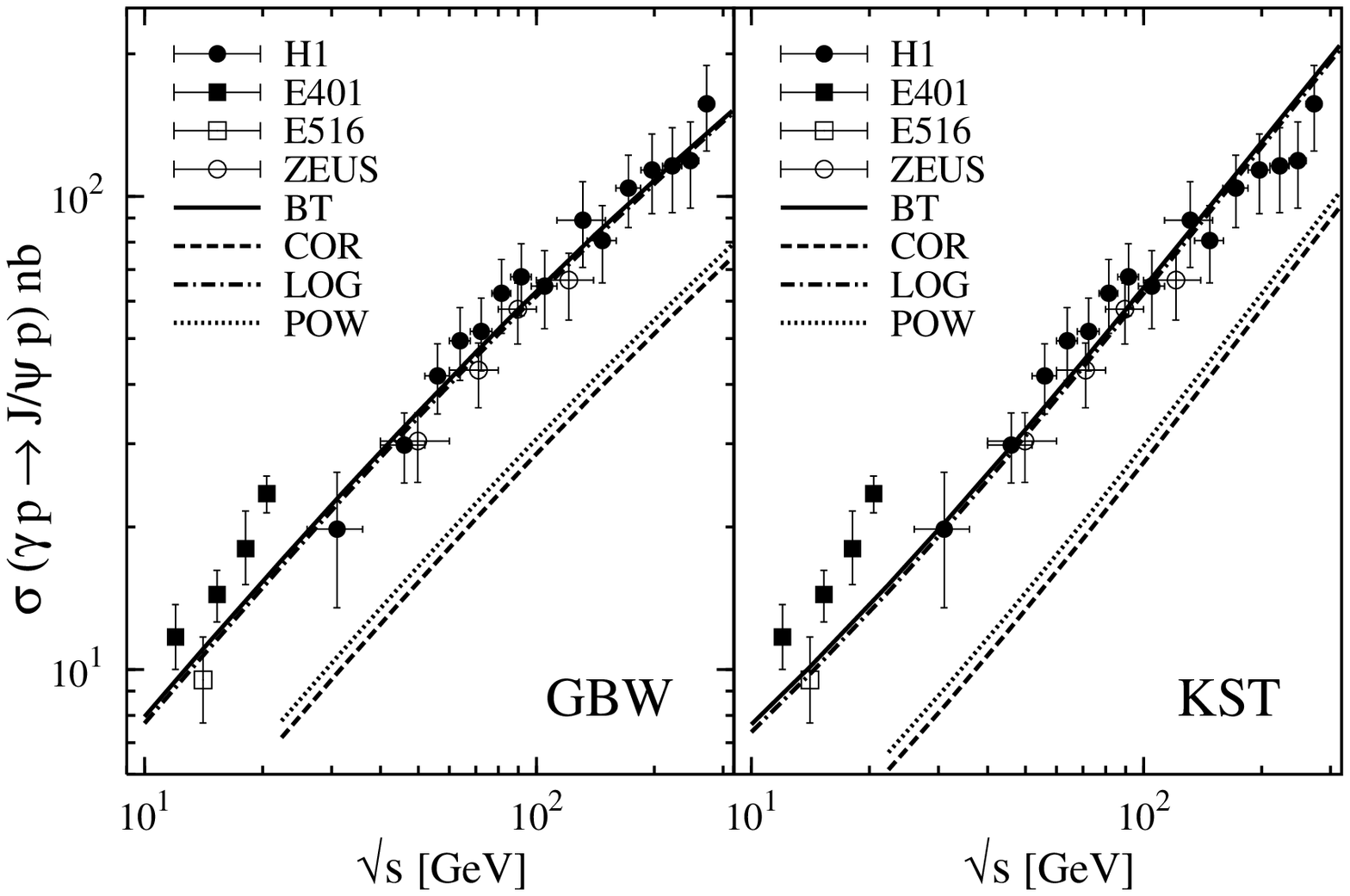}}}
\Caption{
  \label{Fig-s}
  Integrated cross section for elastic photoproduction $\gamma\,p%
  \rightarrow J/\psi\,p$ with real photons ($Q^2=0$) as a function
  of the energy calculated with GBW and KST dipole cross sections
  and for four potentials to generate $\Jpsi$ wave functions.
  Experimental data points from the H1~\Cite{H1-s}, E401~\Cite{E401-s},
  E516~\Cite{E516-s} and ZEUS~\Cite{ZEUS-s} experiments.
}
\EF

Calculations are performed with GBW and KST parameterizations for the
dipole cross section and for wave functions of the $\Jpsi$ calculated
from BT, LOG, COR and POW potentials. One observes
\begin{itemize}
\item There are no major differences for the results using the GBW and
      KST parameterizations.
\item The use of different potentials to generate the wave functions
      of the $\Jpsi$ leads to two distinctly different behaviors. The
      potentials labeled BT and LOG (see sect. \ref{sec-wave}) describe
      the data very well, while the potentials COR and LOG underestimate
      them by a factor of two. The different behavior has been traced
      to the following origin: BT and LOG use $m_c \approx 1.5\GeV$,
      but COR and POW $m_c \approx 1.8\GeV$. While the bound state
      wave functions of $\Jpsi$ are little affected by this difference
      (see Fig.~\ref{Fig-y}), the photon wave function Eq.~\Ref{psi-g}
      depends sensitively on $m_c$ via the argument Eq.~\Ref{eps-Q} of
      the $K_0$ function.
%\vspace*{-0.2cm}
%\end{itemize}
%% --- Figure: Q-dependence for J/psi -----------------------
%\BF
%\centerline{\scalebox{0.45}{\includegraphics{psi-QM.ps}}}
%\vspace*{-0.1cm}
%\Caption{
%  \label{Fig-QM}
%  Integrated cross section for elastic photo production as a function
%  of the photon virtuality $Q^2+M_{J/\psi}$ at energy $\sqrt{s}=90\GeV$.
%  Solid and dashed curves are calculated with GBW and KST dipole cross
%  sections, while thick and thin curves correspond to BT and COR potentials,
%  respectively. Results obtained with LOG and POW potentials are very
%  close to that curves (LOG similar to BT and POW to COR, see also
%  Fig.~\ref{Fig-s}). 
%}
%\EF
%
%\vspace*{-0.5cm}

We compare our calculations also with data for the $Q^2$ dependence of the
cross section. The data are plotted in Fig.~\ref{Fig-QM} at c.m. energy
$\sqrt s=90\GeV$ as a function of $Q^2+M^2_{\Jpsi}$, since in this form
both, data and calculations display an approximate power law dependence. 
%% --- Figure: Q-dependence for J/psi -----------------------
%\BF
%\centerline{\scalebox{0.45}{\includegraphics{psi-QM.ps}}}
%\Caption{
%  \label{Fig-QM}
%  Integrated cross section for elastic photo production as a function
%  of the photon virtuality $Q^2+M_{J/\psi}$ at energy $\sqrt{s}=90\GeV$.
%  Solid and dashed curves are calculated with GBW and KST dipole cross
%  sections, while thick and thin curves correspond to BT and COR potentials,
%  respectively. Results obtained with LOG and POW potentials are very
%  close to that curves (LOG similar to BT and POW to COR, see also
%  Fig.~\ref{Fig-s}). 
%}
%\EF
%
%\vspace*{-11mm}
Such a dependence on $Q^2+M^2_{\Jpsi}$ is suggested by the variable
$\epsilon^2$ in Eq.~(\ref{eps-Q}), which for $\alpha=1/2$ takes the value
$Q^2+(2\,m_c)^2$. It may be considered as an indication that $\alpha=1/2$
is a reasonable approximation for the nonrelativistic charmonium wave function.

\vspace*{-0.2cm}
\end{itemize}
% --- Figure: Q-dependence for J/psi -----------------------
\BF
\centerline{\scalebox{0.45}{\includegraphics{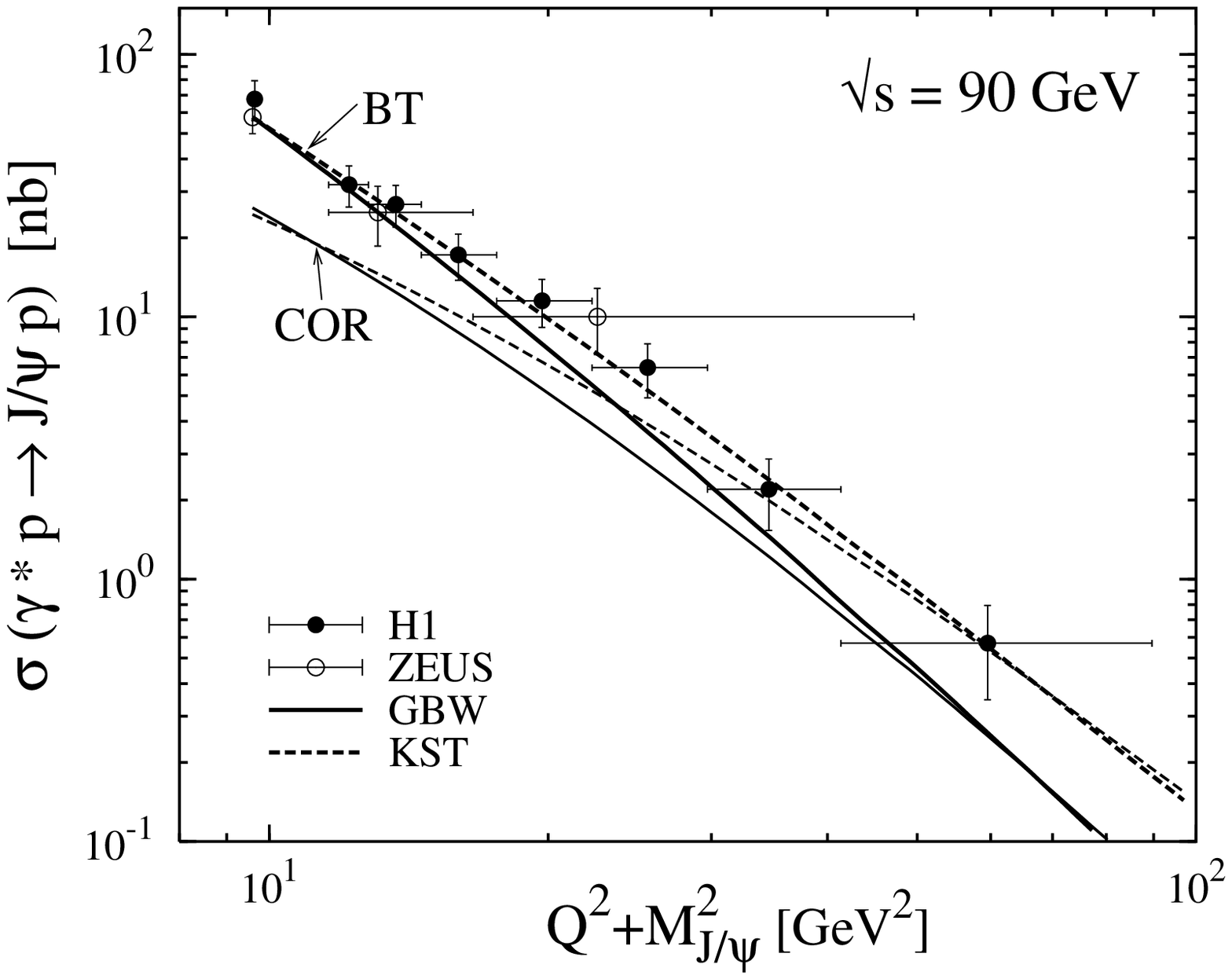}}}
\vspace*{-0.1cm}
\Caption{
  \label{Fig-QM}
  Integrated cross section for elastic photo production as a function
  of the photon virtuality $Q^2+M_{J/\psi}$ at energy $\sqrt{s}=90\GeV$.
  Solid and dashed curves are calculated with GBW and KST dipole cross
  sections, while thick and thin curves correspond to BT and COR potentials,
  respectively. Results obtained with LOG and POW potentials are very
  close to that curves (LOG similar to BT and POW to COR, see also
  Fig.~\ref{Fig-s}). 
}
\EF

\vspace*{-0.5cm}

Our results are depicted for BT and COR potentials and using GBW and KST
cross sections. Agreement with the calculations based on BT potential
is again quite good, while the COR potential grossly underestimate the
data at small $Q^2$. Although the GBW and KST dipole cross sections
lead to nearly the same cross sections for real photoproduction,
their predictions at high $Q^2$ are different by a factor $2-3$.
Supposedly the GBW parameterization should be more trustable
at $Q^2\gg M_{\Psi}^2$.

\vspace*{-2mm}
\subsubsection{Importance of spin effects for the $\psi'$ to $\Jpsi$ ratio}
\label{ratio}

\vspace*{-1mm}
It turns out that the effects of spin rotation have a gross impact on the
cross section of elastic photoproduction $\gamma\,p \to \Jpsi(\psi)p\,$.
We see that these effects add 30-40\% to the $\Jpsi$ photoproduction cross
section. 

The spin rotation effects turn out to have a much more dramatic impact on
$\psi'$ increasing the photoproduction cross section by a factor 2-3.
This spin effects explain the large values of the ratio $R$ observed
experimentally. Our results for $R$ are about twice as large as evaluated
in \cite{Suzuki} and even more than in \cite{Hoyer}.

The ratio of $\psi'$ to $\Jpsi$ photoproduction cross sections is depicted
as function of c.m. energy in Fig.~\ref{Fig-Rs}.
%% --- Figure: psi'(1S) / J/psi(2S) ratio, s-dependence -----
%\vspace*{-1mm}
%\BF
%\centerline{\scalebox{0.5}[0.45]{\includegraphics{psi-Rs.ps}}}
%\Caption{
%\label{Fig-Rs}
%  The ratio of $\psi'$ to $J/\psi$ photoproduction cross sections
%  as a function of c.m. energy calculated for all four potentials
%  with with GBW and KST parameterizations for the dipole cross
%  section. Experimental data points from the SLAC~\Cite{SLAC-R},
%  NA14~\Cite{NA14-R}, E401~\Cite{E401-R}, EMC~\Cite{EMC-R},
%  NMC~\Cite{NMC-R} and H1~\Cite{H1-R} experiments.
%}
%\EF
%\vspace*{-11mm}
Our calculations agree with available data, but error bars are too large to 
provide a more precise test for the theory. Remarkably, the ratio $R(s)$
rises with energy.

%\newpage

% --- Figure: psi'(1S) / J/psi(2S) ratio, s-dependence -----
%\vspace*{-1mm}
\BF
\centerline{\scalebox{0.5}[0.45]{\includegraphics{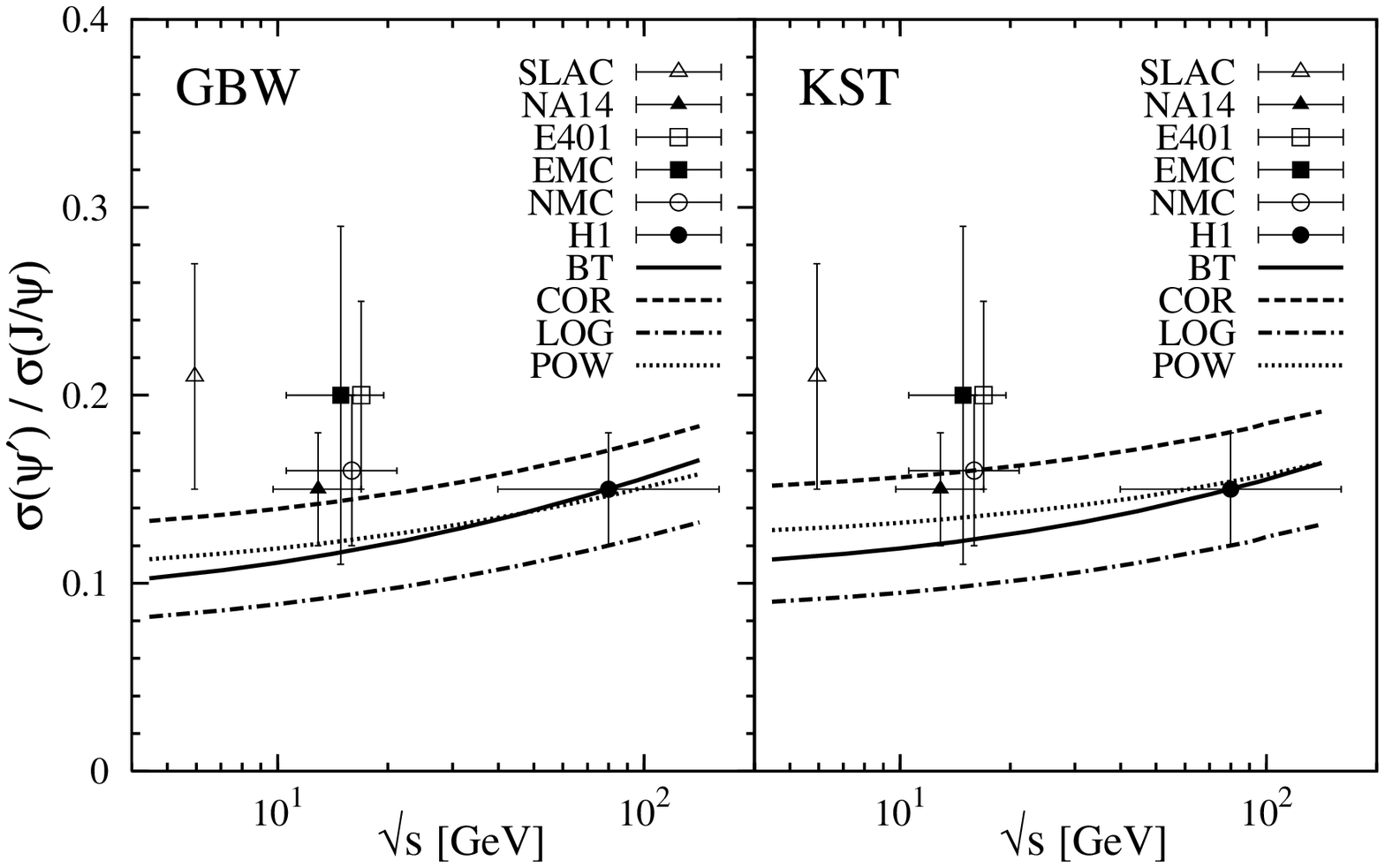}}}
\Caption{
\label{Fig-Rs}
  The ratio of $\psi'$ to $J/\psi$ photoproduction cross sections
  as a function of c.m. energy calculated for all four potentials
  with with GBW and KST parameterizations for the dipole cross
  section. Experimental data points from the SLAC~\Cite{SLAC-R},
  NA14~\Cite{NA14-R}, E401~\Cite{E401-R}, EMC~\Cite{EMC-R},
  NMC~\Cite{NMC-R} and H1~\Cite{H1-R} experiments.
}
\EF
\vspace*{-11mm}

% ------ Charmonium cross sections -------------------------
\subsection{Charmonium-nucleon total cross sections}\label{psi-n}

After the light-cone formalism has been checked with the data for virtual
photoproduction we are in position to provide reliable predictions for
charmonium-nucleon total cross sections. The calculated $\Jpsi$- and 
$\psi'$-nucleon total cross sections are plotted in Fig.~\ref{Fig-S} for
the GBW and KST forms of the dipole cross sections and all four types of
the charmonium potentials. 
% --- Figure: Total J/psi p and psi' p cross sections ------
\BF
\centerline{\scalebox{0.57}[0.5]{\includegraphics{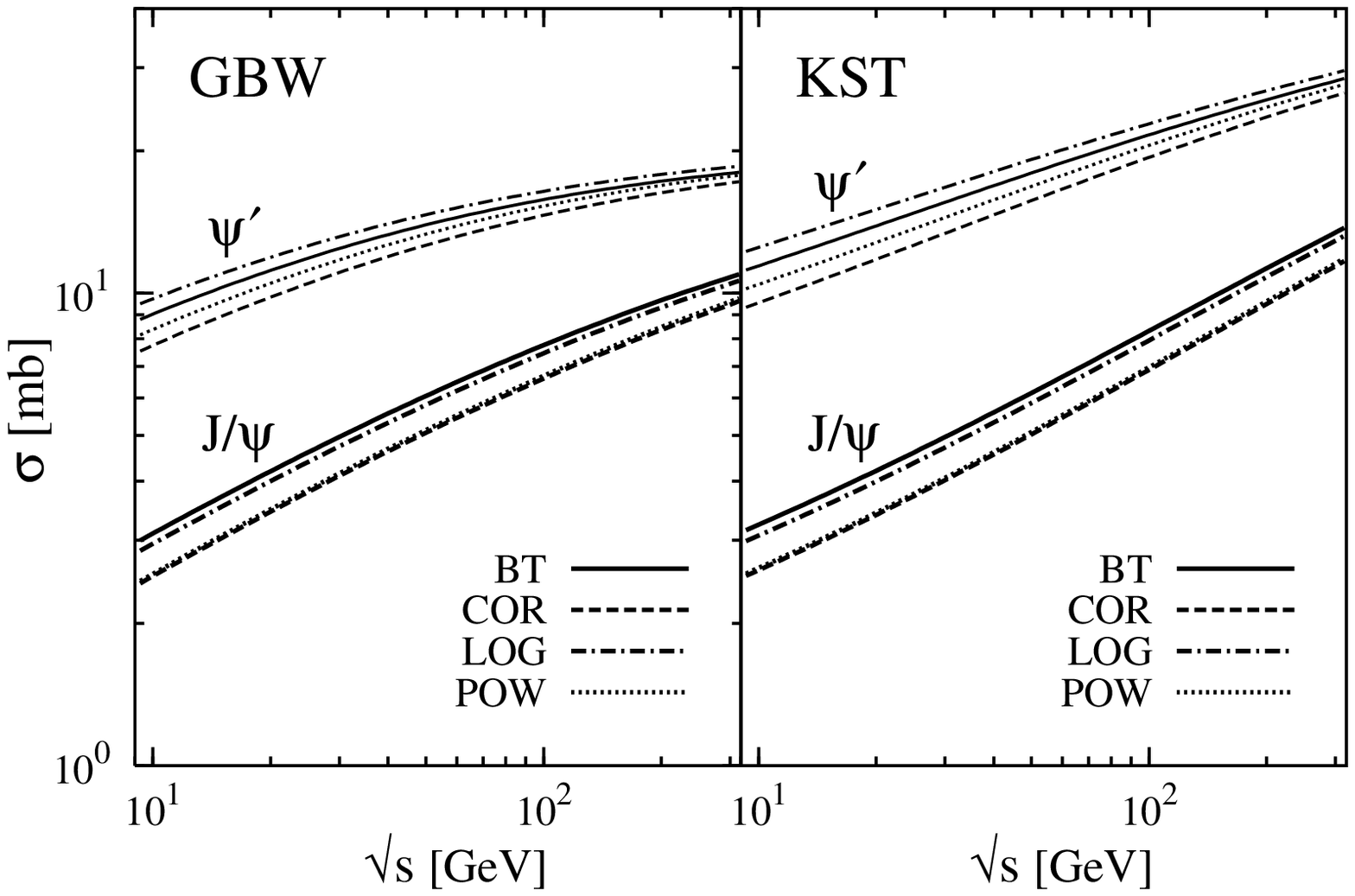}}}
\Caption{
  \label{Fig-S}
  Total $\Jpsi\,p$ (thick curves) and $\psi'\,p$ (thin curves)
  cross sections with the GBW and KST parameterizations for the
  dipole cross section.
}
\EF

According to Fig.~\ref{Fig-S} for the KST parameterization the total cross
sections of charmonia are nearly straight lines as function of $\sqrt{s}$
in a double logarithmic representation, though with significantly different
slopes for the different states. Therefore a parameterization in the form
\BE
  \label{par-sD}
  \sigma^{\psi p}(s) =
  \sigma^{\psi}_0 \cdot \left(\frac{s}{s_0}\right)^{\Delta} \ ,
\EE
seems appropriate, at least within a restricted energy interval. We use
the data shown in Fig. \ref{Fig-S} for the KST parameterization of
$\sigma_{q\bar q}$ and for the BT and LOG potentials and fit them by
the form \Ref{par-sD} with $s_0=1000\GeV$ (see Table~\ref{Tab-sD}).

% --- Table: parameters sigma_0 and Delta ------------------
\noindent
\begin{minipage}{\textwidth}
\begin{table}[htp]
\begin{center}
\begin{tabular}{|c|c|c|c|}
\hline
\vphantom{\bigg\vert}
  & $\la \rT^2 \ra  \,[\fm^2\,]$ 
  & $\sigma^\psi_0\,[\mb  \,]$
  & $\Delta$ \\
\hline &&&\\[-4mm]
$\Jpsi$       & $0.117 \pm 0.003$ & $~5.59 \pm 0.13$ & $0.212 \pm 0.001$ \\
$\chi\,(m=0)$ & $0.181 \pm 0.004$ & $~7.17 \pm 0.07$ & $0.195 \pm 0.001$ \\
$\chi\,(m=1)$ & $0.362 \pm 0.007$ & $13.17 \pm 0.16$ & $0.164 \pm 0.002$ \\
$\psi'$       & $0.517 \pm 0.034$ & $16.63 \pm 0.59$ & $0.139 \pm 0.005$ \\[1mm]
\hline
\end{tabular}
\end{center}
\caption{\label{Tab-sD} Averaged sizes $\la\rT^2 \ra$ for charmonia bound states together with $\sigma_0$ and $\Delta$ in the parameterization \Ref{par-sD} for the $\Jpsi$-, $\psi'$- and $\chi$-proton cross sections.}
\end{table}
\end{minipage}

As expected $\sigma^\psi_0$ rises monotonically with the size $\la\rT^2 \ra$
of the charmonium state, and the cross section for $\psi'\,N$ is about three
times larger that for $\Jpsi$. This deviates from the $r^2$ scaling, since
the mean value $\la r^2\ra$ is 4 times larger for $\psi'$ than for $\Jpsi$.
The exponent $\Delta$ which governs the energy dependence decreases 
monotonically with the size of the charmonium state, demonstrating the usual
correlation between the dipole size and the steepness of energy dependence.
The values of $\Delta$ are larger than in soft interactions of light hadrons
($\sim 0.08$), but smaller than values reached in DIS at high $Q^2$.

Our results at $\sqrt{s}=10\,GeV$ (the mean energy of charmonia
produced in the NA38/NA50 experiments at SPS, CERN) are
\BA
\sigma^{\Jpsi}_{tot}(\sqrt{s}=10\,GeV)     &=& ~3.56 \pm 0.08 \mb\ , 
\label{1s}\\
\sigma^{\psi'}_{tot}(\sqrt{s}=10\,GeV)     &=& 12.2  \pm 0.6  \mb\ ,
\label{2s}\\
\sigma^{``\Jpsi''}_{tot}(\sqrt{s}=10\,GeV) &=& ~5.8  \pm 0.2  \mb\ ,
\label{1s-eff}
\EA
where $``\Jpsi'' = 0.6\cdot\Jpsi+0.3\cdot\chi+0.1\cdot\psi'$.

The cross section Eq.~(\ref{par-sD})) with the parameters in Table~\ref{Tab-sD}
agrees well with $\sigma^{\Jpsi}_{tot}(\sqrt{s}=20\,GeV)=4.4\pm0.6\,\mb$
obtained in the model of the stochastic vacuum \cite{dosch}.

It worth noting that the results for charmonium-nucleon total cross sections
are amazingly similar to what one could get without any spin rotation, or
even performing a simplest integration using the nonrelativistic wave
functions (\ref{Schroed}) in the rest frame of the charmonium:
\BE
  \label{Mpsi-NR}
  \sigma_{tot}^{\psi\,N}(s) \approx
  \int d^3\!r\,\left|\Psi(\vec r)\right|^2\,
  \sigma_{q\bar q}(\rT,s)\ .
\EE

% ------ Conclusion ----------------------------------------
\subsection{Conclusion and discussions}\label{summary}

\vspace*{-2mm}
In this paper we have proposed a simultaneous treatment of elastic
photoproduction $\sigma_{\gamma^*p \to \psi p}(s,Q^2)$ of charmonia
and total cross sections $\sigma^{\psi p}_{tot}(s)$. The ingredients
are (i) the factorized light-cone expressions (\ref{Mgam}) - (\ref{Mpsi})
for the cross sections; (ii) the perturbative light-cone wave functions
for the $c\bar c$ component of the $\gamma^*$; (iii) light cone wave
functions for the charmonia bound states, and (iv) a phenomenological
dipole cross section $\sigma_{q\bar q}(\rT,s)$ for a $c\bar c$ interacting
with a proton.

The dipole cross section rises with energy; the smaller is the transverse
$\bar qq$ separation, the steeper is the growth. The source of the energy
dependence is the expanding cloud of gluons surrounding the $\bar qq$ pair.
The gluon bremsstrahlung is more intensive for small dipoles. The gluon
cloud can be treated as a joint contribution of higher Fock states, 
$|\bar q\,q\,nG\ra$, however, it can be also included into the energy
dependence of $\sigma_{\bar qq}(r_T,s)$, as we do, and this is the full
description. Addition of any higher Fock state would be the double counting.

The effective dipole cross section $\sigma_{q\bar q}(\rT,s)$ is parameterized
in a form which satisfies the expectations $\sigma_{q\bar q}\propto\rT^2$
for $\rT\to0$ (color transparency), but levels off for $\rT\to\infty$. Two
parameterizations for $\sigma_{q\bar q}(\rT,s)$, whose form and parameters
have been fitted to describe $\sigma^{\pi p}_{tot}(s)$ and the structure
function $F_2(x,Q^2)$ are used in our calculations.

While the description of the photon wave function is quite certain, the
light-cone wave function of charmonia is rather ambiguous. We have followed
the usual recipe in going from a nonrelativistic wave function calculated
from a Schr\"odinger equation to a light cone form. We have included the
Melosh spin rotation which is often neglected and found that it is
instrumental to obtain agreement, since no parameter is adjustable.
In particular, it increases the $\psi'$ photoproduction cross section
by a factor 2~-~3 and rises the $\psi'$ to $J/\psi$ ratio to the
experimental value.

At the same time, the charmonium-nucleon total cross sections ($J/\psi,\
\psi',\ \chi(m=0)$ and $\chi(m=1)$) turn out to be rather insensitive to
the way how the light-cone wave function is formed, even applying no 
Lorentz transformation  one arrives at nearly the same results. This is
why we believe that the predicted charmonium-nucleon cross section are
very stable against the ambiguities in the light-cone wave function of
charmonia. A significant energy dependence is predicted which varies
from state to state in accord with our expectations. 

\noindent {\bf Acknowledgment}: This work has been supported by a grant
from the Gesellschaft f\"ur Schwerionenforschung Darmstadt (GSI), grant
no. HD~H\"UFT and by the Federal Ministry BMBF grant no. 06 HD 954,
by the grant INTAS-97-OPEN-31696, and by the European Network: Hadronic
Physics with Electromagnetic Probes, Contract No. FMRX-CT96-0008

% --------- Bibliography -----------------------------------

\newpage

%%\end{document}

%% file: SOURCE/textyura.tex
\newcounter{eqn12}[equation]
\setcounter{equation}{-1}
\stepcounter{equation}

\newcounter{bild12}[figure]
\setcounter{figure}{-1}
\stepcounter{figure}

\newcounter{tabelle12}[table]
\setcounter{table}{-1}
\stepcounter{table}

%\newcounter{kapitel2}[section]
%\setcounter{section}{-1}
%\stepcounter{section}

\newcounter{unterkapitel12}[subsection]
\setcounter{subsection}{-1}
\stepcounter{subsection}
 
\section*{\bf Chiral Lagrangian Approach to the $J/\psi$ Breakup Cross Section}
\addcontentsline{toc}{section}{\protect\numberline{}{Chiral Lagrangian Approach to the $J/\psi$ Breakup Cross Section \\ \mbox{\it V. Ivanov, Yu. Kalinovsky, D. Blaschke, G. Burau}}}
\begin{center}
\vspace*{2mm}
{V.V. Ivanov$^{\ddagger}$, Yu.L. Kalinovsky$^{\ddagger}$}\\
{D.~Blaschke$^{\dagger}$, G. Burau$^{\dagger}$}\\[0.3cm]
{\small\it $^{\ddagger}$ Laboratory of Information Technologies, JINR, 
141980 Dubna, Russia\\
$^{\dagger}$ Fachbereich Physik, Universit\"at Rostock, 18051 Rostock, Germany}
%\date{\today} 
%\maketitle 
\end{center}
\begin{abstract} 
We summarize the results of the $SU(4)$ chiral meson Lagrangian approach to 
the cross section for J/$\psi$ breakup by pion impact. 
The major weakness of this approach is the arbitrariness in the choice of 
hadronic form factors. We suggest to fix this problem by making contact with 
the results of a nonrelativistic quark model for the breakup cross section. 
A model calculation for Gaussian wave functions is presented and the relative 
importance of quark exchange and meson exchange processes is discussed. 
We evaluate the dependence of the cross section on the masses of the final 
D-meson states and compare the result to a parametrization that has been 
employed for the study of in-medium effects on this quantity. 
%\vspace{5mm} 
%\noindent 
%\pacs{PACS number(s): 25.75.-q, 14.40.Gx, 1375.Lb} 
\end{abstract} 
 
\subsection{Introduction} 
The J/$\psi$ meson plays a key role in the experimental search for the 
quark-gluon plasma (QGP) in heavy-ion collision experiments where an anomalous 
suppression of its production cross section relative to the Drell-Yan 
continuum as a function of the centrality of the collision has been found by 
the CERN-NA50 collaboration \cite{na50}. 
An effect like this has been predicted to signal QGP formation \cite{ms86} 
as a consequence of the screening of color charges in a plasma in close 
analogy to the Mott effect (metal-insulator transition) in dense electronic 
systems \cite{rr}. 
However, a necessary condition to explain J/$\psi$ suppression in the static 
screening model is that a sufficiently large fraction of $c\bar c$ pairs after 
their creation have to traverse regions of QGP where the temperature 
(resp. parton density) has to exceed the Mott temperature 
$T^{\rm Mott}_{{\rm J}/\psi}\sim 1.2 - 1.3 T_c$ \cite{kms,rbs} for a 
sufficiently long time interval $\tau>\tau_{\rm f}$,  where 
$T_c\sim 170$ MeV is the critical phase transition temperature and 
$\tau_{\rm f}\sim 0.3 $ fm/c is the J/$\psi$ formation time. 
Within an alternative scenario \cite{rbs2}, J/$\psi$ suppression does not 
require temperatures well above the deconfinement one but can occur already at 
$T_c$ due to impact collisions by quarks from the thermal medium. 
An important ingredient for this scenario is the lowering of the reaction 
threshold for string-flip processes which lead to open-charm meson formation 
and thus to J/$\psi$ suppression. 
This process has an analogue in the hadronic world, where e.g. 
J/$\psi + \pi \rightarrow D^* + \bar D + h.c.$ could occur provided the 
reaction threshold of $\Delta E \sim 640$ MeV can be overcome by pion impact. 
It has been shown recently \cite{bbk} that this process and its in-medium 
modification can play a key role in the understanding of anomalous J/$\psi$ 
suppression as a deconfinement signal. 
Since at the deconfinement transition the $D$- mesons enter the continuum of 
unbound (but strongly correlated) quark- antiquark states (Mott effect), the 
relevant threshold for charmonium breakup is lowered and the reaction rate for 
the process gets critically enhanced. Thus a process which is negligible in 
the vacuum may give rise to additional (anomalous) J/$\psi$ suppression when 
conditions of the chiral/ deconfinement transition and $D$- meson Mott effect 
are reached in a heavy-ion collision but the dissociation of the J/$\psi$ 
itself still needs impact to overcome the threshold which is still present but 
dramatically reduced. 
 
For this alternative scenario as outlined in \cite{bbk} to work the J/$\psi$ 
breakup cross section by pion impact is required and its dependence on the 
masses of the final state $D$- mesons has to be calculated. 
Both, nonrelativistic potential models \cite{mbq95,wsb00} and chiral 
Lagrangian models \cite{mm98,lk00,hg00} have been employed to determine the 
cross section in the vacuum. The results of the latter models appear to be 
strongly dependent on the choice of formfactors for the meson-meson vertices. 
This is considered as a basic flaw of these approaches which could only be 
overcome when a more fundamental approach, e.g. from a quark model, can 
determine these input quantities of the chiral Lagrangian approach. 

In the present paper we would like to reduce the uncertainties of the 
chiral Lagrangian approach by constraining the formfactor from comparison 
with results of a nonrelativistic approach which makes use of meson wave 
functions \cite{wsb00}. Finally, we will obtain a result for the 
off-shell J/$\psi$ breakup cross section which can be compared to the fit 
formula used in \cite{bbk}. This quantity is required for the calculation 
of the in-medium modification of the  J/$\psi$ breakup due to the 
Mott effect for mesonic states at the deconfinement/chiral restoration 
transition which has been suggested \cite{bbk,bbk2} as an explanation of 
the anomalous J/$\psi$ suppression effect observed in heavy-ion 
collisions at the CERN-SPS \cite{na50}. 

\subsection{Effective Chiral Lagrangian} 
 
We start from QCD at low energy. The effective  chiral Lagrangian 
for pseudoscalar (Goldstone) mesons  can be written as 
 
\begin{eqnarray} 
{\cal L}_0 = \frac{F^2_\pi}{8} \mbox{tr} 
\left[\partial_\mu U (x) \partial_\mu U^+ (x)\right]\,, 
\end{eqnarray} 
with $F_\pi = 132 $~{MeV} being the weak pion decay constant, and 
$U (x) =\mbox{exp}\left[ 2i \varphi (x) /F_\pi\right]$. The usual 
multiplet of pseudoscalar mesons is $\varphi = \varphi^a \lambda^a /\sqrt{2}$ 
, $\lambda^a$ are Gell-Mann matrices. 
Notice that $U(x)$ transforms in a so-called non-linear representation of the 
$SU(N_f)_L \times SU(N_f)_R$ group. 
To introduce vector and axial-vector mesons we follow the procedure 
which is connected with a replacement 
\begin{eqnarray} 
{\cal L}_0 \longrightarrow {\cal L} = \frac{F^2_\pi}{8} \mbox{tr} 
\left[{\cal D}_\mu U {\cal D}_\mu U^+ \right]\,, 
\end{eqnarray} 
given by 
\begin{eqnarray} 
{\cal D}_\mu = \partial_\mu U -igA_\mu^L U + ig U A_\mu^R\,. 
\end{eqnarray} 
The left - and right- handed spin-1 fields, $A_\mu^L$ and $A_\mu^R$, 
are  combinations of vector and axial-vector meson fields 
\begin{eqnarray} 
A_\mu^L &=& \frac{1}{2} \left( V_\mu + A_\mu\right)\,, 
\nonumber \\ 
A_\mu^R &=& \frac{1}{2} \left( V_\mu  -  A_\mu\right)\,. 
\end{eqnarray} 
The coupling of these mesons to pseudoscalars is introduced as 
the gauge coupling: $g$ is the gauge coupling constant and  can be determined from the $\rho \longrightarrow \pi \pi$~ decay:~ $g_{\rho\pi\pi}=8.6$~. 
Therefore, the Lagrangian involving spin-1 and spin-0 mesons takes the form 
\begin{eqnarray} 
{\cal L}(\varphi,V,A)&=& 
{1\over 8}F_\pi^2 \mbox{tr}\left[ {\cal D}_\mu U({\cal D}_\mu U)^+ \right] 
+ {1\over 8} F_\pi^2 \mbox{tr} \left[ M(U+U^+ - 2) \right] 
\nonumber \\ && 
 -{1\over 2} \mbox{tr} \left[ (F_{\mu\nu}^L)^2 + (F_{\mu\nu}^R)^2 \right] 
+ m_0^2 \mbox{tr} \left[ (A_\mu^L)^2 + (A_\mu^R)^2 \right] 
\nonumber \\ && 
-i \xi \mbox{tr} \left[ ({\cal D}_\mu U)( {\cal D}_\nu U)^+ F_{\mu\nu}^L 
+ (D_\mu U)^+ (D_\nu U) F_{\mu\nu}^R   \right] 
\nonumber \\ && 
+\gamma \mbox{tr} \left[ F_{\mu\nu}^L U F_{\mu\nu}^R U^+ \right]\,. 
\label{Lagr} 
\end{eqnarray} 
The second term is proportional to $M$(mass matrix) and describes the ``soft'' 
breaking of the chiral $SU(N_f)_L\times SU(N_f)_R$ symmetry. 
The corresponding field strength tensors are given by 
\begin{eqnarray} 
F_{\mu \nu}^{L,R} = 
\partial_\mu A_\nu^{L,R} -  \partial_\nu A_\mu^{L,R} 
-i g \left[A_\mu^{L,R}, A_\nu^{L,R} \right]~. 
\end{eqnarray} 
The third and fourth terms in (5) correspond to the free Lagrangian of the 
spin-1 particles. At this level of the chiral symmetry all spin- 1 mesons have 
the same ``bare'' mass $m_0$. The last terms are so-called non-minimal terms 
since it  contains  of higher order in derivatives. 
It also contains the mixed term ($\partial_\mu\varphi A_\mu $). After the 
diagonalization of (5) we obtain the following Lagrangians 
\begin{eqnarray} 
{\cal L}^{(2)}(\varphi,V,A)&=&{1 \over 2}\mbox{tr}(\partial_\mu \varphi)^2 
-{1 \over 2}\mbox{tr}\left(M\varphi^2\right) 
-{1\over 4}\mbox{tr}(F_{\mu\nu}^V)^2 
+{1\over 2}m_V^2\mbox{tr}\left(V_\mu\right)^2\nonumber\\ 
&&-{1\over 4}\mbox{tr}(F_{\mu\nu}^A)^2 
+{1\over 2}m_A^2\mbox{tr}\left(A_\mu\right)^2~, 
\end{eqnarray} 
 
\begin{eqnarray} 
{\cal L}(\varphi^4)&=& 
-{2\over 3F_\pi^2}\mbox{tr}(\partial_\mu\varphi\partial_\mu(\varphi^3)) 
+{1\over 2F_\pi^2}Z^2\mbox{tr}(\partial_\mu(\varphi^2)\partial_\mu(\varphi^2)) 
\nonumber\\ 
&& +{g^2\over 4m_V^2}\mbox{tr}(\partial_\mu\varphi 
\{\varphi^2,\partial_\mu\varphi\}) 
+{1\over 6F_\pi^2}\mbox{tr}\large(M\varphi^4\large) 
\\ && 
+\left( {1\over 8}g^2(1-\gamma)^2\alpha^4 
-\xi {2g \over F_\pi^2}Z^4\alpha^2\sqrt{1-\gamma}\right)\mbox{tr}(\partial_\mu\varphi\partial_\nu\varphi[\partial_\mu\varphi,\partial_\nu\varphi])~,\nonumber 
\end{eqnarray} 
 
\begin{eqnarray} 
{\cal L}(VV\varphi\varphi)&=& 
-{g^2\over 4Z^4}\mbox{tr}\left((V_\mu\varphi)^2-V_\mu^2\varphi^2\right) 
\nonumber \\ && 
-{1\over F_\pi^2}{\gamma\over {1-\gamma}} 
\mbox{tr}\left(\varphi^2(F_{\mu\nu}^V)^2-(F_{\mu\nu}^V\varphi)^2\right) 
\nonumber \\ && 
+{1\over 16}g^2\alpha^2(1+\gamma)\mbox{tr} 
\left(\large[\partial_\mu\varphi,V_\nu\large] 
+\large[V_\mu,\partial_\nu\varphi\large]\right)^2 
\nonumber \\ && 
+{1\over 8}g^2\alpha^2(1-\gamma)\mbox{tr} 
\left(\left[V_\mu,V_\nu\right][\partial_\mu\varphi,\partial_\nu\varphi]\right) 
\nonumber \\ && 
+{g\alpha \over 2F_\pi}{\gamma \over \sqrt{1-\gamma}}\mbox{tr} 
\left(\varphi[F_{\mu\nu}^V,\left([\partial_\mu\varphi,V_\nu] 
+[V_\mu,\partial_\nu\varphi]\right)]\right) 
\nonumber \\ && 
-{2g\xi\over F_\pi^2}{Z^4\over \sqrt{1-\gamma}}\mbox{tr} 
\left(\partial_\mu\varphi\partial_\nu\varphi[V_\mu,V_\nu]\right) 
\nonumber \\ && 
+{2g\xi\over F_\pi^2}{Z^2 \over \sqrt{1-\gamma}}\mbox{tr} 
\left( (\partial_\mu\varphi[\varphi,V_\nu] 
+[\varphi,V_\mu]\partial_\nu\varphi)F_{\mu\nu}^V\right)~, 
\end{eqnarray} 

\newpage

\begin{eqnarray} 
{\cal L}(A,V,\varphi)&=&-i{g^2F_\pi\over 4Z^2}\sqrt{{1-\gamma}\over{1+\gamma}}\mbox{tr}\left(V_\mu[A_\mu,\varphi]\right) 
\nonumber \\ && 
+i{1\over F_\pi}{\gamma\over \sqrt{1-\gamma^2}}\mbox{tr}\left(\varphi[F_{\mu\nu}^V,F_{\mu\nu}^A]\right) 
\nonumber \\ && 
+i{g^2 F_\pi\over 4m_V^2Z^2}(1-\delta)\sqrt{{1-\gamma}\over{1+\gamma}}\mbox{tr}\left(F_{\mu\nu}^V[A_\mu,\partial_\nu\varphi]\right) 
\nonumber \\ && 
+i{g^2 F_\pi\over 4m_V^2}\sqrt{{1+\gamma}\over {1-\gamma}}\mbox{tr}\left(F_{\mu\nu}^A[V_\mu,\partial_\nu\varphi]\right)~, 
\end{eqnarray} 
\begin{eqnarray} 
{\cal L}(V\varphi\varphi)&=& 
-i{g\over 2}\mbox{tr}\left(V_\mu 
\left( \varphi 
\stackrel{\leftrightarrow}{\partial}_\mu 
\varphi \right) 
\right) 
+i{g\delta\over 2m_V^2}\mbox{tr}\left(F_{\mu\nu}^V\partial_\mu\varphi\partial_\nu\varphi\right)~, 
\end{eqnarray} 
\begin{eqnarray} 
{\cal L}(V^4)&=&{1 \over 16}{g^2\over 1-\gamma}\mbox{tr}\left([V_\mu,V_\nu]^2\right)~, 
\end{eqnarray} 
\begin{eqnarray} 
{\cal L}(V^3)&=&i{g \over 4}\mbox{tr}\left( F_{\mu\nu}[V_\mu,V_\nu] \right)~, 
\end{eqnarray} 
where 
$$ 
\delta=1-Z^2-{2Z^4\over 1-Z^2}{\xi g \over \sqrt{1-\gamma}}~, 
$$ 
and $F_{\mu\nu}^V=\partial_\mu V_\nu - \partial_\nu V_\mu,\, F_{\mu\nu}^A=\partial_\mu A_\nu - \partial_\nu A_\mu,\, (\varphi\stackrel{\leftrightarrow}{\partial}_\mu\varphi)=\varphi\,\partial_\mu\varphi-\partial_\mu\varphi\ \varphi$, which have the standard definition for commutators and anticommutators 
\cite{iz}.\\ 
 To obtain the Lagrangian with the ``hidden'' chiral symmetry 
(vector mesons as dynamic gauge bosons) we choose a gauge where 
left- and right-handed fields in the Lagrangian will be identical to the 
vector field \,$V_\mu$:\, 
$A_\mu^{L'}=A_\mu^{R'}=V_\mu$\, and \,$A'_{\mu}= 0.$ 
This can be done  by a gauge 
transformation which conserves the $SU(N_f)_L\times SU(N_f)_R$ symmetry 
\begin{eqnarray} 
\nonumber \\ && 
A_\mu^L=A_\mu^R=V_\mu~, 
\nonumber \\ && 
U\longrightarrow U_L\ U\ U_R^+~, 
\nonumber \\ && 
A_\mu^L\longrightarrow U_L\ A_\mu^L\ U_L^+\ 
+\ {i \over g}\ U_L\ \partial_\mu U_R^+~, 
\nonumber \\ && 
A_\mu^L\longrightarrow U_R\ A_\mu^L\ U_R^+\ 
+\ {i \over g}\ U_R\ \partial_\mu U_R^+~, 
\end{eqnarray} 
with the specific choice $U_L=U^{1 \over 2}$ and $U_R=U^{-{1 \over 2}}$, 
so that pseudoscalar mesons are gauge parameters. 
Now we can rewrite the Lagrangian (5) as a sum of three Lagrangians 
 
\begin{eqnarray} 
{\cal L}_0&=& \frac{F^2_\pi}{8} \mbox{tr} 
\left({\cal D}_\mu U {\cal D}_\mu U^+ \right)~, 
\end{eqnarray} 
 
\begin{eqnarray} 
{\cal L}_1&=& -{1\over 2} \mbox{tr} \left( (F_{\mu\nu}^L)^2+ 
(F_{\mu\nu}^R)^2\right) 
+\gamma\, \mbox{tr} \left( F_{\mu\nu}^L U F_{\mu\nu}^R U^+ \right)~, 
\end{eqnarray} 
 
\begin{eqnarray} 
{\cal L}_2&=&m_0^2 \mbox{tr} \left( (A_\mu^L)^2+ (A_\mu^R)^2\right) 
+B\,\mbox{tr}\left( A_\mu^L U A_\mu^R U^+ \right) 
\nonumber \\ && 
+C\,\mbox{tr}\left( A_\mu^L A^{R\mu} + A_\mu^R A^{L\mu}\right)~. 
\label{Lag3} 
\end{eqnarray} 
 
Note that we have added two gauge invariant terms to the  Lagrangian (17). 
The second term (with the coefficient B) in (17) plays an important role 
in the description of the width of the $\rho\rightarrow \pi\pi$~decay, 
and the third term (with the coefficient C) maintains the gauge invariance 
of the J/$\psi+\pi\rightarrow D^*+\bar{D}$~decay. Applying the 
substitutions (14) to the Lagrangian (5), we obtain 
 
\begin{eqnarray*} 
{\cal L}_0\rightarrow {\cal L'}_0=0~, 
\end{eqnarray*} 
\begin{eqnarray*} 
{\cal L}_1\rightarrow {\cal L'}_1= 
(\gamma - 1)\,\mbox{tr}\left(F_{\mu\nu}^V\,F^{\mu\nu V}\right)~, 
\end{eqnarray*} 
 
\begin{eqnarray} 
{\cal L}_2\rightarrow {\cal L'}_2&=& 
\frac {m_V^2} {2} \mbox{tr}\left(V_\mu^2\right) 
%\nonumber \\ 
- i{g_{V\varphi\varphi} \over 2}\mbox{tr}\left(V_\mu\left( \varphi 
\stackrel{\leftrightarrow}{\partial}_\mu 
\varphi \right)\right) 
\nonumber \\ 
&+& {8C \over F_\pi^2}\mbox{tr}\left(\left(V_\mu\varphi\right)^2- 
V_\mu^2\varphi^2\right)+{\cal L}(\varphi)~, 
\end{eqnarray} 
by~$m_V^2=2(B+2m_0^2+2C),~g_{V\varphi\varphi}=2(B-2C+2m_0^2)/(gF_\pi^2)$,~ 
where the vector mass and the vector-pseudoscalar-pseudoscalar 
coupling are defined. 
 
\subsection{J/$\psi$ absorption cross sections} 
 
\begin{figure}[htb] 
\parbox{0.45\textwidth}{ 
\center{\epsfig{figure=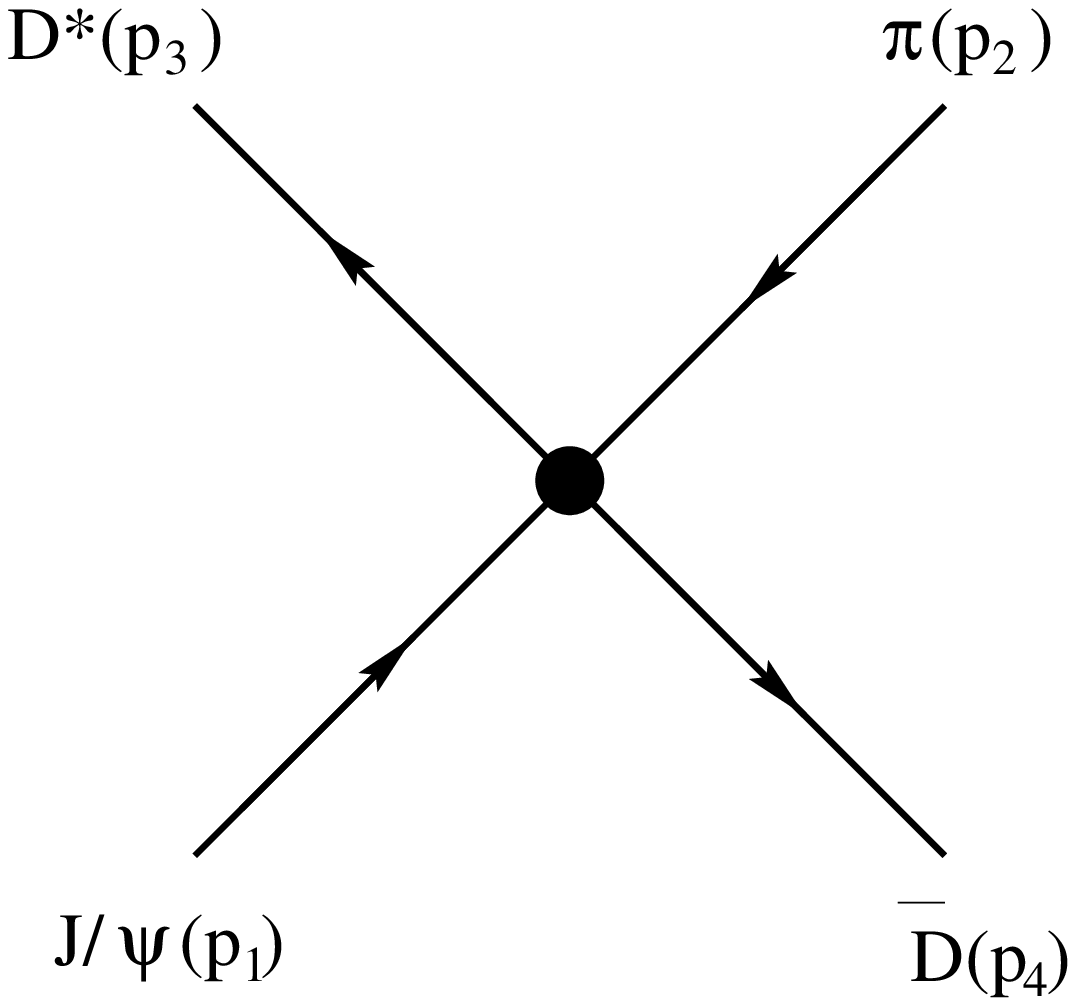,width=0.4\textwidth,angle=-90} 
\\{\bf I}} 
%\centerline{{\bf I}} 
} 
\parbox{0.45\textwidth}{ 
\center{\epsfig{figure=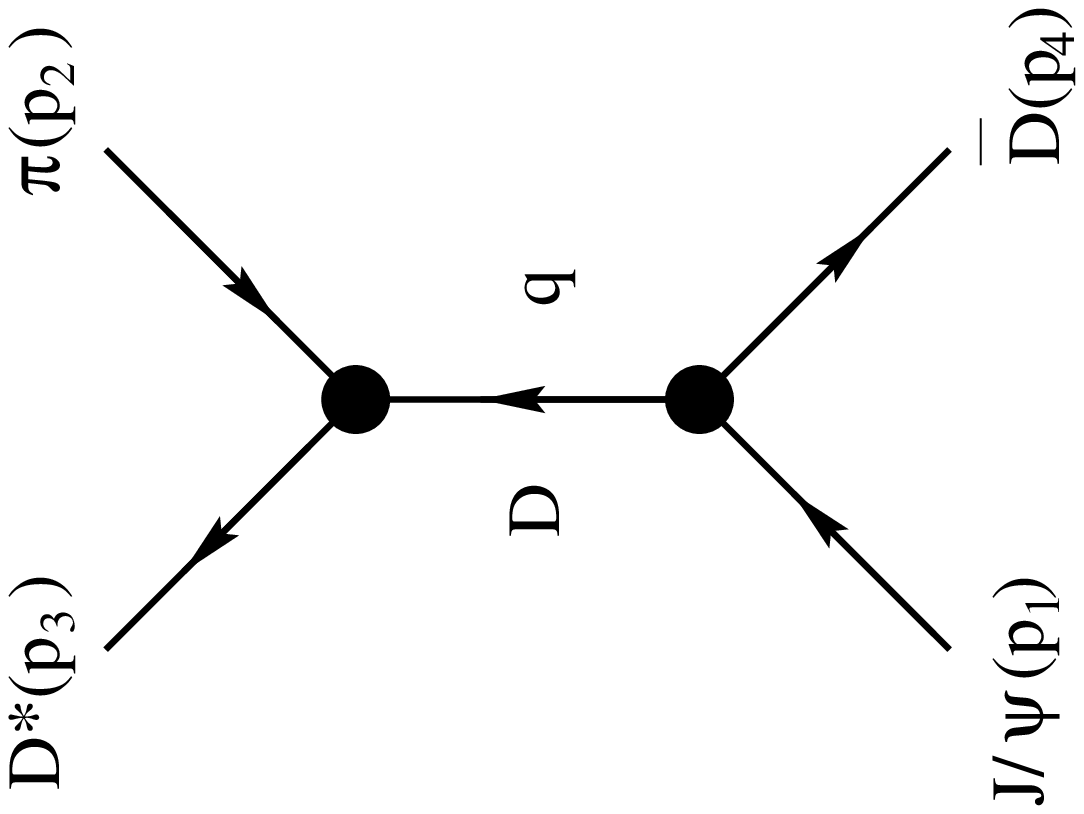,width=0.4\textwidth,angle=-90} 
\\{\bf III}} 
%\centerline{{\bf III}} 
} \vspace{5mm} 
%\parbox{0.45\textwidth}{ 
\center{\epsfig{figure=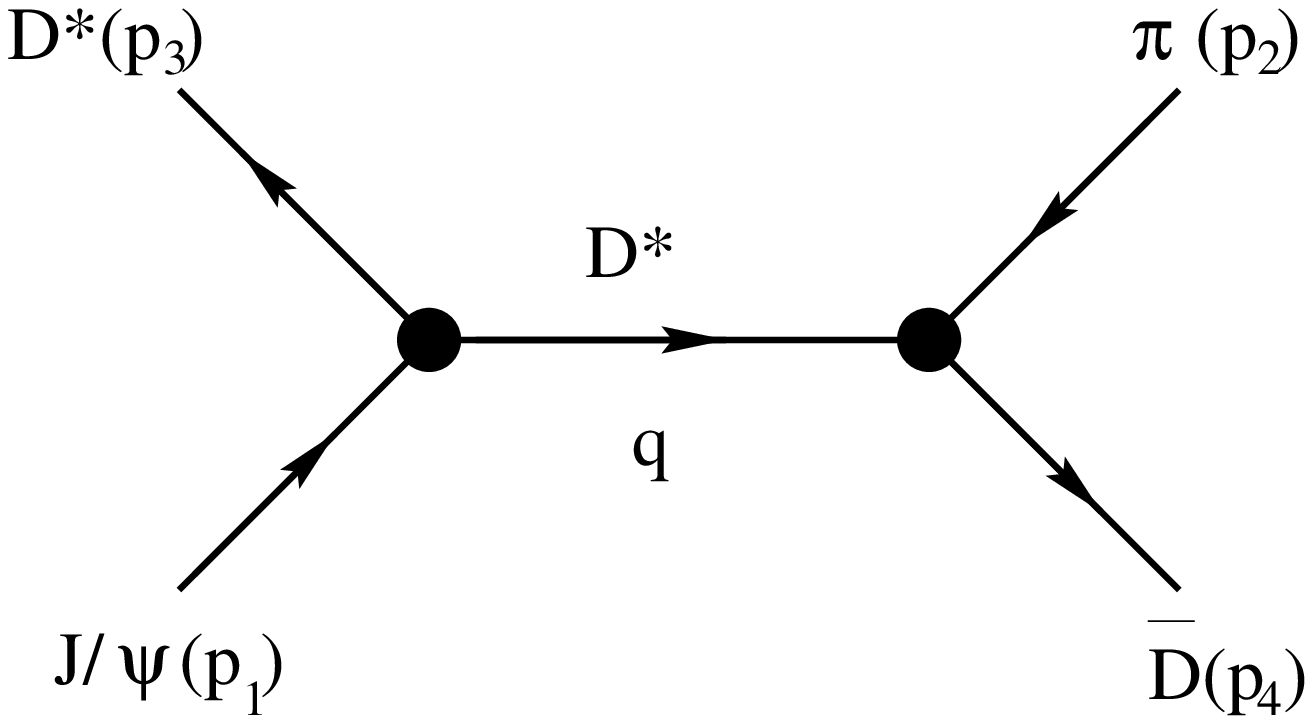,width=0.35\textwidth,angle=-90} 
\\{\bf II}} 
%\centerline{{\bf II}} 
%} 
\vspace{5mm} \caption{Diagrams for J/$\psi$ breakup by pion impact: 
J/$\psi+\pi\rightarrow D^*+\bar{D}$; \qquad 
I - contact term, II+III - D-meson exchange processes.} 
\label{fig1} 
\end{figure}

The above effective Lagrangian allows us to study the following processes for 
{J/$\psi$} absorption by $\pi$ and $\rho$ mesons: 
\begin{eqnarray} 
&& J/\psi + \pi \rightarrow D^* + \bar{D},\ 
J/\psi + \pi \rightarrow D + \bar{D}^*,\ \\ 
&& J/\psi + \rho \rightarrow D + \bar{D},\ 
J/\psi + \rho \rightarrow D^* + \bar{D^*}~. 
\end{eqnarray} 
The corresponding diagrams for the pionic processes, except the process 
{J/$\psi + \pi \longrightarrow D + \bar{D}^*$}, which has the same cross 
section as the process {J/$\psi + \pi \longrightarrow D^* + \bar{D}$}, are 
shown in Fig. \ref{fig1}. 

The full amplitude for the first process 
{J/$\psi + \pi \longrightarrow D^* + \bar{D}$},~without isospin factors and 
before summing and averaging over external spins,~is given by 
\begin{eqnarray} 
{\cal M}_1\equiv{\cal M}_1^{\mu\nu}\varepsilon_{1\mu}\varepsilon_{3\nu}&=&\left(\sum_{i=a,b,c} {\cal M}_{1i}^{\mu\nu}\right)\varepsilon_{1\mu}\varepsilon_{3\nu}~, 
\end{eqnarray} 
with 
\begin{eqnarray*} 
{\cal M}_{1a}^{\mu\nu}= -g_{\pi D D^*}g_{J/\psi DD}(-2p_2+p_3)^\nu\left({1\over u-m_D^2}\right)(p_2-p_3+p_4)^\mu ~, 
\end{eqnarray*} 
\begin{eqnarray*} 
{\cal M}_{1b}^{\mu\nu} &=& g_{\pi D D^*}g_{J/\psi D^* D^*} (-p_2 -p_4)^\alpha \left( {1 \over t-m_{D^*}^2}\right) 
\nonumber \\ 
&\times&\left[ g^{\alpha\beta} - {(p_2 -p_4)^\alpha (p_2 -p_4)^\beta \over m_{D^*}^2}\right] 
\nonumber \\ 
&\times&\left[(-p_1-p_3)^\beta g^{\mu\nu}+(-p_2+p_1+p_4)^\nu g^{\beta\mu}+(p_2+p_3-p_4)^\mu g^{\beta\nu}\right]\ , 
\end{eqnarray*} 
\begin{eqnarray} 
{\cal M}_{1c}^{\mu\nu} = -g_{J/\psi \pi D D^*}\  g^{\mu\nu} \ . 
\end{eqnarray} 

Similarly, the full amplitude for the second  process  J/$\psi + \rho \rightarrow D + \bar{D}$ is given by 
\begin{eqnarray} 
{\cal M}_2\equiv{\cal M}_2^{\mu\nu}\varepsilon_{1\mu}\varepsilon_{2\nu}&=&\left(\sum_{i=a,b,c} {\cal M}_{2i}^{\mu\nu}\right)\varepsilon_{1\mu}\varepsilon_{2\nu} 
\end{eqnarray} 
with 
\begin{eqnarray} 
\nonumber \\ && 
{\cal M}_{2a}^{\mu\nu} = -g_{\rho D D}g_{J/\psi DD}(p_2-2p_3)^\mu\left({1\over u-m_D^2}\right)(p_2-p_3+p_4)^\nu \ , 
\nonumber \\ && 
{\cal M}_{2b}^{\mu\nu} = -g_{\rho D D}g_{J/\psi DD}(-p_2+2p_4)^\mu\left({1\over t-m_D^2}\right)(-p_2-p_3+p_4)^\nu \ , 
\nonumber \\ && 
{\cal M}_{2c}^{\mu\nu} = g_{J/\psi \rho D D}\  g^{\mu\nu} \ . 
\end{eqnarray} 
 
For the third process $ J/\psi + \rho \rightarrow D^* + \bar{D^*}$,~the full amplitude is given by 
 
\begin{eqnarray} 
{\cal M}_3\equiv{\cal M}_3^{\mu\nu\lambda\omega}\varepsilon_{1\mu}\varepsilon_{2\nu}\varepsilon_{3\lambda}\varepsilon_{4\omega}&=&\left(\sum_{i=a,b,c} {\cal M}_{3i}^{\mu\nu\lambda\omega}\right)\varepsilon_{1\mu}\varepsilon_{2\nu}\varepsilon_{3\lambda}\varepsilon_{4\omega}\ , 
\end{eqnarray} 
\begin{eqnarray*} 
{\cal M}_{3a}^{\mu\nu\lambda\omega}&=& 
g_{\rho D^* D^*}g_{J/\psi D^* D^*}\left[(-p_2-p_3)^\alpha g^{\mu\lambda} 
+2p_2^\lambda g^{\alpha\mu}+2p_3^\mu g^{\alpha\lambda} \right] 
\nonumber \\ 
&\times& \left({1 \over u-m_{D^*}^2}\right) 
\left[g^{\alpha\beta}-{(p_2 -p_3)^\alpha (p_2-p_3)^\beta\over m_{D^*}^2}\right] 
\nonumber \\ 
&\times& \left[-2p_1^\omega g^{\beta\nu}+(p_1+p_4)^\beta 
g^{\nu\omega}-2p_4^\nu g^{\beta\omega}\right]~, 
\end{eqnarray*} 
\begin{eqnarray*} 
{\cal M}_{3b}^{\mu\nu\lambda\omega}&=& 
g_{\rho D^* D^*}g_{J/\psi D^* D^*} 
\left[-2p_2^\omega g^{\alpha\mu}+(p_2+p_4)^\alpha g^{\mu\omega}-2p_4^\mu g^{\alpha\omega} \right] 
\nonumber \\ 
&\times&\left({1 \over t-m_{D^*}^2}\right) 
\left[ g^{\alpha\beta} - {(p_2 -p_4)^\alpha (p_2 -p_4)^\beta \over m_{D^*}^2}\right]\nonumber \\ 
&\times&\left[(-p_1-p_3)^\beta g^{\nu\lambda}+2p_1^\lambda g^{\beta\nu}+2p_3^\nu g^{\beta\lambda}\right]\ , 
\end{eqnarray*} 
\begin{eqnarray} 
{\cal M}_{3c}^{\mu\nu\lambda\omega}=g_{J/\psi\rho D^* D^*}(g^{\mu\lambda} g^{\nu\omega} + g^{\mu\omega} g^{\nu\lambda}-2 g^{\mu\nu} g^{\lambda\omega})\ . 
\end{eqnarray} 
In the above,~$p_j$ denotes the momentum of particle~$j$. 
We choose the convention that particle $1$ and $2$ represent 
initial-state mesons while particles $3$ and $4$ represent 
final-state mesons on the left and right sides of 
the diagrams shown in Fig. 1,~respectively. 
The indices $\mu,~\nu,~\lambda$ and $\omega$~ denote 
the polarization components of external particles while 
the indices $\alpha$ and $\beta$ denote those of the exchanged mesons. 

After averaging (summing) over initial (final) spins and including isospin 
factors, the differential cross sections for the three processes are 
given by 

\begin{eqnarray} 
{d\sigma_1 \over dt}= 
{1 \over {96\pi s p_{i,c.m.}^2} }{\cal M}_1^{\mu\nu} 
{\cal M}_1^{* \mu ' \nu '} 
\left( g^{\mu\mu '} - { p_1^\mu p_1^{\mu '} \over m_1^2 }\right) 
\left( g^{\nu\nu '} - { p_3^\nu p_3^{\nu '} \over m_3^2 }\right), 
\end{eqnarray} 
 
\begin{eqnarray} 
{d\sigma_2 \over dt} 
={1 \over 288\pi s p_{i,c.m.}^2} 
{\cal M}_2^{\mu\nu}{\cal M^*}_2^{\mu ' \nu '} 
\left(g^{\mu\mu '}-{p_1^\mu p_1^{\mu '} \over m_1^2}\right) 
\left(g^{\nu\nu '} - {p_2^\nu p_2^{\nu '} \over m_2^2 }\right), 
\end{eqnarray} 
 
\begin{eqnarray} 
{d\sigma_3 \over dt}&=& 
{1 \over 288\pi s p_{i,c.m.}^2} 
{\cal M}_3^{\mu\nu\lambda\omega} 
{\cal M^*}_3^{\mu ' \nu '\lambda '\omega '} 
\left(g^{\mu\mu '}-{p_1^\mu p_1^{\mu '} \over m_1^2}\right) 
 \left(g^{\nu\nu '} - {p_2^\nu p_2^{\nu '} \over m_2^2 }\right) 
\nonumber \\ 
&\times& 
\left(g^{\lambda\lambda '} - {p_3^\lambda p_3^{\lambda '} \over m_3^2 } 
\right) 
\left(g^{\omega\omega '} - {p_4^\omega p_4^{\omega '} \over m_4^2 }\right), 
\end{eqnarray} 
with $s=(p_1+p_2)^2$,~and 
\begin{eqnarray} 
p_{i,c.m.}^2={[s-(m_1+m_2)^2][s-(m_1-m_2)^2]\over 4s}~, 
\end{eqnarray} 
is the squared momentum of initial-state mesons in the center-of-momentum 
(c.m.) frame. The definition of $p_{f,c.m.}$ for the final-state mesons 
is analogous with the replacement $(m_1,m_2)\to(m_3,m_4)$. 
 
\subsection{Hadronic Formfactors} 
 
\begin{figure}[htb] 
\centering{\epsfig{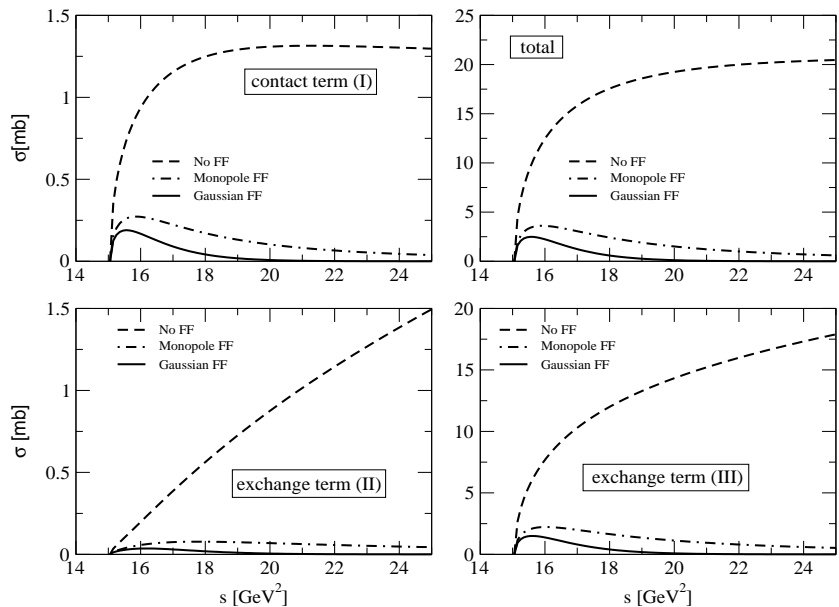}}
\vspace{5mm}
\caption{Upper right panel: total cross section for J/$\psi$ 
breakup by pion impact without formfactor (dashed line), with monopole 
type formfactor (dash-dotted line) and with Gaussian formfactor (solid 
line) as a function of the squared c.m. energy of initial - state mesons. 
The partial contributions from the diagrams I, II, and III of Fig. \ref{fig1} 
shown in the other panels.} 
\label{fig2} 
\end{figure}

The chiral Lagrangian aproach for J/$\psi$ breakup by light meson impact 
makes the assumption that mesons and meson-meson interaction vertices are 
local (four-momentum independent) objects. This neglect of the finite 
extension of mesons as quark-antiquark bound states has dramatic 
consequences: it leads to a monotonously rising behaviour of the cross 
sections for the corresponding processes, see the dashed lines in Fig. 
\ref{fig2}. 
This result, however, cannot be correct away from the reaction threshold 
where the tails of the mesonic wave functions determine the high-energy 
behaviour of the quark exchange (in the nonrelativistic formulation of 
\cite{mbq95,wsb00}) or quark loop (in the relativistic formulation 
\cite{b+00}) diagrams describing the microscopic processes underlying the 
J/$\psi$ breakup by meson impact. As long as the mesonic wave functions 
describe quark-antiquark bound states which have a finite extension in 
coordinate- and momentum space, the J/$\psi$ breakup cross section is 
expected to be decreasing above the reaction threshold and asymptotically 
small at high c.m. energies. This result of the quark model approaches to 
meson-meson interactions \cite{mbq95,wsb00,b+00} can be mimicked within 
chiral meson Lagrangian approaches by the use of formfactors at the 
interaction vertices \cite{lk00,hg00}. We will follow here the definitions 
of Ref. \cite{lk00}, where the formfactor of the four-point vertices of 
Fig.1, i.e. of the box diagram (I) as well as of the meson exchange 
diagrams (II, III) is taken as the product of the triangle diagram 
formfactors 
\begin{equation} 
\label{f3f3} F_4^i({\bf q}^2)=\left[F_3({\bf 
q}^2)\right]^{2}~~,~i=I,II,III~, 
\end{equation} 
with the squared three-momentum ${\bf q}^2$ given by the average value of 
the squared three-momentum transfers in the $t$ and $u$ channels 
\begin{equation} 
{\bf q}^2=\frac{1}{2}\left[({\bf p_1}-{\bf p_3})^2+ ({\bf p_1}-{\bf 
p_4})^2\right]_{\rm c.m.}= p^2_{i,{\rm c.m.}}+p^2_{f,{\rm c.m.}}~. 
\end{equation} 

For the triangle diagrams, we use formfactors with a momentum dependence 
in the monopole form ($M$) 
\begin{equation} 
F_3^M({\bf q}^2)=\frac{\Lambda^2}{\Lambda^2 +{\bf q}^2 }~, 
\end{equation} 
and in the Gaussian ($G$) form 
\begin{equation} 
F_3^G({\bf q}^2)=\exp(-{{\bf q}^2/\Lambda^2})~. 
\end{equation} 

At this point we have to add the comment that this choice, however, is 
obviously not supported by the underlying quark substructure diagrams that 
can provide a justification for the use of formfactors: While the triangle 
diagram is of third order in the wave functions so that the meson 
exchange diagrams are suppressed at large momentum transfer by six wave 
functions, the box diagram appears already at fourth order thus being 
less suppressed than suggested by the ansatz (\ref{f3f3}) of Ref. 
\cite{lk00}. 
A proper analysis of the resulting discrepancy is beyond the 
scope of the present paper and has to be deferred to future work.   
For the cross section of the three diagrams including the effect of 
hadronic formfactors, we multiply the bare expressions with the 
formfactors given above. The results are depicted in Fig. \ref{fig2}. In 
the last section, we want to discuss the results and their possible 
implications for phenomenological applications. 

\begin{figure}[tb]
\centering{\epsfig{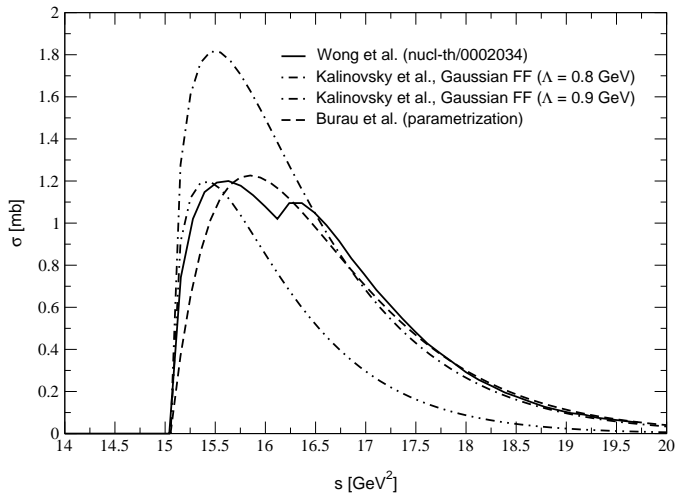}}
\vspace{5mm} 
\caption{Total J/$\psi$ breakup cross section with Gaussian formfactor 
($\Lambda=0.9$ GeV - dot-dashed line, $\Lambda=0.8$ GeV - dot-dot-dashed 
line) compared to the nonrelativistic quark exchange model (Wong et al. 
\protect\cite{wsb00} - solid line) and its parametrization by Burau et al. 
\protect\cite{bbk2} (dashed line).} 
\label{fig3} 
\end{figure} 

\subsection{Results and Discussion} 
 
The J/$\psi$ breakup cross section by $\pi$ and $\rho$ meson impact has 
been formulated within a chiral $U(4)$ Lagrangian approach. Numerical 
results have been obtained for the pion impact processes with the result 
that the D-meson exchange in the t-channel is the dominant subprocess 
contributing to the J/$\psi$ breakup. The use of formfactors at the 
meson-meson vertices is mandatory since otherwise the high-energy 
asymptotics of the processes with hadronic final states will be 
overestimated, see Fig. \ref{fig2}. From a comparison with results of a 
nonrelativistic potential model calculation, we can choose the shape of 
the formfactor to be Gaussian and fix the range $\Lambda=0.9$ GeV from 
the asymptotic high energy behaviour, see Fig. \ref{fig3}.
Within our semi-quantitative discussion, we do not attempt a high accuracy 
description of the nonrelativistic result which accounts for another 
final state D-meson pair, see \cite{mbq95,wsb00}. 

Finally, we want to present an exploration of the influence of a variation 
of the final state D-meson masses on the effective J/$\psi$ breakup cross 
section. Our motivation for considering mesonic states to be off their 
mass-shell is their compositeness which can become apparent in a 
high-temperature (and density) environment at the deconfinement/chiral 
restoration transition, when these states change their character 
qualitatively being resonant quark- antiquark scattering states in the 
quark plasma rather than on-shell mesonic bound states.

\begin{figure}[tb]
\centering{\epsfig{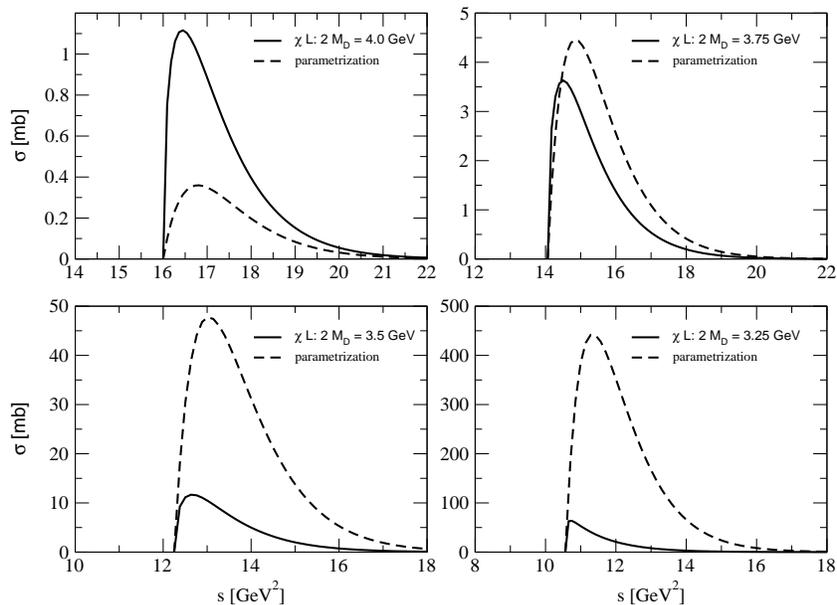}}
\vspace{5mm}
\caption{Total J/$\psi$ breakup cross section in the chiral Lagrangian 
model with Gaussian formfactor ($\Lambda=0.9$ GeV - solid line) compared 
to the parametrization of the nonrelativistic quark exchange model 
\protect\cite{wsb00} by Burau et al. \protect\cite{bbk2} (dashed line). 
The four panels illustrate the differences between both models when the 
final state masses $M_{D_1}=M_{D_2}=M_{D}$ are varied.}
\label{fig4} 
\end{figure}

The consequence 
of this Mott-transition from bound to resonant states for the J/$\psi$ 
breakup has been explored by Burau et al. \cite{bbk2,bbk}, see also these 
proceedings, using a fit formula for the D-mass dependence of the breakup 
cross section which shows a strong enhancement when the process becomes 
subthreshold ($M_D<M_D^{\rm vac}$). This behaviour is qualitatively 
approved within the present chiral U(4) Lagrangian + formfactor model 
although the subthreshold enhancement is more moderate, see Fig. 
\ref{fig4}. 
A more consistent description should include a quark model derivation of 
the formfactors for the meson-meson vertices and their possible medium 
dependence. Such an investigation is in progress. 
 
\subsection*{Acknowledgement} 
V.I. and Yu.K. acknowledge support from the Deutsche 
Forschungsgemeinschaft under grant no. 436 RUS 17/102/00 and from the 
Ministery for Education, Science and Culture of Mecklenburg-Western Pommerania.

%\end{document} 

\newpage

%% file: SOURCE/proceed.tex
\section*{\bf Mott Effect and Anomalous $J/\psi$ Suppression}
\addcontentsline{toc}{section}{\protect\numberline{}{Mott Effect and Anomalous $J/\psi$ Suppression \\ \mbox{\it G. Burau, D. Blaschke, Yu. Kalinovsky}}}
\begin{center}
\vspace*{2mm}
{Gerhard R.G. Burau$^a$, David B. Blaschke$^a$ and Yuri L. Kalinovsky$^b$}\\[0.3cm]
{\small\it $^a$Fachbereich Physik, Universit\"at Rostock, D-18051 Rostock,
Germany\\
$^b$Laboratory of Information Technologies, JINR, 141980 Dubna, 
Russia}
\end{center}
%

%\conference{Workshop on ``Quark Matter in Astro- and Particle Physics'', Rostock, Germany, 27-29 November 2000}

\begin{abstract}
We investigate the in-medium modification of the charmonium break-up 
process due to the Mott effect for light ($\pi$) and open-charm ($D$, $D^*$) 
quark-antiquark bound states at the chiral/deconfinement phase transition.
A model calculation for the process $J/\psi+\pi\to D+\bar D^* + h.c.$ 
is presented which demonstrates that the Mott effect for the D-mesons leads to 
a threshold effect in the thermal averaged break-up cross section. 
This effect is suggested as an explanation of the phenomenon of anomalous 
$J/\psi$ suppression in the CERN NA50 experiment.
\end{abstract}
 
\newcounter{eqn4}[equation]
\setcounter{equation}{-1}
\stepcounter{equation}

\newcounter{bild4}[figure]
\setcounter{figure}{-1}
\stepcounter{figure}

\newcounter{tabelle4}[table]
\setcounter{table}{-1}
\stepcounter{table}

%\newcounter{kapitel4}[section]
%\setcounter{section}{-1}
%\stepcounter{section}

\newcounter{unterkapitel4}[subsection]
\setcounter{subsection}{-1}
\stepcounter{subsection}

%
%\begin{keyword}
%J/$\psi$ suppression, bound state dissociation, Mott effect\\[2mm]
%\end{keyword}
%\end{frontmatter}
%
%--------------------start--------------------
\subsection{Introduction} 
Recent results of the CERN NA50 collaboration on anomalous $J/\psi$ 
suppression \cite{na50} in ultrarelativistic Pb-Pb collisions at 158 AGeV 
have renewed the quest for an explanation of the processes which may cause the 
rather sudden drop of the $J/\psi$ production cross section for transverse 
energies above $E_T\sim 40$ GeV in this experiment.
An effect like this was predicted as a signal for quark gluon plasma 
formation \cite{ms} due to screening of the $c \bar{c}$ interaction.
Soon after that it became clear that for temperatures and densities just above
the deconfinement transition the Mott effect for the $J/\psi$ does not occur
and that a kinetic process is required to dissolve the $J/\psi$ \cite{b} in a 
break-up process by impact of thermal photons \cite{Hansson:1988uk}, 
quarks \cite{rbs}, gluons \cite{Kharzeev:1994pz} 
or mesons \cite{Vogt:1988fj,mbq}.

In this paper, we suggest that at the chiral/deconfinement phase transition 
the charmonium break-up reaction cross sections are critically enhanced 
since the light and open-charm mesonic states of the dissociation processes 
become unbound (Mott effect) so that the reaction thresholds are effectively 
lowered.
We present a model calculation for the particular process 
$J/\psi+\pi\to D+\bar D^* + h.c.$ in a hot gas of resonant (unbound but 
correlated) quark-antiquark states in order to demonstrate that the Mott 
dissociation of the final states (D-mesons) at the chiral phase transition
leads to a threshold effect for the in-medium $J/\psi$ break-up cross section 
and thus the survival propability.

\subsection{In-medium modification of charmonium break-up cross sections}

The inverse lifetime of a charmonium state in a hot and dense many-particle 
system is given by the imaginary part of its selfenergy 
\begin{equation}
\tau^{-1}(p) = \Gamma(p) = \Sigma^>(p) - \Sigma^<(p)~.
\end{equation}
\vspace*{-0.5cm}
\begin{figure}[htb]
\centering{\
\epsfig{figure=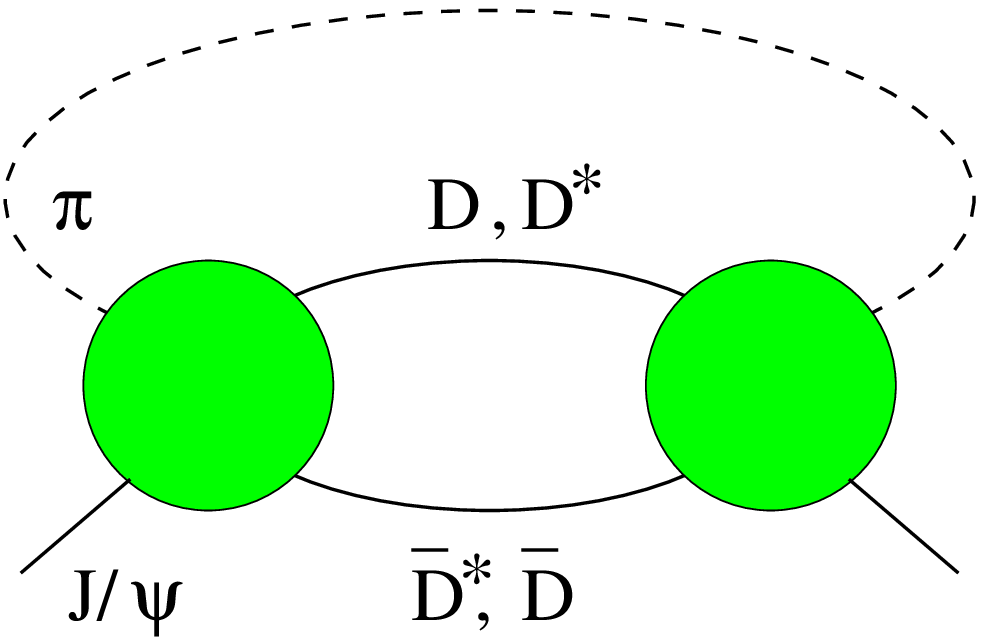,height=5.0cm,width=9.0cm}}
\vspace{0.5cm}
\caption{Diagrammatic representation of the complex selfenergy
for the $J/\psi$ due to break-up in (off-shell) $D$, $\bar{D}^*$ pairs by 
impact of (off-shell) pions from a hot medium.
\label{fig:cex}}
\end{figure}

In the Born collision approximation for the dominant process in a hot gas of 
pion-like correlations, as shown in Fig. \ref{fig:cex}, we have \cite{kb}
\begin{equation}
\Sigma^{\stackrel{>}{<}}(p) = \int\limits_{p'} 
\int\limits_{p_1}\int\limits_{p_2} (2\pi)^4 \delta_{p,p';p_1,p_2} 
\left|{\cal M}\right|^2 
G^{\stackrel{<}{>}}_{\pi}(p')~ 
G^{\stackrel{>}{<}}_{D_1}(p_1)~ 
G^{\stackrel{>}{<}}_{D_2}(p_2)~,
\end{equation}
where the thermal Green functions 
$G^{>}_{i}(p) = [1 + f_i(p)] A_i(p)$ and $G^{<}_{i}(p) = f_i(p) A_i(p)$ 
are defined by the spectral function $A_i(p)$ and the distribution function
$f_i(p)$ of the bosonic state $i$; with the notation 
$\delta_{p,p';p_1,p_2} = \delta(p + p' - p_1 - p_2)$, 
$\int\limits_{p} = \int \frac{d^4p}{(2\pi)^4}$. 

In the low density approximation for the final states ($f_{D_i}(p) \approx 0$),
one can safely neglect $\Sigma^<(p)$ so that 
\begin{equation}
\tau^{-1}(p) = \int\limits_{p'} \int\limits_{p_1} \int\limits_{p_2} 
(2\pi)^4 \delta_{p,p';p_1,p_2} 
\left| {\cal M} \right|^2 
f_\pi(p')~ A_\pi(p')~ A_{D_1}(p_1)~ A_{D_2}(p_2). 
\end{equation}
With the differential cross section 
\begin{equation}
\frac{d\sigma}{dt} = \frac{1}{16\pi} 
\frac{\left| {\cal M}(s,t) \right|^2}
{\lambda(s,M_\psi^2,s')}~, 
\end{equation}
using $s = (p + p')^2$, $t = (p - p_1)^2$, $s' = p'^2$ and 
$\lambda(s, M_{\psi}^2, s') = 
[s - (M_{\psi} + \sqrt{s'})^2][s - (M_{\psi} - \sqrt{s'})^2] 
= 4~ v^2_{\rm rel}~ [{\bf p}^2 + M_{\psi}^2][{\bf p'}^2 + s']$
one can show that the $J/\psi$ relaxation time in a hot pion as well as pionic 
resonance gas is given by 
\begin{equation}
\label{tau}
\tau^{-1}(p) = \int \frac{d^3{\bf p'}}{(2\pi)^3} 
\int ds' f_\pi({\bf p'},s')~ A_\pi(s') 
v_{\rm rel}~ \sigma^*(s)~, 
\end{equation}
where depending on the properties of the medium the pion spectral function 
(as well as the D-meson spectral functions) describes either $q \bar{q}$ 
bound states or resonant (off-shell) correlations. 
The in-medium break-up cross section is given by
\begin{equation}
\label{sig*}
\sigma^*(s) = \int ds_1~ ds_2~ 
A_{D_1}(s_1)~ A_{D_2}(s_2)~ \sigma(s; s_1, s_2)~. 
\end{equation}
Note that there are two kinds of medium effects due to (i) the spectral 
functions of the final states and (ii) the explicit medium dependence of the 
matrix element ${\cal M}$. 
In the following model calculation we will use the approximation 
$\sigma(s; s_1, s_2) \approx \sigma^{\rm vac}(s; s_1, s_2)$ 
justified by the locality of the transition matrix ${\cal M}$ which makes it 
rather inert against medium influence.

\subsection{Model calculation}

The quark exchange processes in meson-meson scattering can be calculated 
within the diagrammatic approach of Barnes and Swanson \cite{bs} which allows 
a generalization to finite temperatures in the thermodynamic Green function
technique \cite{br}. 
This technique has been applied to the calculation of $J/\psi$ break-up cross
sections by pion impact in \cite{mbq}. 
The approach has been extended to excited charmonia states and consideration 
of rho-meson impact recently \cite{wsb}.
The generic form of the resulting cross section (given a band of uncertainty)
can be fit to the form 
\begin{equation}
\label{sig0}
\sigma^{\rm vac}(s; M_{D_1}^2, M_{D_2}^2) = 
\sigma_0 \ln(s/s_0) \exp(-s/\lambda^2)\quad ,\quad s \ge s_0~,
\end{equation}
where $s_0 = (M_{D_1} + M_{D_2})^2$ is the threshold for the process to occur,
$\sigma_0 = 7.5 \cdot 10^{9}$ mb and $\lambda=0.9$ GeV.

Recently, the charmonium dissociation processes have been calculated also in
an effective Lagrangian approach \cite{mm,haglin}, but the freedom of choice 
for the formfactors of meson-meson vertices makes predictions uncertain. 
The development of a unifying approach on the basis of a relativistic confining
quark model is in progress \cite{kb99} and will remove this uncertainty by 
providing a derivation of the appropriate formfactors from the underlying 
quark substructure. 

The major modification of the charmonium break-up process which we expect at 
finite temperatures in a hot medium of strongly correlated quark-antiquark 
states comes from the Mott effect for the light as well as the open-charm 
mesons. 
At finite temperatures when the chiral symmetry in the light quark sector is 
restored, the continuum threshold for light-heavy quark pairs drops below the 
mass of the D-mesons so that they are no longer bound states constrained
to their mass shell, but become rather broad resonant correlations in the 
continuum. 
This Mott effect has been discussed within relativistic quark models 
\cite{b+93} for the light meson sector but can also be generalized to the case 
of heavy mesons 
\cite{gk}. Applying a confining quark model \cite{Blaschke:1998gk} we have 
obtained the critical temperatures $T^{\rm Mott}_{D^*}=130$ MeV, 
$T^{\rm Mott}_{D}=140$ MeV and $T^{\rm Mott}_{\pi}=150$ MeV \cite{kb99}. 

\begin{figure}[htb]
\centering{\
\epsfig{figure=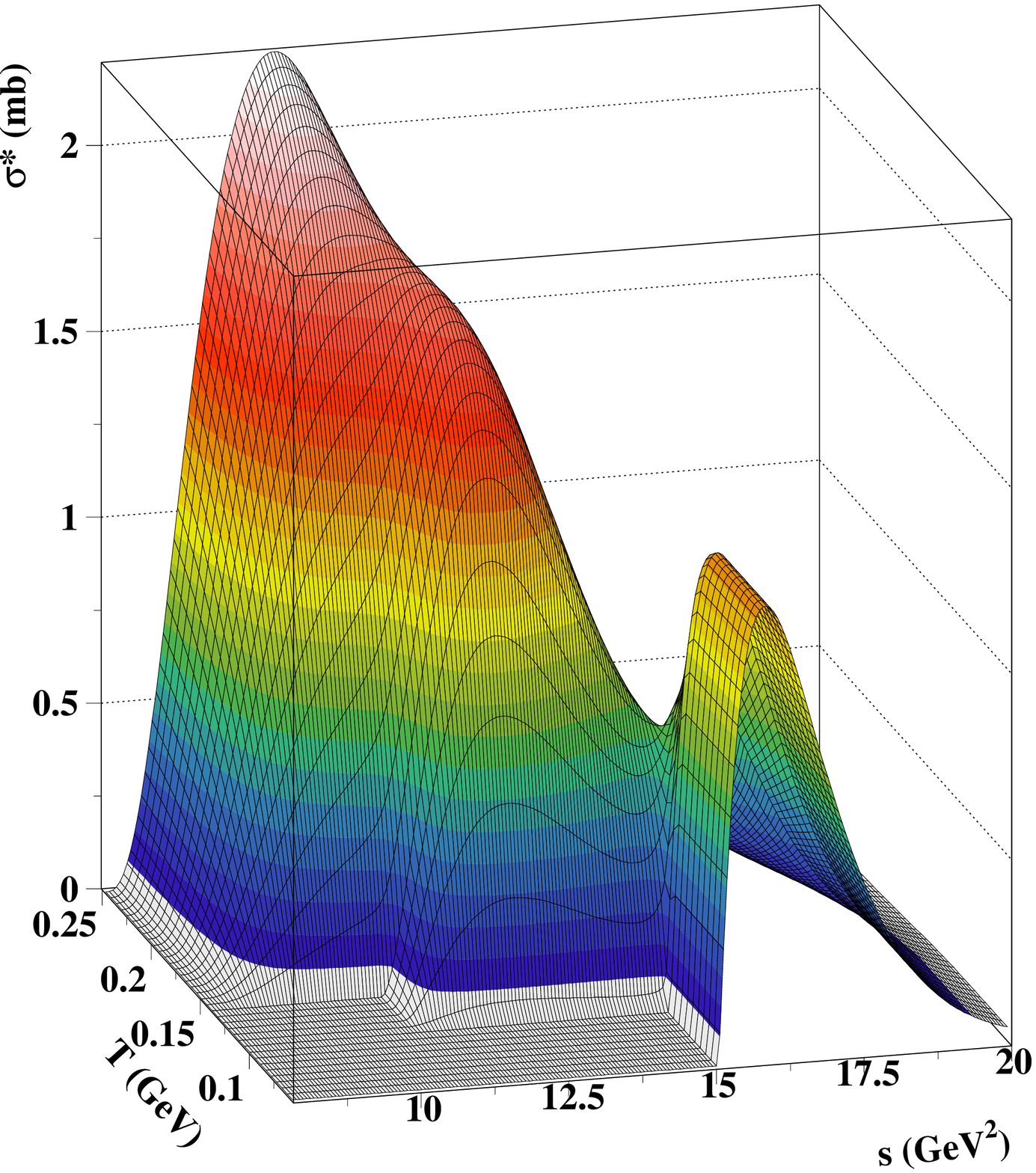,height=7cm,width=10.0cm}}
\vspace{-0.5cm}
\caption{Energy- and temperature dependent in-medium $J/\psi$ break-up cross 
section for pion impact. Thresholds occur at the Mott temperatures for the 
open-charm mesons: $T^{\rm Mott}_{D^*} = 130$ MeV, $T^{\rm Mott}_{D} = 140$ 
MeV.
\label{fig:sig_t}}
\end{figure}

In order to study the implications of the pion and $D$-meson Mott effect for 
the charmonium break-up we adopt here a Breit-Wigner form for the spectral 
functions 
\begin{eqnarray}
\label{ad}
A_i(s) &=& \frac{1}{\pi} 
\frac{\Gamma_i(T)~M_i(T)}{(s - M_i^2(T))^2 + \Gamma_i^2(T) M_i^2(T)}~,
\end{eqnarray} 
which in the limit of vanishing width $\Gamma_i(T)\to 0$ goes over into the 
delta function $\delta(s - M_i^2)$ for a bound state in the channel $i$.
The width of the pions as well as the $D$-mesons shall be modeled by a 
microscopic approach. For our exploratory calculation, we adopt here
\begin{equation}
\Gamma_{\pi,D}(T) = c~ (T - T^{\rm Mott}_{\pi,D})~ \Theta(T - T^{\rm Mott}_{\pi,D})~,
\end{equation}
where the coefficient $c = 2.67$ is assumed to be universal for the pions and 
$D$-mesons and it is obtained from a fit to the pion width above the pion Mott 
temperature, see \cite{b+95}. For the meson masses we have 
$M_{\pi,D}(T) = M_{\pi,D} + 0.75~ \Gamma_{\pi,D}(T)$.
The result for the in-medium $J/\psi$ break-up cross section (\ref{sig*}) is 
shown in Fig. \ref{fig:sig_t}.

%\vspace*{-0.5cm}

With $M_{D^*}=2.01$ GeV and $M_{\bar D}=1.87$ GeV follows for the threshold
$s_0 = 15.05$ GeV$^2$. At a temperature $T = 140$ MeV, where the D-meson can 
still be considered as a true bound state, the $D^*$-meson has already entered 
the continuum and is a resonance with a half width of about 80 MeV.
Due to the Mott effect for the open-charm mesons (final states), 
the charmonium dissociation process become "subthreshold" ones and their 
cross sections which are peaked at threshold rise and spread to lower onset 
with cms energy. 
This is expected to enhance strongly the rate for the charmonium dissociation 
processes in a hot resonance gas. 
To simplify matters we can assume a uniform Mott temperature for the D- and 
D$^*$-meson amounting to $T_{\pi}^{\rm Mott}$. In this scenario we find 
of course only one threshold in the temperature behavior of the in-medium 
$J/\psi$ break-up cross section (\ref{sig*}).

\subsection{$J/\psi$ dissociation in a hot ``pion'' gas}

We calculate the inverse relaxation time for a $J/\psi$ at rest in a hot gas 
of pions (below $T_{\pi}^{\rm Mott}$) and pion-like correlations 
(above $T_{\pi}^{\rm Mott}$) by specifying Eq. (\ref{tau}) for this simplified 
case 
\begin{eqnarray}
\tau^{-1}(T) &=& \int\frac{d^3{\bf p'}}{(2\pi)^3}
\int ds_{\pi} A_{\pi}(s_{\pi}) 
f_\pi({\bf p'}, s_{\pi}; T) \frac{|{\bf p'}|}{E_\pi({\bf p'}, s_{\pi})} 
\sigma^*(s)\\ \nonumber \\
&=& <\sigma^* v_{\rm rel}>n_\pi(T)~, 
\end{eqnarray}
with the dispersion relation 
$E_\pi({\bf p'}, s_{\pi}) = \sqrt{{\bf p'}^2 + s_{\pi}}$, 
the thermal Bose distribution function 
$f_\pi({\bf p'}, s_{\pi}; T) = 3 \left\{\exp[E_{\pi}({\bf p'}, s_{\pi})/T] - 1\right\}^{-1}$ 
and the particle density $n_{\pi}(T)$ for the ``pions''. The cms energy of the 
``pion'' impact on a $J/\psi$ at rest is $s({\bf p'}; s_{\pi}) = s_{\pi} + M_{\psi}^2 + 2M_{\psi}E_\pi({\bf p'}, s_{\pi})$. 

\begin{figure}[htb]
\centering{\epsfig{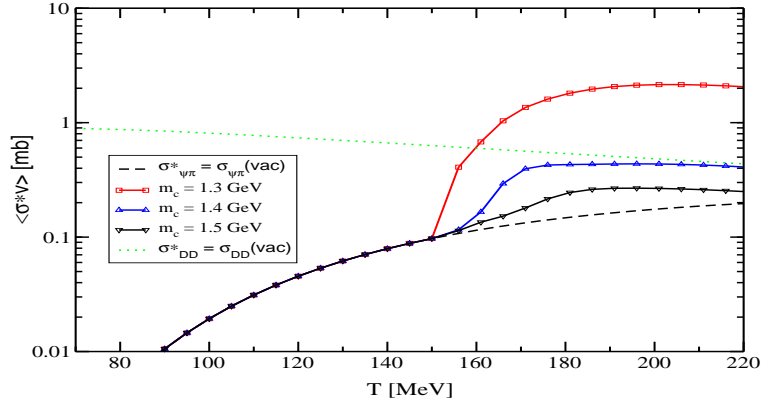}}
\vspace{0.5cm}
\caption{Temperature dependence of the thermal averaged in-medium $J/\psi$ 
break-up cross section for different charm quark masses with off-shell effects 
for $\pi$, $D$ and $D^*$ above $T^{Mott} = T_c = 150$ MeV. 
The dotted line shows the thermal averaged cross section for the back process 
($D D^*$ annihilation) without Mott effect.
\label{fig:sigv}}
\end{figure}

The result for the temperature dependence of the thermal averaged $J/\psi$ 
break-up cross section $<\sigma^* v_{\rm rel}>$ is shown in 
Fig. \ref{fig:sigv}. 
This quantity has to be compared to the nuclear absorption cross section for 
the $J/\psi$ of about 3 mb which has been extracted from charmonium 
suppression data in p-A collisions \cite{hhk}.
%\vspace{0.5cm}
%

It is remarkable that it is practically negligible below the Mott temperature 
$T^{\rm Mott} = 150$ MeV for the open-charm mesons but comparable to the 
nuclear absorption cross section above the chiral/deconfinement temperature of
$T_{\rm crit}\approx 150$ MeV. It is obvious that the transition from D-meson 
bound states to unbound light-heavy quark correlations is responsible for the 
strong increase by one to two orders of magnitude. 
Note that in this calculation the Mott effect for the pion (initial state) 
above $T^{\rm Mott}_{\pi}$ has been also included, but does not alter the 
result obtained previously \cite{Blaschke:2000er} in a calculation neglecting 
this effect.

Therefore we expect the in-medium enhanced charmonium dissociation process  
to be sufficiently effective to destroy the charmonium state on its way 
through the hot fireball of the heavy-ion collision and to provide an 
explanation of the observed anomalous $J/\psi$ suppression phenomenon 
\cite{na50}. 
A detailed comparison with the recent data from the NA50 collaboration
requires a model for the heavy-ion collision. 
The effective in-medium break-up cross section for the $J/\psi$ derived in 
this work provides an input for all calculations which use this quantity, 
e.g. Glauber-type models 
\cite{Blaschke:2000er,wong,mb,Capella:2000zp,Blaizot:2000ev,Hufner:2000nv}, 
more detailed calculations based on a parton cascade model \cite{Bass:1999vz} 
or molecular dynamics \cite{Cassing:2000bj}.

\subsection{Summary and Outlook}

In this contribution we have presented an approach to charmonium break-up 
in a hot and dense medium which is applicable in the vicinity of the 
chi\-ral/de\-con\-fine\-ment phase transition where mesonic bound states get 
dissolved in a Mott-type transition and should be described as resonant 
correlations in the quark plasma. This description can be achieved using the 
concept of the spectral function which can be obtained from relativistic quark 
models in a systematic way. The result of an exploratory calculation employing 
a temperature-dependent Breit-Wigner spectral function for light and 
open-charm mesons presented in this work has demonstrated that heavy-flavor 
dissociation processes are critically enhanced at the QCD phase transition and 
could represent the physical mechanism behind the phenomenon of anomalous 
$J/\psi$ suppression.

In subsequent work we will relax systematically approximations which have been 
made in the present paper and improve inputs which have been used. 
In particular, we will investigate the off-shell behaviour of the charmonium 
break-up cross section in the vacuum (\ref{sig0}) and calculate the spectral 
functions (\ref{ad}) at finite temperature within a relativistic quark model.
Dyson-Schwinger equations provide a nonperturbative, field-\-theo\-retical 
approach which has recently been applied also to heavy-meson observables 
\cite{ikr} and have proven successful for finite-temperature generalization 
\cite{bbkr,Roberts:2000aa}.
Further intermediate open-charm states can be considered; the states in the
dense environment should include rho mesons and nucleons besides of the 
pions which all can be treated as off-shell quark correlations at the QCD 
phase transition. 

In future experiments at LHC the charm distribution in the created fireball 
may be not negligible so that the approximation $f_{D_i}(p) \approx 0$ has to 
be relaxed. In this case, one has to include the gain process ($D \bar{D}$ 
annihilation) encoded in the $\Sigma^{<}$ function. 
In comparision to previous investigations \cite{ko,pbm} the present quantum 
kinetic treatment contains Bose enhancement factors in the $G^{>}_{i}$ 
functions which modify the charm equilibration process.

\subsection*{Acknowledgements}

This work has been supported by the Heisenberg-Landau program for scientific 
collaborations between Germany and the JINR Dubna and by the DFG 
Graduiertenkolleg ``Stark korrelierte Vielteilchensysteme'' at the University 
of Rostock. We thank T. Barnes, P. Braun-Munzinger, J. H\"ufner, M.A. Ivanov, 
C.D. Roberts, G. R\"opke, S.M. Schmidt and P.C. Tandy for their discussions
and stimulating interest in our work.

%______________________________ References ______________________________

%%----------------------------------------
%%\end{document}
\newpage

%% file: SOURCE/procAPP.tex
\newcounter{eqn7}[equation]
\setcounter{equation}{-1}
\stepcounter{equation}

\newcounter{bild7}[figure]
\setcounter{figure}{-1}
\stepcounter{figure}

\newcounter{tabelle7}[table]
\setcounter{table}{-1}
\stepcounter{table}

\newcounter{unterkapitel7}[subsection]
\setcounter{subsection}{-1}
\stepcounter{subsection}

\section*{\bf Few-Body Correlations in Fermi Systems}
\addcontentsline{toc}{section}{\protect\numberline{}{Few-Body Correlations in Fermi Systems \\ \mbox{\it M. Beyer}}}
\begin{center}
\vspace*{2mm}
{M. Beyer}\\[0.3cm]
{\small\it FB Physik, University of Rostock, 18051 Rostock, Germany}
\end{center}
\vspace{1ex}
Interacting quantum systems with strong or long-range
interactions exhibit quite a rich phase structure.  Cluster formation
and superconductivity are examples.  These phenomena are also expected
in the astrophysical context, e.g., during the formation or in the
structure of neutron stars. To describe these phenomena a proper
treatment has to go beyond the simple picture of noninteracting
quasiparticles. An appealing formalism for a systematic approach is
provided by the framework of Dyson equations.  Within an equal
(imaginary) time formalism Dyson equations can be derived for an
arbitrary large cluster embedded in a medium~\cite{duk98}. For
practical use and the sake of simplicity the medium is treated as
uncorrelated to derive the respective $n$-body cluster Green
functions. Further, we neglect ``backward'' propagating particles, so
the Fock spaces for different number of particles $n$ are
disconnected. This way it is possible to derive effective in-medium
$n$-body equations that can be solved rigorously with few-body
techniques~\cite{Beyer:1996rx,Beyer:1997sf,Beyer:1999tm,Beyer:1999zx,Kuhrts:2000jz,Beyer:1999xv,Beyer:2000ds,Kuhrts:2000zs,alt67,san74,alt72,san75}.

These resulting two-, three-, and four-body equations elaborated here
include the dominant medium effects in a systematic way. These are the
self energy corrections for masses and the Pauli blocking that in turn
leads to a change of binding energies, viz. change of the masses of
clusters, and change of reaction rates.  Further, within this approach
the critical temperatures for condensation (of bosons containing two or
four particles) are calculated.

\begin{figure}[htb]
\begin{center}
\psfig{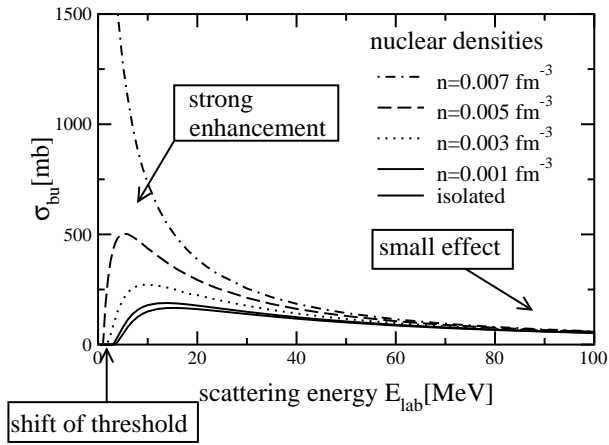}
\vspace*{0.3cm}
\caption{\label{fig:Across} 
    Break-up cross section for different densities of nuclear matter for temperature $T=10$ MeV}
\end{center}
\end{figure}

\begin{figure}[hbt]
\begin{center}
\epsfig{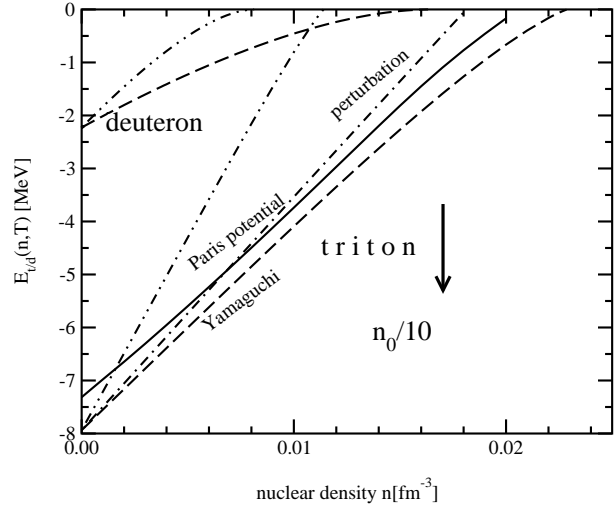}
\vspace*{0.2cm}
\caption{\label{fig:Atri0} 
  Triton binding energy as a function of nuclear matter density.
  Dashed-dot-dot $T=10$ MeV, other $T=20$ MeV.}
\end{center}
\end{figure}

Defining $H_0=\sum_{i=1}^n \varepsilon_i$ with the quasi-particle self
energy 
\begin{equation}
\varepsilon_1 = k^2_1/2m_1+\sum_{2}V_2(12,\widetilde{12})f_2
\end{equation}
 and the Fermi function $f_1\equiv f(\varepsilon_1) =
1/(e^{\gb(\varepsilon_1 - \mu)}+1),$ the $n$-particle cluster
resolvent $G_0$ is
\begin{equation}
G_0(z) = (z-  H_0)^{-1}
\;{N} \equiv R_0(z)\;{ N}.
\end{equation}
Here $G_0$, $H_0$, and $N$ are matrices in $n$ particle space and $z$ denotes the Matsubara frequency~\cite{fet71}. The
Pauli-blocking factors for $n$-particles are
\begin{equation}
N=\bar f_1\bar f_2 \dots \bar f_n
\pm f_1f_2\dots f_n,\qquad\bar f=1-f
\end{equation}
{Note: $NR_0=R_0N$.}  Defining the effective potential $V\equiv
\sum_{\mathrm{pairs}\;\ga} { N_2^{\ga}}V_2^{\ga}$ the full
and the channel resolvents are
\begin{eqnarray}
G(z)&=&(z-H_0- V)^{-1}{N}
\equiv R(z){N},\\
G_\ga(z)&=&(z-H_0- { N_2^{\ga}}V_2^{\ga})^{-1}{ N}
\equiv R_\ga(z){ N},
\end{eqnarray}
Note that $V^\dagger\neq V$ and $R(z)N\neq NR(z)$.
For the scattering problem it is convenient to define the in-medium
AGS operator $U_{\gb\ga}(z)$
\begin{equation}
R(z)=\gd_{\ga\gb}R_\gb(z) + R_\gb(z) { U_{\gb\ga}(z)} R_\ga(z)
\end{equation}
that after some algebra leads to the in-medium AGS equation 
\begin{equation}
 U_{\gb\ga}(z)=\bar\gd_{\gb\ga}R_0(z)^{-1}+\sum_\gc
\bar\gd_{\gb\gc} 
N_2^\gc T_2^\gc(z) R_0(z) U_{\gc\ga}(z),
\end{equation}
where $\bar\gd_{\gb\ga}=1-\gd_{\gb\ga}$.  The square of this
$t$-operator is directly linked to the differential cross section for
the scattering process $\ga\rightarrow\gb$. The driving kernel
consists
 of the two-body $t$-matrix derived in the same
formalism, however given earlier and known as Feynman-Galitskii
equation\cite{fet71}
\begin{eqnarray}\nonumber
T_2^\gamma(z) &=&   V_2^\gamma + 
 V_2^\gamma { N^\gc_2}R_0(z)  T_2^\gamma(z)\nonumber.
\end{eqnarray}
A numerical solution using a coupled Yamaguchi potential has been
given in Ref.~\cite{Beyer:1996rx}. For a temperature $T=10$ MeV and the
three-body system at rest in the medium results a given in
Fig.\ref{fig:Across}
For the bound state problem it is convenient to introduce form
factors
\begin{equation}
|F_\gb\rangle=\sum_\gc\bar\gd_{\gb\gc} { N_2^\gc}  V_2^\gc 
|\psi_{B_3}\rangle.
\end{equation}
Since the potential is nonsymmetric right and left eigenvectors are
different, although the bound state energies are the same,
\begin{eqnarray}
|F_\ga\rangle
&=&\sum_\gb \bar\gd_{\ga\gb}  
{ N_2^\gb} T_2^\gb(B_3) R_{0}(B_3)|F_\gb\rangle,\\
|\tilde F_\ga\rangle
&=&\sum_\gb \bar\gd_{\ga\gb} T_2^\gb(B_3) { N_2^\gb} R_{0}(B_3)
|\tilde F_\gb\rangle.
\end{eqnarray}
The binding energy depends on $\mu,\;T,\;P_{\mathrm c.m.}$. Results
for $P_{\mathrm c.m.}=0$ are given in Fig.\ref{fig:Atri0} for
different potentials and temperatures. Note that the dependence on
density is rather similar for two different potentials studied,
although the binding energies for the isolated triton differs by 10\%
the Mott density is practically at the same place once the binding
energies are renormalized to each other. For helium the Mott density
is smaller due to the Coulomb force, however for asymmetric nuclear
matter, e.g. $N_p/N_n\simeq 0.72$ (for the $^{129}$Xe$+~^{119}$Sn
reaction) this effect is compensated~\cite{mat00}. The dependence of
the Mott effect on the momentum is given in Ref.~\cite{Beyer:1999zx}.

We now turn to the four-body problem in matter. In addition to having
different channels as for the three body system now the channels
appear in different partitions that makes the four-body problem even
more involved. The partitions of the four-body clusters are denoted by
$\rho,\tau,\sigma,\dots$, e.g., $\rho=(123)(4),(234)(1), \dots$ for
$3+1$-type partitions, or $\rho=(12)(43), (23)(41), \dots$ for
$2+2$-type partitions.  The two-body sub-channels are denoted by pair
indices $\ga,\gb,\gc,\dots$, e.g. pairs $(12)$, $(24)$,\dots The two-
and three-body $t$-matrices have to be defined with respect to the
partitions that leads to additional indices. 
The four-body in-medium homogeneous AGS equation are defined for the
form factors
\begin{equation}
        |{\cal F}_\gb^\gs\rangle = \sum_\tau \bar\gd_{\gs\tau}
        \sum_\ga \bar\gd^\tau_{\gb\ga} R_0^{-1}(B_4) |\psi_\ga\rangle,
\end{equation}
where $\bar\gd^\tau_{\gb\ga}=\bar\gd_{\gb\ga}$, if
$\gb,\ga\subset\tau$ and $\bar\gd^\rho_{\gb\ga}=0$ otherwise and
$|\psi_\ga\rangle$ is the $\alpha$-üparticle wave function. They
read~\cite{Beyer:2000ds}
\begin{equation}
|{\cal F}^\gs_\gb\rangle=\sum_{\tau\gc} \bar\gd_{\gs\tau}
U^\tau_{\gb\gc}(B_4)  R_0(B_4) { N_2^\gc} 
T_2^\gc(B_4) 
R_0(B_4) |{\cal F}^\tau_\gc\rangle,
\end{equation}
where $\ga\subset\gs,\gc\subset\tau$.  A numerical solution of this
equation is rather complex. In order to reduce computational time we
introduce a energy dependent pole expansion (EDPE) that has been
proven useful in many application involving the $\alpha$-particle and
is accurate enough for the present purpose.  However, we have to
generalize the original version of the EDPE because of different right
and left eigenvectors. Details will be omitted here,
see~\cite{Beyer:2000ds}

In the {\em two-body} sub-system the EDPE reads
\begin{eqnarray}
T_\gc(z) &\simeq &\sum_n
|\tilde\Gamma_{\gc n}(z)\rangle t_{\gc n}(z)\langle \Gamma_{\gc n}(z)|
\simeq\sum_n |\tilde g_{\gc n}\rangle t_{\gc n}(z)\langle g_{\gc n}|
\nonumber\\
&=&\sum_n
{  N_2^\gc} |g_{\gc n}\rangle t_{\gc n}(z)\langle g_{\gc n}|.
\label{eqn:Tsep2}
\end{eqnarray}
where we have chosen a Yamaguchi ansatz for the form factors for
simplicity.  Inserting this ansatz into the Feynman-Galitskii equation
determines the propagator $t_{\gc n}(z)$. In the {\em three-body}
sub-system the EDPE expansion reads
\begin{equation}
\langle g_{\gb m}(z)| R_0(z) U^\tau_{\gb\gc}(z)R_0(z)| 
\tilde g_{\gc n}(z)\rangle
\simeq \sum_{t,\mu\nu} |\tilde\Gamma^{\tau t, \mu}_{\gb m}(z)\rangle
t^{\tau t}_{\mu\nu}(z)\langle \Gamma^{\tau t, \nu}_{\gc n}(z)|.
\label{eqn:pole3}
\end{equation}
with the three-body EDPE functions 
\begin{equation}
|\tilde\Gamma^{\tau t, \mu}_{\gb m}(z)\rangle
= \langle g_{\ga n}|R_0(z)| \tilde g_{\gb m}\rangle
t_{\gb m}(B_3) |\tilde\Gamma^{\tau t, \mu}_{\gb m}\rangle,
\end{equation}
that we get from solving the proper Sturmian equations
\begin{eqnarray}
\eta_{t,\mu}|\tilde\Gamma^{\tau t, \mu}_{\ga n}\rangle&=&
\sum_{\gb m}
 \langle g_{\ga n}|R_0(B_3)| \tilde g_{\gb m}\rangle
t_{\gb m}(B_3)|\tilde\Gamma^{\tau t, \mu}_{\gb m}\rangle
\label{eqn:sturm}\\
\eta_{t,\mu}|\Gamma^{\tau t, \mu}_{\ga n}\rangle&=&
\sum_{\gb m}
 \langle \tilde g_{\ga n}|R_0(B_3)|  g_{\gb m}\rangle
t_{\gb m}(B_3)|\Gamma^{\tau t, \mu}_{\gb m}\rangle
\end{eqnarray}
Inserting everything into the homogeneous AGS equations allows us to
redefine the form factors that are now operators in the coordinates of
the $2+2$ or $3+1$ system, respectively
\begin{equation} 
|\Gamma^{\gs s}_\mu\rangle 
= \sum_{\gb m}\langle\Gamma^{\gs s}_{\gb m,\nu}(B_4)|t_{\gb m}(B_4)
\langle g_{\gb m}(B_4)| R_0(B_4)|{\cal F}^\gs_\gb\rangle
\end{equation} 
and therefore the final homogeneous equation
\begin{equation}
|\Gamma^{\gs s}_\mu\rangle =
\sum_{\tau t}\sum_{\nu\gk} \sum_{\gc n} \bar\gd_{\gs\tau}
\langle\Gamma^{\gs s, \nu}_{\gc n}(B_4)|t_{\gc n}(B_4)
|\tilde\Gamma^{\gs s, \mu}_{\gc n}(B_4)\rangle\;
t^{\tau t}_{\mu\gk}(B_4)\;|\Gamma^{\tau t}_\gk\rangle,
\label{eqn:coup4}
\end{equation}
is an effective one-body equation with
and effective potential ${\cal V}$ and an effective resolvent
${\cal G}_0$:
\begin{eqnarray}
{\cal V}^{\gs s,\tau t}_{\mu\nu}(z)
&= &\sum_{\gc n} \bar\gd_{\gs\tau}
\langle\Gamma^{\gs s, \mu}_{\gc n}(z)|t_{\gc n}(z)
|\tilde\Gamma^{\gs s, \nu}_{\gc n}(z)\rangle,
\label{eqn:pot4}\\
{\cal G}^{\gs s,\tau t}_{\mu\nu,0}(z)&=& t^{\tau t}_{\mu\nu}(z).
\end{eqnarray}

\begin{figure}[t]
\begin{center}
\psfig{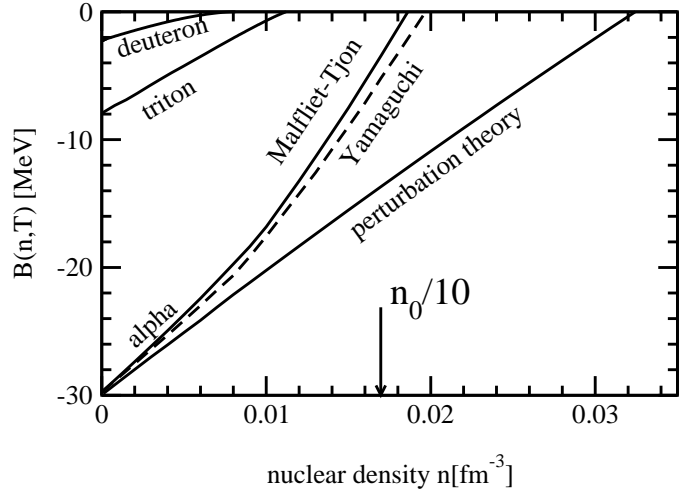}
\caption{\label{fig:amottC} Energy dependence of the binding energy of the
  $\alpha$-particle}
\end{center}
\end{figure}
The binding energies of the two-, three-, and four-body systems are
shown in Fig.\ref{fig:amottC} for a temperature of $T=10$ MeV and a
c.m. momentum of the respective cluster of $P_{\mathrm c.m.}=0$. The
$B=0$ line reflects the respective continuum. Investigating the zeros
of the two-body Joost function the quasi deuteron survives as an
anti-bound state (not resonance) with increasing densities, viz. for
energies above the continuum\cite{bey01}. The fate of the triton and
of the $\alpha$ particle for $B>0$ still needs to investigated as well
as a possible appearance of Efimov states related to $B\rightarrow 0$
of the sub-system. Since the Efimov states are 'excited' states, e.g.
for the three-body system close to the $2+1$ threshold, their blocking
may be smaller since the wave functions contain higher momentum
components. 
Note that the slope of the binding energies as a function of densities
for the larger clusters is also larger. This is a clear indication
that the masses of the clusters change with increasing density. The
Mott density of the $\alpha$ particle appears at
$$
B_4(n_{\mathrm Mott}=0.19 {\mathrm fm}^{-3},T=10\mbox{
  MeV},P_{\mathrm c.m.}=0)=0
$$
\begin{figure}[t]
\begin{center}
\psfig{figure=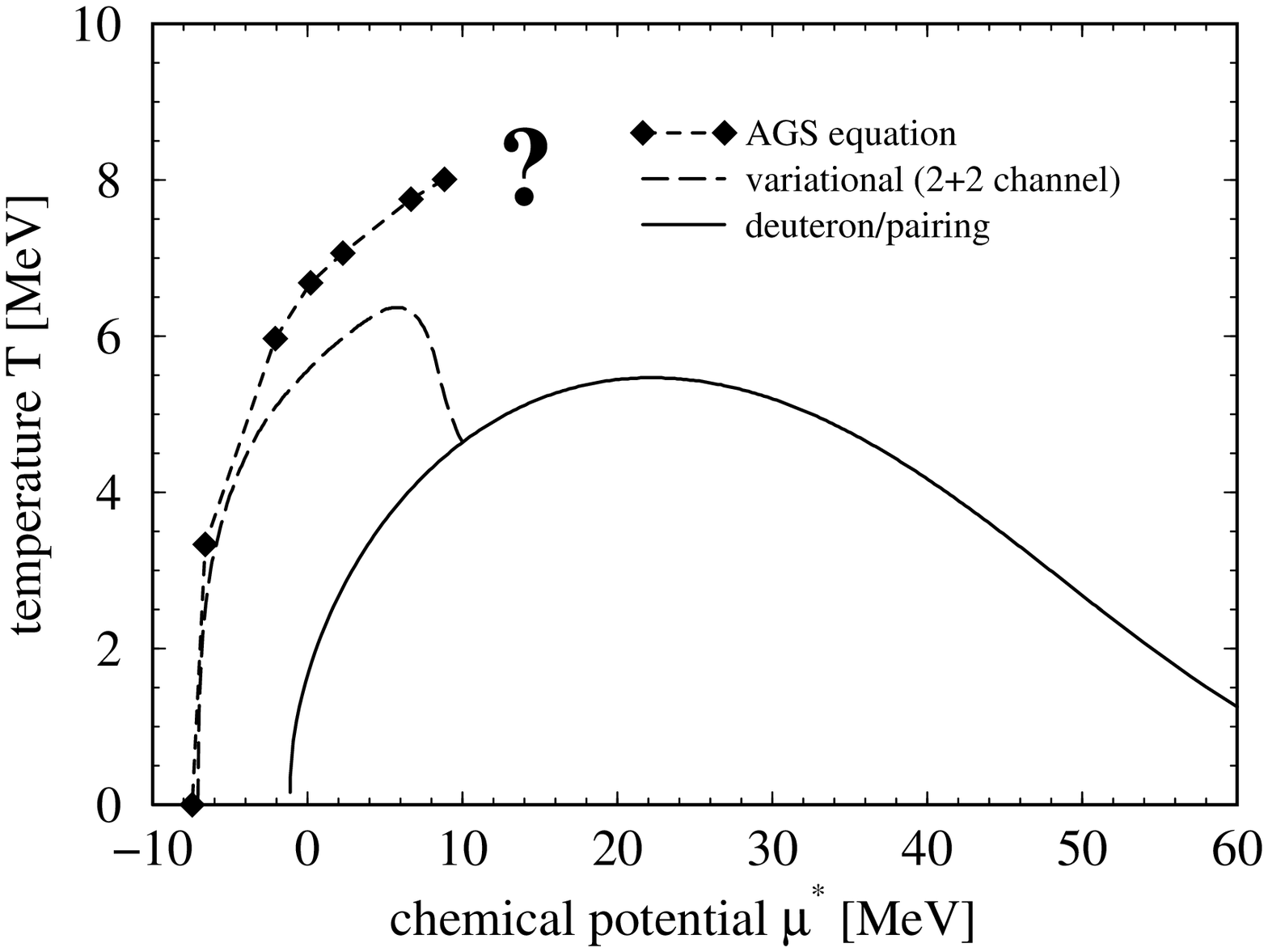,width=0.7\textwidth}
\end{center}
\caption{\label{fig:ProcTc} Temperature vs. effective chemical potential 
  ($\mu^*=\mu-\Sigma(0)$), $\Sigma(0)$ Hartree-Fock shift of the
  single particle energy at zero momentum.  Lines show the critical
  temperature for pairing (solid) and quartetting. The dashed line
  shows a result given in \protect\cite{roe98}, the diamonds show the
  solution of the AGS equation given in (\ref{eqn:coup4})}
\end{figure}
For comparison a perturbative result using Gaussian functions
fitted to the charge radius of the $\alpha$-particle is also given.

%\begin{figure}[t]
%\begin{center}
%\psfig{figure=ProcTc.eps,width=0.7\textwidth}
%\end{center}
%\caption{\label{fig:ProcTc} Temperature vs. effective chemical potential 
%  ($\mu^*=\mu-\Sigma(0)$), $\Sigma(0)$ Hartree-Fock shift of the
%  single particle energy at zero momentum.  Lines show the critical
%  temperature for pairing (solid) and quartetting. The dashed line
%  shows a result given in \protect\cite{roe98}, the diamonds show the
%  solution of the AGS equation given in (\ref{eqn:coup4})}
%\end{figure}
Finally, I address the question of a possible four-particle condensate
or quartetting~\cite{roe98}.  The condition is $B_4(n,T_c,P_{\mathrm
  c.m.}=0)=4\mu $. From Fig.\ref{fig:amottC} we argue that $\alpha$
condensation is likely, i.e.
$$
T_c^{\alpha}>T_c^{\mathrm NN},
$$
where the critical temperature for $\alpha$ condensation turns out
to be higher than for the pairing. However, for $\mu>0$ the situation
seems not so clear, since the four-body AGS equation (\ref{eqn:coup4})
develops poles related to zeros in $4\mu-H_0$. Unlike the two-body
case were these poles disappear because the numerator becomes as well
zero at $2\mu$ and $T_c$ (viz. $1-f_1-f_2\rightarrow 0$). A vanishing
numerator is not obvious for the four-body case because there are more
channels involved. It remains to be clarified, if the rapid fall of
the critical temperature found in Ref.~\cite{roe98} using variational
treatment with square-integrable functions remains, if one uses an
exact treatment of the four-body problem. This is currently
investigated.

{\em Acknowledgment} I am very grateful to S. Mattiello, C. Kuhrts,
G.  R\"opke, P. Schuck, S.A. Sofianos, and W. Schadow, who have
contributed in certain stages to some results presented here, and to
T. Frederico for lively discussions. I gratefully acknowledge the warm
and pleasant atmosphere at the Department of Physics during stays at
UNISA, Pretoria. Work supported by Deutsche Forschungsgemeinschaft
and University of South Africa.

%%\end{document}
\newpage

%% file: SOURCE/main.tex
\section*{\bf Statistical Multifragmentation in Thermodynamical Limit: An Exact Solution for Phase Transitions}
\addcontentsline{toc}{section}{\protect\numberline{}{Statistical Multifragmentation in Thermodynamical Limit: An Exact Solution for Phase Transitions \\ \mbox{\it K.A. Bugaev, M.I. Gorenstein, I.N. Mishustin, W. Greiner}}}
\begin{center}
\vspace*{2mm}
{\underline{K.A. Bugaev$^{1,2}$,} M.I. Gorenstein$^{1,2}$,
I.N. Mishustin$^{1,3,4}$} and\\ {W. Greiner$^{1}$}\\[0.3cm]
{\small\it $^1$ Institut f\"ur Theoretische Physik,
Universit\"at Frankfurt, Germany\\
$^2$ Bogolyubov Institute for Theoretical Physics,
Kyiv, Ukraine\\
$^3$ Kurchatov Institute, Russian Research Center,
Moscow, Russia\\
$^4$ Niels Bohr Institute, University of Copenhagen, Denmark}
\end{center}

%\vspace{1cm}

%\newcounter{eqn}[equation]
%\newcounter{bild}[figure]
%\newcounter{Tabelle}[table]
%\newcounter{Kap}[kap]
%\newcounter{Ueber}[chapter]
%\newcounter{Kapitel}[section]
%\newcounter{Unterkapitel}[subsection]

\begin{abstract}
An exact analytical solution of the 
statistical  multifragmentation model is found in thermodynamic limit.
Excluded volume effects are taken into account in the thermodynamically 
self-consistent way.
The model exhibits a 1-st order 
phase transition of the liquid-gas type.
An extension of the model including the Fisher's term is also studied.
The possibility of the second order phase transition at or above the critical 
point is discussed.
The mixed phase region of the phase diagram, where
the gas of nuclear fragments coexists with the infinite
liquid condensate, is unambiguously identified.
The peculiar thermodynamic properties of the model near the boundary
between the mixed phase
and the pure gaseous phase are studied.
The results for the caloric curve and specific heat are presented
and a physical picture 
of the nuclear liquid-gas phase transition is clarified.
\end{abstract}

\vspace*{0.3cm}

\noindent
\hspace*{1.0cm}\begin{minipage}[t]{11.cm}
{\bf Key words:} Nuclear matter, 
1-st order liquid-gas phase\\
transition, mixed phase thermodynamics
\end{minipage}

\vspace*{0.2cm}

\noindent
\hspace*{1.0cm}
{PACS: 21.65.+f, 24.10. Pa, 25.70. Pq}

%\newpage
\vspace{0.5cm}

Nuclear multifragmentation is one of 
the most interesting and widely discussed phenomena in
intermediate energy nuclear reactions.
The statistical multifragmentation model (SMM) 
(see \cite{Bo:95,Gr:97} and references therein)
was recently applied
to study 
the liquid-gas phase transition
in nuclear matter
\cite{Ch:95,Gu:98,Gu:99,Gu:00}. Numerical calculations
within the canonical ensemble exhibited
many intriguing peculiarities of the finite multifragment
systems. 
However, the investigation of the system's
behavior in the thermodynamic limit was still missing.
Therefore, there was no rigorous proof of the phase
transition existence, and the phase diagram structure
of the SMM remained unclear.
Previous numerical studies for the finite nuclear systems
(the canonical and microcanonical ensembles) led to 
unjustified
(and sometimes wrong) statements concerning  the nuclear liquid-gas phase
transition in the thermodynamic limit.
In our recent paper \cite{Bu:00} an exact analytical solution
of the SMM was found within the grand canonical ensemble
which naturally allowed to study the thermodynamic limit. 
The self-consistent treatment
of the excluded volume effects
was an important part of this study. 
In this letter we investigate
the peculiar thermodynamic properties near the boundary
between the mixed phase
and the pure gaseous phase. 
New results for the caloric curve and the specific heat are presented
and a physical picture
of the nuclear liquid-gas phase transition in SMM is clarified.
This physical picture differs from the one advocated in the previous 
numerical studies.

In the SMM the states of the system  are specified by the multiplicity
sets  $\{n_k\}$
($n_k=0,1,2,...$) of $k$-nucleon fragments.
The partition function of a single fragment with $k$ nucleons is
\cite{Bo:95}:
$
\omega_k =V\left(m T k/2\pi\right)^{3/2}z_k~
$,
where $k=1,2,...,A$ ($A$ is the total number of nucleons
in the system). $V$ and $T$ are, respectively, the  volume
and the temperature of the system,
$m$ is the
nucleon mass.
The first two factors in $\omega_k$ originate
from the
non-relativistic thermal motion 
and the last factor,
 $z_k$, represents the intrinsic partition function of the
$k$-fragment.
For \mbox{$k=1$} (nucleon) we take $z_1=4$ 
(4 internal spin-isospin states) 
and for fragments with $k>1$ we use the expression motivated by the
liquid drop model (see details in \mbox{Ref. \cite{Bo:95}):} 
$
z_k=\exp(-f_k/T),
$ with the fragment free energy 
\begin{equation}\label{fk}
f_k~ = ~- [W_{\rm o}~+~
T^2/\epsilon_{\rm o}~]k~+~\sigma (T)~ k^{2/3}~+~\tau~ T\ln k~.
\end{equation}
Here $W_{\rm o}=16$~MeV is the bulk binding energy per nucleon,
$T^2/\epsilon_{\rm o}$ is the contribution of 
the excited states taken in the Fermi-gas
approximation ($\epsilon_{\rm o}=16$~MeV) and $\sigma (T)$ is the
temperature dependent surface tension which is parameterized 
in the following form:
\begin{equation}\label{Sig}
\sigma (T)=\sigma_{\rm o}
[(T_c^2~-~T^2)/(T_c^2~+~T^2)]^{5/4},
\end{equation}
with $\sigma_{\rm o}=18$~MeV and $T_c=18$~MeV ($\sigma=0$
at $T \ge T_c$). The last Fisher's term in Eq.~(\ref{fk}) with
dimensionless parameter
$\tau$ is introduced for generality.
The canonical partition function (CPF) of the ensemble of nuclear
fragments 
has the following form:
\begin{equation} \label{Zc}
Z^{id}_A(V,T)~=~\sum_{\{n_k\}}~\prod_{k=1}^{A}~\frac{\omega_k^{n_k}}{n_k!}~
\delta(A-\sum_k kn_k)~.
\end{equation}
The model defined by Eqs.(\ref{fk},\ref{Zc}) with $\tau=0$
was studied numerically in Refs.~\cite{Ch:95,Gu:98,Gu:99,Gu:00}.
This is a simplified version of the SMM since the symmetry-energy  and
Coulomb
contributions are neglected. However, its investigation appears to be 
very important for understanding the physics of multifragmentation.

In Eq. (\ref{Zc}) the nuclear fragments are treated as point-like objects.
However, these fragments have non-zero proper volumes and
they should not overlap
in the coordinate space. 
In the 
Van der Waals excluded volume 
approximation 
this is achieved
by replacing
the total volume $V$
in Eq. (\ref{Zc}) by the free (available) volume 
$V_f\equiv V-b\sum_k kn_k$, where
$b=1/\rho_{{\rm o}}$
($\rho_{{\rm o}}=0.16$~fm$^{-3}$ is the normal nuclear density).  
Therefore, the corrected CPF becomes:
$
Z_A(V,T)=Z^{id}_A(V-bA,T)
$.

The calculation of $Z_A(V,T)$
is difficult because of the constraint $\sum_k kn_k =A$.
This difficulty can be partly avoided by calculating the grand canonical
partition function:
\begin{equation} 
{\cal Z}(V,T,\mu)~\equiv~\sum_{A=0}^{\infty}
\exp\left(\mu A/T \right)
~Z_A(V,T)~\Theta (V-bA) \label{Zgc}~, 
\end{equation}
where the chemical potential $\mu$ is introduced.
The calculation of ${\cal Z}$  is still rather
difficult. The summation over the sets $\{n_k\}$
in $Z_A$ cannot be performed analytically because of
the additional \mbox{$A$-dependence}
in the free volume $V_f$ and the restriction
$V_f>0 $.
The problem can be solved 
by introducing the so-called
isobaric partition function (IPF) which is calculated
in a straightforward way (see details 
in Refs.~\cite{Bu:00,Go:81,Bu:00B,Bu:00C}):
\begin{equation} \label{Zs}
\hat{\cal Z}(s,T,\mu) ~ \equiv ~ \int_0^{\infty}dV~\exp(-sV)
~{\cal Z}(V,T,\mu)  
~ = ~ \frac{1}{s~-~{\cal F}(s,T,\mu)}~,
\end{equation}
where
\begin{eqnarray}
{\cal F}(s,T,\mu) &=&
\left( \frac{mT }{2\pi}\right)^{3/2} 
\left[z_1 \exp\left(\frac{\mu-sbT}{T}\right)  
\right.
\nonumber \\
&+& \sum_{k=2}^{\infty} 
\left.
k^{3/2 - \tau} \exp\left(
\frac{(\nu - sbT)k -
\sigma k^{2/3}}{T}\right)\right]\,\,, 
\label{Fs}
\end{eqnarray}
with 
$
\nu \equiv \mu + W_{\rm o}+T^2/\epsilon_{\rm o}
$.
In the thermodynamic limit $V\rightarrow \infty$ the pressure
of the system
is defined by the farthest-right singularity, $s^*(T,\mu)$, of  
the IPF $\hat{\cal Z}(s,T,\mu)$
\begin{equation}\label{ptmu}
p(T,\mu)~\equiv~ T~\lim_{V\rightarrow \infty}\frac{\ln~{\cal Z}(V,T,\mu)}
{V}~=~T~s^*(T,\mu)~.
\end{equation}
The study of the 
system's behavior in the thermodynamic limit
is therefore reduced to the investigation of
the singularities \mbox{of $\hat{\cal Z}$.}

The IPF (\ref{Zs})  has two types of singularities:
\mbox{1) the simple} pole singularity
defined by the following equation
\mbox{$s_g(T,\mu)= {\cal F}(s_g,T,\mu)~$;}
2) the singularity  of the function ${\cal F}$ 
 itself at the point $s_l(T,\mu)=\nu/Tb$ where the coefficient 
in linear over $k$ terms of the exponent in Eq.~(\ref{Fs}) 
is equal to zero.

The simple pole singularity corresponds to the gaseous phase 
where pressure $p_g\equiv Ts_g$ is determined by the 
following transcendental
equation:
$
p_g(T,\mu)=T{\cal F}(p_g/T,T,\mu)
$.
The singularity $s_l(T,\mu)$ of the function ${\cal F}$
defines the liquid pressure:
$
p_l(T,\mu)\equiv Ts_l(T,\mu)=
{\nu}/{b}.
$
Here the liquid is represented by an infinite fragment
(condensate) with $k=\infty$.

In the region of the $(T,\mu)$-plane where $\nu < bp_g(T,\mu)$ the
gaseous phase
is realized ($p_g > p_l$), while  the liquid phase
dominates at $\nu > b p_g(T,\mu)$. The liquid-gas phase transition
occurs when  the two singularities coincide,
i.e. $s_g(T,\mu)=s_l(T,\mu)$.
As ${\cal F}$ in Eq. (\ref{Fs}) 
is a monotonously decreasing
function of $s$ 
the necessary condition for the phase
transition is that the function
${\cal F}$ is finite in its singular
point $s_l$. 
At $\tau =0$ this condition requires $\sigma(T) >0$ and, therefore,
$T<T_c$.
Otherwise, ${\cal F}(s_l,T,\mu)=\infty$ and the system
is always in the gaseous phase as $s_g>s_l$. 
As one can see from Eq.(\ref{Fs}) the convergence properties of
${\cal F}(s,T,\mu)$ depend significantly on the Fisher's exponent
$\tau$ in the vicinity of the critical point where the surface
term vanishes.
%%% MYM
In what follows we mainly concentrate 
on the  case  $\tau =0$. Other possibilities
which appear at $\tau >0$ are discussed shortly. 
Their detail study can be found in Ref.~\cite{Bu:00}.
Here we only note that Eqs. (\ref{Zs}, \ref{Fs}) represent an exact 
solution of the Fisher's droplet model \cite{Fisher:67} 
where additionally the effects of excluded volume are
incorporated.

The baryonic density $\rho$ in the liquid and gaseous phases
is given by the following formulae, 
respectively: 
\begin{equation}\label{rho}
 \rho_l  \equiv  
\left(\partial  p_l/\partial \mu\right)_{T}
=1/b~,~~~~
\rho_g \equiv
\left(\partial  p_g/\partial \mu\right)_{T}=
\rho_{id}/( 1 + b \rho_{id} )~,
\end{equation}
where the function $ \rho_{id}$ is the density of point-like
nuclear fragments with shifted, 
$
\mu \rightarrow \mu -bp_g
$,
chemical potential:
\begin{eqnarray}
\rho_{id}(T,\mu) &=& \left( \frac{mT }{2\pi}\right)^{3/2} 
\left[ z_1 \exp\left(\frac{\mu-bp_g}{T}\right) 
\right.
\nonumber \\
&+& \left. \sum_{k=2}^{\infty} 
k^{5/2} \exp\left(
\frac{(\nu - bp_g)k -
\sigma k^{2/3}}{T} \right) \right]~.
\label{rhoid}
\end{eqnarray}

At $T<T_c$ the system undergoes a 1-st order phase transition
across the line $\mu^*=\mu^*(T)$ defined by
the condition of coinciding singularities:
$ s_l=s_g,  $ i.e., $p_l = p_g$. 
The phase transition line 
$\mu^*(T)$
in the $(T,\mu)$-plane
corresponds to the mixed liquid and gas
states. This line 
is transformed into
the finite mixed-phase region in the $(T,\rho)$-plane
shown in Fig. 1. 
The baryonic density 
in the mixed phase
is a superposition of the liquid and gas baryonic densities:
$
\rho=\lambda\rho_l+(1-\lambda)\rho_g~,
$
where $\lambda$ ($0<\lambda <1$) is the fraction of the system's volume
occupied by the liquid  inside the mixed phase.
Similar linear combinations are also valid for the entropy density $s$
and the energy density $\varepsilon$ 
with $(i=l,g)$
$s_i=\left(\partial p_i/\partial T\right)_{\mu}$\,,
$\varepsilon_i=T \left(\partial p_i/\partial T \right)_{\mu} +
\mu \left(\partial p_i/\partial \mu\right)_{T} -p_i$.

Inside the mixed phase at constant density $\rho$ the
parameter $\lambda$ has a specific temperature dependence
shown in Fig. 2:
from an approximately
constant value $\rho/\rho_{\rm{o}}$ at small $T$ the function 
$\lambda(T)$ drops to zero in a narrow
vicinity of the boundary separating the  mixed phase and 
the pure gaseous phase.
This corresponds to a fast change of the configurations from
the state which is  dominated by one infinite liquid fragment to 
the gaseous multifragment configurations. It happens inside the
mixed phase  without
discontinuities in the thermodynamical functions.

An abrupt decrease of $\lambda(T)$ near this boundary
causes a strong
increase of the energy density as a function of temperature.
This is evident from Fig.~3 which shows the caloric curves at different
baryonic densities. One can clearly see a 
leveling of temperature at energies per nucleon between 10 
and 20 MeV.
As a consequence this  
leads to a sharp peak 
in the specific heat per nucleon at constant density,
$c_{\rho}(T)\equiv (\partial \varepsilon/\partial T)_{\rho}/\rho~$,
presented in Fig. 4.
A finite discontinuity of $c_{\rho}(T)$ arises
at the boundary between the mixed phase and  the gaseous phase.
This finite discontinuity
is caused by the fact that
$\lambda(T)=0$, but
$(\partial\lambda/\partial T)_{\rho} \neq 0$
at this boundary 
(see Fig. 2).

The negative values of the specific heat shown in Fig. 4
appear due to the parameterization of the surface tension (\ref{Sig}):   
its second derivative with respect to  temperature generates
a negative contribution in the vicinity of 
$ T_c $  for all densities. 
This feature of the model  is unphysical and it has to be 
modified. 
%%%MYM1

It should be emphasized that the energy density is continuous
at the boundary of the mixed phase and the gaseous phase, hence
the sharpness of the
peak in $c_{\rho}$ is entirely due to the strong temperature
dependence
of $\lambda(T)$ near this boundary. 
%Therefore,
%in contrast to the expectation in Refs. \cite{Gu:98,Gu:99}, 
Moreover, at any $\rho < \rho_{\rm o}$
the maximum value of $c_{\rho}$ remains finite
and the peak width in $c_{\rho}(T)$ is nonzero in the thermodynamic
limit considered in our study. 
This is in contradiction with the expectation of Refs. \cite{Gu:98,Gu:99}
that an infinite peak of zero width will appear in $c_{\rho}(T)$ in this
limit.
Also a comment about the so-called ``boiling point''
is appropriate here.
This is a discontinuity in the energy density $\varepsilon(T)$ 
or, equivalently, a plateau in the
temperature as a function of the excitation energy. 
Our analysis shows that this type of behavior indeed happens 
at constant pressure, but not at constant density! This is similar to
the usual picture of a liquid-gas phase transition.
In Refs. \cite{Gu:98,Gu:99} a rapid  
increase of the energy density as a function of temperature
at fixed $\rho$ near the boundary of the mixed and gaseous phases
(see Fig.~3)
was misinterpreted as a manifestation of the ``boiling point''.

The results presented in Figs.~1-4 are obtained for $\tau=0$.
New possibilities appear
at non-zero values of the parameter $\tau$.
At $0<\tau \le 5/2$ the qualitative picture remains the same
as discussed above, although there are some
quantitative changes.
For $\tau > 5/2$  the condition ${\cal F}(s,T,\mu) < \infty$ 
is also satisfied
in the singularity point $s_l(T, \mu)$ 
for all  $T>T_c$ where $\sigma(T)=0$.
Therefore, the liquid-gas phase transition
extends now to all temperatures. Its properties
are, however, different for $\tau >7/2$ and for $\tau <7/2$
(see Fig.~5).
If $\tau >7/2$ the
gas density is always lower than $1/b$ as $\rho_{id}$ is finite.
Therefore, the
liquid-gas transition at $T>T_c$ remains
the 1-st order phase transition with discontinuities
of baryonic density, entropy and energy densities.

At $5/2 < \tau < 7/2$ the baryonic density of the gas
in the mixed phase, 
$\rho_g^{mix} \equiv \rho_{id}^{mix}(T)/( 1+ b~\rho_{id}^{mix}(T) )$,
becomes equal to that of the liquid at $T>T_c$, i.e.,  
$\rho_g^{mix}=1/b\equiv \rho_{\rm o}$, since
\begin{eqnarray}
\rho_{id}^{mix}(T)~ \equiv~\rho_{id}(T,\mu^*(T))~
& = &  \left( \frac{ mT }{2 \pi}\right)^{3/2}
\left[z_1 \exp\left(-~\frac{W}{T}\right)\right.
%\right.
\nonumber \\
&+& \sum_{k=2}^{\infty}\left.
k^{\frac{5}{2}- \tau}
\exp\left(-~ \frac{\sigma ~k^{2/3}}{T}\right)\right]~ \rightarrow \infty\,\,,
\label{rhoidmix}
\end{eqnarray}
if surface tension vanishes $\sigma = 0$.
It is easy to prove that the entropy and energy densities
for the liquid and gas phases are also equal to each other.
There are discontinuities only in the derivatives of these densities
over $T$ and $\mu$, i.e., $p(T,\mu)$  has discontinuities
of its second derivatives.
Therefore,  the liquid-gas transition at $T>T_c$ for $5/2 < \tau < 7/2$
becomes the 2-nd order phase transition.
According to standard definition, the point $T=T_c$,
$\rho = 1/b$ separating the first and second order
transitions is the tricritical point.
One can see that this point is now at a finite pressure.

It is interesting to note that at $\tau >0$ the mixed
phase boundary shown in Fig.5 is not so steep function of $T$ as
in the case $\tau=0$ presented in Fig.1. Therefore, the peak in the specific
heat discussed above becomes less pronounced.

\vspace{0.2cm}
In conclusion, the simplified version of the SMM
is solved analytically
in the grand canonical ensemble.
The progress is achieved by 
reducing the description of phase transitions
to the investigation  of the isobaric
partition function singularities. The model clearly 
demonstrates a 1-st order
phase transition of the liquid-gas type.
The considered system has  peculiar
properties near the boundary of the mixed and gaseous
phases.
The rapid change
of the thermodynamical functions with $T$ at fixed $\rho$ 
takes place 
near this  boundary
due to the disappearance of the infinite liquid fragment. 
This leads to leveling of 
the caloric curves shown in Fig.~3
at temperatures between 6 -- 10 MeV depending on the density. 
As a consequence 
a sharp peak and a finite discontinuity
are developed 
in the specific heat $c_{\rho}(T)$ 
at the boundary of the mixed and gaseous phases. 

The phase diagram appears to be rather
sensitive to the value of the parameter $\tau$
in the Fisher's free energy term included in our treatment.
New interesting possibilities for the phase diagram emerge
for $\tau >5/2$  in comparison
with the case when $\tau<5/2$.
The case $5/2<\tau<7/2$ is particularly interesting
because of the appearance of the tricritical point separating the
1-st and 2-nd order phase transitions.

\vspace*{0.2cm}

\begin{center}
{\bf  Acknowledgments.}  
\end{center}

The authors 
are grateful to 
D. Blaschke, J. Polonyi 
and P.T.~Reuter for stimulating discussions. 
We  thank the Alexander von Humboldt Foundation
and DAAD (Germany) for the financial support. 
The research described in this publication was made possible in part by
Award No. UP1-2119 of the U.S. Civilian Research \& Development
Foundation for the Independent States of the Former Soviet Union
(CRDF).
K.A.B. gratefully acknowledges
the warm hospitality of the Particle- and Astrophysics Group
of the University of Rostock, where this talk was given.

\input SOURCE/capts.tex

\newpage

%% file: SOURCE/capts.tex
%%%%%%%%%%%%%%%%%%%%% FIGURECAPTIONS FILE

%\newpage
\clearpage

\begin{figure}
\hspace*{1.1cm}\mbox{\psfig{figure=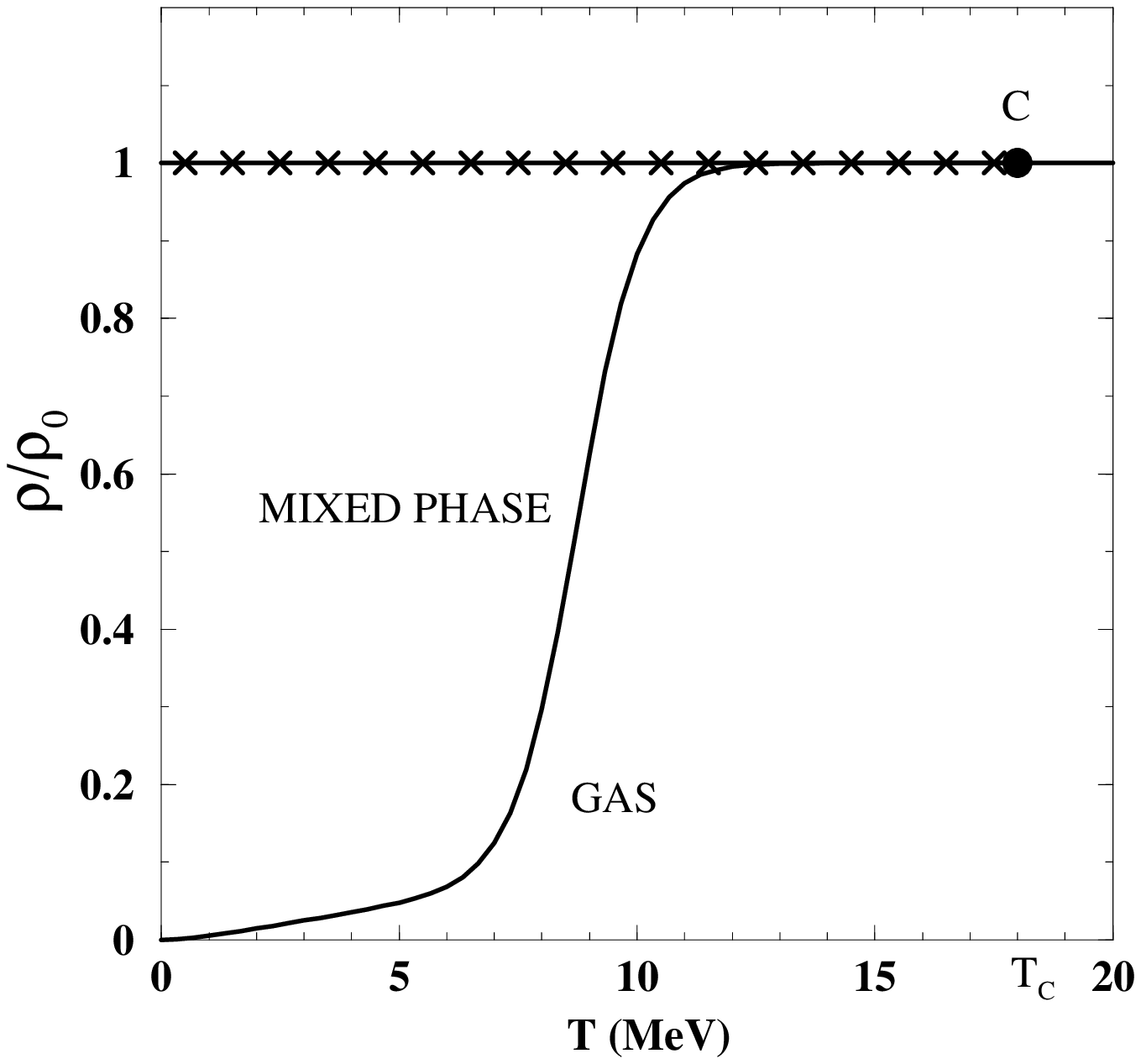,width=12.cm}}
%%%\vspace*{2mm}

\vspace*{-0.5cm}

\caption{\label{fig:one}
Phase diagram in the $(T,\rho)$-plane.
The mixed phase and pure gaseous phase boundary
is shown by the solid line.
The pure liquid phase (shown by crosses) corresponds
to
the fixed density $\rho = \rho_{\rm o}$.
Point $C$ is the critical point,
at $T>T_c$ only the pure gaseous phase
exists.
}
\end{figure}

%\newpage
%\clearpage

\vspace*{-0.5cm}

\begin{figure}   
\hspace*{1.1cm}\mbox{\psfig{figure=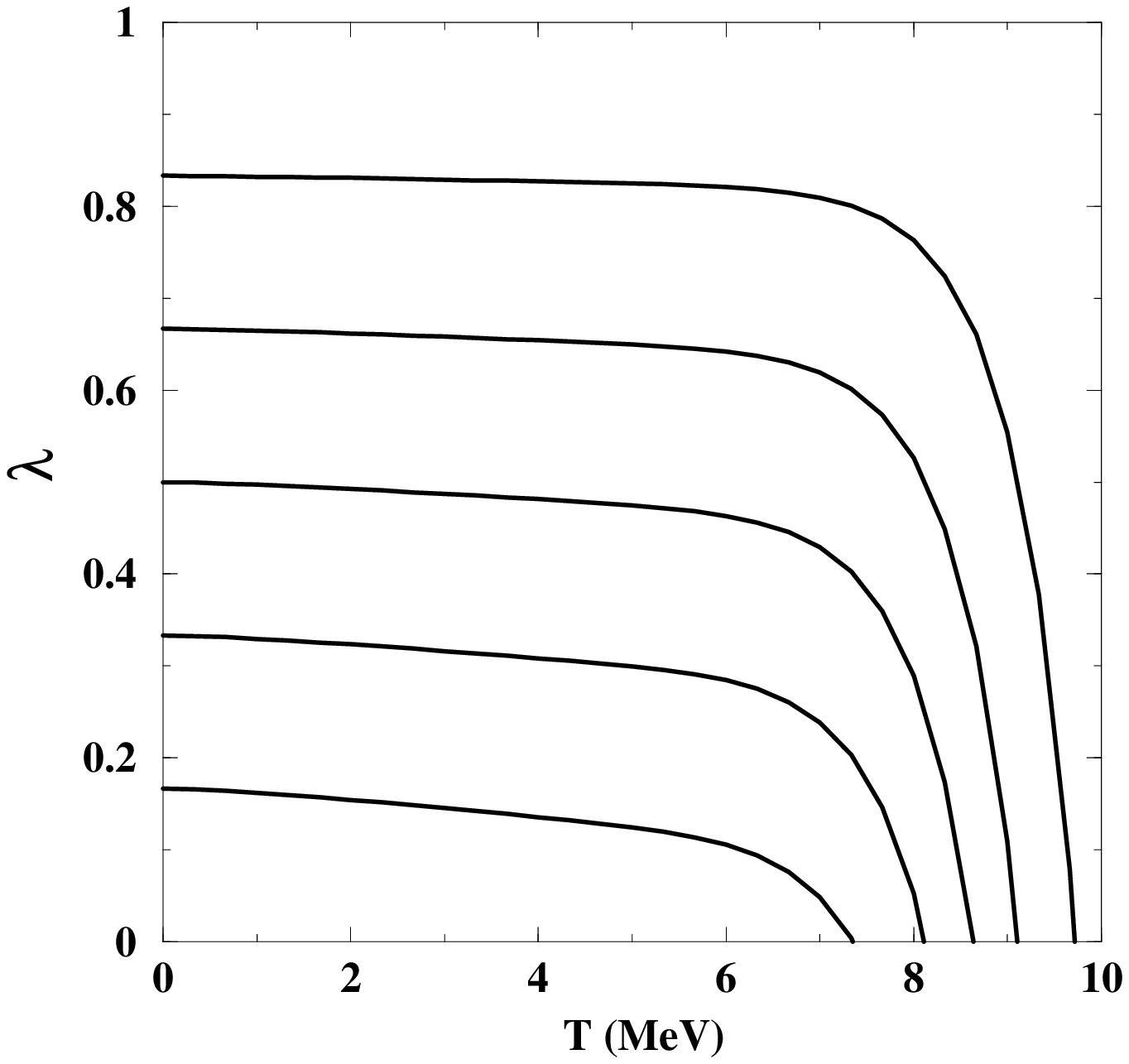,width=12.cm}}
%%%\vspace*{2mm}

\vspace*{-0.5cm}

\caption{\label{fig:two}
Volume fraction $\lambda(T)$ of the liquid
inside the mixed phase is
shown as a function of temperature
for fixed nucleon densities ${\rho}/{\rho_{\rm o}} = 1/6, 1/3, 1/2, 2/3,
5/6$
(from bottom to top).
}
\end{figure}

\newpage
\clearpage

\begin{figure}
\hspace*{1.1cm}\mbox{\psfig{figure=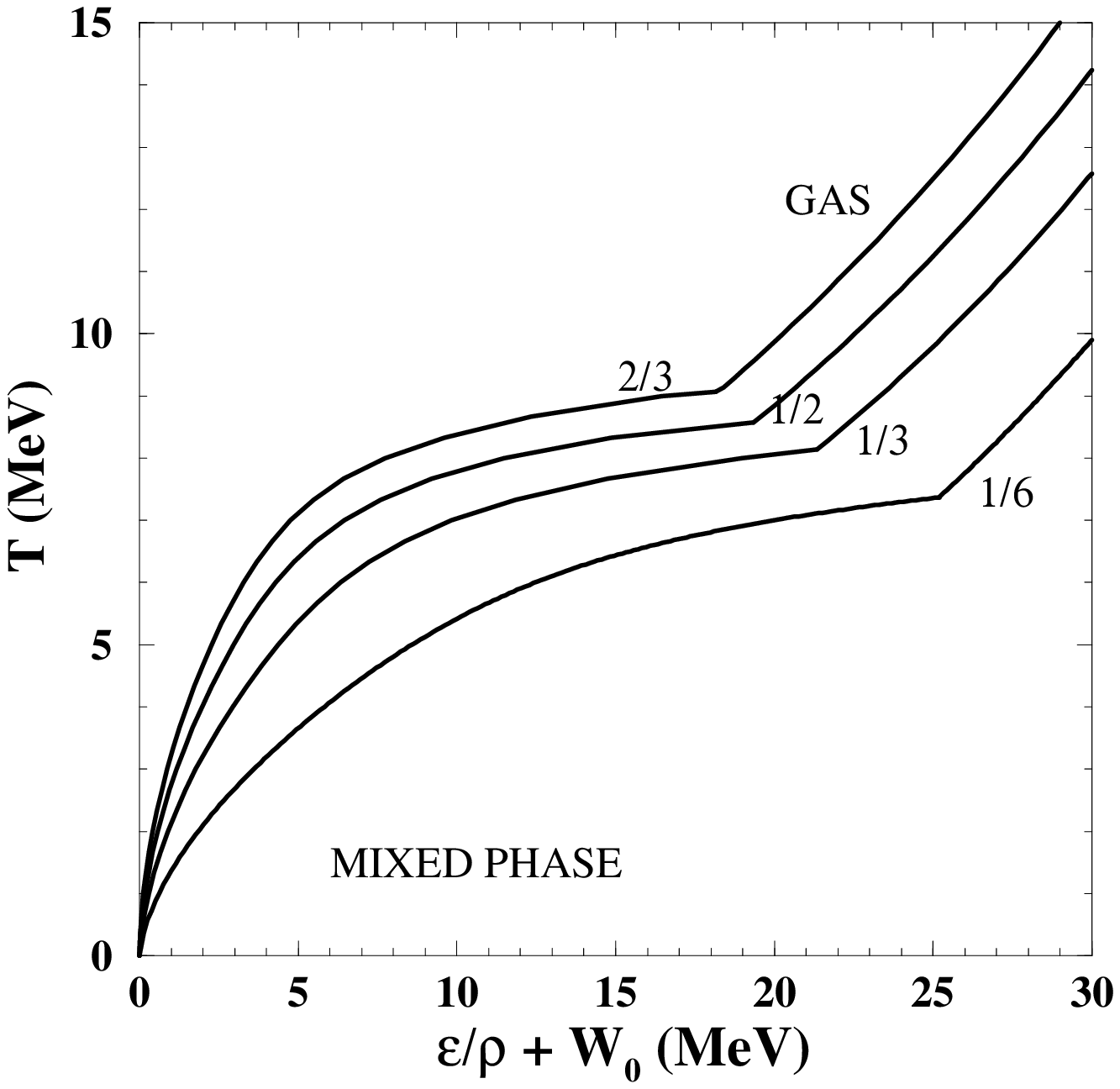,width=12.cm}}
%%%\vspace*{2mm}

\vspace*{-0.5cm}

\caption{\label{fig:three}
Temperature as a function of energy density per nucleon
(caloric curve)
is shown for fixed nucleon densities ${\rho}/{\rho_{\rm o}} = 1/6, 1/3,
1/2, 2/3$.
}
\end{figure}

%\newpage
%\clearpage

\begin{figure}
\hspace*{1.1cm}\mbox{\psfig{figure=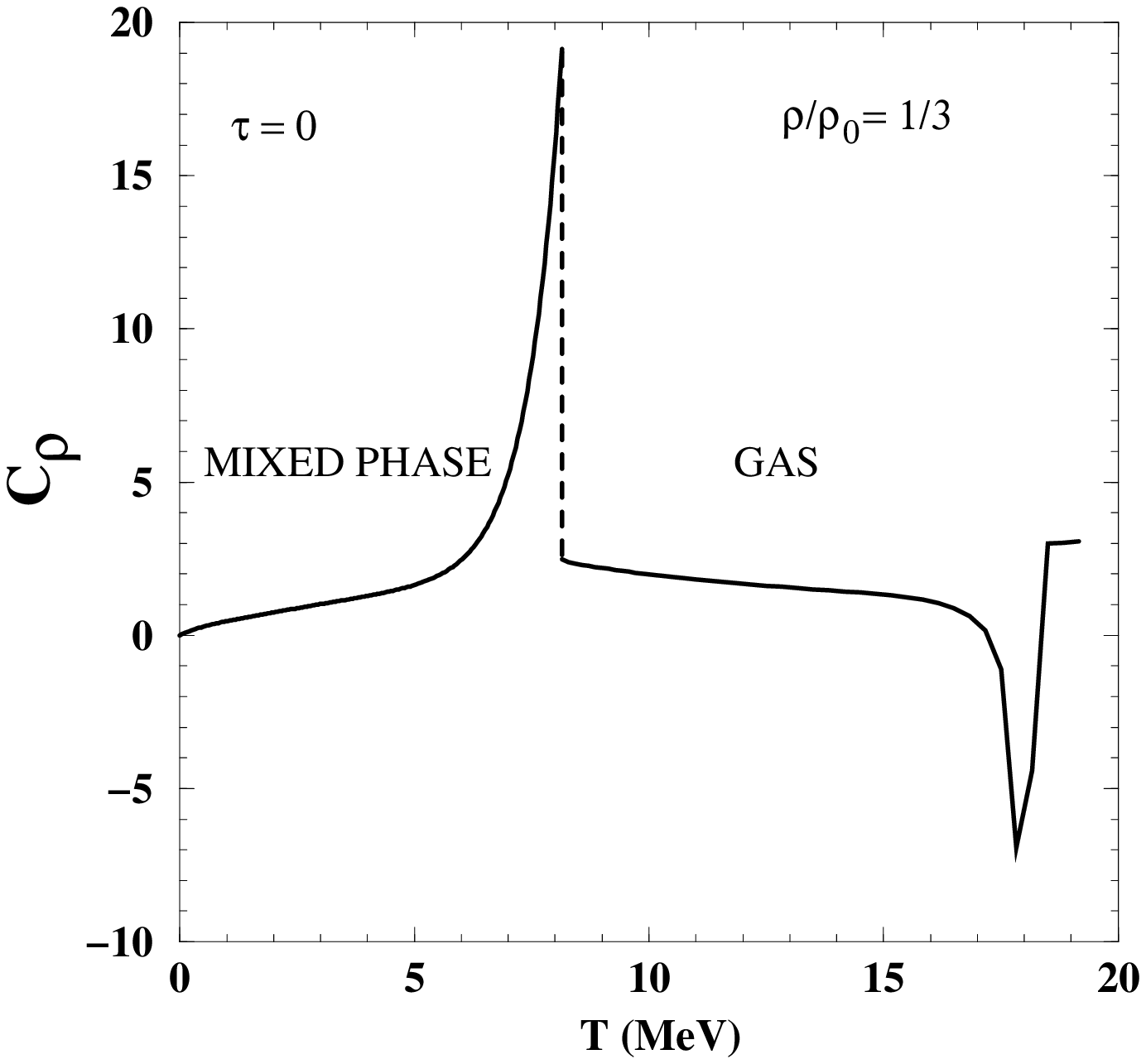,width=12.cm}}
%%%\vspace*{2mm}

\vspace*{-0.5cm}

\caption{\label{fig:four}
Specific heat per nucleon as a function of temperature
at fixed nucleon density ${\rho}/{\rho_{\rm o}} = 1/3$. The dashed line
shows the finite discontinuity of $c_{\rho}(T)$
at the boundary of the mixed and gaseous phases.
The negative specific heat appears in the vicinity of $T = 18$ MeV. 
}
\end{figure}

\newpage
%\clearpage

\vspace*{-1.5cm}

\begin{figure}
\hspace*{1.1cm}\mbox{\psfig{figure=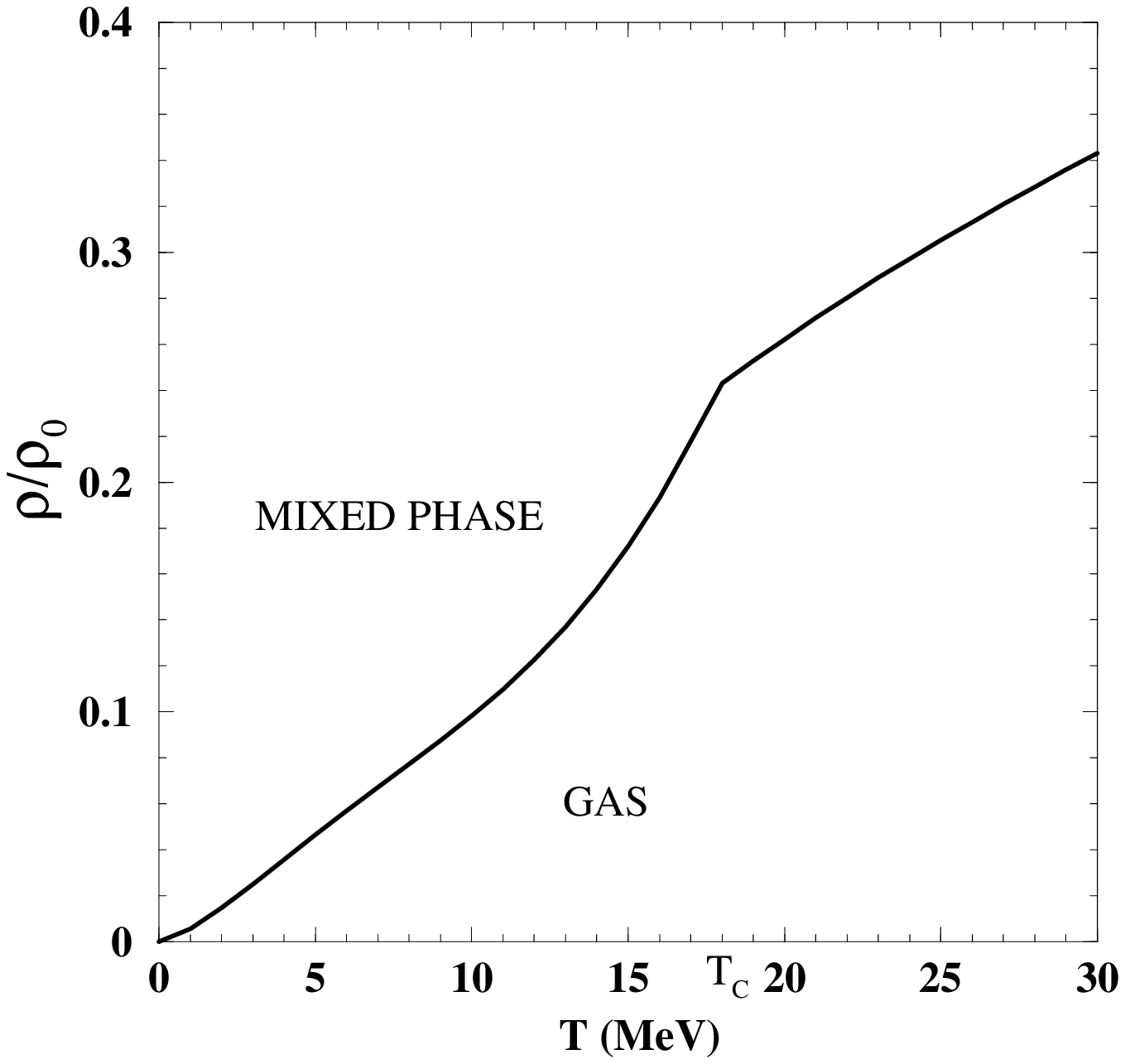,width=10.0cm}}

\vspace*{ -1.5cm}

\hspace*{1.1cm}\mbox{\psfig{figure=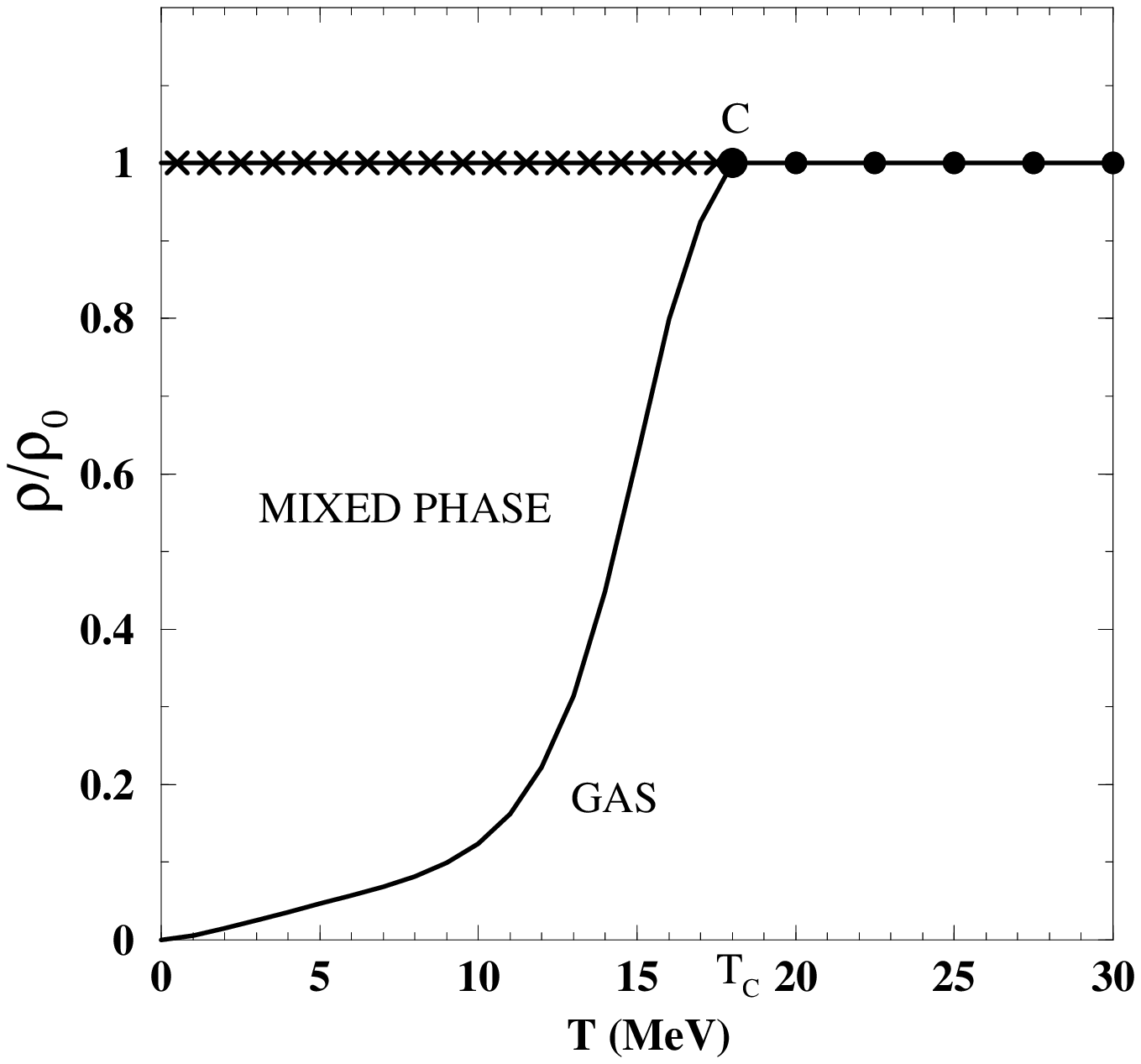,width=10.0cm}}

\vspace*{-0.5cm}

%\noindent
\caption{\label{fig:five}
Phase diagrams in
$T-\rho$ plane for $\tau = 3.6$ (upper panel) and $\tau = 2.6$ (lower panel).
Point $C$ in the lower panel is the tricritical point.   
Crosses correspond to the liquid phase of the first order phase transition
and dots correspond to the states of the second order one.
}
\end{figure}

%% file: SOURCE/DSE.tex
\addcontentsline{toc}{section}{$\left. \right.$\hspace{0.3cm} \bf Chiral Quark Models and Astrophysics}
$\left. \right.$
\vspace*{5cm}
\begin{center}
{\Large Contributions on}
\end{center}
\begin{center}
{\Large {\bf Chiral Quark Models and Astrophysics}}
\end{center}
\newpage

%% file: SOURCE/shortRdr.tex
\section*{\bf On the \mbox{$\eta$--$\eta^\prime$} Complex in 
the SD--BS Approach}
\addcontentsline{toc}{section}{\protect\numberline{}{On the \mbox{$\eta$--$\eta^\prime$} Complex in the SD--BS Approach \\ \mbox{\it D. Klabucar, D. Kekez, M.D. Scadron}}}
\begin{center}
\vspace*{2mm}
{D. Klabucar$^{\dagger}$, D. Kekez$^{\ddagger}$ and M. D. Scadron$^{*}$}\\[0.3cm] 
{\small\it $^\dagger$ Physics Department, Faculty of Science, 
University of Zagreb, \\  Bijenicka c. 32, Zagreb 10000, Croatia\\ 
$^{\ddagger}$ Rudjer Boskovi\'{c} Institute, P.O.B. 180, 10002 Zagreb, Croatia\\ 
$^{*}$ Physics Department, University of Arizona, Tucson, AZ 85721, USA}
\end{center} 

\begin{abstract}
\noindent The bound-state Schwinger-Dyson and Bethe-Salpeter (SD--BS) 
approach is chirally well-behaved and provides a reliable treatment of 
the $\eta$--$\eta^\prime$ complex although a ladder approximation is
employed. Allowing for the effects of the SU(3) flavor symmetry breaking 
in the quark--antiquark annihilation, leads to the improved 
$\eta$--$\eta^\prime$ mass matrix.

\end{abstract}
%\pacs{11.30.Rd, 11.40.-q, 11.40.Ha, 11.10.St}

\newcounter{eqn3}[equation]
\setcounter{equation}{-1}
\stepcounter{equation}

\newcounter{bild3}[figure]
\setcounter{figure}{-1}
\stepcounter{figure}

\newcounter{tabelle3}[table]
\setcounter{table}{-1}
\stepcounter{table}

%\newcounter{kapitel3}[section]
%\setcounter{section}{-1}
%\stepcounter{section}

\newcounter{unterkapitel3}[subsection]
\setcounter{subsection}{-1}
\stepcounter{subsection}

\subsection{\mbox{$\eta$--$\eta^\prime$} phenomenology and Goldstone structure}
\label{Preliminaries}

The physical isoscalar pseudoscalars $\eta$ and $\eta'$ are usually given as
\begin{equation}
|\eta\rangle = \cos\theta\, |\eta_8\rangle
                       - \sin\theta\, |\eta_0\rangle~,
\,\,\,\,\,\,\,
|\eta^\prime\rangle = \sin\theta\, |\eta_8\rangle
                              + \cos\theta\, |\eta_0\rangle~,
\label{eta-etaPrimeDEF}
\end{equation}
{\it i.e.}, as the orthogonal mixture
of the respective octet and singlet isospin zero states, $\eta_8$ and
$\eta_0$. In the flavor SU(3) quark model, they are defined through
quark--antiquark ($q\bar q$) basis states $|f\bar{f}\rangle$ ($f=u,d,s$)
as
\begin{mathletters}
\label{eta8-eta0def}
        \begin{eqnarray}
        |\eta_8\rangle
        &=&
        \frac{1}{\sqrt{6}} (|u\bar{u}\rangle + |d\bar{d}\rangle
                                            -2 |s\bar{s}\rangle)~,
\label{eta8def}
        \\
        |\eta_0\rangle
        &=&
        \frac{1}{\sqrt{3}} (|u\bar{u}\rangle + |d\bar{d}\rangle
                                             + |s\bar{s}\rangle)~.
\label{eta0def}
        \end{eqnarray}
\end{mathletters}
The exact SU(3) flavor symmetry ($m_u = m_d = m_s$) is nevertheless
badly broken. It is an excellent approximation to assume the exact
isospin symmetry ($m_u = m_d$), and a good approximation to take even
the chiral symmetry limit for $u$ and $d$-quark
(where {\it current} quark masses $m_u = m_d = 0$), but
for a realistic description, the strange quark mass $m_s$ must be
significantly heavier than $m_u$ and $m_d$. The same holds for the
{\it constituent} quark masses, denoted by $\hat{m}$ for {\it both} 
$u$ and $d$ quarks since we rely on the isosymmetric limit, and by 
$\hat{m}_s$ for the $s$-quark. They are nonvanishing in the chiral 
limit (CL). In the strange sector, CL is useful only qualitatively, 
as a theoretical limit. (CL would reduce $\hat{m}_s$ to $\hat{m}$, 
on which CL has almost negligible influence.)

Thus, with $|u\bar{u}\rangle$ and $|d\bar{d}\rangle$ being practically
chiral states as opposed to a significantly heavier $|s\bar{s}\rangle$,
Eqs.~(\ref{eta8-eta0def}) do not define the octet and singlet states 
of the exact SU(3) flavor symmetry, but the {\it effective} octet and 
singlet states. Hence, as in Ref. \cite{KlKe2} for example, only in
the sense that the same $q\bar q$ states $|f\bar{f}\rangle$ ($f=u,d,s$)
appear in both Eq.~(\ref{eta8def}) and Eq.~(\ref{eta0def}) do these
equations implicitly assume nonet symmetry (as pointed out by Gilman and 
Kauffman \cite{GilKauf}, following Chanowitz, their Ref.~[8]). However, 
in order to avoid the U$_A$(1) problem, this symmetry must ultimately be
broken at least at the level of the masses. In particular, it must
be broken in such a way that $\eta \to \eta_8$ becomes massless but
$\eta' \to \eta_0$ remains massive (as in Ref. \cite{KlKe2}) when
CL is taken for all three flavors, $m_u, m_d, m_s \to 0$.
Nevertheless, the CL-vanishing octet eta mass $m_{\eta_8}$ is rather 
heavy for the realistically broken SU(3) flavor symmetry;
for the empirical pion and kaon masses $m_\pi$ and $m_K$,
the Gell-Mann-Okubo mass formula $m_\pi^2 + 3m_{\eta_{8}}^2 = 4m_K^2$
yields $m_{\eta_8} \approx 567$ MeV. In that case, and for the 
empirical masses of $\eta (547)$ and $\eta'(958)$, the singlet 
$\eta_0$ mass $m_{\eta_0}$ (nonvanishing even in CL) can 
be found from the mass--matrix trace 
\begin{equation}
        m_{\eta_8}^2 + m_{\eta_0}^2 =
        m_{\eta}^2 + m_{\eta'}^2 \approx
        1.22 ~\w{GeV}^2,\wsep{.1}{giving}
        m_{\eta_0} \approx 947 ~\w{MeV} \, .
        \label{eqno1}
\end{equation}

Alternatively, one can work in a nonstrange ({\it NS})--strange ({\it S})
basis:
% $|\eta_\NSt\rangle$ and $|\eta_\St\rangle$, where
\begin{mathletters}
\label{NS-Sbasis}
        \begin{eqnarray}
        |\eta_\NSt\rangle
        &=&
        \frac{1}{\sqrt{2}} (|u\bar{u}\rangle + |d\bar{d}\rangle)
  = \frac{1}{\sqrt{3}} |\eta_8\rangle + \sqrt{\frac{2}{3}} |\eta_0\rangle~,
\label{etaNSdef}
        \\
        |\eta_\St\rangle
        &=&
            |s\bar{s}\rangle
  = - \sqrt{\frac{2}{3}} |\eta_8\rangle + \frac{1}{\sqrt{3}} |\eta_0\rangle~.
\label{etaSdef}
        \end{eqnarray}
\end{mathletters}
In analogy with Eq. (\ref{eqno1}), in this basis one finds
\begin{equation}
        m_{\eta_{NS}}^2 + m_{\eta_S}^2 = 
        m_\eta^2 + m_{\eta'}^2 \approx
        1.22 ~\w{GeV}^2,
        \label{eqno2}
\end{equation}
whereas the {\it NS--S} mixing relations, diagonalizing the mass matrix, are
\begin{equation}
|\eta\rangle = \cos\phi_P |\eta_\NSt\rangle
             - \sin\phi_P |\eta_\St\rangle~,
\,\,\,\,\,\,\,
|\eta^\prime\rangle = \sin\phi_P |\eta_\NSt\rangle
             + \cos\phi_P |\eta_\St\rangle~.
\label{eqno3}
\end{equation}
The singlet-octet mixing angle $\theta$, defined by
Eqs.~(\ref{eta-etaPrimeDEF}), is related
to the {\it NS--S} mixing angle $\phi$ above as \cite{14} 
$\theta = \phi - \arctan \sqrt{2} =  \phi - 54.74\deg$.

Although mathematically equivalent to the $\eta_8$--$\eta_0$ basis, the
{\it NS--S} mixing basis is more suitable for most quark model considerations,
being more natural in practice when the symmetry between the {\it NS} and {\it S}
sectors is broken as described in the preceding passage.
There is also another important reason to keep in mind the
$|{\eta_\NSt}\rangle$-$|{\eta_S}\rangle$ state mixing angle $\phi$.
This is because it offers the quickest way to show the consistency of
our procedures and the corresponding results obtained using just one 
($\theta$ or $\phi$) {\it state} mixing angle,
with the two-mixing-angle scheme considered in Refs.
\cite{Leutwyler98,KaiserLeutwyler98,FeldmannKroll98EPJC,FeldmannKroll98PRD,FeldmannKrollStech98PRD,FeldmannKrollStech99PLB,Feldmann99IJMPA},
which is defined with respect to the mixing of the decay constants.
For clarification of the relationship with, and our results in
the two-mixing-angle scheme, we refer to Ref. \cite{KeKlSc2000}, 
particularly to its Appendix.
Here, we simply note that our considerations will ultimately lead
us to $\phi \approx 42^\circ$, practically the same as the result of 
Refs. \cite{FeldmannKroll98EPJC,FeldmannKroll98PRD,FeldmannKrollStech98PRD,Feldmann99IJMPA} and in agreement with data ({\it e.g.}, see Table 2 of Feldmann's 
review \cite{Feldmann99IJMPA}).

\begin{figure}[t]
%\rule{5cm}{0.2mm}\hfill\rule{5cm}{0.2mm}
%\vskip 2.5cm
%\rule{5cm}{0.2mm}\hfill\rule{5cm}{0.2mm}
\centerline{\psfig{figure=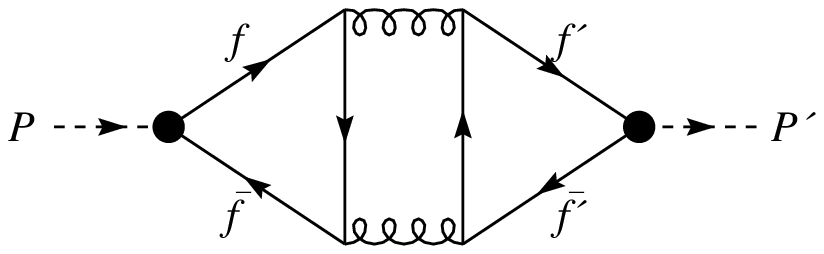,height=1.8in}}
\caption{ Nonperturbative QCD quark annihilation illustrated by the
two-gluon exchange diagram.  It shows the transition of the
$f\bar f$ pseudoscalar $P$ into the pseudoscalar $P^\prime$ having
the flavor content $f^\prime {\bar f}^\prime$. The dashed lines and
full circles depict the $q\bar q$ bound-state pseudoscalars and vertices,
respectively.
%\label{fig:figure1}
}
\end{figure}

As for a theoretical determination of the $\eta$--$\eta^\prime$ mixing 
angle $\phi$ or $\theta$, we 
follow the path of Refs.~\cite{14}. The contribution of the
gluon axial anomaly to the singlet $\eta_0$ mass is 
essentially just parameterized and not really calculated,
but some useful information can be obtained from the isoscalar
$q\bar q$ annihilation graphs of which the ``diamond" one in Fig.\ 1 
is just the simplest example. That is, we can take Fig.\ 1 in the 
nonperturbative sense, where the
two-gluon intermediate ``states'' represent any even number of
gluons when forming a C$^+$ pseudoscalar $\ov{q}q$ meson \cite{17},
and where quarks, gluons and vertices can be dressed nonperturbatively,
and possibly include gluon configurations such as instantons.
Factorization of the quark propagators in Fig.\ 1 characterized by the 
ratio $X \approx \hat{m} /\hat{m}_s$ leads to the $\eta$--$\eta^\prime$ 
mass matrix in the {\it NS--S} basis \cite{14}
\eqnb
        \pip{
                \begin{array}{ll}
        m_\pi^2 + 2 \beta       &  \, \, \,\, \quad \sqrt{2} \beta X \\
                \, \, \sqrt{2}\beta X   & \, 2 m_K^2 - m_\pi^2 + \beta X^2
                \end{array}
        }
%       \to
\begin{array}{c} \vspace{-2mm} \longrightarrow \\ \phi \end{array}
        \pip{
                \begin{array}{ll}
                        m_\eta^2        & 0 \\
                        0               & m_{\eta'}^2
                \end{array}
        },
        \label{eqno6}
\eqne
where $\beta$ denotes the total annihilation strength of the
pseudoscalar $q\bar q$ for the {\it light} flavors $f=u,d$,
whereas it is assumed attenuated by a factor $X$ when a $s\bar s$
pseudoscalar appears. (The mass matrix in the $\eta_8$-$\eta_0$
basis reveals that in the $X\to 1$ limit, the CL-nonvanishing 
singlet $\eta_0$ mass is given by $3\beta$.)
The two parameters on the left-hand-side (LHS) of (\ref{eqno6}), $\beta$ and
$X$, are determined by the two diagonalized $\eta$ and $\eta'$
masses on the RHS of (\ref{eqno6}).  The trace and determinant of the
matrices in (\ref{eqno6}) then fix $\beta$ and $X$ to be \cite{14}
\eqnb
        \beta =
        \frac   { (m_{\eta'}^2 - m_\pi^2) (m_\eta^2 - m_\pi^2) }
                        { 4 (m_K^2 - m_\pi^2) }
        \approx
        0.28 ~\w{GeV}^2~~,\hspace{.3in} X \approx 0.78~,
        \label{eqno7}
\eqne
with the latter value suggesting a {\it S/NS} constituent quark mass ratio
$X^{-1} \sim \hat{m}_s / \hat{m} \sim 1.3$~, near the values in 
Refs.~\cite{17,15,16,18,19}, $\hat{m}_s / \hat{m} \approx 1.45$. 

This fitted nonperturbative scale of $\beta$ in (\ref{eqno7}) depends only on
the gross features of QCD.  If instead one treats the QCD graph of
Fig.\ 1 in the perturbative sense of literally two gluons
exchanged, then one obtains \cite{20} only $\beta_{2g} \sim 0.09
~\w{GeV}^2$, which is about $1/3$ of the needed scale of $\beta$
found in (\ref{eqno7}).
(This indicates that just the perturbative ``diamond" graph 
can hardly represent even the roughest approximation to the effect of 
the gluon axial anomaly operator 
$\epsilon^{\alpha\beta\mu\nu} G^a_{\alpha\beta} G^a_{\mu\nu}$.)
The above fitted quark annihilation (nonperturbative) scale $\beta$ in
(\ref{eqno7}) can be converted to the {\it NS--S} $\eta$--$\eta^\prime$ mixing angle
$\phi$ in (\ref{eqno3}) from the alternative mixing relation 
$\tan2 \phi = 2 \sqrt{2} \beta X(m_{\eta_S}^2-m_{\eta_{NS}}^2)^{-1} 
\approx 9.02$ to \cite{14} 
\eqnb
        \phi = \arctan
        \pib{
                \frac   {(m_{\eta'}^2 - 2m_K^2 + m_\pi^2) (m_\eta^2 -
                                 m_\pi^2)}
                                {(2m_K^2 - m_\pi^2 - m_\eta^2) (m_{\eta'}^2
                                 - m_\pi^2)}
        }^{1/2} \approx
        41.84\deg ~.
        \label{eqno8}
\eqne
This kinematical QCD mixing angle (\ref{eqno8}) or $\theta = \phi -54.74^\circ
\approx -12.9^\circ$ has dynamical analogs \cite{KlKe2,KeKlSc2000}, 
namely the coupled SD-BS approach discussed below, in Sec. \ref{MainSec}.
Since this predicted $\eta$--$\eta^\prime$ mixing angle in (\ref{eqno8}) is 
compatible with the values 
%in (\ref{eqno4}) and (\ref{eqno5}),
repeatedly extracted in various empirical ways
\cite{15,16}, and more recently from the
FKS scheme and theory \cite{FeldmannKroll98EPJC,FeldmannKroll98PRD,FeldmannKrollStech98PRD,FeldmannKrollStech99PLB,Feldmann99IJMPA},
we confidently use the value (\ref{eqno8}) in the mixing angle relations 
(\ref{eqno3}) to infer the nonstrange and strange $\eta$ masses,
\begin{mathletters}
\label{eqno9}
\begin{eqnarray}
        m_{\eta_{NS}}^2 &=& \cos^2 \phi ~m_\eta^2 +
        \sin^2 \phi ~m_{\eta'}^2 \approx (757.9~\w{MeV})^2
        \label{eqno9a}
\\
        m_{\eta_{S}}^2 &=& \sin^2 \phi ~m_\eta^2 + \cos^2 \phi ~m_{\eta'}^2
        \approx (801.5~\w{MeV})^2~~.
        \label{eqno9b}
\end{eqnarray}
\end{mathletters}

Thus it is clear that the true physical masses $\eta (547)$ and
$\eta' (958)$ are respectively much closer to the Nambu-Goldstone
(NG) octet $\eta_8 (567)$ and the non-NG singlet $\eta_0 (947)$
configurations than to the nonstrange $\eta_\NSt (758)$ and strange
$\eta_S (801)$ configurations inferred in Eqs.\ (\ref{eqno9}).  However, the
mean $\eta$--$\eta^\prime$ mass $(548 + 958) /2 \approx 753 ~\w{MeV}$ is
quite near the nonstrange $\eta_\NSt (758)$.  But since $\eta_8
(567)$ appears far from the NG massless limit we must ask: how
close is $\eta_0 (947)$ to the chiral-limiting nonvanishing singlet
$\eta$ mass?

To answer this latter question, return to Fig. 1 and the quark
annihilation strength $\beta \approx 0.28$ GeV$^2$ in Eq.~(\ref{eqno7}).
These $\overline q q$ states presumably hadronize into the U$_A$(1) singlet
state (\ref{eta0def}),
%$|\eta_0 \rangle = 
%|\overline u u + \overline d d + \overline s s \rangle / \sqrt3$,
for effective squared mass in the SU(3) limit with $\beta$ remaining
unchanged \cite{20}:
\eqnb
        m^2_{\eta _0} = 3 \beta \approx (917 ~\w{MeV})^2~.
        \label{eqno10}
\eqne
\noindent
This latter CL $\eta_0$ mass in (\ref{eqno10}) is only 3\% shy of the exact
chiral-broken $\eta_0 (947)$ mass found in Eq.~(\ref{eqno1}). (Such a 3\% CL 
reduction also holds for the pion decay constant $f_\pi \approx 93$ 
MeV $\to 90$ MeV \cite{22} and for $f_+ (0) = 1 \to 0.97$ \cite{23}, the 
$K$--$\pi$ $K_{l3}$ form factor.)

Our $\eta$--$\eta^\prime$ mixing analysis on the basis of phenomenological
mass inputs thus tells us that the physical $\eta (547)$ is 97\% of the
{\it chiral-broken} NG boson $\eta_8 (567)$.  Also the mixing-induced CL
singlet mass of 917 MeV in (\ref{eqno10}) is 97\% of the chiral-broken
singlet $\eta_0 (947)$ in (\ref{eqno1}), which in turn is 99\% of the
physical $\eta'$ mass $\eta' (958)$. This can be viewed as the 
phenomenological resolution of the U$_A$(1) problem of the masses and
(quasi-)Goldstone boson structure of the observed $\eta(547)$ and
$\eta^\prime(958)$ mesons. Or rather, from a more microscopic standpoint,
the above represents phenomenological constraints that microscopic, 
more or less QCD--based studies of the $\eta$--$\eta^\prime$ 
complex must respect.

{\subsection{Bound-state SD--BS approach to \mbox{$\eta$--$\eta^\prime$}}
\label{MainSec}

The coupled Schwinger-Dyson (SD) and Bethe-Salpeter (BS)
approach \cite{Roberts0007054}
can be formulated so that it has strong and clear connections with
QCD, the fundamental theory of strong interactions. In this approach, by
solving the SD equation for dressed quark propagators of various flavors,
one explicitly constructs constituent quarks. They in turn build $q\bar q$
meson bound states which are solutions of the BS equation employing the
dressed quark propagator obtained as the solution of the SD equation.
If the SD and BS equations are so coupled in a consistent approximation,
the light pseudoscalar mesons are simultaneously the $q\bar q$ bound states
and the (quasi) Goldstone bosons of dynamical chiral symmetry breaking
(D$\chi$SB). The resulting relativistically covariant bound-state
model (such as the variant of Ref.~\cite{jain93b}) is consistent with
current algebra because it incorporates the correct chiral symmetry behavior
thanks to D$\chi$SB obtained in an, essentially, Nambu--Jona-Lasinio fashion,
but the SD--BS model interaction is less schematic.
In Refs. \cite{KlKe2,jain93b,munczek92,KeBiKl98,KeKl1,KeKl3} for
example, it is combined nonperturbative and perturbative gluon exchange;
the effective propagator function is the sum of the known perturbative
QCD contribution and the modeled nonperturbative component.
For details, we refer to Refs. \cite{KlKe2,jain93b,munczek92,KeBiKl98,KeKl1},
while here we just note that the momentum-dependent dynamically generated
quark mass functions ${\cal M}_f(q^2)$ ({\it i.e.}, the quark propagator
SD solutions for quark flavors $f$) illustrate well
how the coupled SD-BS
approach provides a modern constituent model which is consistent
with perturbative and nonperturbative QCD. For example, the perturbative
QCD part of the gluon propagator leads to the deep Euclidean behaviors
of quark propagators (for all flavors) consistent with the asymptotic
freedom of QCD \cite{KeBiKl98}.
However, what is important in the present paper, is the behavior
of the mass functions ${\cal M}_f(q^2)$ for {\it low} momenta
[$q^2=0$ to $-q^2\approx (400 \, {\rm MeV})^2$], where ${\cal M}_f(q^2)$
(due to D$\chi$SB) has values consistent with typical values
of the constituent mass parameter in constituent quark models.
For the (isosymmetric) $u$- and $d$-quarks, our concrete model choice
\cite{jain93b} gives us ${\cal M}_{u,d}(0)=356$ MeV in the chiral limit
({\it i.e.}, with vanishing ${\widetilde m}_{u,d}$, the explicit chiral symmetry
breaking bare mass term in the quark propagator SD equation, resulting in
vanishing pion mass eigenvalue, $m_\pi=0$, in the BS equation),
and ${\cal M}_{u,d}(0)=375$ MeV [just 5\% above ${\cal M}_{u,d}(0)$ in the
chiral limit] with the bare mass
${\widetilde m}_{u,d}=3.1$ MeV, leading to a realistically light pion,
$m_\pi=140.4$ MeV. Similarly, for the $s$ quark, ${\cal M}_s(0)=610$ MeV.
The simple-minded constituent masses in both {\it NS} and {\it S} sectors, 
$\hat{m}$ and $\hat{m}_s$ employed in Sec. \ref{Preliminaries}, have thus 
close analogues in the coupled
SD--BS approach which explicitly incorporates some crucial features of QCD,
notably D$\chi$SB. Thanks to D$\chi$SB, this dynamical, bound-state approach 
successfully incorporates the partially Goldstone boson structure of the 
mixed $\eta (547)$ and $\eta' (958)$ mesons \cite{KlKe2}. 

Before addressing its mass matrix, let us briefly recall what the SD--BS 
approach revealed \cite{KlKe2,KeKlSc2000} about the mixing angle inferred 
from $\eta, \eta^\prime \to \gamma\gamma$ decays.
The SD--BS approach incorporates the correct chiral symmetry behavior 
thanks to D$\chi$SB and is consistent with current algebra. Therefore, 
and this gives particular weight to the constraints placed on the mixing angle
$\theta$ by the SD-BS results on $\gamma\gamma$ decays of pseudoscalars,
this approach reproduces
(when care is taken to preserve the vector Ward-Takahashi identity of QED)
analytically and exactly the CL pseudoscalar $\to \gamma\gamma$ decay 
amplitudes ({\it e.g.}, $\pi^0\to \gamma\gamma$), which are fixed by the 
Abelian axial anomaly.
(Note that they are otherwise notoriously difficult to reproduce 
in bound-state approaches, as discussed in Ref.~\cite{KeBiKl98}.) 

General and robust considerations in this chirally well-behaved
approach showed \cite{KlKe2} that, unlike the pion case, 
$\eta_8 , \eta_0 \to \gamma \gamma$
(and therefore also their mixtures $\eta, \eta^\prime \to \gamma \gamma$) 
decay amplitudes cannot be given through their respective axial-current
decay constants $f_{\eta_8}, f_{\eta_0}$, and also gave strong bounds on
these amplitudes with respect to the pion decay constant $f_\pi$
({\it i.e.}, w. r. to the $\pi^0\to \gamma \gamma$ amplitude).
All this says that in models relying on quark degrees of
freedom, reasonably accurate reproduction of the
empirical $\eta,\eta^\prime\to \gamma\gamma$ widths is possible 
only for $\theta$-values less negative than $-15^\circ$. For the 
concrete \cite{munczek92,jain93b} model adopted in Ref. \cite{KlKe2}, 
our calculated $\eta,\eta^\prime\to\gamma\gamma$ widths 
fit the data best for $\theta = -12.0^\circ$.

For the very predictive SD-BS approach to be consistent,
the above mixing angle extracted
from $\eta,\eta^\prime\to\gamma\gamma$ widths, should be
close to the angle $\theta$ predicted by diagonalizing the
$\eta$--$\eta^\prime$ mass matrix.  
In this section, it is given in the quark $f\bar f$ basis: 
\begin{equation}
{\hat M}^2 = \mbox{\rm diag} (M_{u\bar{u}}^2,M_{d\bar{d}}^2,M_{s\bar{s}}^2)
+ \beta 
\left( \begin{array}{ccl} 1 & 1 & 1 \\
                          1 & 1 & 1 \\
                          1 & 1 & 1
        \end{array} \right)~.
\label{M2}
\end{equation}
As in Sec. \ref{Preliminaries}, $3\beta$ (called $\lambda_\eta$
in Ref.~\cite{KlKe2}) is the contribution of the gluon axial anomaly
to $m_{\eta_0}^2$, the squared mass of $\eta_0$.
We denote by $M_{f{\bar f}^\prime}$ the masses obtained as eigenvalues 
of the BS equations for $q\bar q$ pseudoscalars with the
flavor content ${f{\bar f}^\prime}$ ($f, f^\prime = u, d, s$).
However, since Ref.~\cite{KlKe2} had to employ a rainbow-ladder approximation 
(albeit the improved one of Ref.~\cite{jain93b}), it could not calculate the 
gluon axial anomaly contribution $3\beta$. It could only avoid the
U$_A$(1)-problem in the $\eta$--$\eta^\prime$ complex by {\it parameterizing} 
$3\beta$, namely that part of the $\eta_0$ mass squared which remains 
nonvanishing in the CL. Because of the rainbow-ladder approximation (which 
does not contain even the simplest annihilation graph -- Fig. 1), the 
$q\bar q$ pseudoscalar masses $M_{f{\bar f}^\prime}$ {\it do not} contain 
any contribution from $3\beta$, unlike the
nonstrange and strange $\eta$ masses $m_{\eta_{NS}}$ [in Eq.~(\ref{eqno9a})]
and $m_{\eta_S}$ [in Eq.~(\ref{eqno9b})], which do, and which must not
be confused with $M_{u\bar u}=M_{d\bar d}$ and $M_{s\bar s}$.
Since the flavor singlet gluon anomaly contribution $3\beta$ does not 
influence the masses $m_\pi$ and $m_K$ of the non-singlet pion and kaon, 
the realistic rainbow-ladder modeling aims directly at
reproducing the empirical values of these masses: 
$M_{u\bar u}=M_{d\bar d}=m_\pi$ and $M_{s\bar d} = m_K$. In contrast, 
the masses of the physical etas, $m_\eta$ and $m_{\eta^\prime}$, must be 
obtained by diagonalizing the $\eta_8$-$\eta_0$ sub-matrix containing 
both $M_{f\bar f}$ and the gluon anomaly contribution to $m_{\eta_0}^2$.

Since the gluon anomaly contribution $3\beta$ vanishes in the large $N_c$ 
limit as $1/N_c$, while all $M_{f{\bar f}^\prime}$ vanish in CL, our $q\bar q$ 
bound-state pseudoscalar mesons behave
in the $N_c\to\infty$ and chiral limits in agreement with QCD and
$\chi$PT ({\it e.g.}, see \cite{G+L}): as the strict CL
is approached for all three flavors, the SU(3) octet pseudoscalars
{\it including} $\eta$ become massless Goldstone bosons, whereas the
chiral-limit-nonvanishing $\eta^\prime$-mass $3\beta$ is of order $1/N_c$ 
since it is purely due to the gluon anomaly. 
If one lets $3\beta \to 0$ (as the gluon anomaly contribution 
behaves for $N_c\to\infty$), then for any quark masses and 
resulting $M_{f\bar f}$ 
masses, the ``ideal" mixing ($\theta=-54.74^\circ$) takes place so that 
$\eta$ consists of $u,d$ quarks only and becomes 
degenerate with $\pi$, whereas $\eta^\prime$ is the pure $s\bar s$ 
pseudoscalar bound state with the mass $M_{s\bar s}$.

In Ref.~\cite{KlKe2}, numerical calculations of the mass matrix were 
performed for the realistic chiral and SU(3) symmetry breaking, 
with the finite quark masses (and thus also the finite BS $q\bar q$ bound-state 
pseudoscalar 
masses $M_{f\bar f}$) fixed by the fit~\cite{jain93b} to static properties 
of many mesons but excluding the $\eta$--$\eta^\prime$ complex. The mixing 
angle which diagonalizes the  $\eta_8$-$\eta_0$ mass matrix thus depended 
in Ref.~\cite{KlKe2} only on the value of the additionally introduced 
``gluon anomaly parameter" $3 \beta$. Its preferred value 
turned out to 
be $3 \beta=1.165$ GeV$^2$=(1079 MeV)$^2$, leading to the mixing 
angle $\theta=-12.7^\circ$ 
[compatible with $\phi=41.84^\circ$ in Eq.~(\ref{eqno8})]
and acceptable $\eta\to \gamma \gamma$ and 
$\eta^\prime\to \gamma\gamma$ decay amplitudes. Also, the $\eta$ mass was 
then fitted to its experimental value, but such a high value of $3\beta$ 
inevitably resulted in a too high  $\eta^\prime$ mass, above 1 GeV.
(Conversely, lowering $3\beta$ aimed to reduce $m_{\eta^\prime}$, 
would push $\theta$ close to $-20^\circ$, making predictions
for $\eta,\eta^\prime \to \gamma\gamma$ intolerably bad.)
However, unlike Eq.~(\ref{eqno6}) in the present paper, it should be 
noted that Ref.~\cite{KlKe2} did not introduce into the mass matrix
the ``strangeness attenuation parameter" $X$ which should suppress 
the nonperturbative quark $f\bar f \to f^\prime {\bar f}^\prime$ 
annihilation amplitude (illustrated by the ``diamond" graph in Fig. 1)
when $f$ or $f^\prime$ are strange.
Ref. \cite{KeKlSc2000} concluded that it was precisely
the lack of the strangeness attenuation factor $X$
that prevented Ref. \cite{KlKe2} from satisfactorily
reproducing the $\eta^\prime$ mass when it successfully did so
with the $\eta$ mass and $\gamma\gamma$ widths.

One can expect that the influence of this suppression should be
substantial, since $X \approx {\hat m}/\hat{m}_s$ should be a reasonable 
estimate of it, and this nonstrange-to-strange {\it constituent} mass
ratio in the considered variant of the SD-BS approach~\cite{KlKe2} is 
not far from $X$ in Eq.~(\ref{eqno7}) and from the mass ratios in Refs.~\cite{17,18,19}, 
and is even closer to the mass ratios in the Refs. \cite{16}.
Namely, two of us found~\cite{KlKe2} it to be around 
${\cal M}_u(0)/{\cal M}_s(0)=0.615 $ if the constituent mass was
defined at the vanishing argument $q^2$ of the momentum-dependent 
SD mass function ${\cal M}_f(q^2)$.

We therefore introduce the suppression parameter $X$ 
the same way as in the {\it NS--S} mass matrix (\ref{eqno6}), 
whereby the mass matrix in the $f\bar f$ basis becomes 
\begin{equation}
{\hat M}^2 = \mbox{\rm diag} (M_{u\bar{u}}^2,M_{d\bar{d}}^2,M_{s\bar{s}}^2)
    + \beta
        \left( \begin{array}{ccl} 1 & 1 & X \\
                                  1 & 1 & X \\
                                  X & X & X^2
        \end{array} \right)~.
\label{M2wX}
\end{equation}
The very accurate isospin symmetry makes the mixing of
the isovector $\pi^0$ and the isoscalar etas negligible
for all our practical purposes. 
Going to a meson basis of $\pi^0$ and etas enables us
therefore to separate the $\pi^0$ and restrict ${\hat M}^2$ to 
the $2\times 2$ subspace of the etas. In the {\it NS--S} basis,
\eqnb
        \pip{
                \begin{array}{ll}
         m_{\eta_{NS}}^2    &  m_{\eta_{S}\eta_{NS}}^2 \\
            m_{\eta_{NS}\eta_{S}}^2 &    m_{\eta_{S}}^2
                \end{array}
        }
     =
        \pip{
                \begin{array}{ll}
      M_{u\bar{u}}^2 + 2 \beta  & \quad \sqrt{2} \beta X \\
        \,  \sqrt{2} \beta X    & M_{s\bar{s}}^2 + \beta X^2
                \end{array}
        }.
        \label{SD-BS-NS-S}
\eqne
To a very good approximation, Eq. (\ref{SD-BS-NS-S}) recovers 
Eq.~(\ref{eqno6}). This is because not only
$m_\pi=M_{u\bar u}=M_{d\bar d}$, but also because
$M_{s\bar s}^2$ differs from $2 m_K^2 - m_\pi^2$ only by a couple
of percent, thanks to the good chiral behavior of the 
masses $M_{f{\bar f}^\prime}$ calculated in SD-BS approach.
(These $M_{f\bar{f}^\prime}^2$ and the CL model values of $f_\pi$ and 
quark condensate, satisfy Gell-Mann-Oakes-Renner relation to first 
order in the explicit chiral symmetry breaking \cite{munczek92}.)
The SD-BS--predicted octet (quasi-)Goldstone masses $M_{f{\bar f}^\prime}$ 
are known to be empirically successful in our concrete model choice
\cite{jain93b}, but the question is whether the SD-BS approach can also 
give some information on the $X$-parameter.  
If we treat {\it both} $3\beta$ and $X$ as free parameters, we can of 
course fit both the $\eta$ mass and the $\eta^\prime$ mass to their 
experimental values. 
For the model parameters as in Ref. \cite{jain93b}
(for these parameters our independent calculation gives 
$m_\pi=M_{u\bar u}=140.4$ MeV and $M_{s\bar s}=721.4$ MeV),
this 
happens at $3\beta=0.753$ GeV$^2$ =(868 MeV)$^2$
and $X=0.835$.  However,
the mixing angle then comes out as $\theta=-17.9^\circ$, which is 
too negative to allow consistency of the empirically found two-photon
decay amplitudes of $\eta$ and $\eta^\prime$, with predictions of
our SD-BS approach for the two-photon decay amplitudes of $\eta_8$ and 
$\eta_0$ \cite{KlKe2}. 

Therefore, and also to avoid introducing another free parameter in 
addition to $3\beta$, we take the path where the dynamical information 
from our SD-BS approach is used to estimate $X$.
Namely, our $\gamma\gamma$ decay amplitudes $T_{f\bar f}$
can be taken as a serious guide for estimating the $X$-parameter 
instead of allowing it to be free. 
We did point out in Sec. \ref{Preliminaries}
that the attempted treatment \cite{20} of the gluon anomaly contribution
through just the ``diamond diagram" contribution to $3\beta$,
indicated that just this partial contribution is quite insufficient.
This limits us to keeping $3\beta$ as a free parameter,
but we can still suppose that this diagram can help us get the
prediction of the strange-nonstrange {\it ratio} of the complete
pertinent amplitudes $f\bar f \to f^\prime {\bar f}^\prime$ as follows.
Our SD-BS modeling in Ref. \cite{KlKe2} employs an infrared-enhanced
gluon propagator \cite{jain93b,KeBiKl98} weighting the integrand strongly
for low gluon momenta squared.
Therefore, in analogy with Eq.~(4.12) of Kogut and
Susskind \cite{4} (see also Refs. \cite{FrankMeissner97,hep-ph9707210}),
we can approximate the Fig. 1 amplitudes $f\bar f \to {\rm 2 gluons}
\to f^\prime {\bar f}^\prime$, {\it i.e.}, the contribution of the quark-gluon
diamond graph to the element $f f^\prime$ of
the $3\times 3$ mass matrix, by the factorized form
\begin{equation}
{\widetilde T}_{f\bar f}(0,0) 
\, {\cal C} \, \, {\widetilde T}_{f^\prime {\bar f}^\prime}(0,0) \, .
\label{factoriz} 
\end{equation}
In Eq. (\ref{factoriz}), the quantity ${\cal C}$ is given by the 
integral over two gluon propagators remaining after factoring out 
${\widetilde T}_{f\bar f}(0,0)$ and 
${\widetilde T}_{f^\prime {\bar f}^\prime}(0,0)$, the respective 
amplitudes for the transition of the $q\bar q$ pseudoscalar bound 
state for the quark flavor $f$ and $f^\prime$ into two vector bosons, 
in this case into two gluons.
The contribution of Fig. 1 is thereby expressed with the help of
the (reduced) amplitudes ${\widetilde T}_{f\bar f}(0,0)$ calculated
in Ref. \cite{KlKe2} for the transition of $q\bar q$ pseudoscalars 
to two real photons ($k^2={k^\prime}^2=0$), while in general 
${\widetilde T}_{f\bar f}(k^2,k^{\prime 2}) \equiv
{T}_{f\bar f}(k^2,k^{\prime 2})/Q_f^2$ are the ``reduced" two-photon
amplitudes obtained by removing the squared charge factors $Q_f^2$
from ${T}_{f\bar f}$, the $\gamma\gamma$ amplitude of the pseudoscalar
$q\bar q$ bound state of the hidden flavor $f {\bar f}$. 
Although ${\cal C}$ is in principle computable, 
all this unfortunately does not amount to determining $\beta, \beta X$ 
and $\beta X^2$ in Eq. (\ref{M2wX}) since the higher (four-gluon, 
six-gluon, ... , etc.) contributions are clearly lacking. We therefore 
must keep the total (light-)quark annihilation strength $\beta$ as a 
free parameter. However, if we assume
that the suppression of the diagrams with the strange quark in a loop
is similar for all of them, Eq.~(\ref{factoriz}) and the ``diamond" 
diagram in  Fig. 1 help us to at least estimate the parameter $X$ as 
$X \approx {\widetilde T}_{s\bar s}(0,0)/{\widetilde T}_{u\bar u}(0,0)$.
This is a natural way to build in the effects of the SU(3) flavor 
symmetry breaking in the $q\bar q$ annihilation graphs.
(Recall that ${\widetilde T}_{s\bar s}(0,0)/{\widetilde T}_{u\bar u}(0,0)
\approx \hat{m}/\hat{m}_s$ to a good approximation \cite{KeKlSc2000}.)

We get $X=0.663$ from the two-photon amplitudes we obtained in the chosen 
SD-BS model \cite{jain93b}. This value of $X$ agrees well with the other 
way of estimating $X$, namely the nonstrange-to-strange constituent mass 
ratio of Refs. \cite{17,18,19}. 
With $X=0.663$, requiring that 
the $2\times 2$ matrix trace, $m_\eta^2 + m_{\eta^\prime}^2$,
be fitted to its empirical value,
fixes the chiral-limiting nonvanishing singlet 
mass squared to $3\beta=0.832$ GeV$^2$=(912 MeV)$^2$, just 0.5\%
below Eq.~(\ref{eqno10}), while $m_{\eta_{NS}}=757.87$ MeV and 
$m_{\eta_S}=801.45$ MeV, practically the same as Eqs. (\ref{eqno9}). 
The resulting mixing angle and $\eta$, $\eta^\prime$ masses are
\begin{equation}
\phi = 41.3^\circ \wsep{.1}{or} \theta = - 13.4^\circ \, \,  ;
\quad m_\eta = 588 \, \mbox{MeV} \, , 
\quad m_{\eta^\prime} = 933 \, \mbox{MeV} \, .
\label{results1}
\end{equation}
These results are for the original parameters of Ref. \cite{KlKe2}.
Reference \cite{KeKlSc2000} also varied the parameters to check the
sensitivity on SD-BS modeling, but the results changed little.

The above results of the SD-BS approach~\cite{KlKe2} are very satisfactory 
since they agree very well with what was found in Sec. \ref{Preliminaries} 
by different methods. They also agree with the UKQCD lattice results 
\cite{UKQCDetas2000} on $\eta$--$\eta^\prime$ mixing. Their calculated 
mixing parameter $x_{ss}$ corresponds to our $\beta X^2$, and their mixing 
parameters $x_{nn}$ and $x_{ns}$ ($n=u,d$), corresponding respectively
to our $\beta$ and $\beta X$, are aimed to obey $x_{nn} \approx 2 x_{ss}$
and $x_{ns}^2 \approx x_{nn} x_{ss}$. UKQCD prefers \cite{UKQCDetas2000} 
$x_{ss} = 0.13$ GeV$^2$, $x_{nn} = 0.292$ GeV$^2$ and $x_{ns} = 0.218$
GeV$^2$. This, together with their preferred {\it input} values 
$M_{n\bar n}=0.137$ GeV and $M_{s\bar s}=0.695$ GeV, give the 
{\it NS--S} mass matrix (\ref{SD-BS-NS-S}) with elements reasonably 
close to ours, resulting in a rather close mixing angle, 
$\theta = - 10.2^\circ$.

\subsection{Conclusion}
We have shown that the treatment of the $\eta$--$\eta^\prime$ complex in the 
SD--BS approach \cite{KlKe2} is sensible in spite of employing the ladder 
approximation. This is confirmed especially by Ref. \cite{KeKlSc2000} which 
showed its connection and robust agreement with the phenomenological studies 
of the $\eta$--$\eta^\prime$ complex. It is therefore desirable to extend the 
SD--BS studies of the $\eta$--$\eta^\prime$ mass matrix to finite temperatures.
Usually, one has neglected all temperature dependences in the mass matrix, 
except the one of the gluon anomaly contribution $3\beta$ which is assumed 
very strong, which is appropriate if the U$_A$(1) symmetry breaking is due to 
instantons \cite{PisarskiWilczek84,AlkofAmundReinh89,Kapusta+al96,HuangWang96}.
However, rather strong topological arguments of Kogut {\it et al.} 
\cite{Kogut+al98} that the U$_A$(1) symmetry is not restored at critical 
(but only at a higher, possibly infinite) $T$, motivates also the scenario 
where $3\beta(T)\approx const$, while other entries in the mass matrix carry 
the temperature dependence. The inclusion of their $T$--dependence is needed 
also because the scenario with the instanton--induced, strongly $T$--dependent 
$\beta$ should be carefully re-examined, since it has lead to contradicting 
conclusions:
the depletion of $\eta^\prime$ production in Ref. \cite{AlkofAmundReinh89}, 
but $\eta^\prime$--enhancement in Ref. \cite{Kapusta+al96}.

The temperature dependence of $m_\pi = M_{u\bar u} =  M_{d\bar d}$, 
${\cal M}_{u,d}(q^2)$, $f_\pi$ and 
$\langle{u\bar u}\rangle (= \langle{d\bar d}\rangle)$, was already studied 
in various SD--BS models \cite{Maris+alNT0001064,Blaschke+alNT0002024}, so 
that the extension \cite{Blaschke+al01} to the $T$--dependence of 
the remaining needed ingredients, $M_{s\bar s}$, ${\cal M}_{s}(q^2)$, 
$f_{s\bar s}$ and $\langle{s\bar s}\rangle$, should be straightforward.

\vskip 4mm

\noindent {\bf Acknowledgments:} 
D. Kl. thanks D. Blaschke, S. Schmidt and G. Burau, the organizers
of the workshop ``Quark Matter in Astro- and Particle Physics" 
(27.-29. November 2000, Rostock, Germany) for their hospitality and 
for the support which made his participation possible.

%\vspace{-2mm}

%\end{document}
\newpage

%% file: SOURCE/ruivo1NEU.tex
\section*{\bf Pseudoscalar mesons in asymmetric matter}
\addcontentsline{toc}{section}{\protect\numberline{}{Pseudoscalar Mesons in Asymmetric Matter \\ \mbox{\it M.C. Ruivo, P. Costa, C.A. de Sousa}}}
\begin{center}
\vspace*{2mm}
{Maria C. Ruivo, Pedro Costa  and C\' elia  A. de Sousa}\\[0.3cm]
{\small\it Centro de F\'isica Te\' orica, Departamento de F\'isica\\
Universidade de Coimbra
%\thanks{Work supported in part by Centro de F\' isica Te\'orica, and PRAXIS/P/FIS/12247/98, FCT, %Portugal. One of us (MCR) wishes to thank financial support from the University of Rostock, Germany.}\\
P - 3004 - 516 Coimbra, Portugal}
\end{center}
%\lefthead{LEFT head}
%\righthead{RIGHT head}

\begin{abstract}The  behavior of  kaons  and pions in  hot non strange   quark matter, simulating neutron matter, 
is investigated within the SU(3)  Nambu-Jona-Lasinio [NJL] and in the Enlarged Nambu-Jona-Lasinio [ENJL ] (including vector pseudo-vector interaction) models. At zero temperature, it is found that in the NJL model, where the phase transition is first order,   low energy modes with $K^-,  \pi^+$ quantum numbers, which are particle-hole excitations  of the Fermi sea, appear. Such modes are not found  in the ENJL model and in NJL at finite temperatures. The increasing temperature has also the effect of reducing the splitting between the charge multiplets.
\end{abstract}

\newcounter{eqn10}[equation]
\setcounter{equation}{-1}
\stepcounter{equation}

\newcounter{bild10}[figure]
\setcounter{figure}{-1}
\stepcounter{figure}

\newcounter{tabelle10}[table]
\setcounter{table}{-1}
\stepcounter{table}

%\newcounter{kapitel10}[section]
%\setcounter{section}{-1}
%\stepcounter{section}

\newcounter{unterkapitel10}[subsection]
\setcounter{subsection}{-1}
\stepcounter{subsection}

\subsection{Introduction}

During the last few years,  major  experimental and theoretical efforts have been dedicated to heavy-ion collisions aiming at understanding  the properties of hot and dense matter and looking for  signatures of phase transitions to the quark-gluon plasma.  As a matter of fact, it  is believed that, at critical values of the density, $\rho_c$, and/or temperature, $T_c$, the system  undergoes a phase transition,  the QCD vacuum being then des\-cri\-bed by a   weakly interacting  gas of quarks and gluons, with restored chiral symmetry.

 The nature of the phase transition is an important issue   nowadays. Lattice simulations \cite{Kanaya}  provide information  at zero density and finite temperature but for finite  densities no firm lattice results are available and most of our knowledge  comes from model calculations. The Nambu-Jona-Lasinio\cite{NJL} [NJL] type models have been extensively used over the past years to describe low energy features of hadrons and also to investigate  restoration of chiral symmetry with temperature or density \cite{Hatsuda94,Ruivo,SousaRuivo,RSP,Hiller}. 

Recently, it was shown by Buballa\cite{buballa} that, with a convenient parameterization, the SU(2) and SU(3) NJL models exhibit  a first order
phase transition, the system being in a mixed phase between  $\rho\,=\,0$ and $\rho\,=\,\rho_c$, the energy per particle having an absolute minimum at $\rho\,=\,\rho_c$. This suggests an interpretation of the model within the   philosophy of the MIT bag model. The system has two phases, one consisting of   droplets of quarks of high density and low mass surrounded by a non trivial vacuum and  the other one consisting of a quark phase of restored chiral symmetry. Similar concepts appear in NJL inspired models including form factors \cite{raja,david}. The physical meaning of the  mesonic excitations in the medium within this interpretation of the model is an interesting subject that will be analyzed in this paper.

The possible modifications of meson properties in the medium is an important issue nowadays. 
 The study of pseudoscalar mesons, such as kaons and pions, is particularly interesting, since, due to their Goldstone boson nature, they are intimately associated with the breaking of chiral symmetry. Since the  work of Kaplan and Nelson \cite{Kaplan}, the study of medium effects on these mesons in flavor asymmetric media attracted a lot of attention. Indeed, the charge multiplets of those mesons, that are degenerated in vacuum or in symmetric matter, were predicted to have a splitting in flavor asymmetric matter. In particular,  as the density increases, there would be  an increase of the mass of $K^+$ and a decrease of the mass of $K^-$; a similar effect would occur for $\pi^-$ and $\pi^+$ in neutron matter. The mass decrease of one of the multiplets raises, naturally, the issue of meson condensation, a topic specially relevant to Astrophysics.

Most  theoretical approaches  dealing with 
 kaons in flavor asymmetric media, predict  a slight  raising of the $K^+$ mass 
and a pronounced lowering of the $K^-$ mass  \cite{Lutz94,RSP,Cassing,Schaffner}, a conclusion which is supported by the analysis of data on kaonic atoms \cite{Friedmann94}. Experimental results  at GSI  seem to be compatible with
 this scenario \cite{Schr94,Herrman,Barth}.

Studies on pions in asymmetric medium are mainly related with the problem of the $u-d$ asymmetry in a nucleon sea rich in neutrons. Such  flavor asymmetry has been established in SIS and DY experiments and theoretical studies show that there is a significant difference in $\pi^+\,\,,\pi^-$ distribution functions in neutron rich matter \cite{peng}.

From the theoretical point of view, the driving mechanism for the mass splitting  is attributed   mainly to the 
selective effects of the Pauli principle, although, in the case of 
 $K^-$, the interaction with  the
 $\Lambda (1405)$ resonance    plays an important role as well. 
In the study of the effects of the medium on  hadronic behavior, one should have in mind that the medium is a complex system, where a great variety of medium particle-hole excitations  occur, some of them with the same quantum numbers of  the hadrons under study; the interplay of all these  excitations might play a significant role in the modifications of hadron properties. In previous works we have established, within the framework of  NJL models, the presence, in flavor asymmetric media, of  low  energy pseudoscalar modes, which are excitation of the Fermi sea  \cite{RuivoSousa,SousaRuivo}. The combined effect of density and temperature, as well as the effect of vector interaction,  was discussed for the case of kaons in symmetric nuclear matter without strange quarks \cite{RSP,Ruivo}.

This paper addresses the following points: a) analyzes of the phase transition with density and temperature in   neutron matter in the $SU(3)$ NJL model with two different parameterizations  and within the ENJL model ; b) behavior of kaonic and pionic excitations in these models and discussion of the meaning of the Fermi sea excitations, in connection with the nature of the phase transition; c) combined effect of density and temperature.
\section*{Formalism}
We work in a flavor $SU(3)$ NJL type model with  scalar-pseudoscalar and vector-pseudovector pieces, and a determinantal term, the 't Hooft interaction, which breaks the $U_A(1)$ symmetry. We use the following Lagrangian:
\begin{equation}  
\begin{array}{rcl}  
\cal L\,&=& \bar q\,(\,i\, {\gamma}^{\mu}\,\partial_\mu\,-\,\hat m)\, q% \\[4pt]  &+&
+\frac{1}{2}\,g_S\,\,\sum_{a=0}^8\, [\,{(\,\bar q\,\lambda^a\, 
q\,)}^2\,\,+\,\,{(\,\bar q \,i\,\gamma_5\,\lambda^a\, 
q\,)}^2\,] \\[4pt] 
 &-&\frac{1}{2}\,g_V\,\,\sum_{a=0}^8\, [\,{(\,\bar q\,\gamma_\mu\,\lambda^a\, 
q\,)}^2\,\,+\,\,{(\,\bar q \gamma_\mu\,\gamma_5\,\lambda^a\, 
q\,)}^2\,] \\[4pt]  
&+& g_D\,\,  \{\mbox{det}\,[\bar q\,(\,1\,+\,\gamma_5\,)\,q\,] 
%\\[4pt]  
 + \mbox{det}\,[\bar 
q\,(\,1\,-\,\gamma_5\,)\,q \,]\, \} \label{1} 
\\ & &
 \end{array}\label{eq:lag} 
\end{equation}
 %%%%%%%%%%%%%%%%%%%%%%%%%%%%%%%%%%%%%%%%%%%%
In order to discuss the predictions of different models we  consider the cases:  $g_V=0$ with two parametrizations (NJL I and NJL II)  and  $g_V\neq 
0$ (ENJL). The model  parameters, the bare quark masses   $m_d=m_u, m_s$, the 
coupling constants  and the cutoff in 
three-momentum space, $\Lambda$, are essentially fitted to  the experimental values of $m_\pi,\,f_\pi,\,m_K$ and to the phenomenological values of  the quark condensates,  $<\bar uu>,\, <\bar 
dd>,\,<\bar ss>$. The parameter sets used are, for NJL I:  
 $\Lambda=631.4$ MeV, $ g_S\,\Lambda^2=3.658,\,g_D\,\Lambda^5=- 9.40, \, m_u=m_d=5.5$ MeV and 
 $m_s=132.9$ MeV; for ENJL:  $\Lambda=750$ MeV, $ g_S\,\Lambda^2=3.624$, $ g_D\,\Lambda^5=- 9.11,\, g_V\,\Lambda^2=3.842$, $m_u=m_d=3.61$ MeV and
  $m_s=88$ MeV.  For NJL II we use the parametrization of  \cite{RKH}, $\Lambda=602.3$ MeV, $ g_S\,\Lambda^2=3.67$, $g_D\Lambda^5=-12.39$, $m_u=m_d=5.5$ MeV and 
 $m_s=140.7$ MeV, 
which underestimates the pion mass ($m_\pi\,=135\mbox{ MeV}$) and of $\eta$ by about $6\%$.  \vskip0.5cm

The   six quark interaction can be put  in a form suitable to use the bosonization procedure (see \cite{Vogl,Ripka,Ruivo}):

\begin{equation} 
  {\cal L_D}\,=\, \frac{1}{6} g_D\,\, D_{abc} \,(\bar q\, {\lambda}^c\,q\,)\,[\,(\,\bar q\,\lambda^a\, 
q\,)(\bar q\,\lambda^b\, 
q\,)  - 3\,(\,\bar q \,i\,\gamma_5\,\lambda^a\, 
q\,)\,(\,\bar q \,i\,\gamma_5\,\lambda^b\, 
q\,)\,]
\end{equation}

\noindent with: 
$D_{abc}=d_{abc}\,, a,b,c\,\, \epsilon\, \{1,2,..8\}\,, \mbox{(structure constants of SU(3))}\,,D_{000}=\sqrt{\frac{2}{3}}\,,
D_{0ab}=-\sqrt{\frac{1}{6}}\delta_{ab}$.

The usual procedure to obtain a four quark effective interaction from this six quark  interaction is to contract  one bilinear $(\bar q\,\lambda_a\,q)$. Then, from the two previous equations, an effective Lagrangian is obtained: 

\vspace{1cm}

\begin{eqnarray}  
L_{eff}\,&=& \bar q\,(\,i\, {\gamma}^{\mu}\,\partial_\mu\,-\,\hat m)\, q \,\,\nonumber  \\ 
  &+& S_{ab}[\,(\,\bar q\,\lambda^a\, 
q\,)(\bar q\,\lambda^b\, 
q\,)]
+\,P_{ab}[(\,\bar q \,i\,\gamma_5\,\lambda^a\, 
q\,)\,(\,\bar q \,i\,\gamma_5\,\lambda^b\, 
q\,)\,]\nonumber \\
&-&\frac{1}{2}\,g_V\,\,\sum_{a=0}^8\, [\,{(\,\bar q\,\gamma_\mu\,\lambda^a\, 
q\,)}^2\,\,+\,\,{(\,\bar q \gamma_\mu\,\gamma_5\,\lambda^a\, 
q\,)}^2\,]\,\,  
 \end{eqnarray}

\noindent where:
\begin{eqnarray}
S_{ab}\,=\,g_S\,\delta_{ab}\,+\,g_D D_{abc} \,< \bar q\, {\lambda}^c\,q\,>\nonumber \\
P_{ab}\,=\,g_S\,\delta_{ab}\,-\,g_D D_{abc} \,< \bar q\, {\lambda}^c\,q\,>
 \end{eqnarray}

By using the usual methods of bosonization one gets the following effective action:
\begin{eqnarray}\label{action}
 I_{eff}=&-i&Tr\ {\rm ln}(\,i\,\partial_\mu \gamma_{\mu}-\hat m+\sigma_a\,\lambda^a +i\,\gamma_5\,
 {\phi}_a\,\,{\lambda}^a\,+\,\gamma^\mu \,V_\mu\,+\gamma_5\,\gamma^\mu\,A_\mu)\nonumber\\
&-&\frac{1}{2}(\,\sigma_a\,{S_{ab}}^{-1}\,\sigma_b\,+{\phi}_a\, { P_{ab}}^{-1}\,\phi_b\,)\nonumber \\
&+&\frac{1}{2G_V}(\,{V^a_{\mu}}^2\,+\,{A^a_{\mu}}^2\,),
\end{eqnarray}

\noindent from which we obtain the gap equations and meson propagators. 

In order to introduce the finite temperature and density, we use the thermal Green function, which, for a quark  $q_i$ at finite temperature $T$  and 
chemical potential $\mu_i$ reads:

\begin{eqnarray}
S (\vec x -\vec  x',\tau -\tau')& = &\frac{i}{\beta}{\sum}_n e^{-i \omega_n (\tau - \tau')}\nonumber \\
& \int& \frac{d^3p}{(2 \pi)^3} \frac{e^{-i\,\vec p\,(\vec x -\vec x')}}{\gamma_0 (i \omega_n + {\overline \mu_i}) -  \vec \gamma .  \vec p - M_i}\,,
\end{eqnarray}

 \noindent where  $\beta =1/T\,, {\overline \mu_i}\,=\,\mu_i\,-\,\Delta E_i$, $\Delta E_i$ is the energy gap induced by the vector interaction, $M_i$ the mass of the constituent  quarks, 
   $E_i\,=\,(p^2\,+\,M^2_i)^{1/2}$  and   $\omega_n\,=\,(2 n + 1)\,\frac{\pi}{\beta}\,\,\,, n\,=\,0\,,\pm 1\,,\pm 2\,,....,$ are the Matsubara frequencies.
The following gap equations are obtained:

\begin{equation}
M_i\,=\,m_i\,-2\,g_S\,<\bar q_i\, q_i>\,-\,2\,g_D\,<\bar q_j\, q_j><\bar q_k\, 
q_k>
\end{equation}
\begin{equation}
\Delta E_i\,=\,2\,g_V\,<{q_i}^+\, q_i> 
\end{equation}

\noindent with $i\,,j\,,k$ cyclic and
  $<\bar q_i\, q_i>$, $<{q_i}^+\, q_i>$ are respectively the quark condensates and the quark densities at finite $T$ and $\mu_i$.

The condition for the existence  the poles in the propagators of kaons leads to the following dispersion relation:

\begin{equation}
(\,1\,-\,K_P\,\,J_{PP}\,)\,\,(\,1\,-\,K_A\,\,J_{AA}\,)\,-\,K_P\,K_A\,\, 
{J^2}_{PA}\,=\,0
\end{equation}
with: 
 $$
\omega\, J_{PA}\,=\,(M_u + M_s) \,J_{PP}\,+\,2\,(<\bar u u\,>+\,<\bar s s>). $$
 $$
\omega\, J_{AA}\,=\,(M_u + M_s) \,J_{PA}\,+\,2\,(< u^+ u>\,-\,< s^+ 
s>).$$

$$
J_{PP}\,=\,2\,N_c\,\int {d^3 p\over (2\pi)^3}\,\left\{\frac {M_u (M_s - M_u)
  \,- q_0\, E_u}{({E_s}^2 - (q_0+ E_u)^2) E_u} \mbox{ tanh}
\frac{\beta(E_u+{\bar \mu_u)}}{2}\,+\right.
$$
\begin{equation}  
\left.\frac {M_u (M_s - M_u) \ +
q_0 E_u}{({E_s}^2 - (q_0 -E_u)^2) E_u} \mbox{ tanh}
\frac{\beta(E_u- {\bar \mu_u)}}{2}\,\,+\, s \rightarrow u\,,\,q_0
\rightarrow -q_0\right\}
\end{equation}
with  $K_P\,=\,g_S\,+\,g_D \,<\bar d d>$ and  $K_A\,=\,-g_V$,
 ${\bar \mu_u}=\mu_u-\Delta E_u$,
$q_0 =\pm m_{K^{\pm}}- (\Delta E_u-\Delta E_s)$   for $K^{\pm}$ 
  
Similar expressions are obtained for pions in neutron  matter, by replacing $s\leftrightarrow d$ \cite{SousaRuivo}.

\subsection{Phase transitions at finite chemical potential and temperature}

We reanalyze  the problem of the phase transitions in order to establish a connection between the vacuum state and its excitations. We consider here the case of asymmetric quark matter without strange quarks simulating neutron matter: $\rho_u\,=\frac{1}{2}\,\rho_d\,\,,\rho_s\,=\,0$  and calculate the  the energy and pressure.  
At zero temperatures  we found a first order phase transition in NJL model, exhibiting different characteristics according to the parameterization used. Within the parameterization NJL I, the pressure is negative for $0.8 \rho_0\leq \rho \leq 1.65 \rho_0$  and the absolute minimum of the energy per particle is at $\rho=0$ (dashed curves in Fig.1). 

%%\begin{figure}[h]
%%\vspace{5.5cm}
%%\hspace{-0.5cm}\special{pressao.eps}
%%\hspace{5.5cm}\special{energia.eps}
%%\caption{Pressure and energy per particle in NJL I (dashed curves) and NJL II (full curves).}
%%\end{figure}
%%
%\begin{figure}[h]
%\centering{\
%\epsfig{figure=pressao.eps,height=5.5cm,width=5.5cm}}
%\vspace{-0.5cm}
%\caption{Pressure and energy per particle in NJL I (dashed curves) and NJL II (full curves).
%%\label{...}
%}
%\end{figure}
%
\begin{figure}[h]
\centerline{
\epsfig{figure=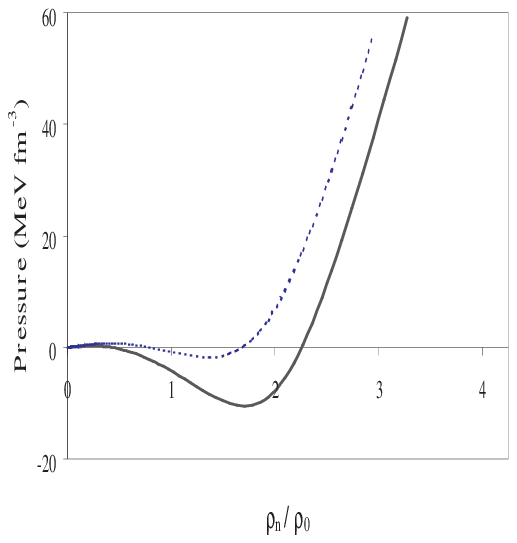,height=5.5cm,width=5.5cm}
\epsfig{figure=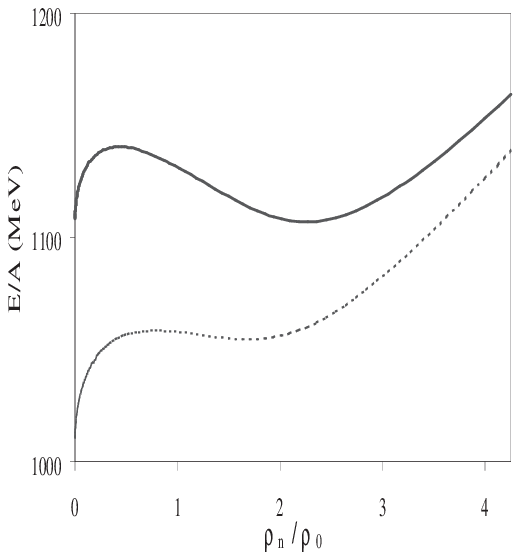,height=5.5cm,width=5.5cm}}
\caption{Pressure and energy per particle in NJL I (dashed curves) and NJL II (full curves).
%\label{...}
}
\end{figure}

The system, within the range of densities indicated above, is in a mixed phase consisting of droplets of massive quarks of low density and  droplets of light quarks of high density   and, for $\rho > 1.65 \rho_0$,   is in a quark phase   with partially restored chiral symmetry (in the SU(2) sector).  These droplets are unstable since the absolute minimum of the energy per particle is at $\rho=0$. With  parameterization II, the mixed phase starts at $\rho\simeq 0$, because, although the zeros of the pressure are at $\rho = 0.44 \rho_0,\,\,\rho_c=2.25 \rho_0$, the compressibility is negative in the low density region for $\rho \simeq 0$; the energy per particle has an absolute minimum at the critical density of $E/A \,= 1102$ MeV, about three times the masses of the constituent non strange quarks in vacuum (Fig 1.   full curves). The model may now be interpreted as having   a hadronic phase  ---  droplets of light $u\,,d$ quarks with  a density  $\rho_c=2.25 \rho_0$ surrounded by a non trivial vacuum --- and, above the critical density, a quark phase with partially restored  SU(2) chiral symmetry. The model is not suitable to describe hadrons for $\rho\,< \rho_c$. Since there is no definition of the density in the mixed phase, we study, in the following, the mesonic excitations only for $\rho > \rho_c$.

The  phase transition becomes second order at finite density and temperatures around $20 MeV$    and also in the 
 ENJL  model, for the set of parameters chosen. In these cases the system has positive pressure but the absolute minimum of the energy  per particle is at zero density.   Although a gas of quarks does not exist at low densities, the model has been used to study the influence of the medium in the mesonic excitations of the vacuum. Of course, an extrapolation of quark matter to hadronic  matter should be made, since we do not have, at low densities, a gas of hadrons.

\subsection{Behavior of pions and kaons in the medium}

At zero temperature we observe a splitting between charge multiplets, both in NJL and  ENJL models (see figs. 2-3).

%%\begin{figure}[h]
%%\vspace{5.5cm}
%%\hspace{-0.5cm}\special{kaao.eps}
%%\hspace{5.5cm}\special{pi.eps}
%%\caption{Masses of kaons and pions in NJL II. $\omega_{up}\, \mbox{and }\, \omega_{low}$ denote the limits of the Fermi sea continuum.}
%%\end{figure}
%%
%\begin{figure}[h]
%\centering{\
%\epsfig{figure=kaao.eps,height=5.5cm,width=5.5cm}}
%\vspace{-0.5cm}
%\caption{Masses of kaons and pions in NJL II. $\omega_{up}\, \mbox{and }\, \omega_{low}$ denote the limits of the Fermi sea continuum.
%%\label{...}
%}
%\end{figure}
\begin{figure}[h]
\centerline{
\epsfig{figure=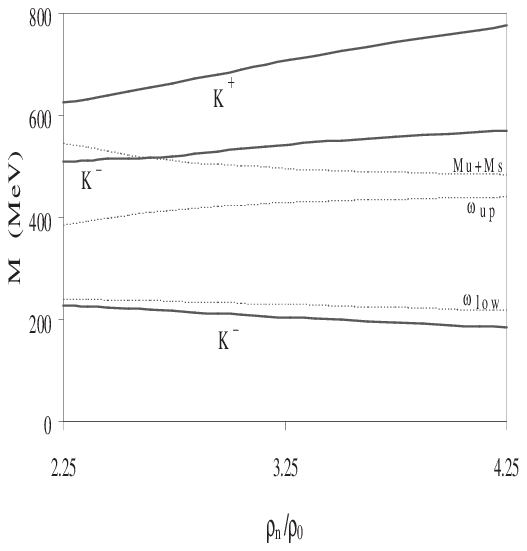,height=5.5cm,width=5.5cm}
\epsfig{figure=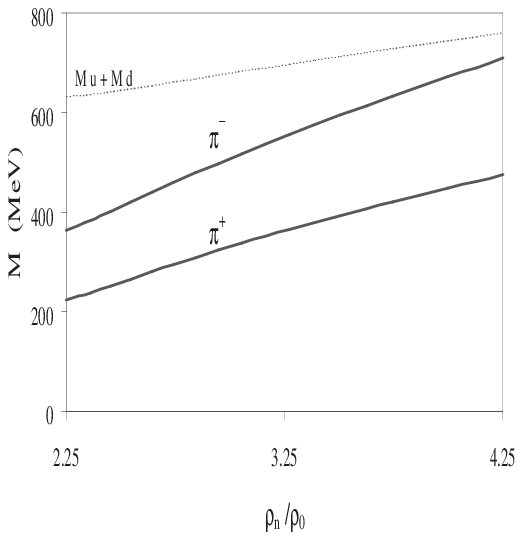,height=5.5cm,width=5.5cm}}
\caption{Masses of kaons and pions in NJL II. $\omega_{up}\, \mbox{and }\, \omega_{low}$ denote the limits of the Fermi sea continuum.
%\label{...}
}
\end{figure}

These modes are excitations of the Dirac sea modified by the presence of the medium. The increase of $K^+ \mbox{and}\, \pi^- $ masses with respect to those of $K^- \, \mbox{and}\, \pi^+$ is due to Fermi blocking and is more pronounced for kaons than for pions because, there are $u$ and $d$ quarks in the Fermi sea and therefore   there are  repulsive effects due to the Pauli principle acting on $\pi^+$. In the case of kaons, we do not have strange quarks at these densities and so there is mora phase space available to create  $\bar u s$ pairs of quarks and there are less repulsive  effects on $K^-$.

An interesting feature in  NJL model is that, besides these modes, low energy modes with quantum numbers of $K^- \, \mbox{and}\, \pi^+$ appear. These are particle-hole excitations of the Fermi sea  which correspond to  $\Lambda\, (1106)$-particle-proton-hole for kaons an to a proton-particle-neutron-hole for the case of pions. A similar effect is found for kaons in symmetric nuclear matter  \cite{SousaRuivo,RSP}. For the case of pions, the low energy modes is less relevant and exist only for $\rho=\rho_c$, merging in the Fermi sea continuum afterwards. We notice that when the 't Hooft interaction is not included \cite{SousaRuivo} or in SU(2) \cite{Hiller} the low energy mode for  pions is more relevant.

The sum rules are a very important tool to analyze the collectivity and relative importance of the modes \cite{RuivoSousa,SousaRuivo,Hiller,RSP}. One can derive a generalization of the PCAC relation in the medium from the Energy Weighted Sum Rule (EWSR), well known from Many Body Theories.
For the mesonic state $|r>$ with energy $\omega_r$ 
associated with the transition operator $\Gamma$
 the strength function $F_r\,=\,\omega_r\,\,{|\,<\,r\,|\,\Gamma\,
|\,0\,>\,|}^2$ 
satisfies 
the EWSR \, which reads
\begin{equation}
m_1\,=\,{\sum}_r\,\omega_r\,\,{|\,<\,r\,|\,\Gamma\,
|\,0\,>\,|}^2\,\,=\,
\frac{1}{2}\,
<\,\Phi_0\,|\,[\,\Gamma\,,\,[\,H\,,\Gamma\,]\,]\,|\,
\Phi_0\,>,
\label{10}
\end{equation}
the transition operator being  defined in the present case by 
$\Gamma\,=\,\Gamma_+\,+\,\Gamma_-$, with  
$\Gamma_\pm\,=\gamma_5\,(\lambda_4\pm i\ \lambda_5)/\sqrt{2}$, for kaons, and , $\Gamma_\pm\,=\gamma_5\,(\lambda_1\pm i\ \lambda_2)/\sqrt{2}$, for pions.
We obtain therefore the GMOR relation  in the medium:

\begin{equation}
\sum_\alpha \,m_{K,\alpha}^2\,f_{K,\alpha}^2\,\,\simeq\,-\,\frac{1}{2}
(m_u\,+\,m_s\,)\,[<\,\overline u\,\,
u\,>\,+\,
<\,\overline s\,\,s\,>\,]\,.\label{11}
\end{equation}
and
\begin{equation}
\sum_\alpha \,m_{\pi,\alpha}^2\,f_{\pi,\alpha}^2\,\,\simeq\,-\,\frac{1}{2}
(m_u\,+\,m_d\,)\,[<\,\overline u\,\,
u\,>\,+\,
<\,\overline d\,\,d\,>\,]\,.\label{12}
\end{equation}

We verified that in the medium the degree of satisfaction of the sum rule is good, provided, naturally, that all the bound state solutions are considered. The strength associated to the low energy mode can not be neglected as the density increases  \cite{RuivoSousa,SousaRuivo}. 

%\begin{figure}[h]
%\vspace{5.5cm}
%\hspace{-0.5cm}\special{kvector.eps}
%\hspace{5.5cm}\special{pivector.eps}
%\caption{Masses of kaons and pions in ENJL model.}
%\end{figure}
%
%\begin{figure}[h]
%\centering{\
%\epsfig{figure=kvector.eps,height=5.5cm,width=5.5cm}}
%\vspace{-0.5cm}
%\caption{Masses of kaons and pions in ENJL model.
%%\label{...}
%}
%\end{figure}
\begin{figure}[h]
\centerline{
\epsfig{figure=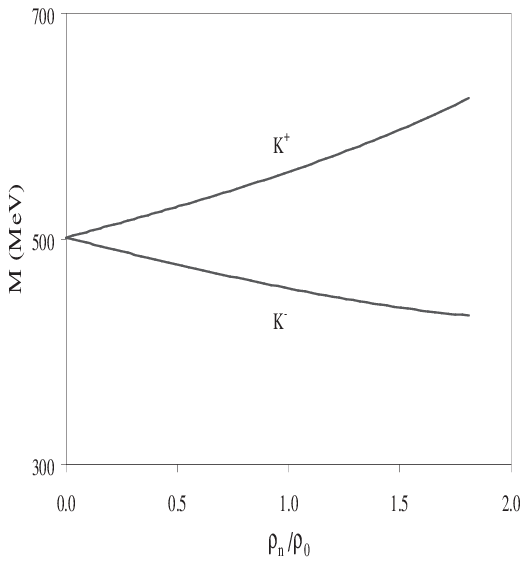,height=5.5cm,width=5.5cm}
\epsfig{figure=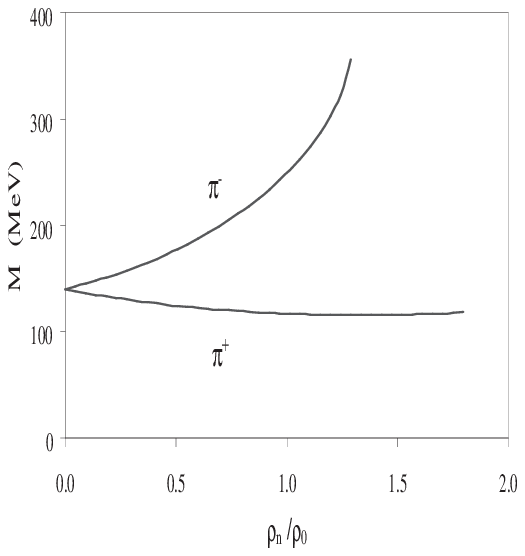,height=5.5cm,width=5.5cm}}
\caption{Masses of kaons and pions in ENJL model.
%\label{...}
}
\end{figure}

In the ENJL model the low energy mode does not appear, which is consistent with the fact that the analysis of the EOS shows that there is no stable Fermi sea.  The attractive effects concentrate on $K^-\,\,,\pi^+$. The splitting  between the charge multiplets is even larger in this model. 

\vspace{0.8cm}

The influence of the temperature is, on one side, to inhibit the occurrence of the low energy mode (this mode is not seen above very low temperatures)  and, on the other side, to reduce the splitting of between the upper energy modes (see Fig. 4).

We notice that temperature has  an effect on the low energy mode similar to the vector pseudovector interaction in vacuum. This  is meaningful since, as it has been mentioned above, as the temperature increases the phase transition is second order and the minimum of energy per particle is at $\rho=0$ and consequently the Fermi sea is not stable. So, it is reasonable that the excitations of the Fermi sea are not seen.

%\begin{figure}[h]
%\vspace{5.5cm}
%\hspace{-0.5cm}\special{kaaot.eps}
%\hspace{5.5cm}\special{pit.eps}
%\caption{Masses of kaons and pions for $T=50$ MeV and $T=100$ MeV, in NJL II.}
%\end{figure}
%
%\begin{figure}[h]
%\centering{\
%\epsfig{figure=kaaot.eps,height=5.5cm,width=5.5cm}}
%\vspace{-0.5cm}
%\caption{Masses of kaons and pions for $T=50$ MeV and $T=100$ MeV, in NJL II.
%%\label{...}
%}
%\end{figure}
\begin{figure}[h]
\centerline{
\epsfig{figure=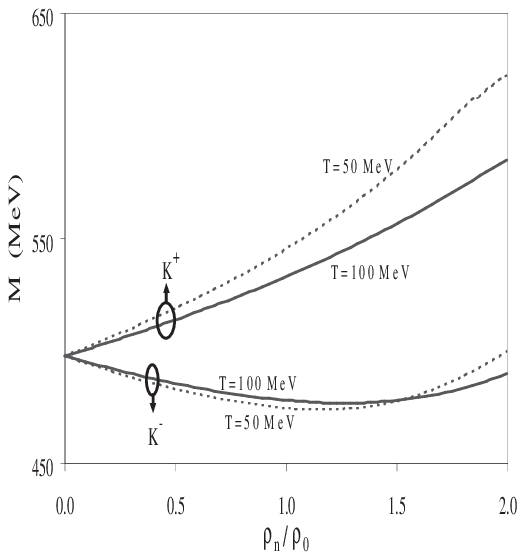,height=5.5cm,width=5.5cm}
\epsfig{figure=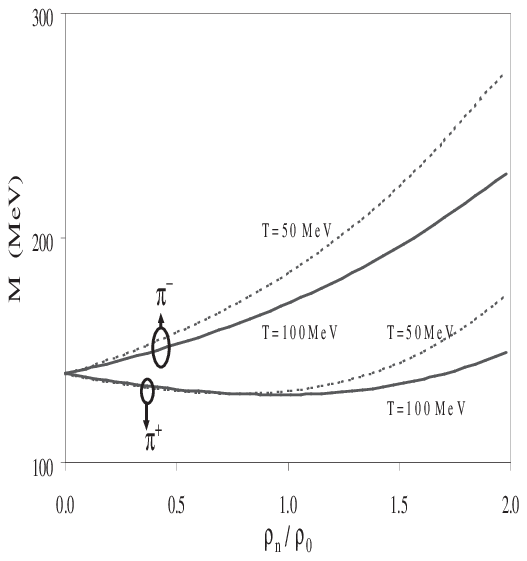,height=5.5cm,width=5.5cm}}
\caption{Masses of kaons and pions for $T=50$ MeV and $T=100$ MeV, in NJL II.
%\label{...}
}
\end{figure}

In conclusion, we have discussed the behavior of kaons and pions in neutron matter in NJL and ENJL model, in connection with the nature of the phase transition, at finite density with zero or non zero temperature. In the NJL model where the phase transition is first order and a stable Fermi sea exists at the critical density, we find low energy particle hole excitations of the Fermi sea, besides the usual splitting of charge multiplets which are excitations of the Dirac sea.  This last effect does not occur in the ENJL model, where the transition is second order. The temperature inhibits this effect and reduces the splitting between the charge multiplets.

\vspace{0.5cm}
\noindent
{\small \bf Acknowledgements}

\noindent
Work supported in part by Centro de F\' isica Te\'orica, and PRAXIS/P/FIS/\\12247/98, FCT, Portugal\\
One of us (MCR) wishes to thank finantial support from the University of Rostock, Germany.

%\vspace{0.8cm}

\newpage

%% file: SOURCE/tre01rev.tex
\section*{\bf Chiral Phase Transition in Covariant Nonlocal NJL Models}
\addcontentsline{toc}{section}{\protect\numberline{}{Chiral Phase Transition in Covariant Nonlocal\\ NJL Models \\
\mbox{\it N.N. Scoccola, I. General, D. Gomez Dumm}}}
\begin{center}
\vspace*{2mm}
{
{I. General$^a$, D. Gomez Dumm$^b\ ^\dagger$ and N.N. Scoccola$^{a,c}$
  \footnote[2]{Fellow of CONICET, Argentina.}}
}\\[0.3cm]
{\small\it $^a$ Physics Department, Comisi\'on Nacional de Energ\'{\i}a At\'omica,\\ Av.Libertador 8250, (1429) Buenos Aires, Argentina\\
$^b$ IFLP, Depto.\ de F\'{\i}sica, Universidad Nacional de La Plata, \\ C.C. 67, (1900) La Plata,

Argentina\\
$^c$ Universidad Favaloro, Sol{\'\i}s 453, (1078) Buenos Aires,
Argentina}
\end{center}

\newcounter{eqn17}[equation]
\setcounter{equation}{-1}
\stepcounter{equation}

\newcounter{bild17}[figure]
\setcounter{figure}{-1}
\stepcounter{figure}

\newcounter{tabelle17}[table]
\setcounter{table}{-1}
\stepcounter{table}

\newcounter{unterkapitel17}[subsection]
\setcounter{subsection}{-1}
\stepcounter{subsection}

\begin{abstract}
The properties of the chiral phase transition at finite
temperature and chemical potential are investigated within an
nonlocal covariant extension of the Nambu-Jona-Lasinio model based
on a separable quark-quark interaction. We consider two types of
non-local regulator functions: a gaussian regulator and the
instanton liquid model regulator. In the first case we study both
the situation in which the Minkowski quark propagator has poles at
real energies and the case where only complex poles appear. We
find that for both regulators the behaviour of the physical
quantities as functions of $T$ and $\mu$ is quite similar. In
particular, for small values of $T$ the chiral phase transition is
always of first order and, for finite quark masses, at certain
``end point" the transition turns into a smooth crossover.
Predictions for the position of this point are presented.

\end{abstract}
\renewcommand{\thefootnote}{\arabic{footnote}}
\setcounter{footnote}{0}

%\pagebreak

\subsection{Introduction}

The behaviour of hot dense hadronic matter and its transition to a plasma
of quarks and gluons has received considerable attention in recent years.
To a great extent this is motivated by the advent of facilities like e.g.\
RHIC at Brookhaven which are expected to provide some empirical information
about such transition. The interest in this topic has been further increased
by the recent suggestions that the QCD phase diagram could be richer than
previously expected (see Ref.\cite{Wil00} for some recent review articles).
Due to the well known difficulties to deal directly with QCD,
different models have been used to study this sort of problems. Among
them the Nambu-Jona-Lasinio model\cite{NJL61} is one of the most popular. In this
model the quark fields interact via local four point vertices which are subject to
chiral symmetry. If such interaction is strong enough the chiral symmetry is spontaneously
broken and pseudoscalar Goldstone bosons appear. It has been shown by many authors
that when the temperature and/or density increase, the chiral symmetry is
restored\cite{VW91}.
Some covariant nonlocal extensions of the NJL model have been studied in the last few years\cite{Rip97}.
Nonlocality arises naturally in the context of several of the most successful approaches
to low-energy quark dynamics as, for example, the instanton liquid model\cite{SS98} and the
Schwinger-Dyson resummation techniques\cite{RW94}. It has been also argued that nonlocal
covariant extensions of the NJL model have several advantages over the local scheme.
Namely, nonlocal interactions regularize the model in such a way that anomalies are
preserved\cite{AS99} and charges properly quantized, the effective interaction is
finite to all orders in the loop expansion and therefore there is not need to introduce
extra cut-offs, soft regulators such as Gaussian functions lead to small NLO
corrections\cite{Rip00}, etc. In addition, it has been shown\cite{BB95} that a proper
choice of the nonlocal regulator and the model parameters can lead to some form of quark
confinement, in the sense of a quark propagator without poles at real energies. Recently, the behaviour
of this kind of models at finite temperature has been investigated\cite{BBKMT00}. In
this work we extend such studies to finite temperature and chemical potential.

\subsection{Non-local NJL models with separable interactions}

We consider a nonlocal extension of the SU(2) NJL model defined by the effective action
\begin{eqnarray}
S &=& \int d^4x \ {\bar \psi}(x) \left( i \rlap/\partial  - m_c \right) {\psi}(x) +
       \int d^4x_1 ... d^4x_4 \ V(x_1,x_2,x_3,x_4)  \nonumber \\
 \! \!\! \! \! \times \! \!\! & \!\!\!\! & \!\!\!\!
  \left( {\bar \psi}(x_1) \psi(x_3) {\bar \psi}(x_2) \psi(x_4) -
         {\bar \psi}(x_1)  \gamma_5 \vec \tau \psi(x_3) {\bar \psi}(x_2) \gamma_5 \vec \tau \psi(x_4)
         \right) , \! \!
\end{eqnarray}
where $m_c$ is the (small) current quark mass responsible for the explicit chiral
symmetry breaking. The interaction kernel in Euclidean momentum space is given by
\begin{equation}
V(q_1,q_2,q_3,q_4) = \frac{G}{2} \ r(q_1^2)  r(q_2^2) r(q_3^2) r(q_4^2)
\ \delta(q_1\!+\!q_2\!-\!q_3\!-\!q_4)\,,
\end{equation}
where $r(q^2)$ is a regulator normalized in such a way that $r(0) = 1$. Some general
forms for this regulator like Lorentzian or Gaussian functions have been used in the
literature. A particular form is given in the case of instanton liquid models.

Like in the local version of the NJL model, the chiral symmetry is spontaneously
broken in this nonlocal scheme for large enough values of the coupling $G$. In
the Hartree approximation the self-energy $\Sigma(q^2)$ at vanishing temperature
and chemical potential is given by
\begin{equation}
\Sigma(q^2) =  m_c + (\Sigma(0) - m_c) r^2(q^2)\,,
\end{equation}
where the zero-momentum self-energy $\Sigma(0)$ is a solution of the gap equation
\begin{equation}
\frac{2\pi^4}{G \ N_c} \left( \Sigma(0) - m_c \right) =
\int d^4q \ \frac{ \left[ m_c + (\Sigma(0) - m_c) r^2(q^2)\right] r^2(q^2)}
               {q^2 + \left[ m_c + (\Sigma(0) - m_c)
               r^2(q^2)\right]^2}\,.
\label{gapeq0}
\end{equation}
In general, the quark propagator might have a rather complicate structure of poles
and cuts in the complex plane. As already mentioned,
the absence of cuts and purely real poles in the Minkowski quark propagator might
be interpreted as a realization of confinement\cite{BB95}. In that case, the poles
will appear as quartets located at $\alpha_p = R_p \pm i \ I_p$,
$\alpha_p = - R_p \pm i \ I_p$. On the other hand, if purely real poles exist they
will show up as doublets $\alpha_p = \pm R_p$.  It is clear that the number and
position of the poles and cuts depend on the details
of the regulator. For example, if we assume it to be a step function as in the standard
NJL model only two purely real poles at $\pm \ M$ appear, with $M$ being the
dynamical quark mass. For a Gaussian interaction,
three different situations might occur. For values of $\Sigma(0)$ below a certain critical value
$\Sigma(0)_{crit}$ two pairs of purely real simple poles and an infinite set of
quartets of complex simple poles appear. At $\Sigma(0) = \Sigma(0)_{crit}$, the two
pairs of real simple poles turn into a doublet of double poles with
$I_p=0$, while for $\Sigma(0) > \Sigma(0)_{crit}$ only an infinite set of quartets of
complex simple poles is obtained. For the Lorentzian interactions there is
also a critical value above which purely real poles cease to exist. However, for
this family of regulators the total number of poles is always
finite. As yet another example, in the case of the instanton model
regulator one has to deal with a cut in the complex plane, together with a
number of complex poles which depends on the magnitude of $\Sigma(0)$.

\subsection{Extension to finite temperature and chemical potential}

To introduce finite temperature and chemical potential we follow the imaginary time formalism.
Thus, we replace the fourth component of the Euclidean quark momentum by $\omega_n - i \mu$,
where $\omega_n = (2 n + 1) \pi T$ are the discrete Matsubara frequencies and $\mu$ is
the chemical potential. In what follows we will assume that the model parameters $G$ and $m_c$,
as well as the shape of the regulator, do not change with $T$ or $\mu$.
Performing this replacement in the gap equation we obtain after some
calculation \cite{GDS00}
\begin{eqnarray}
&\!\!\! &
\frac{\pi^4}{G \ N_c} \left( \Sigma(0) - m_c \right) = \int d^4 q
\ \frac{ \Sigma (q^2) r^2(q^2)}
               { q^2 + \Sigma^2(q^2)} \quad - \nonumber \\
& &  -  \int d^3\vec q \
 \sum_{\alpha_p} \ \!\! ' \
 \gamma_p \ {\rm Re} \left[ \left. \frac{\Sigma(z) r^2(z)}
{1+ \partial_{z} \Sigma^2(z) }\right|_{z=-\alpha_p^2}
\ \frac{\epsilon_p \left( n_+ + n_- \right) }{\epsilon_p^2 + i R_p I_p} \right]\, ,
\label{sum}
\end{eqnarray}
where we have expressed the sum over the Matsubara frequencies in terms of a sum over
the quark propagator poles $\alpha_p = R_p + i I_p$  by
introducing the auxiliary function $f(z) = 1/(1+\exp(z/T))$ and using
standard finite temperature field theory techniques.
In Eq.(\ref{sum}) $\epsilon_p$ is given by
\begin{equation}
\epsilon_p = \sqrt{ \frac{ R_p^2 - I_p^2 + \vec q\ ^2 +
\sqrt{  (R_p^2 - I_p^2 + \vec q\ ^2)^2 + 4 R_p^2 I_p^2 }}{2}} \, ,
\end{equation}
and the prime implies that the sum runs over all the poles
$\alpha_p = R_p + i I_p$ with $R_p > 0$ and $I_p \ge 0$. Moreover,
$\gamma_p= 1/2$ for $I_p=0$ and $\gamma_p= 1$ otherwise and the generalized
occupation numbers $n_\pm$ are
\begin{equation}
n_\pm = \left[ 1+ \exp\left( \frac{\epsilon_p  \mp \mu + i R_p I_p/\epsilon_p}{T} \right) \right]^{-1} \, .
\end{equation}
In deriving Eq.(\ref{sum}) we have assumed that the quark
propagator has no cuts in the complex plane. In the general case
some extra terms containing integrals along the cuts have to
be included in such equation.

\subsection{Results}

Having introduced the formalism needed to extend the model to finite temperature
and chemical potential we turn now to our numerical calculations. In this work we
explicitely study two types of non-local regulators: the
gaussian regulator and the instanton liquid model regulator.

The gaussian regulator has the form
\begin{equation}
r(q^2) = \exp\left( - q^2/2\Lambda^2 \right)\,.
\end{equation}
We consider two sets of values for the parameters of the model. Set I corresponds to
$G~=~50$~GeV$^{-2}$, $m_c~=~10.5$~MeV and $\Lambda~=~627$~MeV, while for Set II the
respective values are $G~=~30$~GeV$^{-2}$, $m_c~=~7.7$~MeV and $\Lambda~=~760$~MeV.
Both sets of parameters lead to the physical values of the pion mass and decay
constant. For Set I the calculated value of the chiral quark condensate at zero
temperature and chemical potential is $- (200$~MeV$)^3$ while for Set II it
is $- (220$~MeV$)^3$. These values are similar in size to those determined from
lattice gauge theory or QCD sum rules. The corresponding results for the self-energy
at zero momentum are $\Sigma (0) = 350$~MeV for Set I and $\Sigma (0) = 300$~MeV for
Set II. It is possible to check that Set I corresponds to a situation in which there are
no purely real poles of the quark propagator and Set II to the case in which
there are two pairs of them. Following Ref.\cite{BB95}, Set I might be interpreted as
a confining one since quarks cannot materialize on-shell in Minkowski space.

The behaviour of the zero-momentum self-energy $\Sigma (0)$ as function of the
chemical potential for some values of the temperature is shown in Fig.\ 1.
\vspace*{-0.5cm}
\begin{figure}[h]
\centerline{\psfig{figure=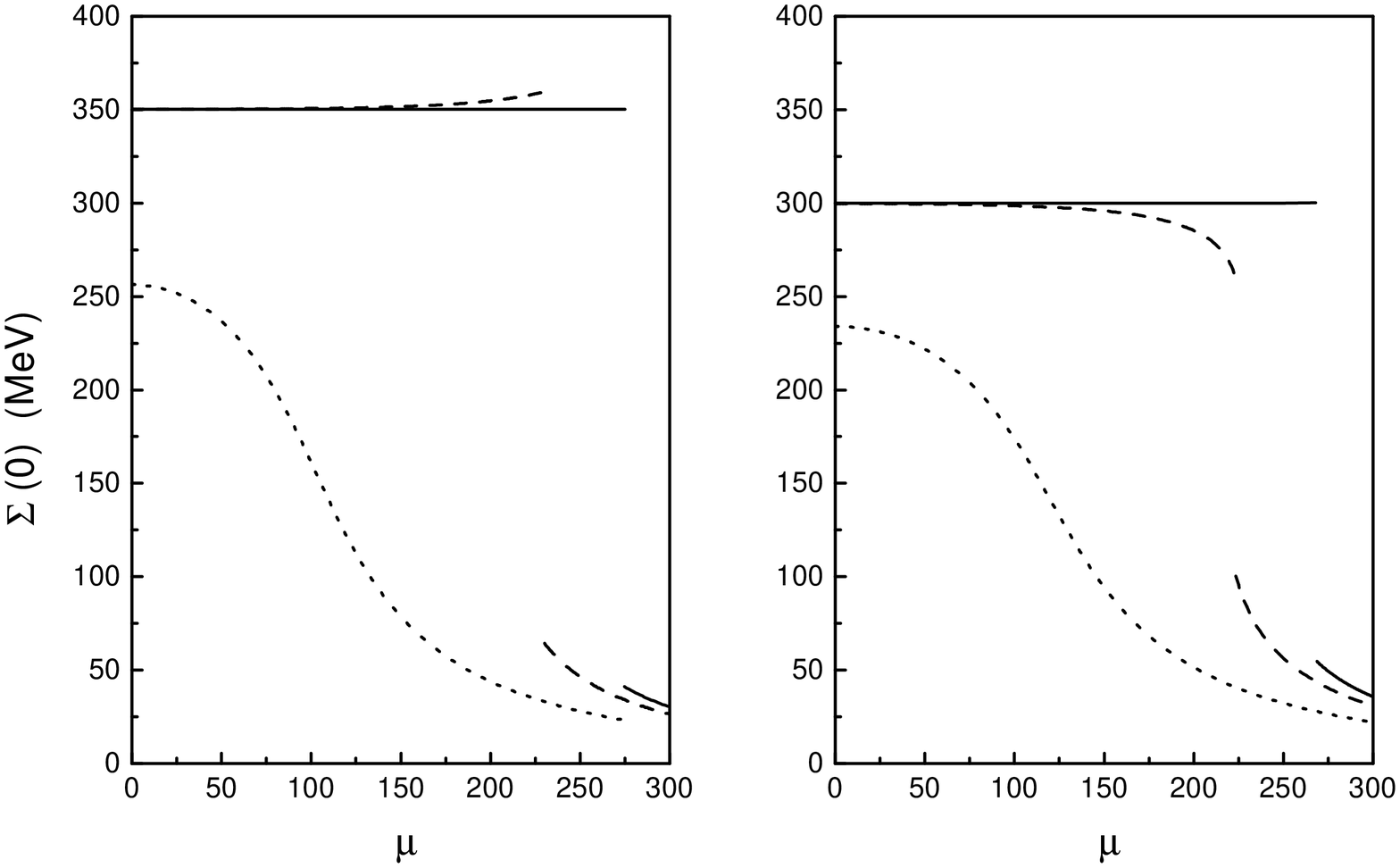,height=8.cm}}
%\vspace{0.5cm}
\protect\caption{\it Behaviour of the self-energy as a function of the chemical potential for three
representative values of the temperature. Full line corresponds to T=0, dashed
line to $T=50$ {\rm MeV} and dotted line to $T=100$ {\rm MeV}. The left panels display
the results for Set I and the right panels those for Set II.}
\label{tmdep}
\end{figure}
We observe that at $T=0$ there is a first order phase transition for both
the confining and the non-confining sets of parameters.
As the temperature increases, the
value of the chemical potential at which the transition shows up decreases. Finally,
above a certain value of the temperature the first order phase transition does not
longer exist and, instead, there is a smooth crossover. This phenomenon is clearly shown
in the right panel of Fig.\ 2, where we display the critical temperature at which the
phase transition occurs as a function of the chemical potential. The point at which
the first order phase transition ceases to exist is usually called ``end point".
\begin{figure}[ht]
%\vspace{-.5cm}
\centerline{\psfig{figure=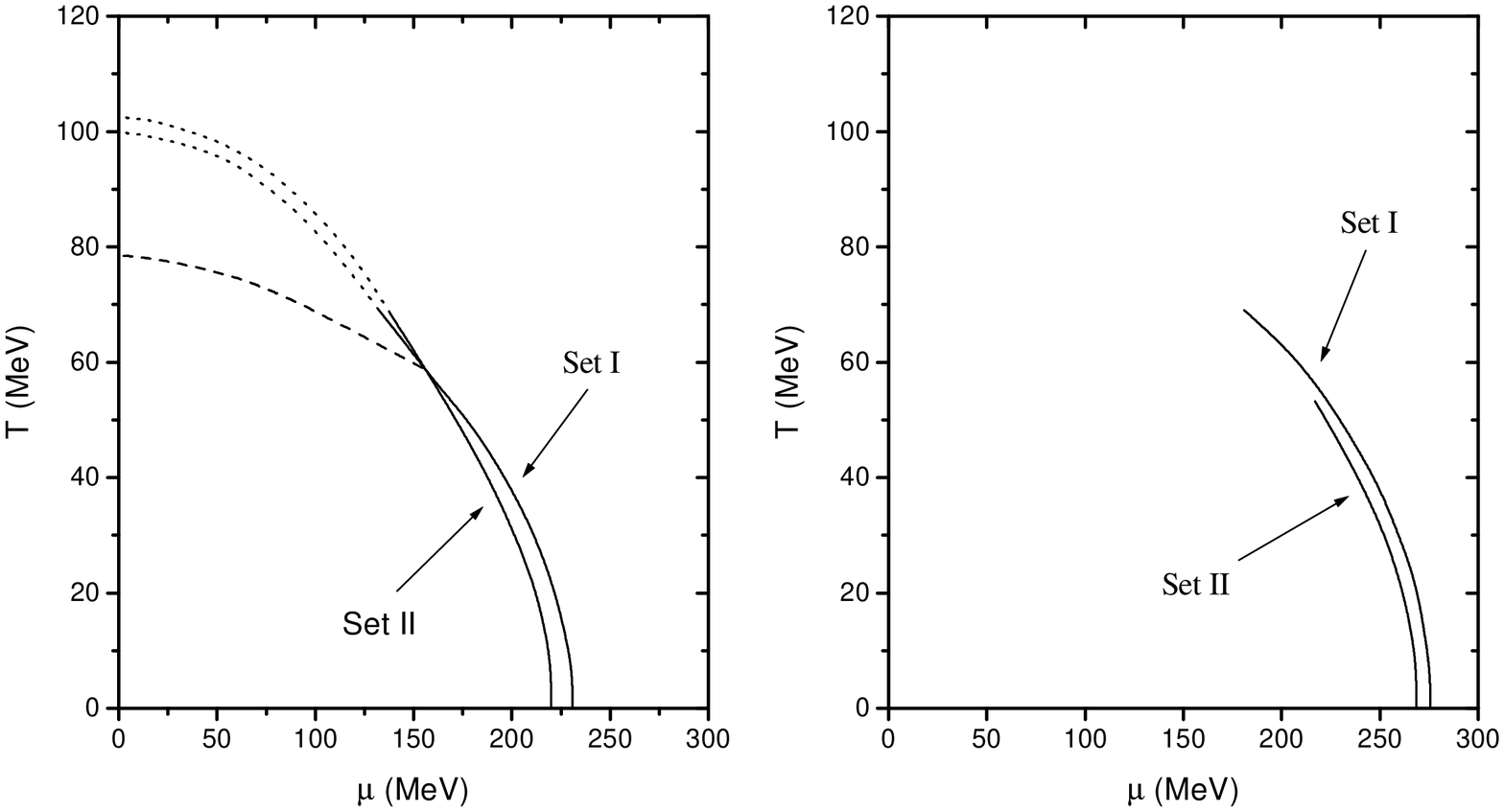,height=8.cm}}
%\vspace{0.5cm}
\protect\caption{\it Critical temperatures as a function of the chemical potential.
The left panel corresponds to the chiral limit and the right panel
to the case of finite quark masses. In both panels the full line stands for
a first order phase transition while in the left panel the dotted line  indicates
that the transition is of second order. The dashed line in the left panel indicates
the temperature at which, for Set I, complex poles of the propagator turn into
real poles.}
\label{trans}
\end{figure}

In the chiral limit, the ``end point'' is expected to turn into
a so-called ``tricritical point'', where the second order
phase transition expected to happen in QCD with two massless quarks
becomes a first order one. Indeed, this is what happens within the present
model for $m_c=0$, as it is shown in the left panel of Fig.\ 2. Some
predictions about both the position of the ``tricritical point'' and its possible
experimental signatures exist in the literature\cite{BR98}. In our case
this point is located at $(T_P,\mu_P) = $(70~MeV$,\,130$~MeV) for
Set I and (70~MeV$,\,140$~MeV) for Set II, while the ``end points'' are placed
at $(T_E,\mu_E) = $(70~MeV$,\,180$~MeV) and (55~MeV$,\,210$~MeV), respectively.

It is interesting to discuss in detail the situation concerning the confining set.
In this case we can find, for each temperature, the chemical potential
$\mu_{d}$ at which confinement is lost. Following the proposal of Ref.\cite{BB95},
this corresponds to
the point at which the self-energy at zero momentum reaches the value
$\Sigma(0)_{crit}$.
Using the values of $m_c$ and $\Lambda$ corresponding to
Set I we get $\Sigma(0)_{crit} = 267$ MeV.  For low temperatures,
$\mu_{d}$ coincides with the the chemical potential at which the
chiral phase transition takes place. However, for a temperature close enough to
that of the ``end point", $\mu_{d}$ starts to be slightly smaller than the value of
$\mu$ that corresponds to the chiral restoration. Above $T_E$ it is difficult to make
an accurate comparison since, for finite quark masses, the chiral restoration
proceeds through a smooth crossover. However, we can still study the situation in
the chiral limit. In this case we find that, in the region where the chiral
transition is of second order, deconfinement always occurs, for fixed $T$,
at a lower value of $\mu$ than the chiral transition. The corresponding
critical line is indicated by a dashed line in the left panel of Fig.\ 2. In
any case, as we can see in this figure, the departure of the line of chiral
restoration from that of deconfinement is in general not too large. This
indicates that within the present model both transitions tend to happen at,
approximately, the same point.

We turn now to the results for the instanton liquid model
regulator. In this case we have
\begin{equation}
r(q^2) = - z \, \frac{d}{dz} \left[ I_0(z) K_0(z) -
I_1(z) K_1(z) \right]
\end{equation}
\noindent
where $I$ and $K$ are the modified Bessel functions and
\begin{equation}
z = \frac{\sqrt{p^2} \ \rho}{2}
\end{equation}
Here, $\rho$ stands for the instanton size. We take the standard
value $\rho = 1/3$~fm and fix the coupling constant
to $G= 36.7$~GeV$^{-2}$ so as to reproduce the typical value
for the instanton density $n\approx 1$~fm$^{-4}$\cite{SS98}. Using
these values, together with $m_c= 4.9$~MeV, it is possible to reproduce
the empirical values of the pion mass and decay constant. The
resulting value of the chiral quark condensate at zero
temperature and chemical potential is $- (256$~MeV$)^3$.
The behaviour of the quark self-energy as a function of the
chemical potential is given in Fig.\ 3 for some representative
values of the temperature.
\vspace*{1.cm}
\begin{figure}[ht]
\centerline{\psfig{figure=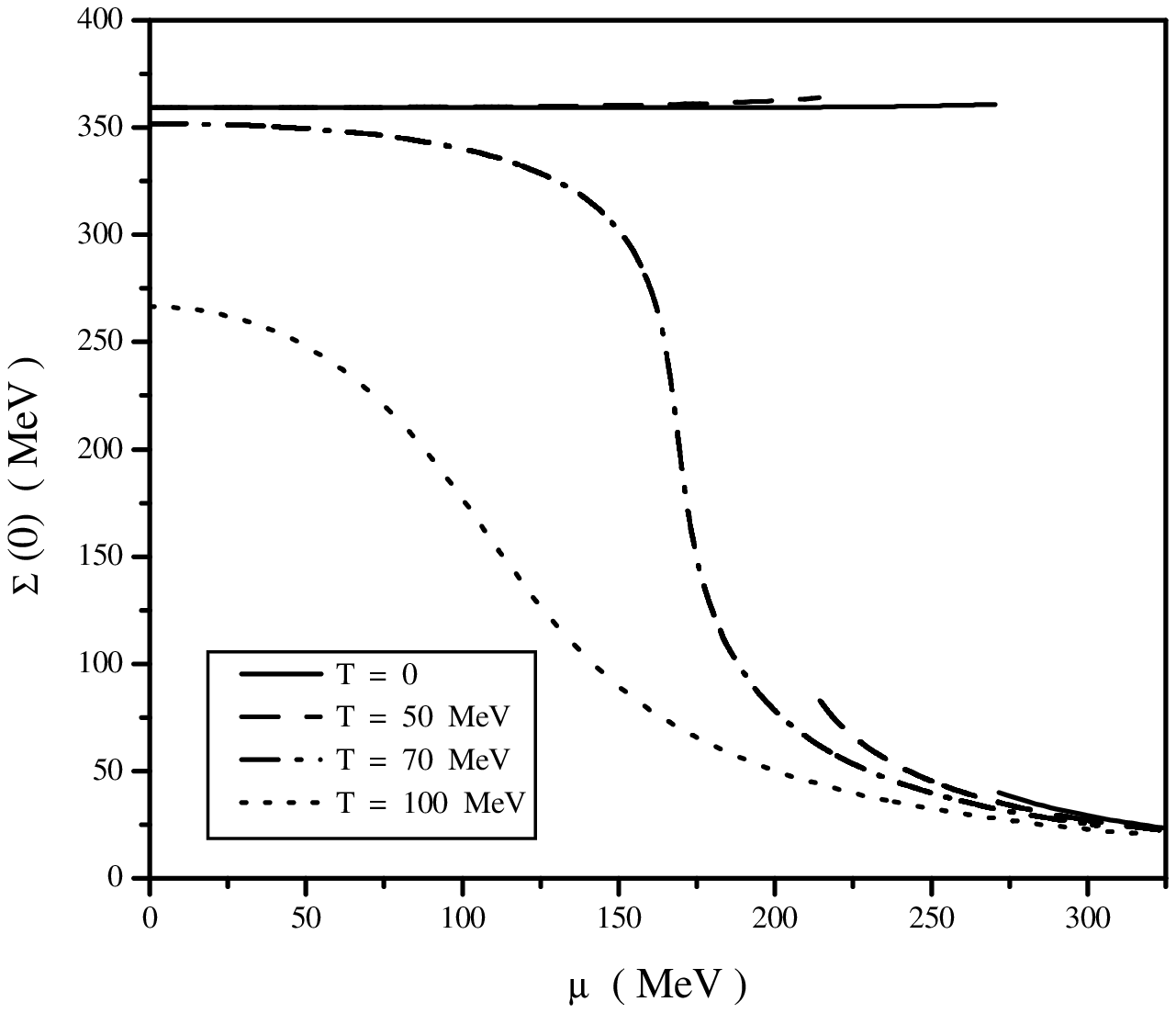,height=6cm}}
\vspace{0.2cm}
\protect\caption{\it Quark self-energy in the instanton liquid model
as a function of the chemical potential
for some representative values of $T$.}
\label{tmdepins}
\end{figure}
As in the case of the gaussian regulator
we observe that for low values of $T$ the chiral phase transition is
of first order while for values above a certain $T_E$ the transition
becomes a smooth crossover. The corresponding values of the critical
temperature as a function of the chemical potential are displayed
in Fig.4. The predicted position of the ``end point" is in this
case  $(T_E,\mu_E) = $(65~MeV$,\,180$~MeV)
\vspace{1.cm}
\begin{figure}[ht]
\centerline{\psfig{figure=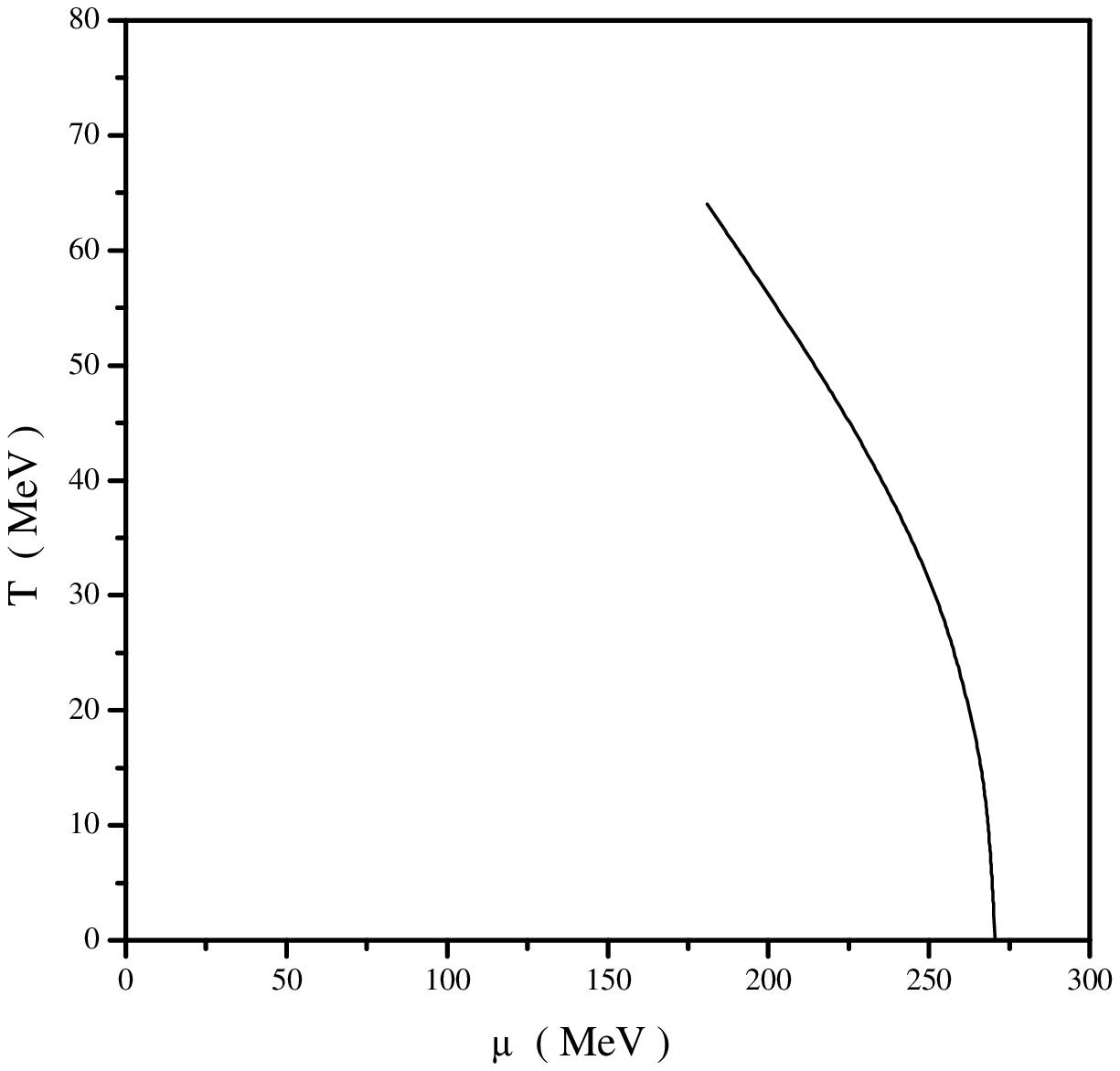,height=6. cm}}
\vspace{0.5cm}
\protect\caption{\it Critical temperatures as a function of the chemical potential in
the instanton liquid model.}
\label{tcins}
\end{figure}

\subsection{Conclusions}

In this work we investigate the features of the
chiral phase transition in some non-local extensions of the
NJL with covariant separable interactions. We consider two
types of regulators: the Gaussian regulator and the instanton
liquid model regulator. We find that in both cases the
phase diagram is quite similar. In particular, we obtain
that for two light flavors the transition is of first order at
low values of the temperature and becomes a smooth crossover
at a certain ``end point''. Our predictions for the
position of this point are very similar for both types of regulators and
slightly smaller than the values in Refs.\cite{BR98},
$T_P~\approx~100$~MeV and $\mu_P \approx 200 - 230$~MeV.
In this sense, we
should remark that our model predicts a critical temperature at $\mu = 0$ of
about 100~MeV, somewhat below the values obtained in modern lattice
simulations which suggest $T_c \approx 140 - 190$~MeV. In any case,
our calculation seems to indicate that $\mu_P$ might be smaller than previously
expected even in the absence of strangeness degrees of freedom.

Several extensions of this work are of great interest. For example, it would be
very important to investigate the impact of the introduction of strangeness
degrees of freedom and flavor mixing on the main features of the chiral phase
transition. In addition, the competition between chiral symmetry breaking
and color superconductivity at large chemical potential deserves further
studies. Work along these lines is under way.

\vspace{0.5cm}

\noindent
{\it NNS would like to thank the organizers of the meeting ``Dynamical aspects of the
QCD phase transition'' for their warm hospitality during his stay in Trento}.

\newpage

%% file: SOURCE/sqm_sepNEU.tex
\section*{\bf Equation of State for Strange Quark Matter in a Separable Model}
\addcontentsline{toc}{section}{\protect\numberline{}{Equation of State for Strange Quark Matter in a\\ Separable Model \\ \mbox{\it C. Gocke, D. Blaschke, A. Khalatyan, H. Grigorian}}}
\begin{center}
\vspace*{2mm}
{Christian Gocke$^{\dagger}$, David Blaschke$^{\dagger}$, \\
Arman Khalatyan$^{\ddagger}$, Hovik Grigorian$^{\ddagger}$}\\[0.3cm]
{\small\it $^{\dagger}$ Department of Physics, Rostock University, 
D-18051 Rostock, Germany\\
$^{\ddagger}$ Department of Physics, Yerevan State University, 
375 025 Yerevan, Armenia}
\end{center}

\newcounter{eqn21}[equation]
\setcounter{equation}{-1}
\stepcounter{equation}

\newcounter{bild21}[figure]
\setcounter{figure}{-1}
\stepcounter{figure}

\newcounter{tabelle21}[table]
\setcounter{table}{-1}
\stepcounter{table}

%\newcounter{kapitel21}[section]
%\setcounter{section}{-1}
%\stepcounter{section}

\newcounter{unterkapitel21}[subsection]
\setcounter{subsection}{-1}
\stepcounter{subsection}

\begin{abstract}
We present the thermodynamics of a nonlocal chiral quark model with separable 
4-fermion interaction for the case of $U(3)$ flavor symmetry within a 
functional integral approach. 
The four free parameters of the model are fixed by the chiral condensate, and 
by the pseudoscalar meson properties (pion mass, kaon mass, pion decay 
constant).
We discuss the $T=0$ equation of state (EoS) which describes quark confinement 
(zero quark matter pressure) below the critical chemical potential 
$\mu_c=333$ MeV. The new result of the present approach is that the 
strange quark deconfinement is separated from the light quark one and occurs
only at a higher chemical potential of $\mu_{c,s}=492$ MeV.  
We compare the resulting EoS to bag model ones for two and three quark flavors,
which have the phase transition to the vacuum with zero pressure also at 
$\mu_c$.

We study quark matter stars in general relativity theory assuming 
$\beta$-equilibrium with electrons and show that for configurations  
with masses close to the maximum of stability at $M=1.62 \div 1.64 ~M_{\odot}$
strange quark matter can occur.   
\end{abstract}

\subsection{Introduction}
Since the discovery of the parton substructure of nucleons and its 
interpretation within the constituent quark model, much effort has been spent 
to explain the properties of these particles. 
The phenomenon of {confinement}, i.e. the property of quarks to exist only 
in bound states as {mesons} and {baryons} in all known systems, poses 
great difficulties for a describing theory. 
So far the problem has been solved by introducing a {color} interaction 
that binds all colored particles to ``colorless'' states.  

However, it is believed and new experimental results \cite{cernlb} underline 
it, that at very high temperatures exceeding $150$ MeV, or densities higher 
than three times nuclear matter density, a transition to deconfined 
quark matter can occur. Besides of the heavy ion collisions performed in
particle physics, the existence of a {deconfinement phase} and 
its properties is of high importance for the understanding of 
{compact stars} \cite{glenden} in astrophysics. 
These, that are popular as Neutron stars, imply core densities above 
three times the nuclear saturation density \cite{nsdata} so that quark matter 
is expected to occur in their interior and several suggestions have been made 
in order to detect signals of the deconfinement transition \cite{gpw,pgb}.

Unfortunately, rigorous solutions of the fundamental theory of color 
interactions ({Quantumchromodynamics (QCD)}) for the EoS at finite baryon 
density could not be obtained yet, even Lattice gauge theory simulations have
serious problems in this domain \cite{udq}. 
To describe interacting quark matter it is therefore necessary to find 
approximating models. The best studied
one is the {Nambu-Jona-Lasinio (NJL)} model that was first developed to 
describe the interaction of nucleons \cite{njl1} and has later been applied
for modeling low-energy QCD \cite{volkov,sandy,hatsuda} with particular 
emphasis on the dynamical breaking of chiral symmetry and the occurence of 
the pion as a quasi Goldstone boson.  
The application of the NJL model for studies of quark matter thermodynamics is 
problematic since it has no confinement and free quarks appear well below the 
chiral phase transition \cite{klevansky,brunodis}. This contradicts to 
results from lattice gauge theory simulations of QCD thermodynamics where
the critical temperatures for deconfinement and chiral restoration coincide. 
That can be helped by using a separable model which can be treated similarly 
to the NJL model but includes a momentum dependence fo the interaction via 
formfactors. It has been shown \cite{sepmod} that in the chiral limit the 
model has no free quarks below the chiral transition. 

Looking again at the densities of compact star cores and comparing with the 
results of NJL model calculations \cite{brunodis,buballa} it seems reasonable 
to include strange-flavor quarks in the model because the energy density is 
sufficiently high for their creation in weak processes. 
Therefore, we extend in the present work the separable model to the case of 
three quark flavors, assuming for simplicity $U(3)$ symmetry. 
We will calculate the partition function
using the method of bosonisation and applying the mean-field approximation. 
Finally we will formulate the resulting thermodynamics of three-flavor quark 
matter and obtain numerical results for the quark matter EoS and compact star 
structure.

%=============================================================================
% THEORY OF QUARK FIELDS
%=============================================================================
%\section*{Theory of Quark Fields}
\subsection{The separable quark model}

The starting point of our approach is an effective chiral quark model 
action with a four-fermion interaction in the current-current form 
%
%\newpage
\begin{eqnarray}
  \label{eq:4}
  {\cal S}[q, \bar q] &=& 
\int \frac{d^4\!k}{(2\pi)^4} \big[\bar q(k)i(\! \not k + \hat m)q(k)
              \nonumber \\
           &+& D_0 \int \frac{d^4\!k^\prime}{(2\pi)^4}\sum_{\alpha=0}^{8} 
                 [(\bar q(k)\lambda_\alpha f(k)q(k))
                   (\bar q(k^\prime)\lambda_\alpha f(k^\prime)q(k^\prime)) \nonumber \\
           \hspace{2cm} &+& (\bar q(k) i\gamma_5\lambda_\alpha f(k)q(k))
   (\bar q(k^\prime) i\gamma_5\lambda_\alpha f(k^\prime)q(k^\prime))]\big] ~. 
\end{eqnarray}
We restrict us here to the {scalar} and {pseudoscalar} currents in Dirac space
which is invariant under chiral rotations of the quark fields and neglect 
color correlations (global color model). 
Furthermore we do not include one of the possible models to account for the 
$U_A(1)$ anomaly since, at least in the quark representation of the Di Vecchia
- Veneziano model \cite{vv}, it can be shown that there is no contribution
to the quark thermodynamics on the mean-field level \cite{alkofer}.
The generalization to the three-flavor case is done using the $U(3)$ symmetry  
where $\lambda_\alpha$ are the Gell-Mann matrices and 
$\lambda_0=\sqrt{\frac{2}{3}}I$. 
For the quark mass matrix in flavor space we use the notation 
\begin{equation}
\label{hat}
\hat m=\sum_f m_f P_f~,
\end{equation}
where $m_f$ are the current quark masses and the projectors $P_f$ on the 
flavor eigenstate $f=u,d,s$ are defined as
\begin{eqnarray}\label{project} 
P_u&=&\frac{1}{\sqrt{6}}\lambda_0+\frac{1}{2}\lambda_3
+\frac{1}{2\sqrt{3}}\lambda_8,  \\
P_d&=&\frac{1}{\sqrt{6}}\lambda_0-\frac{1}{2}\lambda_3
+\frac{1}{2\sqrt{3}}\lambda_8,  \\
P_s&=&\frac{1}{\sqrt{6}}\lambda_0-\frac{1}{\sqrt{3}}\lambda_8~~.  
\end{eqnarray}
Since the Matsubara frequencies in the $T\to 0$ limit become quasicontinuous
variables, the summation over the fourth component $k_4$ of the 4-momentum 
has been replaced by the corresponding integration. 
According to the Matsubara formalism the calculations are performed 
in Euclidean space rather than in Minkowski space where we use 
$\gamma^4= i\gamma^0$.
The partition function in Feynman's path integral representation is given by
\begin{equation}
  \label{eq:5}
  {\cal Z}[T,\hat\mu] = \int {\cal D}\bar q {\cal D}q \exp\left(
         {\cal S}[q, \bar q]  - \int\frac{d^4\!k}{(2\pi)^4} 
 i\hat\mu\gamma^4\bar qq \right)~,
\end{equation}
where the constraint of baryon number conservation is realized by the diagonal
matrix of chemical potentials $\hat\mu$ (Lagrange multipliers) using the 
notation of the hat symbol analogous to (\ref{hat}).

In order to perform the functional integrations over the quark fields 
$\bar q$ and $q$ we use the formalism of bosonisation (see \cite{ripka} and 
references therein) which is based on the Hubbard-Stratonovich transformation 
of the four-fermion interaction terms employing the identity
\begin{eqnarray}
\exp\left\{ D_0 \int_k \int_{k^\prime} \sum_{\alpha=0}^{8} 
j_s^\alpha (k)j_s^\alpha (k^\prime)\right\}
= \nonumber \\
 {\cal N} \prod_\alpha\int d\sigma^\alpha 
\exp\left[\frac{(\sigma^\alpha)^2}{4 D_0} + 
\int_k \int_{k^\prime} j_s^\alpha(k) \sigma^\alpha(k^\prime)\right]~, 
\end{eqnarray}
for the scalar and a similar one for the pseudoscalar channel, where the 
abbreviations $\int_k=\int\frac{d^4\!k}{(2\pi)^4}$ for the phase space integral
and $j_s^\alpha(k)=\bar q(k)\lambda_\alpha f(k)q(k)$ for the scalar current of
the component $\alpha$ have been used.
Now the generating functional is Gaussian with respect to the quark field and 
can be evaluated.
We arrive at the transformed generating functional in terms of bosonic 
variables 
\begin{eqnarray}  \label{eq:6}
{\cal Z}[T,\hat\mu] &=& \int\!\!\!\int{\cal D}\sigma^\alpha{\cal D}\pi^\alpha 
   \exp\left\{\cal{S}[\sigma^\alpha,\pi^\alpha]\right\}
\end{eqnarray}
with the action functional
\begin{eqnarray}
  \label{eq:7}
  {\cal S}[\sigma^\alpha,\pi^\alpha]=
&-&\int\frac{d^4\!k}{(2\pi)^4} \ln \left( \det_{DFC}  
              [\not{\tilde k} + \hat{m} +
              \hat{\sigma} f(\tilde k) + i\gamma_5\hat{\pi} f(\tilde k)] \right) \nonumber \\ 
&+& \frac{\bar\sigma_\alpha\bar\sigma^\alpha}{4D_0} + \frac{\bar\pi_\alpha\bar\pi^\alpha}{4D_0}~. 
\end{eqnarray}
with analogous use of the already known hat symbol and
the 4-vector $\tilde k_f = \left({\vec k},{k_4 +  i\mu_f} \right)$. 
In order to further evaluate the integral over the auxiliary bosonic 
fields $\sigma_\alpha$ and $\pi_\alpha$ we expanded them
around their mean values $\bar\sigma_\alpha$ and $\bar\pi_\alpha$ that 
minimize the action
\[\begin{array}{r@{\:=\:}l}
  \sigma_\alpha & \bar\sigma_\alpha + \tilde\sigma_\alpha(k) \\
  \pi_\alpha & \bar\pi_\alpha + \tilde\pi_\alpha(k) 
\end{array}\]
and neglect the fluctuations $\tilde\sigma_\alpha(k)$ and 
$\tilde\pi_\alpha(k)$ in the following. 
The mean values of the pseudoscalar field vanish for symmetric reasons 
\cite{alkofer}.
The indices $DFC$ refer to the determinant in Dirac-, flavor- and color-
space. So we end up with the mean-field action 
\begin{eqnarray}
  \label{eq:8}
  {\cal S}_{\rm{MF}}[T,\{\mu_f\}] &=& 
         \sum_f\left(-2N_c \int\frac{d^4\!k}{(2\pi)^4}[
             \ln(\tilde k_f^2 + M_f^2) ] + \frac{\Delta_f^2}{8D_0}\right)~,
\end{eqnarray}
with the effective quark masses $M_f=M_f(\tilde k)=m_f + \Delta_f f(\tilde k)$. 
The flavor dependent mass gaps $\Delta_f$ are defined by 
$\hat\sigma=\sum_f\Delta_fP_f$. 
\subsection{Quark matter thermodynamics in mean field approximation}
In the mean field approximation, the grand canonical thermodynamical potential
is given by
\begin{eqnarray}
  \label{eq:9}
\Omega(T,\{\mu\}) &=& 
\beta^{-1}\ln\left\{{\cal Z}[T,\{\mu_f\}]/{\cal
    Z}[0,\{0\}]\right\}\nonumber \\ 
&=& \beta^{-1} \left\{{\cal S}_{\rm{MF}}[T,\{\mu_f\}]-
{\cal S}_{\rm{MF}}[0,\{0\}]\right\}~,
\end{eqnarray}
Where the divergent vacuum contribution has been subtracted.
In what follows we consider the case $T=0$ only. 
In order to interpret our result, we want to represent it as a sum of three 
terms
\begin{eqnarray}
  \label{eq:12}
  \Omega(0,\{\mu\}) &=& \sum_f\left( -2N_c \int\frac{d^4\!k}{(2\pi)^4}
          \ln\left(\frac{\tilde k_f^2 + M_f^2}{\tilde k_f^2 + m_f^2}\right) 
                  + \frac{\Delta_f^2}{8D_0}\right) \nonumber \\
               && -2N_c \sum_f\left(\int\frac{d^4\!k}{(2\pi)^4}
                  \ln\left(\frac{\tilde k_f^2 + m_f^2}{k_f^2 + m_f^2}\right)\right) \\
               && +\sum_f\left( 2N_c \int\frac{d^4\!k}{(2\pi)^4}
                  \ln\left(\frac{k^2 + (M_f^0)^2}{k^2 + m_f^2}\right) 
                  - \frac{(\Delta_f^0)^2}{8D_0}\right)~, \nonumber
\end{eqnarray}
where $M_f^0=m_f+\Delta_f^0f(k^2)$ are the effective quark masses in the vacuum.
The second term on the r.h.s. of this equation is the renormalized 
thermodynamical potential of an ideal fermion gas \cite{kapusta}.
The third term of Eq.\ (\ref{eq:12}) is independent of $T$ and $\mu$, i.e.\ 
it is a (thermodynamical) constant for the chosen model. Refering to the MIT bag model 
we call this term the {bag-constant} $B$. 
The remaining term includes the effects of quark interactions in the mean field
approximation and can be evaluated numerically.

All thermodynamical quantities can now be derived from Eq.\ (\ref{eq:12}). 
For instance,
pressure, density, energy density and the chiral condensate are given by:
\begin{equation}
  \label{eq:13}
  pV = -\Omega\;,\;n = -\frac{\partial\Omega}{\partial\mu}\;,\;
  \varepsilon = -p+\mu n\;,\;
<\!\bar q_fq_f\!> = \frac{\partial\Omega}{\partial m_q}~.
\end{equation}

Still the quark mass gaps $\Delta_f$ have to be determined. This is done by 
solving the gap equations which follow from the minimization conditions
$\frac{\partial\Omega}{\partial\Delta_f}=0$. 
The gap equations read
\begin{equation}
  \label{eq:14}
  \Delta_f = 4D_0(-2N_c)\int\frac{d^4\!k}{(2\pi)^4}\frac{2M_ff(\tilde k)}
{\tilde k_f^2+M_f^2}~.
\end{equation}
As can be seen from (\ref{eq:14}), for the chiral $U(3)$ quark model 
the three gap equations for $\Delta_u,~\Delta_d,~\Delta_s$ are decoupled
and can be solved separately. 
%
%===========================================================
% NUMERICAL RESULTS
%===========================================================
\subsection{Results for the Gaussian formfactor}
\subsubsection{Parametrization of the model}
In the nonlocal separable quark model described above the formfactor of the interaction was not yet specified.
In the following numerical investigations we will employ a simple Gaussian 
\begin{equation}
f(k)=\exp(-k^2/\Lambda^2)~,
\end{equation}
which has been used previously for the description of meson \cite{birse} and 
baryon \cite{golli} properties in the vacuum as well as for those of 
deconfinement and mesons at finite temperature \cite{bkt,bbkmt}. 
A systematic extension to other choices of formfactors is in preparation
\cite{bubsepmod}.
 
The Gaussian model has five free parameters to be defined: the coupling  
constant $D_0$, the interaction range $\Lambda$, and the three current quark
masses $m_u$, $m_d$, $m_s$. Setting $m_u=m_d=:m_q$ we restrict ourselves to 
four free parameters.
These are fixed by the three well known observables:  
pion mass $m_\pi=140$ MeV, kaon mass $m_K=494$ MeV and pion decay 
constant $f_\pi=93$ MeV. 
The formulas for the meson masses and the decay constant 
are calculated as approximations of the {Bethe-Salpeter equation} 
including the generalized {Goldberger-Treiman relation} \cite{bubsepmod}.

The fourth condition comes from values
for the chiral condensate that are conform with phenomenology. 
The resulting parametrisations of the quark model are shown in 
Tab.\ \ref{tbl:1}.

%\vspace*{-0.2cm}
\noindent
\begin{minipage}{\textwidth}
\begin{table}[bth]
  \begin{tabular}{c||c|c|c|c||c|c}
 $-<\!\bar uu + \bar dd\!>^{1/3}$ & $\Lambda$ & $D_0$ & $m_q$ & $m_s$ & $\Delta_q^0$ & $\Delta_s^0$ \\
{[MeV]}&[MeV]&[GeV$^{-2}$]&[MeV]&[MeV]&[MeV]&[MeV]\\
    \hline
    $230$ & $659.2$ & $29.32$ & $6.8$ & $143.5$ & $549.9$ & $767.8$\\
    $235$ & $697.6$ & $23.88$ & $6.4$ & $136.1$ & $497.0$ & $719.3$\\   
    $240$ & $736.5$ & $19.88$ & $6.0$ & $129.5$ & $453.8$ & $682.1$\\   
    $245$ & $775.5$ & $16.85$ & $5.6$ & $123.4$ & $419.7$ & $653.1$\\   
    $250$ & $814.9$ & $14.49$ & $5.3$ & $118.0$ & $391.1$ & $630.0$\\   
    $255$ & $853.8$ & $12.66$ & $5.0$ & $112.9$ & $368.7$ & $611.4$\\   
    $260$ & $894.2$ & $11.11$ & $4.7$ & $108.1$ & $349.1$ & $596.1$\\ 
  \end{tabular}
  \vspace*{0.2cm}
  \caption{Parameter sets for the Gaussian separable model for different values of the chiral condensate $<\!\bar uu + \bar dd\!>$. }\label{tbl:1}
\end{table}
\end{minipage}

\subsubsection{Thermodynamics for quark matter without $\beta$ equilibrium}
\begin{figure}[bth]
  \begin{center}
    \includegraphics[width=0.6\linewidth]{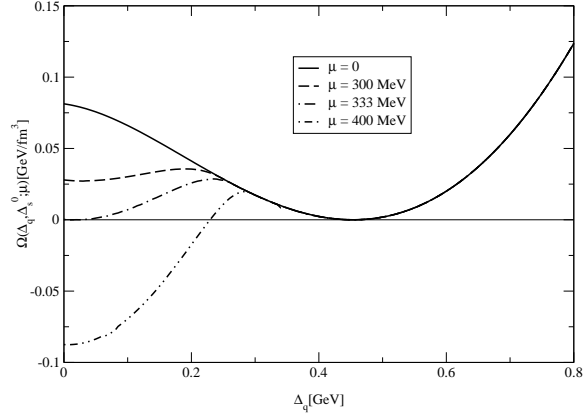} 
    \vspace*{0.2cm}
    \caption{Dependence of the thermodynamical potential on the light flavor 
        gap $\Delta_q=\Delta_u=\Delta_d$
        (order parameter) for different values of the chemical potential,
        $\Delta_s=682$ MeV.
        }
    \label{fig1}
  \end{center}
\end{figure}

\begin{figure}[bth]
  \begin{center}
    \includegraphics[width=0.6\linewidth]{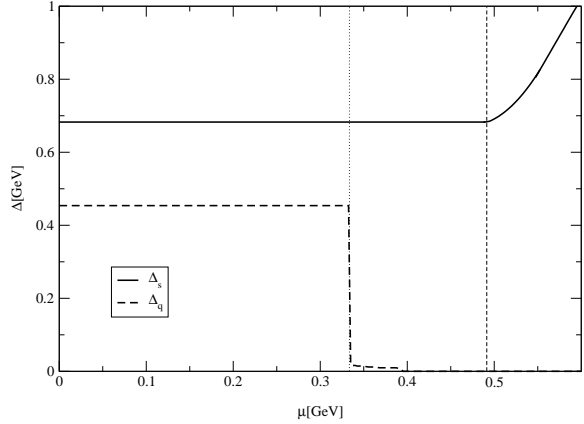} 
    \vspace*{0.3cm}
    \caption{Solutions of the gap equations that minimize the potential}
    \label{fig2}
  \end{center}
\end{figure}

This case is relevant for systems which are considered for time scales larger 
than the typical strong interaction time of about $1$ fm/c but smaller than 
the weak interaction time of several minutes, so that the presence of leptons  
(electrons) does not influence on the composition of quark matter and we can 
choose the chemical equilibrium with $\mu_u=\mu_d=\mu_s=\mu$.
For the numerical calculations we choose the parameter set for the light quark
condensate $-\langle\bar uu + \bar dd\rangle^{1/3}=240$ MeV which is a typical
value known from phenomenology. 
We consider the behavior of thermodynamical quantities at $T=0$ with 
respect to the chemical potential. 
As we set $m_u=m_d$ earlier there is no difference between up- 
and down quarks and both are refered to as light quarks.  
%\newpage
%
Fig.\ \ref{fig1} visualizes the behavior of the thermodynamical potential as a function of the light quark gap $\Delta_q=\Delta_u=\Delta_d$ for different 
values of the chemical potential $\mu$. 
For $\mu<\mu_c=333$ MeV the argument and the value of the global minimum is 
independent of $\mu$ which corresponds to a vanishing quark density 
(confinement). 
At the critical value $\mu=\mu_c=333$ MeV a phase transition occurs from the 
massive, confining phase to a deconfining phase negligibly small mass gap. 
From the solution of the gap equation shown in Fig. \ref{fig2}
\vspace*{-0.2cm}

one can see that the strange quark gap of $\Delta_s=682$ MeV still remains 
unchanged. Thus the strange quarks are confined until a higher value 
of the chemical potential $\mu_{c,s}=492$ MeV is reached. This value is much 
bigger than the current strange quark mass. 
In Fig.\ \ref{fig2} we separate by vertical lines the regions 
of full confinement, two-flavor deconfinement and full deconfinement.
Thus, in the present model, the onset of strange quark deconfinement is 
inhibited. Moreover, the onset of a finite strange quark density is not 
determined by a drop in the strange gap which remains constant and even 
starts to rise for large $\mu$ values.
This result of the present model drastically differs from those of 
bag models or NJL models.
The reason is the 4-momentum dependence of the dynamical quark mass function
which results in complex mass poles for the quark propagators and makes  
the naive identification of the mass gap with a real mass pole impossible
\cite{bubsepmod}.  

\begin{figure}[bth]
  \begin{center}
    \includegraphics[width=0.6\linewidth]{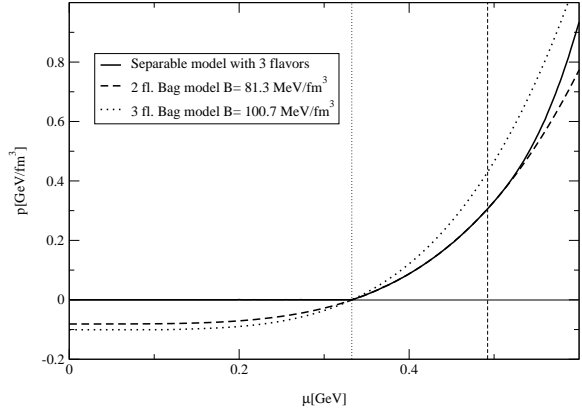} 
    \vspace*{0.3cm}
    \caption{Pressure of the quark matter as a function of the chemical 
potential for the separable model (solid line) compared to a three-flavor 
(dotted line) and a two-flavor (dashed line) bag model. All models have the
same critical chemical potential $\mu_c=333$ MeV for (light) quark 
deconfinement.}
    \label{fig3}
  \end{center}
\end{figure}

The effect on thermodynamical quantities can be understood if we look at the 
pressure. 
In Fig.\ \ref{fig3} we show for comparison the resulting equation of state for 
the pressure of the present separable model together with a two-flavor and a 
3-flavor bag model. Both bag models are chosen such that the critical chemical 
potential for the deconfinement coincides with that of the separable model.
The pressure of the present three-flavor separable model can be well described
by a two-flavor bag model with a bag constant $B=81.3$ MeV/fm$^3$ in the 
region of chemical potentials $333$ MeV $\le \mu \le 492$ MeV where the third 
flavor is still confined.
For comparison, the 3 flavor bag model has a the bag constant $B=100.7$
MeV/fm$^3$ and is considerably harder that the separable one due to the 
additional relatively light strange quark flavor. 

\subsubsection{Inclusion of $\beta$ equilibrium with electrons}
\begin{figure}[bth]
  \begin{center}
    \includegraphics[width=0.6\linewidth]{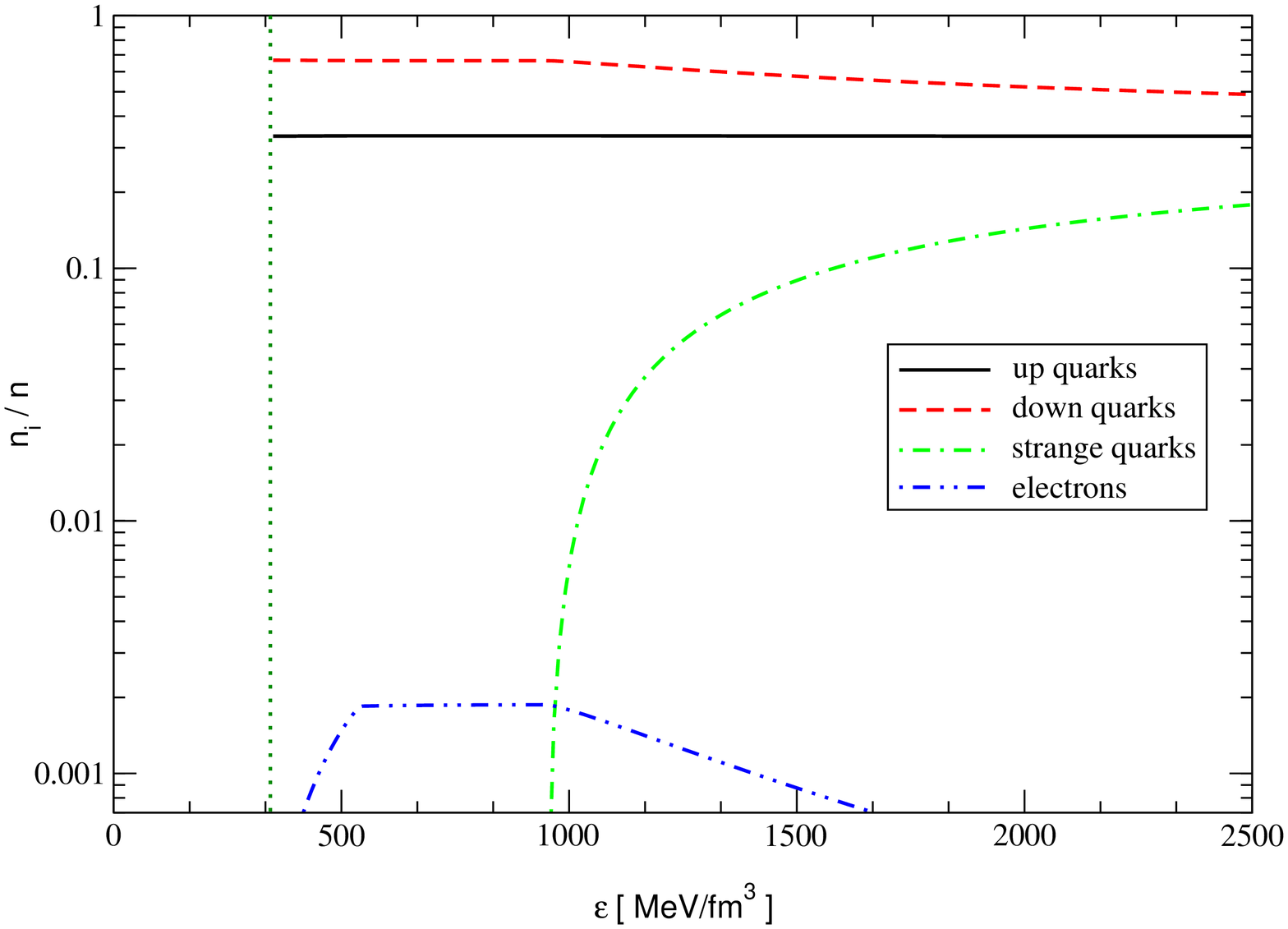} 
    \vspace*{0.3cm}
    \caption{Composition of three-flavor quark matter in $\beta$ equilibrium 
        with electrons.}
    \label{fig4}
  \end{center}
\end{figure}

\begin{figure}[bth]
  \begin{center}
\vspace{1cm}

    \includegraphics[width=0.6\linewidth]{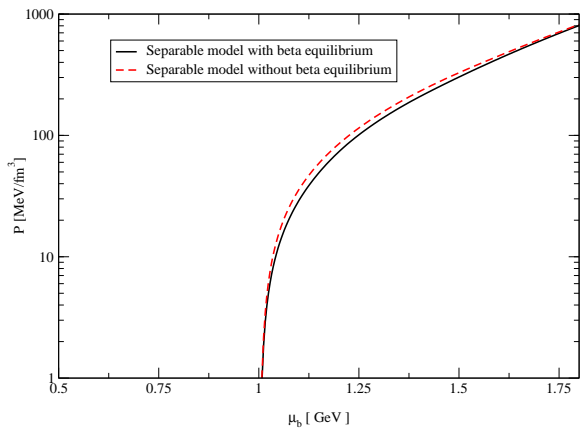} 
    \vspace*{0.3cm}
    \caption{Pressure of three-flavor quark matter with $\beta$ equilibrium 
        and without.}
    \label{fig5}
  \end{center}
\end{figure}

Quark matter in $\beta$-equilibrium is to be supplemented with the two 
relations for conservation of baryon charge and electric charge. 
In the deconfined phase there are quarks and leptons (in our model case up, 
down, strange quarks and electrons) with vanishing net electric charge 
\begin{eqnarray}\label{neutral}
Q_{{\rm q}}(\mu^u,\mu^d,\mu^s) +Q_{{\rm L}}(\mu^e) = {\frac{{2}}{{3}}} n_u -
{\frac{{1}}{{3}}}( n_d + n_s) - n_e = 0~.
\end{eqnarray}
Taking into account the energy balance in weak interactions
\begin{eqnarray}
d &\leftrightarrow& u + e^- +\bar \nu_e\\
s &\leftrightarrow& u + e^- +\bar \nu_e
\end{eqnarray}
and introducing the average quark chemical potential  
$\mu=\frac{1}{3}(\mu_u+\mu_d+\mu_s)$
we can write the $\beta$ equilibrium conditions as
\begin{eqnarray}\label{beta}
\mu_u =  \mu - {\frac{{2}}{{3}}} \mu^e\,
,\qquad \mu_d = \mu_s = \mu + {\frac{{1}}{{3}}} \, \mu^e \ .
\end{eqnarray}
Solving the equation of charge neutrality (\ref{neutral}) one can find the
chemical potential of electrons as a function of $\mu$ and using Eqs. 
(\ref{beta}) the equation of state can be given in terms of a single chemical 
potential $\mu$. 
In Fig. \ref{fig4} we show the composition of the three-flavor quark matter
for the Gaussian separable model in the case of $\beta$ equilibrium with 
electrons as a function of the energy density.
As for the case without $\beta $ equilibrium we can define also in Fig. 
\ref{fig4} the regions of quark confinement 
($\varepsilon<\varepsilon_c=350$ MeV/fm$^3$) and three-flavor deconfinement 
($\varepsilon>\varepsilon_{c,s}=930$ MeV/fm$^3$). 
In the region of two-flavor deconfinement the concentrations of electrons,
up- and down- quarks coincide with those of the two-flavor bag model except 
for the relatively small energies close to $\varepsilon_c$ where the effect of 
a small dynamical quark mass leads to a density dependence of the composition.
In Fig. \ref{fig5} we demonstrate the influence of the $\beta$ equilibrium on 
the equation of state. It can be seen that the difference between pressures 
with and without  $\beta$ equilibrium is limited to the region of intermediate 
densities, where the electron fraction reaches its maximum value 
$x_e\simeq 0.002$.
 
\subsection{Applications for compact stars}

One of the main goals for studying the strange quark matter equation of state
is the possible application for compact stars. In particular, the hypothesis 
that strange quark matter might be more stable than ordinary nuclear matter 
\cite{strange} has lead to the investigation of possible consequences for 
properties of compact stars made thereof \cite{sstar}.
Most of these applications use the bag model equation of state where the 
result depends on the value of the bag constant as a free parameter.
Recently, first steps have been made towards a description of strange quark 
matter within dynamical quark models such as the NJL model \cite{snjl}, 
where the parameters are fixed from hadron properties. 
The non-confining quark dynamics of this model, however, leads to
predictions for dynamical quark masses and critical parameters of the
chiral phase transition which differ from those of confining models
\cite{mnstar,sepmod} and might be quantitatively incorrect.
Here we want to extend previous studies of compact star properties with 
dynamically confining quark models to the strange quark 
sector and find the characteristics of stable compact star configurations 
with the equation of state derived above. 

For the calculation of the self-bound configuration for the quark matter with 
gravitational interaction one needs the condition of mechanical equilibrium of 
the thermodynamical pressure with the gravitational force.
This condition is given by the Tolman-Oppenheimer-Volkoff-Equation
\begin{eqnarray}
\frac{dP}{dr}=- G (\varepsilon(r) + P(r)) \frac{m(r) 
+ 4\pi G r^3 P(r)}{r(r-2Gm(r))}~
\end{eqnarray} 
and defines the profiles for all thermodynamical quantities in the case of nonrotating spherically symmetric distributed matter configurations in general relativity.
In this equation $m$ denotes the accumulated mass in the sphere with radius 
$r$ given by
\begin{equation}
m(r)=4 \pi \int_0^r \varepsilon(r') r'^2 dr
\end{equation}
The gravitational constant is denoted by $G$. 
The radius $R$ of the star is defined by the condition that the pressure 
becomes zero on the surface of the star $P(R)=0$.
The total mass of the star is $M=m(R)$.

\begin{figure}[bth]
  \begin{center}
    \includegraphics[width=0.6\linewidth,angle=-90]{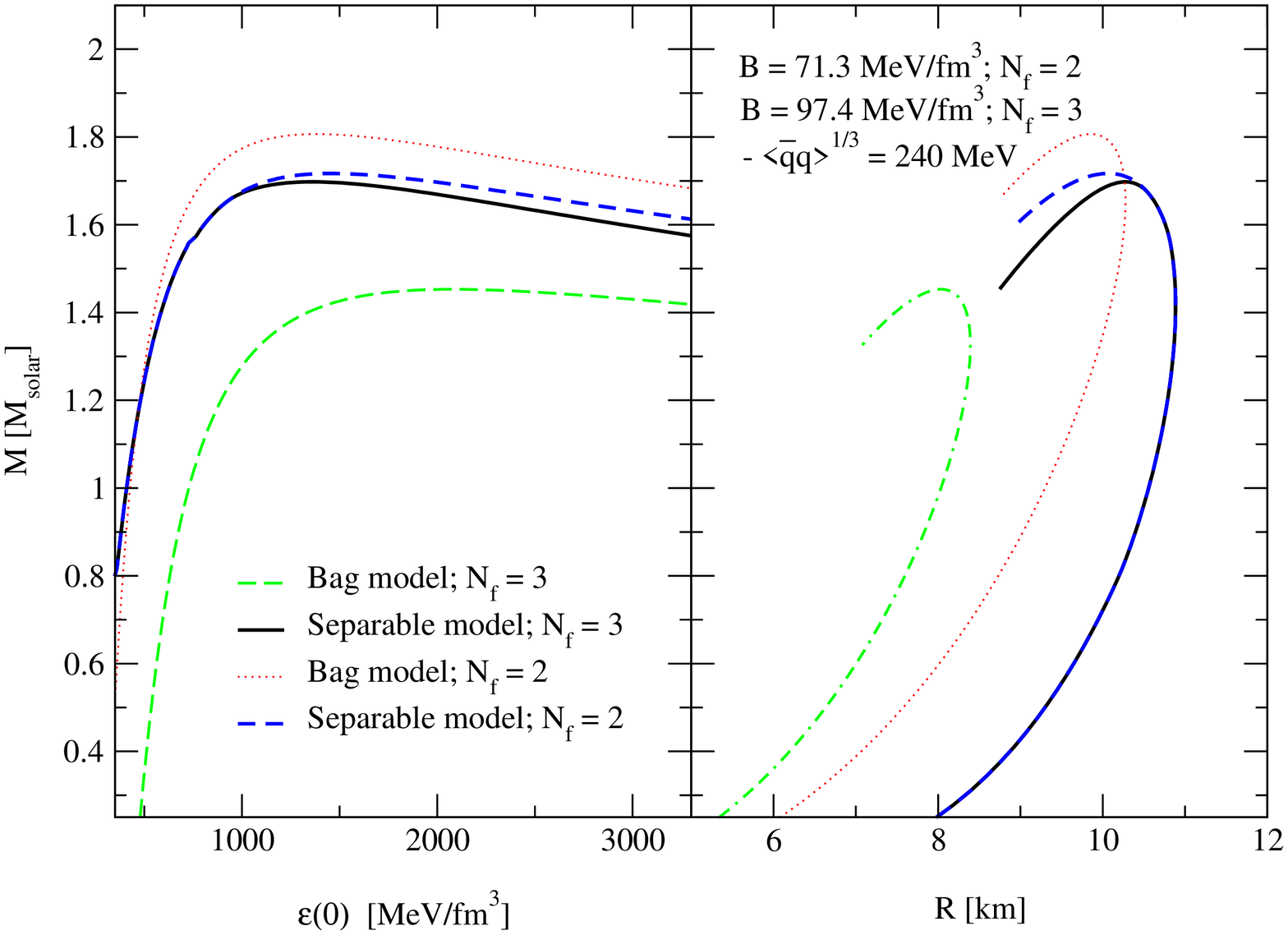} 
    \vspace*{0.3cm}
    \caption{Stability for compact stars composed of quark matter.}
    \label{fig6}
  \end{center}
\end{figure}

Each configuration has one independent parameter which could be chosen to be 
$\varepsilon(0)$, the central energy density. 
In Fig. \ref{fig6} we show the dependence of the total mass of the 
configuration as a function of the central density and the radius for the 
separable quark model and for the bag model in the
cases of two and three flavors respectively. 

The rising branches of the mass-radius or mass-density relations correspond to 
the families of stable compact stars.
The maximum possible mass for the separable model is $1.64~M_\odot$ for the 
three-flavor and $1.71~M_\odot$ for the two-flavor case.
The maximal central density is about $1350$ MeV/fm$^3$ which allows for the 
three-flavor case to have strange quark matter in the core of the quark star. 
The comparison with the corresponding bag model strange stars shows that the 
latter are more compact, their maximum radius is about $8$ km, and less massive
with a maximum mass of about $1.5~M_\odot$.
The maximum radius of stars within the separable model is 11 km and thus 
exceeds the radii for both two and three flavor bag model quark stars. 
The origin
of this difference is the behavior of the pressure in the low density region.  

%==================================================================
%CONCLUSION
%==================================================================
\subsection{Conclusion}
For neutron stars it is relevant to include the effects of strange flavor in a 
model for quark matter. In the simplest case considering $U(3)$ symmetry this
can be done without increasing the complexity of the generating functional.  
We showed that in our separable model the gap equations decouple and can be 
solved separately. The resulting thermodynamics can be solved numerically and
gives the equations of state for interacting quark matter. Unlike the well
known NJL model the separable model is able to express the effects of 
confinement in the thermodynamical quantities.
The new result obtained within the present approach is the separation of the
deconfinement of light quark flavors at $\mu_{c}=333$ MeV from that of
strange quarks which occurs only at a higher chemical potential of 
$\mu_{c,s}=502$ MeV.
A consequence for the application of the EoS presented here in compact
star calculations is that strange quarks do occur only close to the maximum 
mass of $1.64 ~M_{\odot}$, i.e. that for masses below $1.62 ~M_{\odot}$,
only two-flavor quark matter can occur. 
\subsection*{Acknowledgments}
We like to thank our colleagues  M.\ Buballa, Y.\ Kalinovsky, M.\ Ruivo, 
N.\ Scoccola and P.C.\ Tandy. 
Their studies of separable quark models have helped us to 
formulate the present one. We are grateful to R. Alkofer for pointing out 
Ref. \cite{alkofer} to us.
We acknowledge the support of DAAD and DFG for scientist exchange between
the Universities of Rostock and Yerevan.  
\vspace*{1cm}

%%%%%%%%%%%%%%%%%%%%%%%%%%%

%

%% file: SOURCE/dgh2.tex
\newcounter{eqn25}[equation]
\setcounter{equation}{-1}
\stepcounter{equation}

\newcounter{bild25}[figure]
\setcounter{figure}{-1}
\stepcounter{figure}

\newcounter{tabelle25}[table]
\setcounter{table}{-1}
\stepcounter{table}

\newcounter{unterkapitel25}[subsection]
\setcounter{subsection}{-1}
\stepcounter{subsection}

\section*{\bf Quark Matter Effects in the Cooling and Spin Evolution of Neutron Stars}
\addcontentsline{toc}{section}{\protect\numberline{}{Quark Matter Effects in the Cooling and Spin\\ Evolution of Neutron Stars \\ \mbox{\it D. Blaschke, G. Poghosyan, H. Grigorian}}}
\begin{center}
\vspace*{2mm}
{David Blaschke$^{\dagger}$, Gevorg Poghosyan$^{\dagger}$,\\
Hovik Grigorian$^{\ddagger}$}\\[0.3cm]
{\small\it $^{\dagger}$ Department of Physics, University of Rostock,
  D-18051 Rostock, Germany\\
$^{\ddagger}$ Department of Physics, Yerevan State University, Alex Manoogian 1, \\
375025 Yerevan, Armenia}
\end{center}

%\vspace{-4.7cm}

%\hfill
%\parbox{4cm}{\mbox{ECT*-2001-05}\\ \mbox{MPG-VT-UR 221/01}}
%\vspace{4.7cm}

\begin{abstract}
In this contribution we consider 
two aspects of the evolution of a neutron star: cooling and rotational 
evolution. We present recent results of investigations of possible signals
for a deconfinement phase transition in the stars interior from 
observations of the surface temperatures of young (less than $10^3$ yr) pulsars
and of the spin evolution of old (larger than $10^6$ yr) stars.
We have obtained the temperature profiles of young pulsars taking into account 
heat transport and color superconductivity on the evolution of the surface 
temperature in comparison with the observational data.    
For old pulsars we suggest a phase diagram spanned by baryon number $N$ and 
angular velocity $\Omega$ of the configurations and show that there are 
characteristic changes in the rotational evolution of star configurations when 
they cross the critical line which separates the region of quark core stars 
from that of hadronic ones.  
For accreting compact stars in low-mass X-ray binaries a clustering of their 
population along this critical line is suggested to be a detectable signal 
for the occurence of quark matter. 
\end{abstract}

%%%%%%%%%%%%%%%%%%%%%%%%%%%%%%%%%%%%%%%%%%%%%%%%%%%%%%%%%%%%%%%%%%

\subsection{Introduction}

Quantum Chromodynamics (QCD) as the fundamental theory for strongly
interacting matter predicts a deconfined state of quarks and gluons under
conditions of sufficiently high temperatures and/or densities which occur,
e.g., in heavy-ion collisions, a few microseconds after the Big Bang or in
the cores of pulsars. The unambiguous detection of the phase transition from
hadronic to quark matter (or vice-versa) has been a challenge to particle
and astrophysics over the past two decades \cite{qm99,bkr}. While the
diagnostics of a phase transition in experiments with heavy-ion beams faces
the problems of strong nonequilibrium and finite size, the dense matter in a
compact star forms a macroscopic system in thermal and chemical equilibrium
for which signals of a phase transition shall be more pronounced.
Particularly interesting systems for investigating the possible occurence of 
quark matter in neutron star interiors are accreting compact stars in low-mass 
X-ray binaries. Due to their mass accretion flow these systems are 
candidates for the most massive compact stars in nature at the limit of black 
hole formation. Therefore, if the possible deconfinement phase transition in
compact stars exists at all, we expect it to occur in these systems. 
Typical time scales for the deconfinement transition due to either the 
accretion of a fraction of the solar mass or due to frequency changes by 
a fraction of a kHz are larger than $10^8$ yr.  

These timescales for rotational evolution seem well separated from those of
thermal processes which leave traces from the composition of the neutron star 
interior during the time which a heat wave travels from the center to the 
surface of the star, i.e. less than $10^3$ yr. 

A completely new situation might arise if the scenarios suggested for
(color) superconductivity \cite{ARW98dim,RSSV98} with large diquark pairing
gaps ($\Delta _{q}\sim 50\div 100$ MeV) in quark matter are applicable to
neutron star interiors. Then, fast temperature drops could occur between 
$10 \div 500$ yr after the birth of the compact star depending on the size of 
the pairing gaps. The diquark pairing changes the specific heat of quark 
matter and therefore a different photon cooling asymptotics results in the 
case of color superconductivity.
Unfortunately, the photon cooling era (after $10^6 \div 10^8$ yr) does not 
allow to distinguish between details of the internal structure such as sizes 
of gaps and of the quark core.

The question arises whether there are candidates of compact objects which are 
young enough to reveal their internal composition by their cooling behaviour 
and which could traverse the critical phase transition line within less than 
$100$ yr. For the latter condition to be fulfilled, these objects should 
rotate fast enough (period $P\sim 2$ ms) and the braking torque acting on them
should be large enough to make their spindown age small enough. 

It is interesting to note the recently reported $2.14$ ms
optical pulsar candidate in SN 1987A \cite{middleditch} would fulfill these
requirements and it could be interesting to investigate the combined effects 
of deconfinement on rotational and cooling evolution more in detail 
and develop speculations about the mysterious disappearance of this source
$6.5$ yr after the supernova explosion.  

In this contribution we will review the status of the separate investigations 
of quark matter effects in the cooling and rotational evolution of neutron 
stars. 

\subsection{Cooling of young neutron stars}

Let us consider how the color superconducting quark matter, if it
exists in interiors of massive neutron stars, may affect the neutrino cooling 
of hybrid neutron stars (HNS), see \cite{BGV00}. 
Various phases are possible. The two-flavor (2SC) or
the three-flavor (3SC) superconducting phases allow for unpaired quarks of
one color whereas in the color-flavor locking (CFL) phase all the quarks are
paired.

Estimates of the cooling evolution have been performed \cite{BKV00} for a
self-bound isothermal quark core neutron star (QCS) which has a crust but
no hadron shell, and for a quark star (QS) which has neither crust nor
hadron shell. It has been shown there in the case of the 2SC (3SC) phase of a
QCS that the consequences of the occurrence of gaps for the cooling curves
are similar to the case of usual hadronic neutron stars (enhanced cooling).
However, for the CFL case it has been shown that the cooling is extremely
fast since the drop in the specific heat of superconducting quark matter
dominates over the reduction of the neutrino emissivity. As has been pointed
out there, the abnormal rate of the temperature drop is the consequence of
the approximation of homogeneous temperature profiles the applicability of
which should be limited by the heat transport effects. Page et al. (2000)
estimated the cooling of HNS where heat transport
effects within the superconducting quark core have been disregarded.
Neutrino mean free paths in color superconducting quark matter have been
discussed in \cite{cr00} where a short period of cooling delay at the
onset of color superconductivity for a QS has been conjectured in accordance
with the estimates of \cite{BKV00} in the CFL case for small gaps.

A detailed discussion of the neutrino emissivity of quark
matter without taking into account of the possibility of the color
superconductivity has been given first in Ref. \cite{I82}. 
In this work the quark direct Urca (QDU) reactions $d\rightarrow ue\bar{\nu}$ 
and $ ue\rightarrow d{\nu }$ have been suggested as the most efficient 
processes.
In the color superconducting matter the corresponding expression for the
emissivity modifies as 
\[
\epsilon _{\nu }^{{\rm QDU}}\simeq 9.4\times 10^{26}\alpha _{s}(\varrho
/\varrho _{0})Y_{e}^{1/3}\zeta _{{\rm QDU}}~T_{9}^{6}~{\rm erg~cm^{-3}~s^{-1}%
},
\]
where due to the pairing the emissivity of QDU processes is suppressed by a
factor, very roughly given by $\zeta _{{\rm QDU}}\sim \mbox{exp}(-\Delta
_{q}/T)$. At $\varrho /\varrho _{0}\simeq 2$ the strong coupling constant is 
$\alpha _{s}\approx 1$ decreasing logarithmically at still higher densities, 
$Y_{e}=\varrho _{e}/\varrho $ is the electron fraction. If for somewhat
larger density the electron fraction was too small ($Y_{e}<Y_{ec}\simeq
10^{-8}$), then all the QDU processes would be completely switched off \cite
{DSW83} and the neutrino emission would be governed by two-quark reactions
like the quark modified Urca (QMU) and the quark bremsstrahlung (QB)
processes $dq\rightarrow uqe\bar{\nu}$ and $q_{1}q_{2}\rightarrow
q_{1}q_{2}\nu \bar{\nu}$, respectively. The emissivities of QMU and QB
processes are smaller than that for QDU being suppressed by factor $\zeta _{%
{\rm QMU}}\sim \mbox{exp}(-2\Delta _{q}/T)$ for $T<T_{{\rm crit},q}\simeq
0.4~\Delta _{q}$. For $T>T_{{\rm crit},q}$ all the $\zeta $ factors are
equal to unity. The modification of $T_{{\rm crit},q}(\Delta _{q})$ relative
to the standard BCS formula is due to the formation of correlations as,
e.g., instanton- anti-instanton molecules. The contribution of the reaction $%
ee\rightarrow ee\nu \bar{\nu}$ to the emissivity is very small \cite{KH99}, 
%\begin{equation} 
$\epsilon _{\nu }^{ee}\sim 10^{12}\,Y_{e}^{1/3}(\varrho /\varrho
_{0})^{1/3}T_{9}^{8}~{\rm erg~cm^{-3}~s^{-1}}$, %\label{eq:4} 
%\end{equation} 
but it can become important when quark processes are blocked out for large
values of $\Delta _{q}/T$ in superconducting quark matter.

For the quark specific heat \cite{BGV00} used expression of \cite{I82} being
however suppressed by the corresponding $\zeta$ factor due to color
superfluidity. Therefore gluon-photon and electron contributions play
important role.

The heat conductivity of the matter is the sum of partial contributions 
%\begin{equation} 
% \label{conductivity2} 
$\kappa = \sum_{i}\kappa_{i},\,\,\, \kappa_i^{-1} =
\sum_{j}\kappa_{ij}^{-1}~,$ %\end{equation} 
where $i,j$ denote the components (particle species). For quark matter $%
\kappa$ is the sum of the partial conductivities of the electron, quark and
gluon components %\begin{equation} 
% \label{total} 
$\kappa = \kappa_{e} + \kappa_{q}+\kappa_{g}$, %\end{equation} 
where $\kappa_{e}\simeq \kappa_{ee}$ is determined by electron-electron
scattering processes since in superconducting quark matter the partial
contribution $1/\kappa_{eq}$ (as well as $1/\kappa_{gq}$ ) is additionally
suppressed by a $\zeta_{{\rm QDU}}$ factor, as for the scattering on
impurities in metallic superconductors. Due to very small resulting value of 
$\kappa$ the typical time necessary for the heat to reach the star surface
is large, delaying the cooling of HNS.

The equation of state (EoS) used in Ref. \cite{BGV00} included a model for  
hadronic matter, and regions of mixed phase and of pure quark matter. 
A hard EoS for the hadron matter
was used, finite size effects were disregarded in description of the mixed
phase, and the bag constant $B$ was taken to be rather small as motivated 
from confining chiral quark models \cite{bt,gbkg}. 
That led to the
presence of a wide region of mixed and quark phases already for the HNS of
the mass $M = 1.4~M_\odot$. On the other hand, the absence of a dense hadronic
region within this EoS allowed to diminish uncertainties in the description of
in-medium effects in hadronic matter suppressing them relative to that used
in the "standard scenario".

\begin{figure}[tbp]
\centerline{\includegraphics[height=8cm,clip=true]{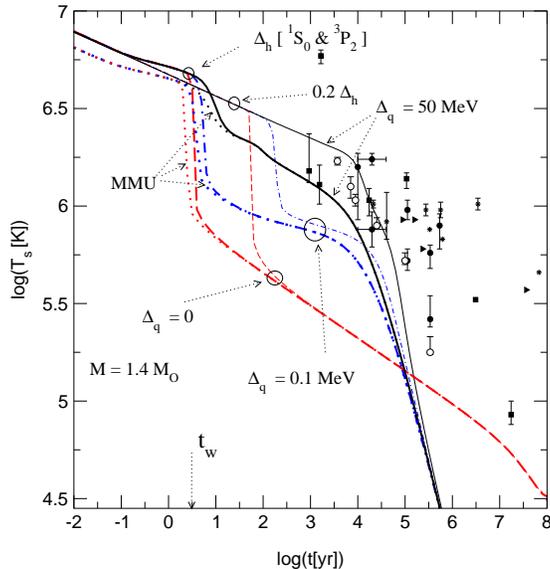}}
\caption{ Evolution of the surface temperature $T_s$ of HNS with $M=1.4
M_{\odot}$ for $T_s =(10 T_m )^{2/3}$, where $T$ is in K, see 
\protect\cite{T79}.
Data are from \protect\cite{SVSWW97} (full symbols) and from 
\protect\cite{YLS99} (empty symbols), 
$t_w$ is the typical time which is necessary for the cooling wave
to pass through the crust.}
\label{fig:tbp}
\end{figure}

With the above inputs, the evolution equation for
the temperature profile has been solved in Ref. \cite{BGV00}. 
In order to demonstrate the influence of the size
of the diquark and nucleon pairing gaps on the evolution of the temperature
profile solutions were performed with different values of the quark and
nucleon gaps. Comparison of the cooling evolution ($\mbox{lg}~T_s$ vs. $%
\mbox{lg}~t$) of HNS of the mass $M = 1.4~M_\odot$ is given in
Fig. \ref{fig:tbp}.
The curves for $\Delta_q \stackrel{\scriptstyle >}{\phantom{}_{\sim}} 1$
MeV are very close to each other demonstrating typical large gap behaviour. 
%As representative example it was taken $\Delta_q=50$ MeV. 
The behaviour of the cooling curve for $t \leq 50 \div 100$ yr is in
straight correspondence with the heat transport processes. The subsequent
time evolution is governed by the processes in the hadronic shell and by a
delayed transport within the quark core with a dramatically suppressed
neutrino emissivity from the color superconducting region. In order to
demonstrate this feature a calculation was performed with the nucleon gaps ($%
\Delta_i (n), ~i= n,p$) being artificially suppressed by a factor 0.2. Then
up to $\mbox{lg}(t[{\rm yr}]) \stackrel{\scriptstyle <}{\phantom{}_{\sim}} 4$
the behaviour of the cooling curve is analogous to the one would be obtained
for pure hadronic matter. The curves labelled "MMU" show the cooling of
hadron matter with inclusion of appropriate medium modifications in the $NN$
interaction. 
%%%%%%%%%%%%%%%%%%%%%%%%%%%%%%%%% Figure 2 %%%%%%%%%%%%%%%%%%%%%%%%%%%%%%% 
%%%%%%%%%%%%%%%%%%%%%%%%%%%%%%%%%%%%%%%%%%%%%%%%%%%%%%%%%%%%%%%%%%%%%%%%% 
These effects have an influence on the cooling evolution only for $\mbox{lg}%
(t[{\rm yr}]) \stackrel{\scriptstyle <}{\phantom{}_{\sim}} 2$ since the
specific model EQS used does not allow for high nucleon densities in the
hadron phase at given example of HNS of $M = 1.4~M_\odot$. The effect would
be much more pronounced for larger star masses, a softer EQS for hadron
matter and smaller values of the gaps in the hadronic phase. Besides,
incorporation of finite size effects within description of the mixed phase
reducing its region should enlarge the size of the pure hadron phase.

The unique asymptotic behaviour at $\mbox{lg}(t[{\rm yr}]) \geq 5$ for all
the curves corresponding to finite values of the quark and nucleon gaps is
due to a competition between normal electron contribution to the specific
heat and the photon emissivity from the surface since small exponentials
switch off all the processes related to paired particles. This tail is very
sensitive to the interpolation law $T_s =f(T_m)$ used to simplify the
consideration of the crust. The curves coincide at large times due to the
uniquely chosen relation $T_s \propto T_m^{2/3}$.

The curves for $\Delta_q = 0.1$ MeV demonstrate an intermediate cooling
behaviour between those for $\Delta_q=50$ MeV and $\Delta_q=0$. Heat
transport becomes not efficient after first $5 \div 10$ yr. The subsequent $%
10^4$ yr evolution is governed by QDU processes and quark specific heat
being only moderately suppressed by the gaps and by the rates of NPBF
processes in the hadronic matter (up to $\mbox{lg}(t[{\rm yr}]) \leq 2.5$).
At $\mbox{lg}(t[{\rm yr}]) \geq 4$ begins the photon cooling era.

The curves for normal quark matter ($\Delta_q =0$) are governed by the heat
transport at times $t \stackrel{\scriptstyle <}{\phantom{}_{\sim}} 5$ yr and
then by QDU processes and the quark specific heat. The NPBF processes are
important up to $\mbox{lg}(t[{\rm yr}])\leq 2$, the photon era is delayed up
to $\mbox{lg}(t[{\rm yr}])\geq 7$. For times smaller than $t_w$
(see Fig. \ref{fig:tbp})
the heat transport is delayed within the crust area \cite{LPPH91dim}. Since
for simplicity this delay was disregarded in the heat transport treatment,
for such small times the curves should be interpreted as the $T_m(t)$
dependence scaled to guide the eye by the same law $\propto T_m^{2/3}$, as $%
T_s$.

For the CFL phase with large quark gap, which is expected to exhibit the most
prominent manifestations of color superconductivity in HNS and QCS
configurations, \cite
{BGV00} thus demonstrated an essential delay of the cooling during the first 
$50 \div 300$ yr (the latter for a QCS) due to a dramatic suppression of the
heat conductivity in the quark matter region. This delay makes the cooling
of HNS and QCS systems not as rapid as one could expect when ignoring the heat
transport. In HNSs compared to QCSs (large gaps) there is an additional delay
of the subsequent cooling evolution which comes from the processes in pure
hadronic matter.

In spite of that we find still too fast cooling for those objects compared
to ordinary NS. Therefore, with the CFL phase of large quark gap it seems
rather difficult to explain the majority of the presently known data both in
the cases of the HNS and QCS, whereas in the case of pure hadronic stars
the available data are much better fitted even within the same simplified
model for the hadronic matter. For 2SC (3SC) phases one may expect analogous
behaviour to that demonstrated by $\Delta_q =0$ since QDU processes on
unpaired quarks are then allowed, resulting in a fast cooling. It is however
not excluded that new observations may lead to lower surface temperatures
for some supernova remnants and will be better consistent with the model
which also needs further improvements. On the other hand, if future
observations will show very large temperatures for young compact stars they
could be interpreted as a manifestation of large gap color superconductivity
in the interiors of these objects.

\subsection{Spin evolution of old neutron stars}

\begin{figure}[bht]
\begin{center}
\includegraphics[width=1.0\textwidth]{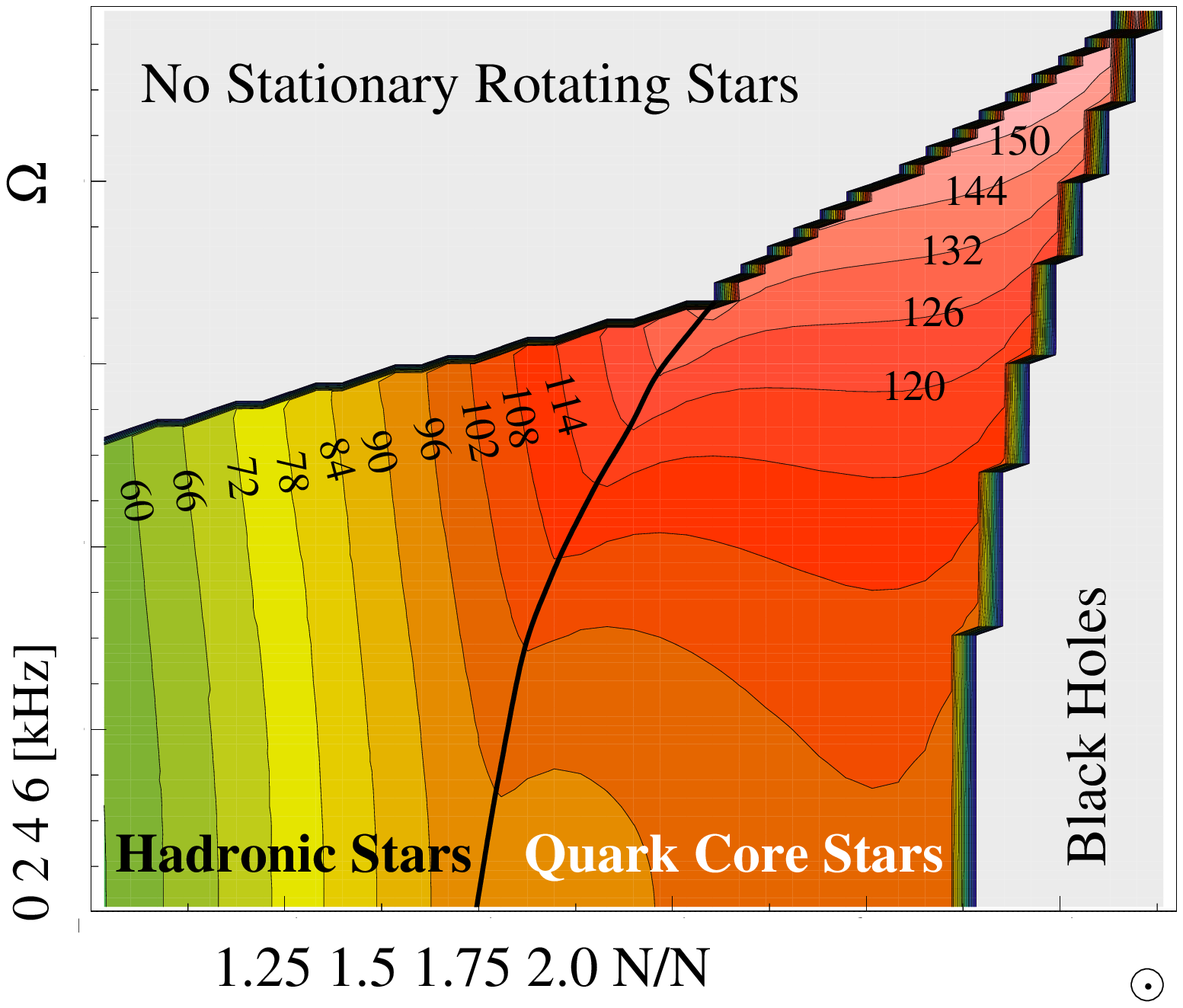}
\end{center}
\caption{Phase diagram for configurations of rotating compact objects in the
plane of angular velocity $\Omega$ and mass (baryon number $N$). Contour
lines show the values of the moment of inertia in M$_\odot$km$^2$. The line $%
N_{{\rm crit}}(\Omega)$ which separates hadronic from quark core stars
corresponds the set of configurations with a central density equal to the
critical density for the occurence of a pure quark matter phase..}
\label{fig:phdiag}
\end{figure}

It is the aim of the present paper to investigate the conditions for an
observational verification of the existence of the critical line 
$N_{{\rm crit}}(\Omega)$ which separates the QCS configurations from hadronic 
ones.  
We will show evidence that in principle such a measurement
is possible since this deconfinement transition line corresponds to a
maximum of the moment of inertia, which is the key quantity for the
rotational behavior of compact objects, see Fig. \ref{fig:phdiag}.

In the case of rigid rotation the moment of inertia is defined by 
\begin{equation}  \label{momi}
I(\Omega, N) = J(\Omega, N)/\Omega~,
\end{equation}
where the angular momentum $J(\Omega, N)$ of the star can be expressed in
invariant form as 
\begin{equation}
J(\Omega, N)=\int T_\phi ^t\sqrt{-g}dV~,  \label{moment}
\end{equation}
with $T_\phi ^t$ being the nondiagonal element of the energy momentum
tensor, $\sqrt{-g}dV$ the invariant volume and $g=\det||g_{\mu\nu}||$ the
determinant of the metric tensor. We assume that the superdense compact
object rotates stationary as a rigid body, so that for a given time-interval
both the angular velocity as well as the baryon number can be considered as
global parameters of the theory. The result of our calculations for the
moment of inertia (\ref{momi}) can be cast into the form 
\begin{equation}
I= I^{(0)}+\sum_{\alpha}\Delta I_{\alpha},
\end{equation}
where $I^{(0)}$ is the moment of inertia of the static configuration with
the same central density and $\Delta I_{\alpha}$ stands for contributions to
the moment of inertia from different rotational effects which are labeled by 
$\alpha$: matter redistribution, shape deformation, and changes in the
centrifugal forces and the gravitational field \cite{cgpb}. 
%They all can be expressed by integrals of the angular
%averaged modifications of the corresponding physical quantities \cite{cgpb}.

In Fig. \ref{fig:phdiag} we show the resulting phase diagram for compact
star configurations which exhibits four regions: (i) the region above the
maximum frequency $\Omega _{{\rm K}}(N)$ where no stationary rotating
configurations are found, (ii) the region of black holes for baryon numbers
exceeding the maximum value $N_{{\rm max}}(\Omega )$, and the region of
stable compact stars which is subdivided by the critical line $N_{{\rm crit}%
}(\Omega )$ into a region of (iii) quark core stars and another one of (iv)
hadronic stars, respectively. The numerical values for the critical lines
are model dependent. For this particular model EoS due to the hardness of
the hadronic branch (linear Walecka model \cite{gbook}) there is a maximum
value of the baryonic mass on the critical line $N_{{\rm crit}}(\Omega
_{k})=1.8N_{\odot }$, such that for stars more massive than that one all
stable rotating configurations have to have a quark core. 

The simple case of the spindown evolution of isolated (non-accreting) pulsars 
due to magnetic dipole radiation would be described by vertical lines in Fig.
\ref{fig:phdiag}.
All other possible trajectories correspond to processes with variable baryon
number (accretion). In the phase of hadronic stars, $\dot{\Omega}$ first
decreases as long as the moment of inertia monotonously increases with $N$.
When passing the critical line $N_{{\rm crit}}(\Omega )$ for the
deconfinement transition, the moment of inertia starts decreasing and the
internal torque term $K_{{\rm int}}$ changes sign. This leads to a narrow
dip for $\dot{\Omega}(N)$ in the vicinity of this line. As a result, the
phase diagram gets populated for $N\stackrel{<}{\sim }N_{{\rm crit}}(\Omega )
$ and depopulated for $N\stackrel{>}{\sim }N_{{\rm crit}}(\Omega )$ up to
the second maximum of $I(N,\Omega )$ close to the black-hole line $N_{{\rm %
max}}(\Omega )$. The resulting population clustering of compact stars at the
deconfinement transition line is suggested to emerge as a signal for the
occurence of stars with quark matter cores. In contrast to this scenario, in
the case without a deconfinement transition, the moment of inertia could at
best saturate before the transition to the black hole region and
consequently $\dot{\Omega}$ would also saturate. This would entail a smooth
population of the phase diagram without a pronounced structure.

The clearest scenario could be the evolution along lines of constant $\Omega 
$ in the phase diagram. These trajectories are associated with processes
where the external and internal torques are balanced. A situation like this
has been described, e.g. by \cite{bildsten} for accreting binaries emitting
gravitational waves.

We consider the spin evolution of a compact star under mass accretion from a
low-mass companion star as a sequence of stationary states of configurations
(points) in the phase diagram spanned by $\Omega $ and $N$. The process is
governed by the change in angular momentum of the star

\begin{equation}  \label{djdt}
\frac{d}{dt} (I(N,\Omega)~ \Omega)= K_{{\rm ext}}~,
\end{equation}
where 
\begin{equation}
K_{{\rm ext}}= \sqrt{G M \dot M^2 r_0}- N_{{\rm out}}  \label{kex}
\end{equation}
is the external torque due to both the specific angular momentum transfered
by the accreting plasma and the magnetic plus viscous stress given by 
$N_{{\rm out}}=\kappa \mu^2 r_c^{-3}$, $\kappa=1/3$ \cite{lipunov}. 
For a star with radius $R$ and magnetic field strength $B$, 
the magnetic moment is given by $\mu=R^3~B$. 
The co-rotating radius $r_c=\left(GM/\Omega^2 \right)^{1/3}$ is very large 
($r_c\gg r_0$) for slow rotators. 
The inner radius of the accretion disc is 
\[
r_0 \approx \left\{ 
\begin{array}{cc}
R~, & \mu < \mu_c \\ 
0.52~r_A~, & \mu \geq \mu_c
\end{array}
\right. 
\]
where $\mu_c$ is that value of the magnetic moment of the star for which the
disc would touch the star surface. The characteristic Alfv\'en radius for
spherical accretion with the rate 
$\dot M=m \dot N$ is 
$r_A=\left(2\mu^{-4}G M \dot M^2\right)^{-1/7}$. 
Since we are interested in the case of fast
rotation for which the spin-up torque due to the accreting plasma in Eq. 
(\ref{kex}) is partly compensated by $N_{{\rm out}}$, eventually leading to a
saturation of the spin-up, we neglect the spin-up torque in $N_{{\rm out}}$
which can be important only for slow rotators \cite{gl},

From Eqs. (\ref{djdt}), (\ref{kex}) one can obtain the first order
differential equation for the evolution of angular velocity

\begin{equation}  \label{odoto}
\frac{d \Omega}{d t}= \frac{K_{{\rm ext}}(N,\Omega)- K_{{\rm int}}(N,\Omega)
} {I(N,\Omega) + {\Omega}({\partial I(N,\Omega)}/{\partial \Omega})_{N}}~,
\end{equation}
where 
\begin{equation}  \label{kint}
K_{{\rm int}}(N,\Omega)=\Omega\dot N ({\partial I(N,\Omega)}/{\partial N}
)_{\Omega}~.
\end{equation}

Solutions of (\ref{odoto}) are trajectories in the $\Omega - N$ plane
describing the spin evolution of accreting compact stars. 
Since $I(N,\Omega)$ exhibits characteristic functional
dependences \cite{phdiag} at the deconfinement phase transition line 
$N_{{\rm crit}}(\Omega)$ we expect observable consequences in the 
$\dot P - P$
plane when this line is crossed.

In our model calculations we assume that both the mass accretion and the
angular momentum transfer processes are slow enough to justify the
assumption of quasistationary rigid rotation without convection. The moment
of inertia of the rotating star can be defined as $I(N,\Omega)=
J(N,\Omega)/\Omega~$, where $J(N,\Omega)$ is the angular momentum of the
star. For a more detailed description of the method and analytic results we
refer to \cite{cgpb} and the works of \cite{hartle,thorne}, as well as \cite
{chubarian,sedrakian}.

The time dependence of the baryon number for the constant accreting rate $%
\dot N$ is given by 
\begin{equation}
N(t)=N(t_0)+ (t-t_0)\dot N~.
\end{equation}
For the magnetic field of the accretors we consider the exponential decay 
\cite{heuvel} 
\begin{equation}
B(t)=[B(0) - B_{\infty}]\exp(-t/\tau_B)+ B_{\infty}~.
\end{equation}
We solve the equation for the spin-up evolution (\ref{odoto}) of the
accreting star for decay times $10^7\le \tau_B {\rm [yr]} \le 10^9$ and
initial magnetic fields in the range $0.2 \leq B(0){\rm [TG]}\leq 4.0 $. The
remnant magnetic field is chosen to be $B_\infty=10^{-4}$TG\footnote[1]
{1 TG= $10^{12}$ G} \cite{page}.

At high rotation frequency, both the angular momentum transfer from
accreting matter and the influence of magnetic fields can be small enough to
let the evolution of angular velocity be determined by the dependence of the
moment of inertia on the baryon number, i.e. on the total mass. This case is
similar to the one with negligible magnetic field considered in \cite
{cgpb,shapirov,colpi} where $\mu \leq \mu_c$ in Eq. (\ref{odoto}), so
that only the so called internal torque term (\ref{kint}) remains.

\subsubsection{Waiting time and population clustering}

The question arises whether there is any characteristic feature in the spin
evolution which distinguishes trajectories that traverse the critical phase
transition line from those remaining within the initial phase.

\begin{figure}[bht]
\centerline{\includegraphics[width=0.75\textwidth,angle=-90]{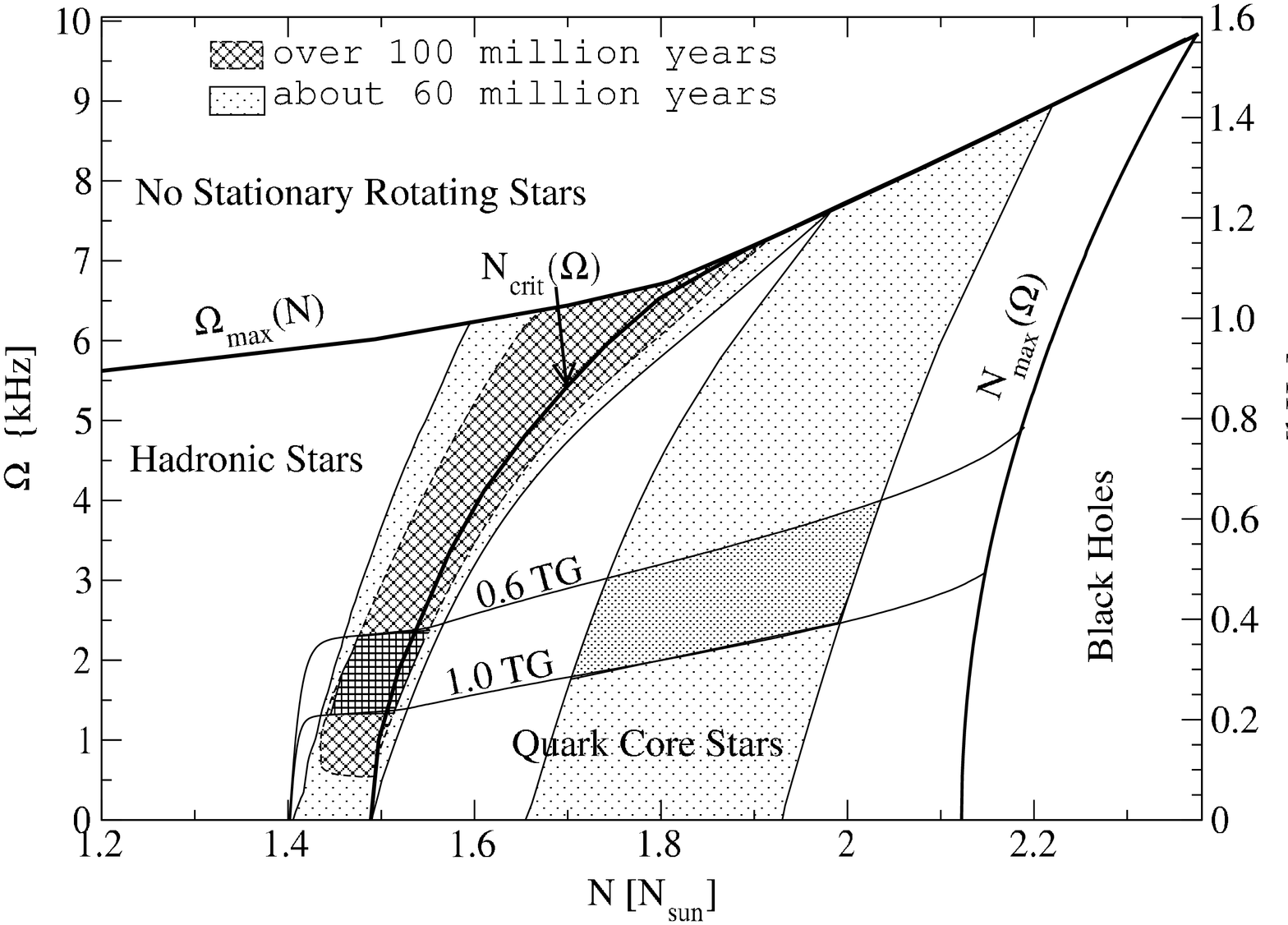}}
\caption{Regions of {waiting times} in the phase diagram for compact hybrid
stars for the (9,-8) scenario. For an estimate of a population statistics we
show the region of evolutionary tracks when the interval of initial magnetic
field values is restricted to $0.6\leq B(0)[{\rm TG}] \leq 1.0$. Note that
the probability of finding a compact star in the phase diagram is enhanced
in the vicinity of the critical line for the deconfinement phase transition 
$N_{{\rm crit}}(\Omega)$ by at least a factor of two relative to all other
regions in the phase diagram. }
\label{fig:ONTcntr}
\end{figure}

The results of Ref. \cite{apjl} show that the waiting time for accreting
stars along their evolution trajectory is larger in a hadronic configuration
than in a QCS, after a time scale when the mass load onto the star becomes
significant. 
This suggests that if a hadronic star enters the QCS region,
its spin evolution gets enhanced thus depopulating the higher frequency
branch of its trajectory in the $\Omega - N$ plane.
%\vspace*{-2cm}
%\begin{figure}[bht]
%\centerline{\includegraphics[width=0.8\textwidth,angle=-90]{ontcontr.ps}}
%\caption{Regions of {waiting times} in the phase diagram for compact hybrid
%stars for the (9,-8) scenario. For an estimate of a population statistics we
%show the region of evolutionary tracks when the interval of initial magnetic
%field values is restricted to $0.6\leq B(0)[{\rm TG}] \leq 1.0$. Note that
%the probability of finding a compact star in the phase diagram is enhanced
%in the vicinity of the critical line for the deconfinement phase transition 
%$N_{{\rm crit}}(\Omega)$ by at least a factor of two relative to all other
%regions in the phase diagram. }
%\label{fig:ONTcntr}
%\end{figure}
%
In Fig. \ref{fig:ONTcntr} we show contours of waiting time regions in the
phase diagram. The initial baryon number is $N(0)=1.4 N_\odot$ and the
initial magnetic field is taken from the interval $0.2\leq B(0)[{\rm TG}]
\leq 4.0$ .

The region of longest waiting times is located in a narrow branch around the
phase transition border and does not depend on the evolution scenario after
the passage of the border, when the depopulation occurs and the probability
to find an accreting compact star is reduced. Another smaller increase of
the waiting time and thus a population clustering could occur in a region
where the accretor is already a QCS. For an estimate of a population
statistics we show the region of evolutionary tracks when the values of
initial magnetic field are within $0.6\leq B(0)[{\rm TG}] \leq 1.0$ as
suggested by the observation of frequency clustering in the narrow interval 
$375 \geq \nu [{\rm Hz}] \geq 225$.

As a strategy of search for QCSs we suggest to select from the LMXBs
exhibiting the QPO phenomenon those accreting close to the Eddington limit 
\cite{heuvel} and to determine simultaneously the spin frequency and the
mass \cite{LM00} for sufficiently many of these objects. The emerging
statistics of accreting compact stars should then exhibit the population
clustering shown in Fig. \ref{fig:ONTcntr} when a deconfinement transition
is possible. If a structureless distribution of objects in the $\Omega - N$
plane will be observed, then no firm conclusion about quark core formation
in compact stars can be made.

%%\vspace*{-2cm}
%
%%\vspace*{-2cm}
%\begin{figure}[bht]
%\centerline{\includegraphics[width=1.05\textwidth]{ontcontr.eps}}
%\caption{Regions of {waiting times} in the phase diagram for compact hybrid
%stars for the (9,-8) scenario. For an estimate of a population statistics we
%show the region of evolutionary tracks when the interval of initial magnetic
%field values is restricted to $0.6\leq B(0)[{\rm TG}] \leq 1.0$. Note that
%the probability of finding a compact star in the phase diagram is enhanced
%in the vicinity of the critical line for the deconfinement phase transition 
%$N_{{\rm crit}}(\Omega)$ by at least a factor of two relative to all other
%regions in the phase diagram. }
%\label{fig:ONTcntr}
%\end{figure}
%%
For the model equation of state on which the results of our present work are
based, we expect a baryon number clustering rather than a frequency
clustering to be a signal of the deconfinement transition in the compact
stars of LMXBs. The model independent result of our study is that a
population clustering in the phase diagram for accreting compact stars shall
measure the critical line $N_{{\rm crit}}(\Omega)$ which separates hadronic
stars from QCSs where the shape of this curve can discriminate between
different models of the nuclear EoS at high densities.

\subsection{Conclusion}

The investigations of cooling and rotational evolution of compact stars with 
quark matter cores show that the presence of quark matter in these objects 
could be an observable phenomenon. 
In order to develop the prognosis of signals for the occurence of quark matter
in neutron stars further, and to identify candidate objects for this 
phenomenon, we need to improve the scenarios presented here in several ways.

On the one hand the models for the EoS need to be improved so that one goes 
beyond the simple mean-field level of description. The equation of state could
be further constrained not only by the saturation properties of nuclear matter
but also by comparison with informations about excitation properties and
softness obtained from relativistic heavy-ion collisions.

On the other hand the scenarii for quark matter occurence should be 
developed. It is conceivable that already in the hot and dense
protoneutron star stage immediately after a supernova explosion quark matter
can occur and modify the propagation of neutrinos \cite{prakash}. 
Effects on the hydrodynamical, gravitational and cooling evolution are 
expected. 

A particularly interesting aspect is the possibility of 
cross-effects between: rotation and cooling (see introduction), accretion and 
cooling, magnetic field and rotation, etc. which leads to the prediction 
of correlations between observables. Such stronger constraints will help 
to prove or disprove the hypothesis of quark matter occurence in neutron stars.

\subsubsection*{Acknowledgement}

The results for the cooling evolution of neutron stars reported here are based
on an approach which has been developed together with D.N. Voskresensky. 
This work is a result of the collaboration between the
Universities of Rostock and Yerevan which is supported by a DAAD program.
G.P. acknowledges support by DFG under grant No. ARM 436/17/00.
D.B. is grateful for hospitality and support at the ECT* Trento during the 
collaboration meetings on {\it Color Superconductivity} and on 
{\it Dynamical aspects of QCD phase transitions}.

\newpage

%% file: SOURCE/proc_1412.tex
\newcounter{eqn2}[equation]
\setcounter{equation}{-1}
\stepcounter{equation}

\newcounter{bild2}[figure]
\setcounter{figure}{-1}
\stepcounter{figure}

\newcounter{tabelle2}[table]
\setcounter{table}{-1}
\stepcounter{table}

\newcounter{unterkapitel2}[subsection]
\setcounter{subsection}{-1}
\stepcounter{subsection}

\section*{\bf Conformal Cosmology and Supernova Data}
\addcontentsline{toc}{section}{\protect\numberline{}{Conformal Cosmology and Supernova Data \\ \mbox{\it D. Behnke, D. Blaschke, V. Pervushin, D. Proskurin}}}
\begin{center}
\vspace*{2mm}
{Danilo Behnke$^{\dagger}$, David Blaschke$^{\dagger}$,  \\ 
Victor Pervushin$^{\ddagger}$, Denis Proskurin$^{\ddagger}$}\\[0.3cm]
{\small\it $^\dagger$ Fachbereich Physik, Universit\"at Rostock, 18051 Rostock, Germany\\ 
$^{\ddagger}$Joint Institute for Nuclear Research, 141980 Dubna, Russia} 
\end{center} 
%\medskip 
%PACS number(s):11.25.-w 
%04.60.-m, 04.20.Cv, 98.80.Hw 
%\medskip 
%
\begin{abstract} 
We define the cosmological parameters $H_{c,0}$, $\Omega_{m,c}$ and 
$\Omega_{\Lambda ,c}$ in the Conformal Cosmology as obtained by the 
homogeneous approximation in a conformal-invariant unified theory 
which is mathematically equivalent to Einstein's gravity but  
given in space with the geometry of similarity. 
We show how the age of the universe depends on 
them, followed by the evolution of the scale parameter of the universe 
and of the density parameters. Possible 
explanations of the recent supernova data of type 1a are discussed. 
\end{abstract} 
%%%%%%%%%%%%%%%%% 
\subsection{Introduction} 
%%%%%%%%%%%%%%%%%
Now there is a very interesting situation in the modern observational 
cosmology stimulated by new data on the distance-redshift 
relation published by the supernova cosmology project (SCP) \cite{snov} and on the 
large-scale structure of the microwave background radiation \cite{10}. 
The SCP data point to an accelerated 
expansion of the universe and have stimulated new developments within 
the standard cosmological model. 
The old naive version of this model with the dust dominance was not sufficient 
to explain these new data. New fits of the modern data in the framework of the 
standard Friedmann-\-Robertson-\-Walker (FRW) model were forced to introduce 
a nonvanishing $\Lambda$-term \cite{9,wett}. The occurence of this term has also been 
interpreted as due to a new form of matter called "quintessence"
\cite{9}, a time dependent speed of light~\cite{barrow} or the fine
structure constant~\cite{moffat}. 
What is the origin of the "quintessence" and why does its density approximately 
coincide with that of matter (luminous plus dark one) at the present stage? 
Present theories - containing a scalar field - can describe the data by 
fitting its effective potential \cite{12} but cannot answer these questions. 
 
One of the interesting alternatives to the standard FRW cosmological model is 
the Jordan-Brans-Dicke (JBD) scalar-tensor theory \cite{ps1,13} with two homogeneous 
degrees of freedom, the scalar field and the scale factor. 
Another alternative is the conformal-invariant version of General
Relativity (GR) 
based on the scalar dilaton field and the geometry of similarity (following 
Weyls ideas \cite{we}) developed in \cite{ps1,pr,grg,plb,pct}. 
This dilaton version of GR (considered also as a particular case of the 
Jordan-Brans-Dicke scalar tensor theory of gravitation~\cite{jbd}) 
is the basis of some speculations on the unification of Einstein's gravity 
with the Standard Model of electroweak and strong interactions~\cite{pr,plb,kl} 
including modern  theories of supergravity~\cite{kl}. 
In the conformal-invariant Lagrangian of matter, the dilaton scales the 
masses of the elementary particles in order to conserve scale invariance of 
the theory.
However, in the current literature ~\cite{kl} a peculiarity of the 
conformal-invariant version of Einstein's dynamics has been overlooked. 
The conformal-invariant version of Einstein's dynamics 
is not compatible with the absolute standard of 
measurement of lengths and times given by an interval in the 
Riemannian geometry as this interval is not conformal-invariant. 
As it has been shown by Weyl in 1918, conformal-invariant 
theories correspond to the relative standard of measurement of 
a conformal-invariant ratio of two intervals, 
given in the geometry of similarity as a manifold of Riemannian geometries 
connected by conformal transformations~\cite{we}. 
The geometry of similarity is characterized by a measure of changing 
the length of a vector in its parallel transport. In the considered 
dilaton case, it is the gradient of the dilaton $\Phi$~\cite{plb}. 
In the following, we call the scalar conformal-invariant theory 
the conformal general relativity (CGR) to distinguish it from the original 
Weyl theory where the measure of changing the length 
of a vector in its parallel transport is a vector field 
(that leads to the defect of the physical ambiguity of the arrow of time 
pointed out by Einstein in his comment to Weyl's paper~\cite{we}). 
 In the present paper we will apply this approach to a description of  
the Hubble diagram (m(z)-relation) including recent data from the SCP \cite{snov} at 
$z\sim 1$.
We make a prediction for the behaviour at $z > 1$ which drastically deviates 
from that of the standard FRW cosmology with a $\Lambda$ - term. 
We suggest this as a test which could discriminate between alternative 
cosmologies when new data are present in near future. 
 
The present paper is devoted to the definition of the cosmological 
parameters in the Conformal Cosmology by the analogy with the 
standard cosmological model~\cite{pal}.  
To emphasize the mathematical equivalence 
of both cases we try to repeat the standard model definitions
restricting ourselves by the consideration 
of the dust-, curvature-, and $\Lambda$-terms.  
% 
%%%%%%%%%%%%%%%%%%%%%%%%
\subsection{Theory and Geometry} 
%%%%%%%%%%%%%%%%%%%%%%%% 
%
We start from the conformal-invariant theory described 
by the sum of the dilaton action and the matter action 
\be \label{cut} 
W=W_{\rm CGR}+W_{\rm matter}. 
\ee 
The dilaton action is the Penrose-Chernikov-Tagirov one for a scalar 
(dilaton) field with the negative sign 
\be 
\label{wgr} 
W_{\rm CGR}(g|\Phi)= \int d^4x\left[-\sqrt{-g}\frac{\Phi^2}{6} R(g)+ 
\Phi \partial_{\mu}(\sqrt{-g}g^{\mu\nu}\partial_{\nu}\Phi )\right]~. 
\ee 
The conformal-invariant action of the matter fields can be chosen in the form 
\be 
\label{smc} 
W_{\rm matter}=\int d^4x\left[{\cal L}_{(\Phi=0)} 
+\sqrt{-g}(-\Phi F+\Phi^2B-\lambda \Phi^4)\right]~, 
\ee 
where $B$ and $F$ are the mass contributions to the Lagrangians of the vector 
($v$) and spinor ($\psi$) fields, respectively, 
\be \label{67} 
B=v_i (y_v)_{ij}v_j~;~~ 
F=\bar\psi_{\alpha} (y_s)_{\alpha\beta}\psi_{\beta}~, 
\ee 
with $(y_v)_{ij}$, and $(y_s)_{\alpha\beta}$ being the mass matrices of 
vector bosons and fermions coupled to the dilaton field. 
The massless part of the Lagrangian density of 
the considered vector and spinor fields is denoted by ${\cal L}_{(\Phi=0)}$. 
The class of theories of the type ~(\ref{cut}) includes the 
superconformal theories with supergravity~\cite{kl} and the standard model 
with a massless Higgs field \cite{plb} as the mass term would violate the 
conformal symmetry. 
% 
%%%%%%%%%%%%%%%%%%%%%%%%%%%%%
\subsection{Homogeneous Approximations} 
%%%%%%%%%%%%%%%%%%%%%%%%%%%%% 
% 
In the Conformal Cosmology, the evolution of a universe is described 
by the scalar dilaton field which can be decomposed into a homogeneous, time 
dependent component and fluctuations, 
$\Phi (T,x)=\vh(T)+\chi(T,x)$. 
In the homogeneous approximation we neglect the fluctuations and start with 
the line element of a homogeneous and 
isotropic universe, which is described by the conformal version of the 
Friedmann-Robertson-Walker (FRW) metric without the scale factor, 
as it disappears due to conformal invariance: 
\begin{eqnarray} 
(ds)^2_c  = g_{00}(t) dt^2 -  \Bigg[ {dr^2 \over 1-k_cr^2/r_0^2} +  
r^2 \Big( d\theta^2 
+ \sin^2\theta \, d\phi^2 \Big) \Bigg]~. 
\label{FRWmetric} 
\end{eqnarray} 
We define 
\be 
dT = \sqrt{g_{00}}dt 
\ee 
 as the conformal time interval. 
From the constraint-type equation $\delta W/\delta g_{00} = 0$ 
we get 
\be\label{phi} 
\vh'^2 = \rho_c\vh + \lambda\vh^4 - \frac{k_c\vh^2}{r_0^2}=\rho_{C}~, 
\ee 
see also \cite{bbp}. 
 
There is a direct correspondence between the conformal cosmology and  
the standard model obtained by the conformal transformations 
\begin{eqnarray} 
dt_f&=&\frac{\vh(T)}{\vh(T_0)} dT = \frac{a(T)}{a(T_0)}dT~,\\ 
\rho_f&=&\frac{\rho_C}{a^4(T)}~, \\
\Lambda&=&\lambda \vh(T_0)^4~, 
\end{eqnarray} 
where $a(T_0)=1$, $\vh(T_0)=M_{\rm Planck}\sqrt{{3}/{(8\pi)}}$. 
The Friedmann time and density are denoted by $t_f$ and $\rho_f$, respectively,  
$a(T)$ is the scale factor and $T_0$ the present value of the conformal time. 
% 
%%%%%%%%%%%%%%%%%%%%%%%%%%%%%%%%%%%%%%%%%%%%%%%%
\subsection{Determination of the Conformal Cosmological Parameters} 
%%%%%%%%%%%%%%%%%%%%%%%%%%%%%%%%%%%%%%%%%%%%%%%% 
% 
We can define the conformal Hubble-constant 
\be 
H_c = \frac{\vh'}{\vh} = \frac{1}{\vh} \frac{d\vh}{dT}~. 
\ee 
and rewrite~(\ref{phi}) to 
\be\label{phi_1} 
H_c^2(T) = \frac{\rho_c}{\vh(T)} + \lambda\vh(T)^2 - \frac{k_c}{r_0^2}. 
\ee 
Applying it to the present time ($T=T_0$), we can write 
\begin{eqnarray} 
1 = \Omega_{m,c} + \Omega_{\Lambda ,c} + \Omega_{k,c}~, 
\label{sum=1} 
\end{eqnarray}  
with the dimensionless parameters 
\bea 
\Omega_{m,c} &\equiv & {\rho_c \over \vh_0 H_{c,0}^2}~,\nonumber\\* 
\Omega_{\Lambda ,c} &\equiv & {\lambda \vh_0^2 \over H_{c,0}^2}~,\nonumber\\* 
\Omega_{k,c} &\equiv & -\; {k\over r_0^2H_{c,0}^2}~, 
\label{Omegas} 
\eea 
where $H_{c,0}$ is  the value of $\vh'/\vh$ at the present time. 
So far, we have discussed only one of the independent equations that 
arises among the set given in Einstein's field equations. In order to 
proceed, we need the other. This is in fact equivalent to the 
statement of conservation of matter, which means that the quantity 
$\rho_c$ is constant in time, and the present-day value of the 
dust matter density is 
\begin{eqnarray} 
\rho_o = \rho_c \vh_0~. 
\end{eqnarray} 

From now on we will use dimensionless variables 
instead of $\vh$ and $T$. We define  
\begin{eqnarray} 
y \equiv \vh/ \vh_0 \,, \qquad \tau_c \equiv H_{c,0}(T-T_0) \,. 
\end{eqnarray} 
Using these variables, we can rewrite Eq.\ (\ref{phi_1}) in the 
following form: 
\begin{eqnarray} 
\left(\frac{dy}{d\tau_c} \right)^2 &=& 
\frac{1}{H^2_{c,0}}\left[ 
{\rho_c y\over \vh_0} + {\lambda y^4 \vh_0^2} - {k_c y^2 \over r_0^2} \right] 
\nonumber\\ 
&=&y^2\left[{1\over y}\Omega_{m,c}+y^2 \Omega_{\Lambda,c}+\Omega_{k,c}\right].
\end{eqnarray} 
Eliminating $\Omega_{k,c}$ now using Eq.\ (\ref{sum=1}), we obtain 
\begin{eqnarray} 
\left( {dy \over d\tau_c} \right)^2 = y^2 \Big[1 + \Big( \frac 1y - 1 \Big) 
\Omega_{m,c} + \Big( y^2-1 \Big) \Omega_{\Lambda ,c} \Big] \,, 
\end{eqnarray} 
or 
\begin{eqnarray} 
d\tau_c = {dy \over y\sqrt{1 + \Big( {1\over y} - 1 \Big) 
\Omega_{m,c} + \Big( y^2-1 \Big) \Omega_{\Lambda ,c}}} \,. 
\label{yeqn} 
\end{eqnarray} 
If there was a big bang, $y$ was zero at the time of the bang, i.e., 
at $T=0$. On the other hand, $y=1$ now, by definition.  Integrating 
Eq.\ (\ref{yeqn}) between these two limits, we obtain 
\begin{eqnarray} 
H_{c,0}T_0 &=& \int_0^1 {dy \over y\sqrt{1 + \Big( {1\over y} - 1 \Big) 
\Omega_{m,c} + \Big( y^2-1 \Big) \Omega_{\Lambda ,c}}} \,. 
\label{yage} 
\end{eqnarray} 
This is the equation which shows that the age of the universe is {\em 
not}\/ independent, but rather is determined by $H_{c,0}$, $\Omega_{m,c}$ and 
$\Omega_{\Lambda ,c}$. 
For the special case of a flat, dust universe without cosmological term 
($\Omega_{m,c}=1$, $\Omega_{\Lambda ,c}=0$), we have $T_0=2~H_{c,0}^{-1}$. 
 
Conventionally, one does not use the dimensionless parameter $y$, but 
rather uses the {\em red-shift parameter}\/ $z$, defined by  
\begin{eqnarray} 
1+z \equiv {\vh_0\over \vh} = {1\over y} \,. 
\label{z} 
\end{eqnarray} 
Using this variable, Eq.\ (\ref{yeqn}) becomes 
\begin{eqnarray} 
d\tau_c =  {dz\over \sqrt{(1+z)^2 (1+\Omega_{m,c} 
z) - z(2+z) \Omega_{\Lambda ,c}}} \,, 
\label{zeqn} 
\end{eqnarray} 
so that 
Eq.\ (\ref{yage}) can be written in the following 
equivalent form: 
\begin{eqnarray} 
H_{c,0}T_0 
&=& \int_0^\infty {dz \over \sqrt{(1+z)^2 (1+\Omega_{m,c} 
z) - z(2+z) \Omega_{\Lambda ,c}}} \,. 
\end{eqnarray} 
Later we will discuss what sort of evolution does this equation 
represent. 
 
In order to discuss the evolution of the universe, let us not 
integrate Eq.\ (\ref{yeqn}) all the way to the initial singularity, 
but rather to any arbitrary time $T$. This gives 
\begin{eqnarray} 
H_{c,0} (T - T_0) 
= \int_0^{y} 
{dy' \over y'\sqrt{1 + \Big( {1\over y'} - 1 \Big) 
\Omega_{m,c} + \Big( y'^2-1 \Big) \Omega_{\Lambda ,c}}} \,. 
\label{intyeqn} 
\end{eqnarray} 
Equivalently, using the red-shift variable, we can write 
\bea\label{lookback} 
H_{c,0} (T_0 - T) 
&=& \int_0^{z} {dz' \over \sqrt{(1+z')^2 (1+\Omega_{m,c} 
z') - z'(2+z') \Omega_{\Lambda ,c}}} \,. 
\eea 
%%%%%%%%%%%%%%%%%%%%%%%%%%%%% 
\subsection{Distance vs redshift relation} 
%%%%%%%%%%%%%%%%%%%%%%%%%%%%%
%
A light ray traces a null geodesic, i.e., a path for which $(ds)_c^2=0$ 
in~(\ref{FRWmetric}). Thus, a light ray coming to us satisfies the 
equation 
\bea 
\frac{dr}{dT} = \sqrt{1-k_c r^2/r_0^2} ~,
\eea 
where $r_0$ is the dimensionless co-ordinate distance introduced in Eq.\ 
(\ref{FRWmetric}). Using Eqs.\ (\ref{z}) and (\ref{zeqn}), we can 
rewrite it as  
\begin{eqnarray} 
{dr\over \sqrt{1 + \Omega_{k,c} H_{c,0}^2 r^2}} &=& dT\nonumber\\ 
&=& {1\over H_{c,0}} \; {dz\over \sqrt{(1+z)^2 (1+\Omega_{m,c} 
z) - z(2+z) \Omega_{\Lambda ,c}}} \,, 
\end{eqnarray} 
where on the left side, we have replaced $k_c$ by $\Omega_{k,c}$ using the 
definition of Eq.\ (\ref{Omegas}).  Integration of this equation 
determines the co-ordinate distance as a function of $z$: 
\bea
\label{sinn}
&H_{c,0}& r(z) = {1\over \sqrt{|\Omega_{k,c}|}}\times\nonumber\\ 
 &{\rm sinn}& \Bigg[ 
\sqrt{|\Omega_{k,c}|} \int_0^z {dz'\over \sqrt{(1+z')^2 (1+\Omega_{m,c} 
z') - z'(2+z') \Omega_{\Lambda ,c}}} \Bigg]~, 
\eea 
where ${\rm sinn}(x)={\rm sinh}(x)$ for $\Omega_{k,c}>0$, 
${\rm sinn}(x)={\rm sin}(x)$ for $\Omega_{k,c}<0$ and 
${\rm sinn}(x)=x$ for the flat universe with $\Omega_{k,c}=0$. 
The equation~(\ref{sinn}) coincides with the similar relation between the 
coordinate distance and the redshift in Standard Cosmology~\cite{pal}. 
The physical distance to a certain object can be defined in various 
ways. 
For what follows, we will need what is called 
the ``luminosity distance'' $\ell_f$, which is defined in a way that the 
apparent luminosity of any object goes like $1/\ell_f^2$. 
In Standard Cosmology we have 
\begin{eqnarray} 
\ell_f(z) = a_0^2 r(z) / a(z) = (1+z) a_0 r \,. 
\end{eqnarray} 

\begin{figure}[bth]
\begin{center}
%\leavevmode
\includegraphics[width=0.2\linewidth,width=9cm,]
{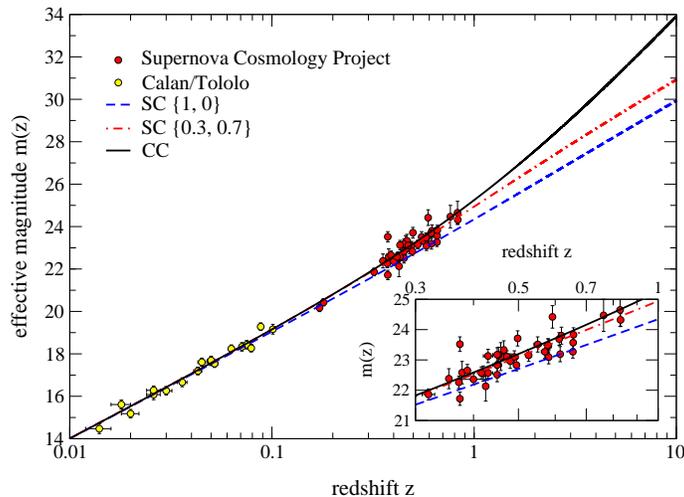} 
\caption{$m(z)$- relation for a flat universe model in SC and CC. The data points include those
from 42 high-redshift Type Ia supernovae~\protect\cite{snov}.
An optimal fit to these data within the Standard Cosmology requires a 
cosmological constant $\Omega_{\Lambda}=0.7$, whereas in the Conformal 
Cosmology presented here no cosmological constant is needed~\protect\cite{bbpp2}.}
\label{fig1}\end{center}\end{figure}

Thus, 
\bea
&H_0& \ell_f(z) = {1+z\over \sqrt{|\Omega_k|}}\times\nonumber\\
&{\rm sinn}& \Bigg[ 
\sqrt{|\Omega_k|} \int_0^z {dz'\over \sqrt{(1+z')^2 (1+\Omega_m 
z') - z'(2+z') \Omega_\Lambda}} \Bigg] \,. 
\eea 
Any observable distances $\ell_f(z)$ in the Standard Cosmology can be converted into 
observable distances $\ell_c(z)$ in the Conformal Cosmology by the 
conformal transformation 
\be 
d \ell_c = (1+z)d \ell_f(z)~. 
\ee
Considering the flat universe with $\Omega_\Lambda = 0$ in the dust stage, 
we get in the Standard Cosmology 
\be 
\ell_f(z) = 2 [(1+z) - \sqrt{1+z}]~. 
\ee 
In Fig.~\ref{fig1} we compare the results of the Standard Cosmology and for the dust case in the Conformal Cosmology 
%\be 
%\ell_c(z) = (2z+z^2)-2/3[(1+z)^{3/2}-1]~, 
%\ee
for the well known effective magnitude-redshift relation $m(z)-$ given by
$m(z) = 5 \log{[H_0\ell(z)]} + {\cal M}$~, where ${\cal M}$ is a constant. 
%\vspace{-2cm}
\vspace*{-1cm}
%%%%%%%%%%%%%%%%%%%%%%%
\subsection{Conclusion} 
%%%%%%%%%%%%%%%%%%%%%%%
We have defined the cosmological parameters $H_{c,0}$, $\Omega_{m,c}$ and 
$\Omega_{\Lambda ,c}$ in the Conformal Cosmology. 
We have shown how the age of the universe depends on 
them. 
The important result of the above derivation of the cosmological
parameters in the Conformal Cosmology is the fact, that we can fit the 
recent Supernova data at $z\sim 1$ in the Hubble diagram (m(z)-relation) in a
simple dust case very well and therefore do 
not need a cosmological constant~\cite{bbpp2}. Furthermore we gave a prediction for 
the behaviour at $z>1$ which deviates from the standard FRW cosmology
with a non-vanishing $\Lambda$-term. 
Within this model we suggest as a possible explanation of the recent data 
from the Supernova Cosmology Project a static (nonexpanding) flat universe 
where the apparent ``acceleration'' stems from the evolution of the scalar 
dilaton field in Conformal General Relativity, when applied to
cosmology. This scenario provides alternative views on the origin of standard
cosmological data (such as the Cosmic Microwave Background Radiation~\cite{bppvg}), which we begin to explore. 
There is another way to realize the mixing of the dilaton with the
standard model Higgs field~\cite{beken}
which was considered recently within the conformal cosmology
approach~\cite{bbpp2} and which slightly modifies the predictions for
the magnitude-redshift relation, Fig. 1.
However, the conclusion of the present paper, that no cosmological
constant is needed for a description of the recent high-redshift
supernova data, remains.

%\newpage
%$\left. \right.$
\newpage

%% file: SOURCE/PARTLIST.tex
\addcontentsline{toc}{section}{$\left. \right.$\hspace{0.3cm} \bf List of Participants}
$\left. \right.$
\vspace*{3cm}
\begin{center}
{\Large List of}
\end{center}
\begin{center}
{\Large {\bf Participants of the Workshops on}}\\
\vspace{1cm}
{\large 
{\bf Quark Matter in Astro- and Particle Physics}\\
\vspace{0.5cm}
{\rm Rostock (Germany), November 2000}\\
\vspace{1cm}
{\bf Dynamical Aspects of the QCD Phase Transition }\\
\vspace{0.5cm}
{\rm Trento (Italy), March 2001}\\
}
\end{center}
\newpage

%% file: SOURCE/participant.tex
%\begin{center}
%\section*{\bf List of Participants}
%\vspace*{2mm}
%\end{center}
%
\newcounter{kapitel13}[section]
\setcounter{section}{-1}
\stepcounter{section}
$\left. \right.$\\
{\bf Reinhard Alkofer}, 
{\tt Reinhard.Alkofer@uni-tuebingen.de},\\
Universit\"at T\"ubingen, Institut f\"ur Theoretische Physik, 
Auf der Morgenstelle 14, D-72076 T\"ubingen, Germany 
\\[2mm]
{\bf Danilo Behnke}, 
{\tt danilo.behnke@physik.uni-rostock.de},
Universit\"at Rostock, Fachbereich Physik, Universit\"atsplatz 3,
D-18051 Rostock, Germany
\\[2mm]
{\bf Michael Beyer}, 
{\tt michael.beyer@physik.uni-rostock.de},
Universit\"at Rostock, Fachbereich Physik, Universit\"atsplatz 3,
D-18051 Rostock, Germany
\\[2mm]
{\bf David Blaschke}, 
{\tt david.blaschke@physik.uni-rostock.de},
Universit\"at Rostock, Fachbereich Physik, Universit\"atsplatz 3,
D-18051 Rostock, Germany 
\\[2mm]
{\bf Jacques Bloch}, 
{\tt bloch@alpha10.tphys.physik.uni-tuebingen.de},
Universit\"at T\"ubingen, Institut f\"ur Theoretische Physik, 
Auf der Morgenstelle 14, D-72076 T\"ubingen, Germany 
\\[2mm]
{\bf Kyrill Bugaev}, 
{\tt bugaev@th.physik.uni-frankfurt.de},
Universit\"at Frankfurt, Institut fuer Theoretische Physik, Robert-Mayer-Str. 8-10, D- 60325 Frankfurt, Germany
\\[2mm]
{\bf Gerhard Burau}, 
{\tt gerhard.burau@physik.uni-rostock.de},
Universit\"at Rostock, Fachbereich Physik, Universit\"atsplatz 3, 
D-18051 Rostock, Germany 
\\[2mm]
{\bf Sergey Dorkin}, 
{\tt dorkin@phys.dvgu.ru}, 
Far Eastern State University, Vladivostok 690000, Russia
\\[2mm]
{\bf Alessandro Drago}, 
{\tt drago@fe.infn.it},
Universita' di Ferrara, Dipartimento di Fisica, Via Paradiso 12, 
44100 Ferrara, Italy 
\\[2mm]
{\bf Christian Gocke}, 
{\tt chris@darss.mpg.uni-rostock.de},
Universit\"at Rostock, Fachbereich Physik, Universit\"atsplatz 1,
D-18051 Rostock, Germany 
\\[2mm]
{\bf Kevin L. Haglin}, 
{\tt haglin@stcloudstate.edu},
Saint Cloud State University,  Department of Physics and Astronomy, 
720 Fourth Avenue South, MS 313, St. Cloud, MN 56301, USA 
\\[2mm]
{\bf Arne H\"oll}, 
{\tt hoell@darss.mpg.uni-rostock.de},
Universit\"at Rostock, Fachbereich Physik, Universit\"atsplatz 3,
D-18051 Rostock, Germany 
\\[2mm]
{\bf J\"org H\"ufner}, 
{\tt joerg.huefner@urz.uni-heidelberg.de},
Universit\"at Heidelberg, Institut f\"ur Theoretische Physik, 
Philosophenweg 19, D-69120 Heidelberg, Germany 
\\[2mm]
{\bf Valery Ivanov}, 
{\tt vivanov@cv.jinr.ru},
Bogoliubov Laboratory of Theoretical Physics, 
Joint Institute for Nuclear Research, 141980 Dubna, Russia 
\\[2mm]
{\bf Yura Kalinovsky}, 
{\tt kalinov@cv.jinr.dubna.su},
Laboratory of Information Technologies, 
Joint Institute for Nuclear Research, 141980 Dubna, Russia 
\\[2mm]
{\bf Dubravko Klabucar}, 
{\tt klabucar@phy.hr},
Zagreb University, Physics Department of Faculty of Science, 
Bijenicka c. 32, HR-10000 Zagreb, Croatia 
\\[2mm]
{\bf Kurt Langfeld}, 
{\tt kurt.langfeld@uni-tuebingen.de},
Institut f\"ur Theoretische Physik, Universit\"at T\"ubingen, 
Auf der Morgenstelle 14, D-72076 T\"ubingen, Germany  
\\[2mm]
{\bf Stefan Leupold}, 
{\tt stefan.leupold@theo.physik.uni-giessen.de},
Universit\"at Giessen, Institut fuer Theoretische Physik I, 
Heinrich-Buff-Ring 16, D-35392 Giessen, Germany
\\[2mm]
\newpage
\noindent
{\bf Gouranga C. Nayak}, 
{\tt nayak@th.physik.uni-frankfurt.de},
Universit\"at Frankfurt, Institut fuer Theoretische Physik, 
Robert-Mayer-Str. 8-10, D-60325 Frankfurt, Germany
\\[2mm]
{\bf Viktor Pervushin}, 
{\tt pervush@thsun1.jinr.ru},
Bogoliubov Laboratory of Theoretical Physics, 
Joint Institute for Nuclear Research, 141980 Dubna, Russia 
\\[2mm]
{\bf Janos Polonyi}, 
{\tt polonyi@fresnel.u-strasbg.fr},
University Strasbourg, 3 rue de l'Universite, 67084 Strasbourg Cedex, France
\\[2mm]
{\bf Sasha Prozorkevich}, 
{\tt smol@ns.ssu.runnet.ru},
Saratov State University 410026, Moskovskaya, 155 Physics Department, Russia
\\[2mm]
{\bf Hugo Reinhardt}, 
{\tt reinhardt@uni-tuebingen.de},
Universit\"at T\"ubingen, Institut f\"ur Theoretische Physik, 
Auf der Morgenstelle 14, D-72076 T\"ubingen, Germany 
\\[2mm]
{\bf Gerd R\"opke}, 
{\tt gerd@darss.mpg.uni-rostock.de},
Universit\"at Rostock, Fachbereich Physik, Universit\"atsplatz 3,
D-18051 Rostock, Germany 
\\[2mm]
{\bf Maria Ruivo}, 
{\tt Maria@teor.fis.uc.pt},
Departamento de Fisica, Universidade de Coimbra, P-3004-516 Coimbra, Portugal
\\[2mm]
{\bf Sebastian Schmidt},\\ 
{\tt basti@pion20.tphys.physik.uni-tuebingen.de},
Universit\"at T\"ubingen, Institut f\"ur Theoretische Physik, 
Auf der Morgenstelle 14, D-72076 T\"ubingen, Germany 
\\[2mm]
{\bf Norberto N. Scoccola}, 
{\tt scoccola@tandar.cnea.gov.ar},
Physics Dept., Comision Nac de Energia Atomica, 
Physics Dept. - Lab. TANDAR Comision Nac. de Energia Atomica,
Av. Libertador 8250, (1429) Ciudad de Buenos Aires, Argentina 
\\[2mm]
{\bf Julien Serreau},
{\tt Julien.Serreau@th.u-psud.fr},
Universit\'e Paris-Sud, Laboratoire de Physique Theorique, B\^{a}timent 210, 
91405 Orsay Cedex, France
\\[2mm]
{\bf Jonivar Skullerud}, 
{\tt jonivar@mail.desy.de},
DESY Theory Group, Notkestra{\ss}e 85, D--22603 Hamburg, Germany
\\[2mm]
{\bf S. A. Sofianos}, 
{\tt sofiasa@kiaat.unisa.ac.za}, 
University of South Africa, Dept. of Physics, Pretoria 0003, South Africa
\\[2mm]
{\bf Markus Thoma}, 
{\tt markus.thoma@cern.ch},
CERN, Theory Division, CH-1211 Geneva 23, Switzerland
\\[2mm]
{\bf Pengfei Zhuang}, 
{\tt zhuang@tphys.uni-heidelberg.de},
Universit\"at Heidelberg, Institut f\"ur Theoretische Physik, 
Philosophenweg 19, D-69120 Heidelberg, Germany 
\\[2mm]

%\end{document}